\documentclass[12pt]{article}
\pdfoutput=1
\usepackage{putex}
\usepackage{comment}
\usepackage{graphicx}
\usepackage{caption}
\usepackage{amsmath}
\usepackage{array}
\usepackage{subcaption}
\usepackage{epstopdf}
\usepackage{enumerate}
\usepackage{cite}
\usepackage{youngtab}
\usepackage{tensor}
\usepackage{slashed}
\usepackage[export]{adjustbox}
\usepackage[aligntableaux=center]{ytableau}
\usepackage[utf8]{inputenc}
\usepackage{hyperref}
\hypersetup{colorlinks=true,
      linkcolor=blue,
      urlcolor=blue,
      filecolor=black,
      citecolor=red}

\usepackage{braket}

\usepackage{tikz}

\usepackage{youngtab}

\newcommand{\RNum}[1]{\uppercase\expandafter{\romannumeral #1\relax}}

\newcommand {\be} {\begin {equation}}
\newcommand {\ee} {\end {equation}}
\newcommand {\nn} {\nonumber}
\newcommand {\bes} {\begin {equation*}}
\newcommand {\ees} {\end {equation*}}

\newcommand{\R}{\mathbb{R}}
\newcommand{\C}{\mathbb{C}}

\newcommand{\ellK}{\mathbb{K}}
\newcommand{\ellE}{\mathbb{E}}

\newcommand{\beq}{\begin{equation}}
\newcommand{\eeq}{\end{equation}}

\def\eqref#1{(\ref{#1})}

\def\ie{\begin{equation}\begin{aligned}}
\def\fe{\end{aligned}\end{equation}}

\numberwithin{equation}{section}

\def\<{\langle}
\def\>{\rangle}

\def\comma{\,,}
\def\period{\,.}
\def\fn#1{\footnote{#1}}
\def\del{\partial}

\def\pmatrix#1#2{\left(\begin{array}{#1}#2\end{array}\right)}

\begin{document}

\preprint{PUPT-2632 \\CERN-TH-2022-017}
%\preprint{CERN-TH-2022-017}
\institution{PU}{Joseph Henry Laboratories, Princeton University, Princeton, NJ 08544, USA}
\institution{CERN}{Department of Theoretical Physics, CERN, 1211 Meyrin, Switzerland
}

%\title{{\LARGE Large Charges on the Wilson Line in $\mathcal{N}=4$ SYM:\\II. Quantum Fluctuations}}
\title{{\LARGE Large Charges on the Wilson Loop in $\mathcal{N}=4$ SYM: \\ II. Quantum Fluctuations, OPE, and Spectral Curve}}
%, Spectral Curve and dCFT}}

\authors{Simone Giombi\worksat{\PU}, Shota Komatsu\worksat{\CERN}, Bendeguz Offertaler\worksat{\PU}
}
\abstract{We continue our study of large charge limits of the defect CFT defined by the half-BPS Wilson loop in planar $\mathcal{N}=4$ supersymmetric Yang-Mills theory. In this paper, we compute $1/J$ corrections to the correlation function of two heavy insertions of charge $J$ and two light insertions, in the double scaling limit where the charge $J$ and the 't Hooft coupling $\lambda$ are sent to infinity with the ratio $J/\sqrt{\lambda}$ fixed. Holographically, they correspond to quantum fluctuations around a classical string solution with large angular momentum, and can be computed by evaluating Green's functions on the worldsheet. We derive a representation of the Green's functions in terms of a sum over residues in the complexified Fourier space, and show that it gives rise to the conformal block expansion in the heavy-light channel. This allows us to extract the scaling dimensions and structure constants for an infinite tower of non-protected dCFT operators. We also find a close connection between our results and the semi-classical integrability of the string sigma model. The series of poles of the Green's functions in Fourier space corresponds to points on the spectral curve where the so-called quasi-momentum satisfies a quantization condition, and both the scaling dimensions and the structure constants in the heavy-light channel take simple forms when written in terms of the spectral curve. These observations suggest extensions of the results by Gromov, Schafer-Nameki and Vieira on the semiclassical energy of closed strings, and in particular hint at the possibility of determining the structure constants directly from the spectral curve.}

\maketitle

\tableofcontents

\section{Introduction}\label{sec:introduction}
Wilson loops are fundamental observables in any gauge theory. They provide a natural basis of gauge-invariant operators, play the role of order parameter for the confinement-deconfinement transition, and satisfy a set of non-perturbative Schwinger-Dyson equations called the loop equations \cite{Makeenko:1979pb,Makeenko:1980vm}, which are especially useful in two dimensions\footnote{Intersecting $1/8$ BPS Wilson loops in $\mathcal{N}=4$ SYM can also be computed using the loop equation of two-dimensional Yang-Mills theory as shown in \cite{Giombi:2020pdd}.} \cite{Kazakov:1980zi,Kazakov:1980zj}.

In the maximally supersymmetric gauge theory in four dimensions known as $\mathcal{N}=4$ supersymmetric Yang-Mills (SYM) theory, one can define  generalizations of the usual Wilson loop that couple to scalar fields and preserve a fraction of supersymmetry. Of particular importance among them is the half-BPS Wilson loop, which couples to a single scalar field and is defined on a circular (or infinite straight line) contour. Thanks to extensive research in the past few years, it has become clear that the half-BPS Wilson loop provides an ideal testing ground for various non-perturbative approaches in quantum field theory. 

Firstly, the half-BPS Wilson loop preserves a one-dimensional superconformal group $OSp(4^{\ast}|4)$  \cite{Bianchi:2002gz, Kapustin:2005py, drukker2006small} and provides a canonical example of a one-dimensional defect conformal field theory (CFT) \cite{Billo:2016cpy}. This enables one to study the correlation functions of insertions on the Wilson loop using both analytical and numerical bootstrap techniques \cite{liendo2018bootstrapping,Ferrero:2021bsb,Barrat:2021yvp,Cavaglia:2021bnz}. Secondly, there exists a ``topological'' subsector of this defect CFT (dCFT) in which the correlation functions become position-independent and can be computed analytically as nontrivial functions of the 't Hooft coupling $\lambda (\equiv g_{\rm YM}^2 N)$ using the method of supersymmetric localization \cite{giombi2018exact,Giombi:2018hsx,Giombi:2020amn,giombi2010correlators,giombi2013correlators}. This led to a rigorous determination of an infinite set of defect conformal data on the half-BPS Wilson loop, which provided important inputs for the conformal bootstrap analysis. Thirdly, the half-BPS Wilson loop in the fundamental representation is holographically dual to an open string minimal surface extending in the AdS$_2$ subspace of AdS$_5\times S^{5}$. Using this dual representation, one can study the correlation functions of insertions at strong coupling via perturbation theory of the string sigma model \cite{giombi2017half}. Finally, the operator insertions on the half-BPS Wilson loop can be mapped to states in an integrable open spin chain and their spectrum can in principle be determined exactly using integrability techniques \cite{drukker2006small,Drukker:2012de,Correa:2012hh,Gromov:2016rrp,Grabner:2020nis}. The three- and higher-point functions also seem amenable to the integrability machinery \cite{Kim:2017phs,Kiryu:2018phb}, in particular to the so-called hexagon formalism \cite{Basso:2015zoa,Fleury:2016ykk,Eden:2016xvg}, although more work is needed to fully develop the formalism.

The study of the half-BPS Wilson loop also allows one to explore the cross-fertilization of different techniques. For instance, the correlation functions in the topological subsector computed from supersymmetric localization in \cite{giombi2018exact,Giombi:2018hsx} can be recast into an integral of {\it Q-functions}, which are the most basic quantities in the integrability formalism \cite{Gromov:2013pga}. This strongly hints at the applicability of integrability to correlation functions and also suggests a deep connection between integrability and supersymmetric localization. In addition, a recent study \cite{Cavaglia:2021bnz} demonstrated that one can determine conformal data to remarkable numerical precision by combining the numerical conformal bootstrap and the spectral data computed from integrability. Alternatively, one can use the conformal bootstrap to extend the results from perturbation theory: this has been demonstrated explicitly in \cite{Ferrero:2021bsb}, which computed the three-loop corrections at strong coupling by imposing the crossing symmetry of the four-point functions. A similar analysis at weak coupling has not been fully developed but a few direct perturbative results, which would provide starting points of such computations, are available in the literature \cite{Kim:2017sju,Cooke:2017qgm,Kiryu:2018phb,Barrat:2020vch,Barrat:2021tpn}.
Furthermore, the half-BPS Wilson loop provides a simple example of defect renormalization group flow, which connects the ordinary Wilson loop without scalar couplings to the half-BPS loop \cite{Polchinski:2011im}. The defect renormalization group flow can be studied both at weak and strong couplings \cite{Beccaria:2017rbe,Beccaria:2019dws} (see also \cite{Beccaria:2021rmj, Beccaria:2022bcr,Cuomo:2022xgw} for related recent works) and it allows one to explicitly check the monotonicity of the defect entropy, which was proven recently in \cite{Cuomo:2021rkm}.

The goal of this paper and its companion \cite{Giombi:2021zfb} is to explore a connection to yet another non-perturbative approach---the large charge expansion of conformal field theory. Starting from the seminal works \cite{Hellerman:2015nra,Monin:2016jmo}, general properties of the large charge sector in interacting CFTs with global symmetries have been actively explored in recent years using effective field theory techniques and treating the inverse of the charge as a small expansion parameter \cite{Alvarez-Gaume:2016vff,Hellerman:2017veg,Hellerman:2017sur,Gaume:2020bmp,Jafferis:2017zna}. In the Wilson loop defect CFT, the simplest analog of the large charge sector is given by the correlation functions of two insertions with R-charge $J$ and several light insertions in the limit
\begin{align}\label{eq:doublescaling}
J\to\infty\comma\qquad \lambda\to \infty\comma\qquad \frac{J}{\sqrt{\lambda}}:\text{ fixed}\period
\end{align}
In this regime, the role of the large charge effective field theory is played by a probe string action in AdS$_5\times S^5$, which becomes classical in the large $J$ limit. As we demonstrated in the previous paper \cite{Giombi:2021zfb}, the leading large charge answer for the correlation functions can be computed by evaluating light vertex operators on a nontrivial classical string solution with large angular momentum, which was constructed in \cite{drukker2006small,Miwa2006HolographyOW,Gromov:2012eu}. In special kinematics, we can compare the results from holography with exact results from supersymmetric localization \cite{giombi2018exact} and verify the agreement of the two approaches.

\begin{figure}[t]
\centering
\includegraphics[clip,height=7cm]{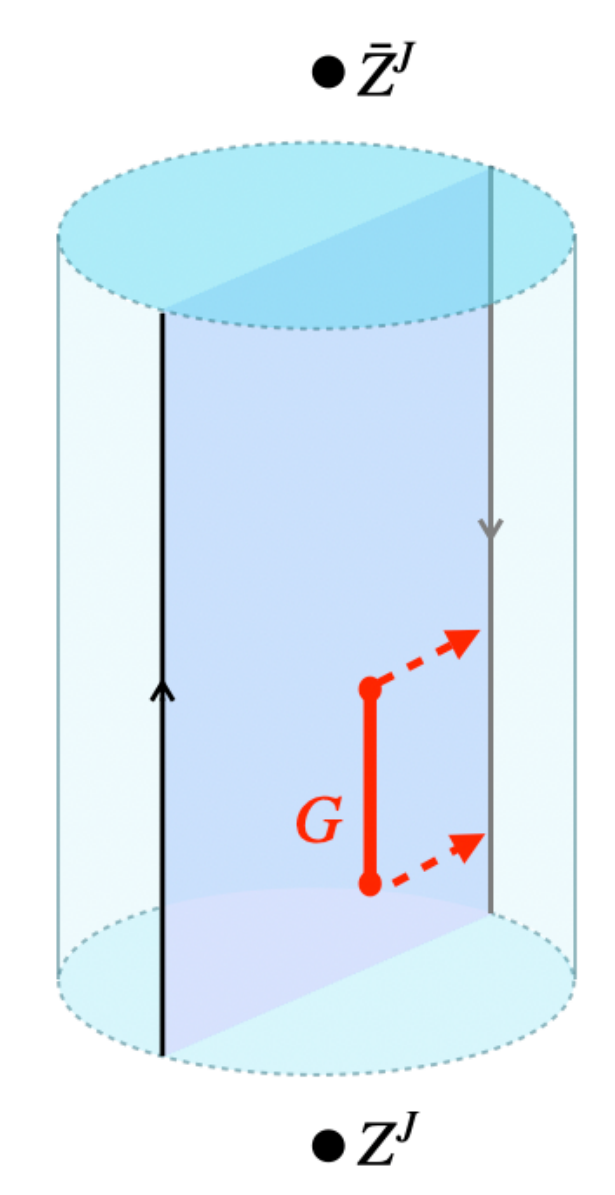}
\caption{A sketch of the setup. In the radial quantization, a single BPS Wilson loop is mapped to two straight lines in $R\times S^{3}$ (which are connected with each other at future and past infinities) while the operator insertions with large charges ($Z^{J}$ and $\bar{Z}^{J}$ in the figure) are mapped to a state defined on $S^3$ in the presence of the Wilson lines. This setup is holographically dual to a string worldsheet anchored at the Wilson lines at the boundary (the blue strip in the figure). To compute the four-point function, we evaluate the Green's function (the red segment in the figure) and push the endpoints to the boundary. This gives the correlation function of two heavy insertions and two light fluctuations on the worldsheet. The picture shows the case where the endpoints are sent to the same boundary on the strip, which gives the four-point function in the ``heavy-heavy-light-light" ordering. One may also send the endpoints to opposite boundaries, which gives the ``heavy-light-heavy-light" ordering of the correlation function. See Figure \ref{fig:extrapolating bulk-to-boundary} for more details.}\label{Fig:Intro}
\end{figure}

In this paper, we consider the leading $1/J$ corrections to the results computed in the previous paper. In the double scaling limit \eqref{eq:doublescaling}, this is equivalent to studying $1/\sqrt{\lambda}$ corrections, which correspond to quantum fluctuations on the string worldsheet. For simplicity, we focus on the four-point functions--- namely, the correlation functions of two large charge insertions and two light insertions. We compute them by evaluating the Green's functions of light fluctuations around the classical string solution with large angular momentum,  and sending the endpoints of the Green's functions to the boundaries of the worldsheet (see Figure~\ref{Fig:Intro}). The results for the Green's functions are given by integrals over Fourier modes, which take the following schematic form,
\beq\label{eq:greenintro}
G(r,\tau;r^{\prime},\tau^{\prime})\overset{r>r^{\prime}}{\sim} \int_{-\infty}^{\infty} dk\, e^{ik (\tau-\tau^{\prime})}\frac{g^{R}(r;k)g^{L}(r^{\prime};k)}{\left<g^{R},g^{L}\right>}\period
\eeq
Here, $G$ is the Green's function, $r$ and $\tau$ are the coordinates on the worldsheet, $g^{L}$ and $g^{R}$ are solutions to the same second-order differential equation (which turns out to be of the Lam\'e type) that satisfy $g^{L}\to 0$ $(g^{R}\to 0)$ at the left (right) boundary of the worldsheet, and, finally, $\left<g^{R},g^{L}\right>$ is the Wronskian of the two solutions (which is a position-independent function of $k$). The integrand has poles in the upper-half $k$ plane where the Wronskian vanishes.\fn{At the positions of the poles, the two solutions $g^{R,L}$ become linearly dependent and there exists a solution to the differential equation that vanishes at both boundaries of the worldsheet. Physically, such a solution corresponds to a normalizable excitation on the worldsheet.} Picking up the residues from those poles, we can rewrite \eqref{eq:greenintro} as a discrete sum, which turns out to precisely reproduce the conformal block expansion in the heavy-light channel. This allows us to read off conformal data of the Wilson loop defect CFT, including the conformal dimensions of (infinitely many) non-protected heavy operators and the ``heavy-heavy-light'' structure constants.

In addition, the discrete sum representation has a tantalizing connection to integrability, and in particular to the {\it spectral curve} and the so-called {\it quasi-momentum}. The spectral curve is one of the fundamental concepts in the classical integrability of the string sigma model: it encodes infinitely many conserved charges as its period integrals and it arises as the semiclassical limit of the Bethe equations, in which the {\it Bethe roots}---i.e.,~the solutions to the Bethe equations---clump up and form branch cuts \cite{Kazakov:2004qf,Beisert:2005bm}. The quasi-momentum $p(x)$ is a function on the spectral curve that satisfies certain analyticity properties and the integral of $p(x)d(x+1/x)$ gives the period integrals on the spectral curve.

As we will show below, each term in the discrete sum corresponds to a point on the spectral curve satisfying a ``quantization condition'' $p(x_n)=n\pi$ with $n\in\mathbb{N}$, and the defect CFT data in the heavy-light channel is given by simple functions on the spectral curve. A similar observation was made by Gromov, Schafer-Nameki and Vieira \cite{Gromov:2008ec}, who developed an efficient method to compute the semiclassical energy of closed strings by expressing it as a function on the spectral curve. A novelty of our results is that we find that the structure constants, not just the energy of the string (or, equivalently, the conformal dimension), simplify when written in terms of the spectral parameter $x$. This hints at an extension of the analysis of \cite{Gromov:2008ec} to the three-point functions and suggests it may be possible to compute them directly from the spectral curve and classical integrability, without explicit reference to string solutions.

\paragraph{Outline of the paper.} The rest of the paper is organized as follows: In Section \ref{sec: setup}, we summarize the basics of the four-point functions of insertions on the half-BPS Wilson loop, including the superconformal Ward identities and the definition of the large charge limit. 
We also present the main results for the four point functions that we will derive in later sections. 
In Section \ref{sec:fluctuations about the classical string}, we review the classical string solution describing the large charge insertions and compute quadratic fluctuations around it. We also explain how to extract the four-point function of two large charge insertions and two light insertions from Green's functions of light fluctuations. In Section \ref{sec:series representations of 4pt functions}, we derive an integral representation of Green's functions and later recast it into a discrete sum by picking up the residues of the poles in the complexified Fourier space. The discrete sum can be identified with the conformal block expansion in the heavy-light channel. Using this fact, in Section \ref{sec:extract OPE data} we read off the conformal dimensions of infinitely many heavy operators and the ``heavy-heavy-light'' structure constants. We also study the behavior of the correlators and OPE data at small and large $J/\sqrt{\lambda}$. 
%derive concrete representations.
In Section \ref{sec:integrability}, we discuss a connection to integrability. We show that each term in the sum corresponds to a point on the spectral curve satisfying a quantization condition and derive simple expressions for the conformal dimensions and the structure constants as functions of the spectral parameter. In Section \ref{sec:vkOMSPdS3m}, we conclude and discuss future directions. Several appendices are included to explain technical details.

\section{Four-point defect correlators with large charge}\label{sec: setup}
\subsection{Preliminaries}\label{sec:preliminaries}
This paper continues the analysis started in \cite{Giombi:2021zfb}. Let us briefly review the setup. We start with the (Maldacena-)Wilson operator in $\mathcal{N}=4$ SYM:
\begin{align}\label{eq:WL}
    \mathcal{W}&\equiv \frac{1}{N}\text{Tr P}\text{ exp}\left(\oint \left(iA_\mu(x) \dot{x}^\mu +|\dot{x}|Y^I\Phi_I(x)\right)dt\right).
\end{align}
Here $x^\mu(t)$ is a closed contour in $\mathbb{R}^4$ (we work in Euclidean signature), $Y^I(t)$ is a closed contour in $S^5$ (i.e., $\delta_{IJ}Y^IY^J=1$, where $I,J=1,\ldots,6$), and the trace is taken in the fundamental representation of the gauge group, which we take to be $U(N)$. The gauge field $A_\mu$ and the scalars $\Phi_I$ transform in the adjoint representation of $U(N)$. We are interested in the special case of the half-BPS Wilson line, for which the spacetime contour is an infinite straight line and $Y^I$ is a point in $S^5$ (equivalently, one may consider a circular contour after a conformal transformation). For concreteness, we let $x^\mu(t)=(t,0,0,0)$ and $Y^I\Phi_I=\Phi_6$. The symmetries preserved by the Wilson line form the one-dimensional superconformal group, $OSp(4^*|4)\subset PSU(2,2|4)$, which includes $16$ supercharges and the bosonic subgroup $SL(2,\R)\times SO(3)\times SO(5)$. Here, $SL(2,\mathbb{R})$ are the conformal symmetries of the Wilson line, $SO(3)$ are the spacetime rotations about the line, and $SO(5)$ is the R-symmetry subgroup that rotates the scalars not coupled to the Wilson line.

The Wilson line defines a one-dimensional defect CFT, in which correlation functions of defect local operators are obtained by inserting local adjoint operators along the spacetime contour \cite{drukker2006small,Cooke:2017qgm,Kim:2017sju,giombi2017half,Beccaria:2017rbe,giombi2018exact,Kiryu:2018phb,Giombi:2018hsx,liendo2018bootstrapping,Beccaria:2019dws,Giombi:2020amn,Grabner:2020nis,Ferrero:2021bsb,Cavaglia:2021bnz,Barrat:2021tpn}. Explicitly, the correlation function of $n$ defect operators $O_i(t_i)\equiv O_i(x(t_i))$ inserted in order on the line (i.e., $t_m<t_{m'}$ if $m<m'$) is defined by
\begin{align}
    \braket{O_1(t_1)\ldots O_n(t_n)}&\equiv \Big\langle\frac{1}{N}\text{Tr }\left[W_{n+1,n}O_n(t_n)W_{n,n-1}\ldots W_{21}O_1(t_1)W_{10}\right]\Big\rangle_{\mathcal{N}=4\text{ SYM}},\label{eq:defect correlator}\\
    W_{ji}&\equiv \text{P}e^{\int_{t_i}^{t_{j}}(A_0+\Phi_6) dt},
\end{align}
where $t_0\equiv -\infty$, $t_{n+1}\equiv \infty$. These correlators have the normalization $\braket{1}= \braket{\mathcal{W}}_{\mathcal{N}=4{\text{ SYM}}}=1$. One may also consider correlation functions involving insertions of 
local gauge invariant operators away from the Wilson line, but in this paper we focus on correlators involving only defect insertions.  

The Wilson line defect correlators satisfy the axioms of a 1d CFT (see Appendix A of \cite{Qiao:2017xif}). For instance, the two- and three-point functions of primary operators $O_1$, $O_2$ and $O_3$ take the form:
\begin{align}
    \braket{O_1(t_1)O_2(t_2)}&= \frac{\mathcal{N}_{O_1O_2}}{t_{21}^{2\Delta_1}}\delta_{\Delta_1\Delta_2},\label{eq:1d CFT 2-pt function}\\
    \braket{O_1(t_1)O_2(t_2)O_3(t_3)}&= \frac{\mathcal{C}_{O_1O_2O_3}}{t_{21}^{\Delta_1+\Delta_2-\Delta_3}t_{32}^{\Delta_2+\Delta_3-\Delta_1}t_{31}^{\Delta_3+\Delta_1-\Delta_2}}, &&t_1<t_2<t_3.\label{eq:1d CFT 3-pt function}
\end{align}
Here, $t_{ij}\equiv t_i-t_j$ is the signed Euclidean distance on the line and $\mathcal{N}_{O_1O_2}$ and $\mathcal{C}_{O_1O_2O_3}$ are the two-point and three-point (i.e., OPE) coefficients. The normalized OPE coefficients are given by $\mathcal{C}_{O_1O_2O_3}/(\mathcal{N}_{O_1O_1^\dagger}\mathcal{N}_{O_2O_2^\dagger}\mathcal{N}_{O_3O_3^\dagger})^{\frac{1}{2}}$.\footnote{One can always rescale the primaries to have unit norm (i.e., $\mathcal{N}_{OO^\dagger}=1$). We do not adopt this convention because some (protected) operators have natural normalizations that contain information about the CFT.}

In 1d CFTs, because operators on a line cannot be moved continuously around each other without becoming coincident,  three-point and higher-point functions generically depend on the circle-ordering of the operators.\footnote{A discussion of operator ordering and discrete symmetries in a 1d defect CFT--- the twist defect in the 3d Ising model--- can be found in Section 2 of \cite{Gaiotto:2013nva}.} Thus, the OPE coefficient in \eqref{eq:1d CFT 3-pt function} is defined with a particular order. By circular permutation, it satisfies $\mathcal{C}_{O_1O_2O_3}=\mathcal{C}_{O_2O_3O_1}=\mathcal{C}_{O_3O_1O_2}$ and likewise $\mathcal{N}_{O_1O_2}=\mathcal{N}_{O_2O_1}$. Given that the Wilson line defect CFT is parity invariant and unitary, there are also relations between configurations of correlators with different circle-orderings. For instance, assuming the primaries are parity eigenstates, the OPE coefficients with different orderings will differ at most by a minus sign:  $\mathcal{C}_{O_1O_2O_3}=(-1)^{P_1+P_2+P_3}\mathcal{C}_{O_3O_2O_1}$, where $(-1)^{P_i}$ is the parity of $O_i$. Furthermore, the time-reversal property of the adjoint map, $\braket{O_n^\dagger(t_n)\ldots O_1^\dagger(t_1)}=\braket{O_1(-t_1)\ldots O_n(-t_n)}^*$, implies $\mathcal{C}_{O_1O_2O_3}^*=\mathcal{C}_{O_3^\dagger O_2^\dagger O_1^\dagger}$.\footnote{There are additional constraints on the OPE data from the reflection positivity of the adjoint, which says $\braket{O_1(t_1)\ldots O_n(t_n)O_n^\dagger(-t_n)\ldots O_1^\dagger(-t_1)}\geq 0$. For example, for the two-point function, it implies $\mathcal{N}_{O^\dagger O}\geq 0$.} The relations between different configurations  of higher-point functions will in general be more complicated.

There are two classes of ``elementary operators'' on the Wilson line dCFT that we will work with. The first class consists of the chiral primaries of the form $(\epsilon\cdot \Phi)^L$ where $\Phi=(\Phi_1,\ldots,\Phi_5)$ are the scalars that do not couple to the Wilson line, $L$ is a non-negative integer, and $\epsilon\in \mathbb{C}^5$ is a polarization vector satisfying $\epsilon^2=0$. This operator transforms in the rank $L$ symmetric traceless representation of $SO(5)$ and its dimension is protected and equal to its R-charge, $\Delta=L$. For the rank-$1$ symmetric traceless (i.e., the fundamental) representation, we can also denote the operators by $\Phi_a$, $a=1,\ldots,5$. For concreteness, we will often phrase our discussion in terms of the specific chiral primaries
\begin{align}\label{eq:Z Zbar def}
    Z&\equiv \Phi_4+i\Phi_5, & \bar{Z}&\equiv\Phi_4-i\Phi_5.
\end{align}

The second class of operators we will work with are the displacement operators, $\mathbb{D}_a$, $a=1,2,3$. Displacement operators exist in any dCFT due to the breaking of translational symmetry \cite{Billo:2016cpy}. They generate local orthogonal translations of the defect. On the Wilson line, the displacement operators take the explicit form $\mathbb{D}_a\equiv iF_{0a}+D_a\Phi_6$, where $F_{0a}$ is the gauge field strength and $D_a$ is the covariant derivative. They transform in the fundamental representation of $SO(3)$ and have protected dimension $\Delta=2$. We can also use the alternative notation $\mu\cdot \mathbb{D}$, where $\mu\in \mathbb{C}^3$ is a polarization vector satisfying $\mu^2=0$.

\paragraph{Previous results in the large charge sector of the Wilson line dCFT.}
In \cite{Giombi:2021zfb}, we studied correlators of chiral primaries in the dCFT in which two of the primaries have ``large'' R-charge. More precisely, we chose the R-charges of the two distinguished primaries to be $J$ and took the following sequence of double-scaling limits,
\begin{equation}\label{eq:large charge limit}
\begin{aligned}
    &1) && N\to \infty \text{ with $g$ and $J$  held fixed}\\
    &2) && J,g\to \infty \text{  with  $\mathcal{J}$  held fixed} 
\end{aligned}
\hspace{2cm} \text{(large charge limit),}
\end{equation}
where $g$ and $\mathcal{J}$ are defined by
\beq
g\equiv \frac{\sqrt{\lambda}}{4\pi}\comma\qquad \qquad\mathcal{J}\equiv \frac{J}{g}\period
\eeq
We call operators whose R-charges scale in proportion with the coupling $g$ in the large charge limit ``heavy'' and operators whose quantum numbers do not scale with $g$ ``light.''

In \cite{Giombi:2021zfb}, we studied the leading large charge behavior of the two-point function $\braket{Z^J\bar{Z}^J}$ and higher-point functions $\braket{Z^J\bar{Z}^J\prod_i Z^{\ell_i}\prod_j \bar{Z}^{\bar{\ell}_j}}$ using AdS/CFT. The two-point function was found to be
\begin{align}\label{eq: Z^JZbar^J}
\braket{Z^J(t_L)\bar{Z}^J(\bar{t}_L)}&=\frac{\mathcal{N}_{Z^J\bar{Z}^J}}{(t_L-\bar{t}_L)^{2J}}, &\mathcal{N}_{Z^J\bar{Z}^J}= (2gc)^{2J} e^{8g(\ellE(c^2)-\ellE(0))},
\end{align}
where the parameter $c^2$ is related to $\mathcal{J}$ by:
\begin{align}\label{eq:J/g and c^2}
\frac{\mathcal{J}}{4}&=\ellK(c^2)-\ellE(c^2),
\end{align}
and $\ellK(c^2)$ and $\ellE(c^2)$ are the complete elliptic integrals of the first and second kind. Meanwhile, the higher-point functions take the simplest form consistent with conformal symmetry:
\begin{align}\label{eq:higher point}
    \frac{\braket{Z^J(t_L)\bar{Z}^J(\bar{t}_L)\prod_{i=1}^m Z^{\ell_i}(t_i)\prod_{j=1}^n\bar{Z}^{\bar{\ell}_j}(\bar{t}_j)}}{\braket{Z^J(t_L)\bar{Z}^J(\bar{t}_L)}}&=\frac{(t_L-\bar{t}_L)^{2\ell_{\rm tot}}(2gc)^{2\ell_{\rm tot}}}{\prod_{i=1}^m (t_i-\bar{t}_L)^{2\ell_i} \prod_{j=1}^n(\bar{t}_i-t_L)^{2\bar{\ell}_i}}\delta_{\ell_{\rm tot}\bar{\ell}_{\rm tot}},
\end{align}
where $\ell_{\rm tot}\equiv \sum_{i=1}^m \ell_i$ and $\bar{\ell}_{\rm tot}\equiv \sum_{j=1}^n \bar{\ell}_j$. 

We also studied the large charge limit of the two-point function $\braket{\tilde{\Phi}^J\tilde{\Phi}^J}$ and higher-point functions $\braket{\tilde{\Phi}^J\tilde{\Phi}^J \prod_i \tilde{\Phi}^{\ell_i}}$, where $\tilde{\Phi}^\ell\equiv (\epsilon(t)\cdot \Phi(t))^\ell$ is a ``topological'' chiral primary whose polarization vector $\epsilon(t)$ is correlated with its position in such a way that its correlation functions are independent of position.\footnote{In one possible realization of the topological primaries on the Wilson line, the polarization vector is $\epsilon(t)=(0,0,1-t^2,2t,i(1+t^2))$. It satisfies $\epsilon(t_i)\cdot \epsilon(t_j)\equiv -2t_{ij}^2$.} These observables are related to the ones in \eqref{eq: Z^JZbar^J} and \eqref{eq:higher point} because, at leading order in large charge, the topological primary truncates to a sum of powers of $Z$ and $\bar{Z}$ only--- namely, $\tilde{\Phi}^\ell\sim (\epsilon_Z(t)Z(t)+\epsilon_{\bar{Z}}(t)\bar{Z}(t))^\ell$ for an appropriate choice of $\epsilon_Z,\epsilon_{\bar{Z}}$. Correlators of the topological operators can be studied using localization\cite{giombi2018exact,Giombi:2018hsx} and, accordingly, in \cite{Giombi:2021zfb} we expressed the topological correlators in terms of an ``emergent'' $J\times J$ matrix model that we could analyze in the large charge limit using the usual saddle point techniques. The result for $\braket{\tilde{\Phi}^J\tilde{\Phi}^J}$ agreed with \eqref{eq: Z^JZbar^J}, and the topological higher-point functions reproduced the appropriate linear combinations of the correlators in \eqref{eq:higher point}.

Finally, also using localization, we were able to relate certain topological correlators to the generalized Bremsstrahlung function, $B_J$, whose leading and subleading behavior in the large charge limit was determined in \cite{Gromov:2012eu,Sizov:2013joa}. As a special case, this allowed us to determine the following large charge OPE coefficient to subleading order:
\begin{align}\label{eq:OPE coeff from localization}
    \frac{\mathcal{C}_{Z^\ell Z^J\bar{Z}^{J+\ell}}^2}{\mathcal{N}_{Z^\ell\bar{Z}^\ell}\mathcal{N}_{Z^J\bar{Z}^J}\mathcal{N}_{Z^{J+\ell}\bar{Z}^{J+\ell}}}=\frac{(g\pi c^2)^\ell}{\ell!}\left(1+\frac{\ell(\ell+1)}{4g}\left[\frac{3}{4\pi}+\frac{1-c^2}{c^2\ellE(c^2)}\right]+O(1/g^2)\right).
\end{align}
These OPE coefficients satisfy $\mathcal{C}_{ZZ^J\bar{Z}^{J+\ell}}=\mathcal{C}_{Z^JZ\bar{Z}^{J+\ell}}=\mathcal{C}_{\bar{Z}\bar{Z}^JZ^{J+\ell}}$ and are real, as required by parity, R-symmetry and time-reversal.

\paragraph{Four-point correlators in the large charge sector.} In this work, we continue to study defect correlators in which two chiral primaries have large R-charge, now extending the analysis to subleading order in the large charge expansion. We will use AdS/CFT to compute the following four-point functions:
\begin{align}
    &\frac{\braket{\epsilon_1\cdot\Phi(t_1)\text{ }\epsilon_2\cdot \Phi(t_2)(\epsilon_3\cdot\Phi(t_3))^J(\epsilon_4\cdot \Phi(t_4))^J}}{\braket{(\epsilon_3\cdot\Phi(t_3))^J(\epsilon_4\cdot \Phi(t_4))^J}}=\frac{2g}{\pi}\frac{\epsilon_1\cdot \epsilon_2}{t_{12}^2}\left[G_1(\chi)+\zeta G_2(\chi)+\xi G_3(\chi)\right],\label{eq:scalar large charge correlator}\\
    &\frac{\braket{\mu_1\cdot\mathbb{D}(t_1)\text{ }\mu_2\cdot \mathbb{D}(t_2)(\epsilon_3\cdot\Phi(t_3))^J(\epsilon_4\cdot \Phi(t_4))^J}}{\braket{(\epsilon_3\cdot\Phi(t_3))^J(\epsilon_4\cdot \Phi(t_4))^J}}=\frac{12g}{\pi}\frac{\mu_1\cdot \mu_2}{t_{12}^4}G_4(\chi).\label{eq:displacement large charge correlator}
\end{align}
Here, $\chi$ is the conformally invariant cross-ratio given by
\begin{align}\label{eq:chi}
    \chi&\equiv \frac{t_{12}t_{34}}{t_{13}t_{24}},&\frac{1}{1-\chi}&=\frac{t_{13}t_{24}}{t_{14}t_{23}},&\frac{1-\chi}{\chi}&=\frac{t_{14}t_{23}}{t_{12}t_{34}},
\end{align}
and $\xi$ and $\zeta$ are the $SO(5)$ invariant cross-ratios of the polarization vectors given by
\begin{align}\label{eq:xi zeta}
    \xi&\equiv \frac{\epsilon_1\cdot \epsilon_3\text{ }\epsilon_2\cdot \epsilon_4}{\epsilon_1\cdot \epsilon_2\text{ }\epsilon_3\cdot \epsilon_4}, & \zeta &\equiv \frac{\epsilon_1\cdot \epsilon_4\text{ }\epsilon_2\cdot \epsilon_3}{\epsilon_1\cdot \epsilon_2\text{ }\epsilon_3\cdot \epsilon_4}.
\end{align}
The dependence of the conformally invariant functions, $G_i(\chi)$, on $N$, $g$ and $J$ is left implicit in our notation. In the large charge expansion (after taking the planar limit), each $G_i(\chi)$ is written as a series in $1/g$ with a general functional dependence on $\mathcal{J}$ at each order. Note that, since $\mathcal{J}=J/g$, this large charge scaling limit effectively resums an infinite number of terms in the ordinary perturbation theory in powers of $1/g$ with $J$ finite. 

The general form of \eqref{eq:scalar large charge correlator} and \eqref{eq:displacement large charge correlator} is fixed by the $SL(2,\mathbb{R})$, $SO(5)$ and $SO(3)$ symmetries. The normalizations are chosen based on the observation that when either $J\to 0$ or $\chi\to 0$, then the normalized four-point functions reproduce the two-point functions of the unit chiral and displacement operators. In particular, $\chi\to 0$ corresponds to the OPE limit $t_1\to t_2$ or $t_3\to t_4$, which is dominated by the exchange of the identity operator between the two light and the two heavy operators. The two-point functions are known \cite{correa2012exact}:
\begin{align}\label{eq:phi-phi D-D two pt}
    \braket{\epsilon_1\cdot \Phi(t_1)\epsilon_2\cdot \Phi(t_2)}=n_1(g)\frac{\epsilon_1\cdot \epsilon_2}{t_{12}^2}, && \braket{\mu_1\cdot \mathbb{D}(t_1)\mu_2\cdot \mathbb{D}(t_2)}=6n_1(g)\frac{\mu_1\cdot \mu_2}{t_{12}^4}.
\end{align}
The relative normalization of these correlators is fixed by supersymmetry (because $\Phi_i$ and $\mathbb{D}_a$ are in the same supermultiplet of $OSp(4^*|4)$) and the coefficient, $n_1(g)$, is known exactly from localization \cite{correa2012exact,giombi2018exact}. In the planar limit, one finds $n_1(g)=\frac{2g}{\pi}\frac{I_2(4\pi g)}{I_1(4\pi g)}$. For our purposes, we will only need the first two terms in the planar strong coupling limit ($N\to \infty$, $g\gg 1$): 
\begin{align}\label{eq:n_1(g)}
    n_1(g)&=\frac{2g}{\pi}\left(1-\frac{3}{8\pi g}+\ldots\right).
\end{align}
Matching \eqref{eq:phi-phi D-D two pt} with \eqref{eq:scalar large charge correlator}-\eqref{eq:displacement large charge correlator}, it follows that $G_1,G_4\to 1-\frac{3}{8\pi g}+O(1/g^2)$ and $G_2,G_3\to 0$, as either $\chi\to 0$ (for any $J$), or $J\to 0$.

It will often be convenient to let $Z^J$ and $\bar{Z}^J$ serve as the charge $J$ operators in \eqref{eq:scalar large charge correlator} and \eqref{eq:displacement large charge correlator}, where $Z$ and $\bar{Z}$ were defined in \eqref{eq:Z Zbar def}. Thus, instead of the general correlators in \eqref{eq:scalar large charge correlator}-\eqref{eq:displacement large charge correlator}, we will typically work with the four correlators 
\begin{align}
\frac{\braket{\Phi_i(t_1)\Phi_j(t_2)Z^J(t_3)\bar{Z}^J(t_4)}}{\braket{Z^J(t_3)\bar{Z}^J(t_4)}}&=\frac{2g}{\pi}\frac{1}{t_{12}^2}G_1(\chi)\delta_{ij},\label{eq:PhPhZ^JZb^J}\\
\frac{\braket{Z(t_1)\bar{Z}(t_2)Z^J(t_3)\bar{Z}^J(t_4)}}{\braket{Z^J(t_3)\bar{Z}^J(t_4)}}&=\frac{4g}{\pi}\frac{1}{t_{12}^2}G_{Z\bar{Z}}(\chi),\label{eq:ZZbZ^JZb^J}\\    \frac{\braket{\bar{Z}(t_1)Z(t_2)Z^J(t_3)\bar{Z}^J(t_4)}}{\braket{Z^J(t_3)\bar{Z}^J(t_4)}}&=\frac{4g}{\pi}\frac{1}{t_{12}^2}G_{\bar{Z}Z}(\chi),\label{eq:ZbZZ^JZb^J}\\
\frac{\braket{\mathbb{D}_a(t_1)\mathbb{D}_b(t_2)Z^J(t_3)\bar{Z}^J(t_4)}}{\braket{Z^J(t_3)\bar{Z}^J(t_4)}}&=\frac{12g}{\pi}\frac{1}{t_{12}^4}G_4(\chi)\delta_{ab},\label{eq:DDZ^JZb^J}
\end{align}
where we define
\begin{align}\label{eq:def GZZb GZbZ}
    G_{Z\bar{Z}}(\chi)&\equiv G_1(\chi)+G_2(\chi), &G_{\bar{Z}Z}(\chi)\equiv G_1(\chi)+G_3(\chi).
\end{align}
The leading order behavior of \eqref{eq:ZZbZ^JZb^J}-\eqref{eq:ZbZZ^JZb^J} is included in \eqref{eq:higher point}. In the present work, we will determine the first subleading correction of \eqref{eq:ZZbZ^JZb^J}-\eqref{eq:ZbZZ^JZb^J} and the leading behavior of \eqref{eq:PhPhZ^JZb^J} and \eqref{eq:DDZ^JZb^J}.

\begin{figure}[t!]
    \centering
    \begin{minipage}{0.66\hsize}
\centering
\vspace{1cm}
\includegraphics[clip, height=1.4cm]{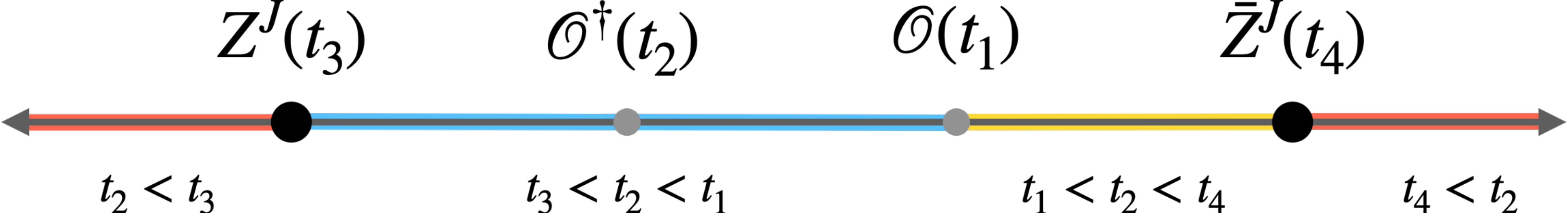}\\
\vspace{1cm}
{\bf a.} Operators on the line...
\end{minipage}
\begin{minipage}{0.33\hsize}
\centering
\includegraphics[clip, height=3.4cm]{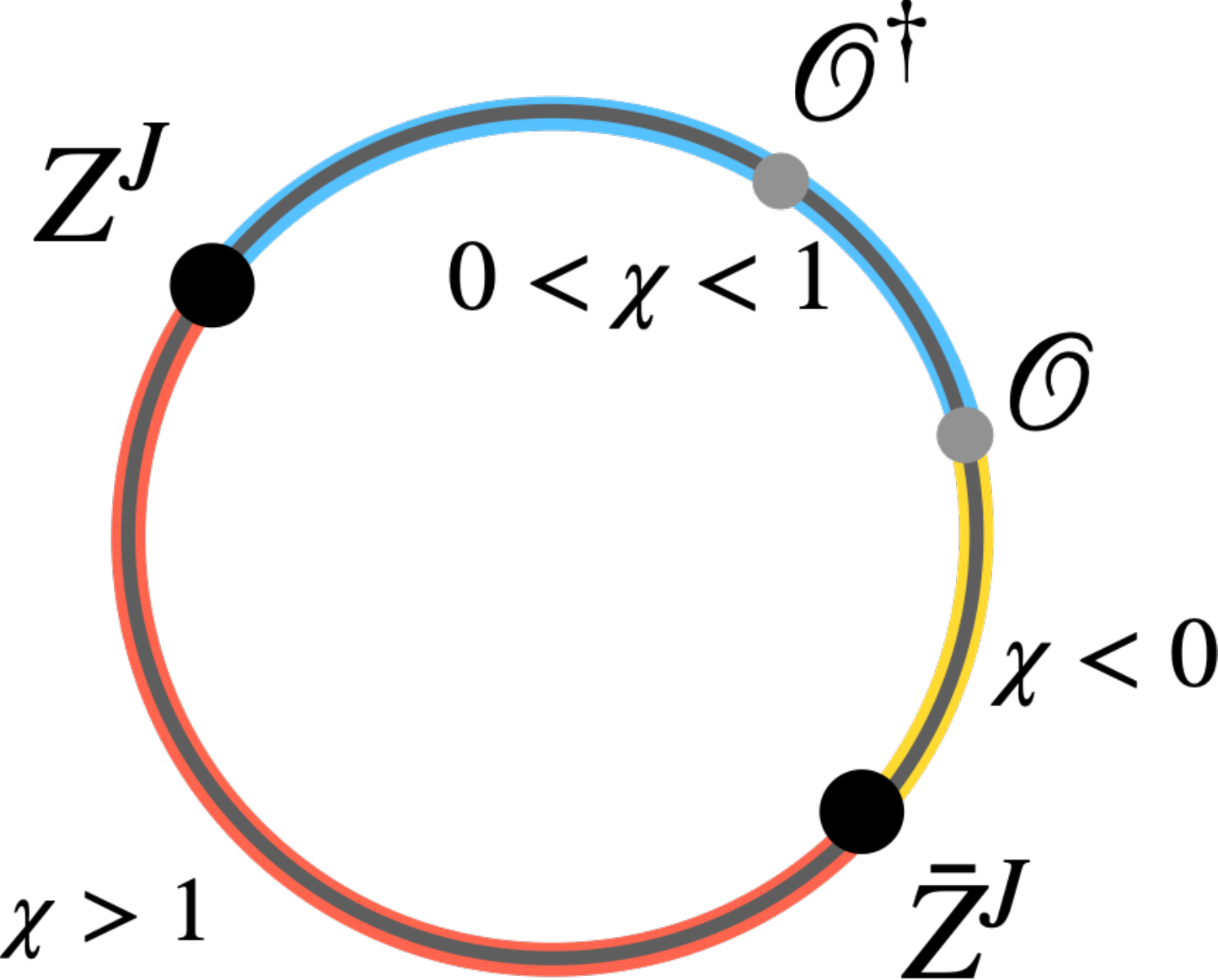}\\
{\bf b.} ... and circle
\end{minipage}
    \caption{The four operators on the Wilson line can be in one of three distinct configurations. \textbf{a.} Using the conformal transformations and parity, we can put $Z^J(t_3)$, $\bar{Z}^J(t_4)$ and the first light operator $\mathcal{O}(t_1)$ at any three points on the line satisfying $t_3<t_1<t_4$. The three configurations of the four-point function are then distinguished by the location of the second light operator: $t_3<t_2<t_1$, $t_1<t_2<t_4$, and $t_2<t_3$ or $t_4<t_2$. In terms of the cross-ratio, these correspond to $0<\chi<1$, $\chi<0$, and $\chi>1$. \textbf{b.} The different configurations can also be visualized by compactifying the line to a circle.}
    \label{fig:four-point configurations}
\end{figure}

We make a few additional comments about \eqref{eq:PhPhZ^JZb^J}-\eqref{eq:DDZ^JZb^J}: The scalars $\Phi_i$, $i=1,2,3$, which are orthogonal to $Z$, $\bar{Z}$ and $\Phi_6$, possess a residual $SO(3)\subset SO(6)$ R-symmetry. Furthermore, the four operators of interest are all parity even and under the adjoint map satisfy $\Phi_i^\dagger=\Phi_i$, $\mathbb{D}_a^\dagger=\mathbb{D}_a$ and $(Z^L)^\dagger=\bar{Z}^L$. Finally, it should be emphasized that there are three inequivalent configurations the four operators on the line can be in, $\chi\in (-\infty,0)$,  $\chi\in(0,1)$, and $\chi\in(1,\infty)$, as illustrated in Figure~\ref{fig:four-point configurations}. Via analytic continuation from these configurations, each four-point function defines three generically distinct multi-valued complex functions with singularities at $\chi=0,1,\infty$ .

\paragraph{Superconformal Ward identities.} Before we commence the analysis of the defect correlators, we note that they are not independent. Specifically, the functions $G_i(\chi)$ are related by crossing symmetry and supersymmetry. Firstly, interchanging $1\leftrightarrow 2$ in \eqref{eq:chi} sends $\chi\leftrightarrow \chi/(\chi-1)$ and it therefore follows from \eqref{eq:PhPhZ^JZb^J}-\eqref{eq:DDZ^JZb^J} that
\begin{align}
    G_1(\chi)&=G_1\left(\frac{\chi}{\chi-1}\right),&  G_{\bar{Z}Z}(\chi)&=G_{Z\bar{Z}}\left(\frac{\chi}{\chi-1}\right),& G_4(\chi)&=G_4\left(\frac{\chi}{\chi-1}\right).\label{eq:crossing Gs}
\end{align}
Thus, one can study the $G_i(\chi)$ on the restricted interval $\chi\in (0,2)$ and extend them to $\chi\in\mathbb{R}$ using \eqref{eq:crossing Gs}, or study $G_i(\chi)$ on $\chi\in \mathbb{R}$ and use \eqref{eq:crossing Gs} as a consistency check.

Less trivially, the correlator in \eqref{eq:scalar large charge correlator} satisfies superconformal Ward identities, which may succinctly be written \cite{liendo2018bootstrapping}:
\begin{align}
    0&=\left(\frac{\partial\mathcal{A}}{\partial \zeta_1}+\frac{1}{2}\frac{\partial \mathcal{A}}{\partial \chi}\right)\biggr\rvert_{\zeta_1=\chi}=\left(\frac{\partial\mathcal{A}}{\partial \zeta_2}+\frac{1}{2}\frac{\partial \mathcal{A}}{\partial \chi}\right)\biggr\rvert_{\zeta_2=\chi}.\label{eq:Ward identity}
\end{align}
Here, $\mathcal{A}\equiv G_1(\chi)+\frac{1}{\zeta_1\zeta_2}G_3(\chi)+\frac{(1-\zeta_1)(1-\zeta_2)}{\zeta_1\zeta_2}G_2(\chi)$ is the conformally invariant part of the RHS of \eqref{eq:scalar large charge correlator} and $\zeta_1$ and $\zeta_2$ are an alternative parametrization of the $SO(5)$ invariants related to $\xi$ and $\zeta$ by $\xi\equiv \frac{1}{\zeta_1\zeta_2}$ and $\zeta\equiv \frac{(1-\zeta_1)(1-\zeta_2)}{\zeta_1\zeta_2}$. Evaluating \eqref{eq:Ward identity} explicitly, we get the following two ODEs,
\begin{align}
\left(\chi\frac{d}{d\chi}-2\right)G_3(\chi)=\left(\chi(\chi-1)\frac{d}{d\chi}+2\right)G_2(\chi)=-\chi^2\frac{dG_1}{d\chi}.\label{eq:J1cWCLsjEW}
\end{align}
These equations allow us to solve for $G_2(\chi)$ and $G_3(\chi)$, and therefore for $G_{Z\bar{Z}}$ and $G_{\bar{Z}Z}$, in terms of $G_1(\chi)$, after using input from the OPE limit and the localization result (\ref{eq:OPE coeff from localization}) to fix the initial conditions. The details of this calculation are given in Section \ref{sec:computing G2 and G3}.  
\begin{comment}
\begin{align}
    G_2(\chi)&=\frac{\chi^2}{(1-\chi)^2}\left(C_2(\chi_0)+\int_{\chi_0}^{\chi}  dx\left(\frac{1}{x}-1\right)\frac{dG_1}{dx}\right),\label{eq:RKeEazh2eB}\\
    G_3(\chi)&=\chi^2\left(C_3(\chi_0)-\int_{\chi_0}^{\chi} dx\frac{1}{x}\frac{dG_1}{dx}\right).\label{eq:Kluz7kJ9dy}
\end{align}

We can further fix the constants of integration, $C_2(\chi_0)$ and $C_3(\chi_0)$, by taking OPE limits of the four-point functions in \eqref{eq:ZZbZ^JZb^J}-\eqref{eq:ZbZZ^JZb^J}. Since \eqref{eq:RKeEazh2eB} and \eqref{eq:Kluz7kJ9dy} automatically satisfy $G_2,G_3\to 0$ as $\chi\to 0$, the useful OPE limits are $\chi\to 1$ and $\chi\to \pm\infty$, in which the light operators each become coincident with one of the heavy operators. The lightest primaries appearing in the OPEs of $Z$ and $Z^J$ and of $Z$ and $\bar{Z}^{J}$ are simply the protected operators s$Z^{J+1}$ and $\bar{Z}^{J-1}$, respectively, and the relevant OPE data is given in \eqref{eq:OPE coeff from localization}. With this extra input from localization, we can fully determine $G_2$ and $G_3$ given $G_1$.
\end{comment}

Since the unit scalar $\Phi_i$ and displacement operator $\mathbb{D}_a$ are in the same superconformal multiplet, the correlators in \eqref{eq:scalar large charge correlator} and \eqref{eq:displacement large charge correlator} are also related by Ward identities \cite{liendo2018bootstrapping}. In principle, one can also determine $G_4$ in terms of $G_1$, just like $G_{Z\bar{Z}}$ and $G_{\bar{Z}Z}$. We will instead study the scalar and displacement correlators independently. This will serve as a test of the consistency of our analysis via the dual string.

\subsection{Summary of explicit results for the four-point functions}

We close this preliminary section by collecting our final results for the four-point functions. They are accessible without a detailed understanding of the derivations in Sections~\ref{sec:fluctuations about the classical string} and \ref{sec:series representations of 4pt functions}. 

The conformally invariant functions $G_i(\chi)$ are naturally expressed as series over the ``fluctuation energies'' $E_n$. These are determined by the quantization condition
\begin{align}\label{eq:quantization condition}
    \int_1^{E_n} dE \rho(E) =|n|,
\end{align}
where the ``energy density'', $\rho(E)$, is  
\begin{align}\label{eq:fluctuation energy density}
    \rho(E)\equiv \frac{2}{\pi}\frac{\mathbb{K}(c^2)E^2-\mathbb{E}(c^2)}{\sqrt{E^2-1}\sqrt{E^2+c^2-1}}.
\end{align}
Note that the integral in \eqref{eq:quantization condition} converges at $E=1$ for both the edge case $c=0$, for which $\rho(E)=1$, and the general case $c\in(0,1)$, for which $\rho(E)\sim 1/\sqrt{E-1}$ as $E\to 1^+$. It is also convenient to define the ``form factor'', $f(E)$:
\begin{align}\label{eq:form factor}
    f(E)\equiv \frac{\pi}{4}\frac{E(E^2+c^2-1)}{\ellK(c^2)E^2-\ellE(c^2)}.
\end{align}

From the semiclassical analysis of the fluctuations of the dual string in Sections~\ref{sec:fluctuations about the classical string}-\ref{sec:series representations of 4pt functions}, we will find that $G_1(\chi)$ and $G_4(\chi)$ are given by
\begin{align}\label{eq:G1 series}
    G_1(\chi)&= \frac{\chi^2}{|1-\chi|}\sum_{n\in \mathbb{Z}}\text{sgn}(1-\chi)^n f(E_n) e^{-E_n|\log|1-\chi||},\\
    G_4(\chi)&= \frac{\chi^4}{(1-\chi)^2}\sum_{n\in \mathbb{Z}}\text{sgn}(1-\chi)^{n+1}f(E_n)\frac{E_n^2-1}{6}e^{-E_n|\log|1-\chi||}.\label{eq:G4 series}
\end{align}
Furthermore, $G_{Z\bar{Z}}(\chi)$ and $G_{\bar{Z}Z}(\chi)$ are obtained from $G_1(\chi)$ and the Ward identities; the explicit results are given in \eqref{eq:G2 series}-\eqref{eq:s7VYqCQAUD}. 

As discussed in detail in Section~\ref{sec:extract OPE data}, these series representations are directly related to the conformal block expansion in the heavy-light channel, and the energies $E_n$ and form factors $f(E_n)$ encode the anomalous dimensions and OPE coefficients of the exchanged operators. The expressions \eqref{eq:G1 series}-\eqref{eq:G4 series} are valid for all $\chi\in \mathbb{R}$, except at $\chi=2$, where the $G_i(\chi)$ are smooth but their series representations do not converge (as explained in Section~\ref{sec:computing G1}, this is related to the radius of convergence of the OPE). It is interesting to note that consistency with the limiting behavior $G_1(\chi),G_4(\chi)\to 1$ as $\chi\to 0$, as required by the fact that in this limit the OPE of the two light or two heavy operators is dominated by the exchange of the identity, implies that we should have
\begin{equation}
\begin{aligned}
&\sum_{n\in \mathbb{Z}}f(E_n) e^{-E_n|\log(1-\chi)|}\stackrel{\chi\rightarrow 0}{\sim} \frac{1}{\chi^2}\,, \\
&\sum_{n\in \mathbb{Z}}f(E_n)\frac{E_n^2-1}{6} e^{-E_n|\log(1-\chi)|} \stackrel{\chi\rightarrow 0}{\sim} \frac{1}{\chi^{4}}\,.
\end{aligned}
\end{equation}
We will explicitly check towards the end of Section~\ref{sec:computing G1} that these indeed hold, based on the large $n$ behavior of the energies $E_n$ and form factors $f(E_n)$. This is a non-trivial test of the crossing symmetry of our results. 

The four-point functions and the OPE data can also be studied analytically at both small $\mathcal{J}$ ($c^2\to 0^+$) and large $\mathcal{J}$ ($c^2\to 1^-$), as discussed in Section~\ref{sec:small and large J/g}. The expansion of the four-point functions in small $c^2$ involves polylogarithms.\footnote{The expansion in powers of $c^2$ is equivalent to an expansion in powers of $\frac{J}{g}$, and is therefore related to the standard $1/g$ perturbation theory in the string sigma model. The appearance of polylogarithms is expected for loop diagrams in AdS$_2$, see e.g. \cite{Mazac:2018ycv}.}  Meanwhile, in the  $c^2\to 1$ (${\cal J}\to \infty$) limit, the behavior of the four-point functions depends significantly on whether they are in the ``heavy-heavy-light-light" ($\chi<1$) or the ``heavy-light-heavy-light" ($\chi>1$) configurations: the correlators in the latter configuration vanish while the correlators in the former configuration attain finite limits given by Bessel functions. See Figures~\ref{fig:G1 vs chi}, \ref{fig:G1+G2 vs chi} and \ref{fig:G4 vs chi} for plots of $G_1(\chi)$, $G_{Z\bar{Z}}(\chi)$ and $G_4(\chi)$ as functions of $\chi$ for representative values of $c^2$, including the edge cases $c^2=0$ and $c^2=1$.

\section{Semiclassical analysis of the dual string}\label{sec:fluctuations about the classical string}

We will compute the next-to-leading order terms in the large charge expansion of the four point functions in \eqref{eq:PhPhZ^JZb^J}-\eqref{eq:DDZ^JZb^J} by studying the semiclassical fluctuations of the string that is holographically dual to the Wilson line with $Z^J$ and $\bar{Z}^J$ inserted. We first sketch the basic idea in Section~\ref{sec:preview}, and then fill in the details in Sections~\ref{sec:quadratic action}-\ref{sec:diciontary} and Section~\ref{sec:series representations of 4pt functions}.

\subsection{Preview}\label{sec:preview}
Invoking AdS/CFT, we can schematically write the defect four-point function of two heavy operators $Z^J$, $\bar{Z}^J$ and two light operators, $\mathcal{O}$ ($\equiv Z$, $\bar{Z}$, $\Phi_i$, or $\mathbb{D}_a$) and $\mathcal{O}^\dagger$, as a string path integral:
\begin{align}\label{eq:string path integral}
\braket{\mathcal{O}(t_1)\mathcal{O}^\dagger(t_2)Z^J(t_3)\bar{Z}^J(t_4)}=\frac{\int \mathcal{D}\Psi e^{-S[\Psi]}v^J_{Z}(t_3)v^J_{\bar{Z}}(t_4)v_{\mathcal{O}}(t_1)v_{\mathcal{O}^\dagger}(t_2)}{\int \mathcal{D}\Psi e^{-S[\Psi]}}.
\end{align}
Here, $\int \mathcal{D}\Psi$ denotes integration over the fields of the superstring sigma model (which we denote collectively by $\Psi$) whose bosonic components are the coordinates of the string in AdS$_5\times S^5$, $S[\Psi]$ is the string action, and $v_{Z}$, $v_{\bar{Z}}$, $v_{\mathcal{O}}$ and $v_{\mathcal{O}^\dagger}$ are vertex operators dual to $Z$, $\bar{Z}$, $\mathcal{O}$ and $\mathcal{O}^\dagger$. In accordance with the ``extrapolate dictionary", we define the vertex operator $v_\mathcal{O}$ corresponding to the Wilson line defect operator $\mathcal{O}$ by evaluating the dual field $\Psi_\mathcal{O}$ at the point on the boundary of the worldsheet where $\mathcal{O}$ is located. Schematically,
\begin{align}\label{eq:schematic vertex operator}
    v_{\mathcal{O}}(t)\equiv \lim_{z\to 0^+}\frac{2g}{z^{\Delta}} \Psi_{\mathcal{O}}(t,z),
\end{align}
where $z$ is a particular bulk coordinate that together with $t$ parametrizes the string worldsheet (with boundary at $z=0$), and $\Delta$ is the dimension of $\mathcal{O}$ in the dCFT.

Taking the planar limit in \eqref{eq:string path integral} picks out the disk topology and taking the large $g$ limit means the path integral is dominated by its saddle point. Without the two large charge insertions, the saddle point would be a classical string extending in an AdS$_2$ subspace of AdS$_5$ and sitting at a point on $S^5$. With the two large charge insertions $Z^J$ and $\bar{Z}^J$, the saddle point solution is a classical string, $\Psi_{\rm cl}$, that carries angular momentum $J$ along the circle in $S^5$ dual to $Z$. This solution extremizes the ``total'' action, which includes the contribution of the vertex operators:
\begin{align}\label{eq:total action}
    S_{\rm tot}[\Psi]\equiv S[\Psi]-J\log\left(v_Z(t_3)v_{\bar{Z}}(t_4)\right).
\end{align}
The string tension is $2g$, so the two terms are the same order in the large charge limit.

We reviewed the classical string dual to the Wilson loop with $Z^J$ and $\bar{Z}^J$ in \cite{Giombi:2021zfb}, which was discussed previously in \cite{drukker2006small,Miwa2006HolographyOW,Gromov:2012eu}. We computed the action and the vertex operators dual to $Z$ and $\bar{Z}$ on the classical solution, which determined the leading large charge behavior of the two-point and higher-point functions in \eqref{eq: Z^JZbar^J} and \eqref{eq:higher point}. In this work, we go beyond the leading order and therefore need to take into account the fluctuations about the classical string. Letting $\Psi=\Psi_{\rm cl}+\delta \Psi$, we expand $S_{\rm tot}$ to quadratic order in the fluctuation modes:
\begin{align}\label{eq:schematic quadratic action}
    S_{\rm tot}[\Psi_{\rm cl}+\delta \Psi]&=S_{\rm tot}[\Psi_{\rm cl}]+\underbrace{\frac{1}{2}\frac{\delta^2 S_{\rm tot}}{\delta \Psi\delta \Psi}\biggr\rvert_{\Psi_{\rm cl}}\delta \Psi\delta\Psi}_{\equiv S_2[\delta \Psi]}+O(\delta \Psi^3).
\end{align}
We will see in Section~\ref{sec:quadratic action} that there are four distinct bosonic fluctuation modes, corresponding to the four types of defect operators on the Wilson line appearing in \eqref{eq:PhPhZ^JZb^J}-\eqref{eq:DDZ^JZb^J}. We will not study the fermionic modes.

From \eqref{eq:string path integral}, it follows that the four-point function normalized by the large charge two-point function is given by
\begin{align}\label{eq:schematic 4 pt}
&\frac{\braket{\mathcal{O}(t_1)\mathcal{O}^\dagger(t_2)Z^J(t_3)\bar{Z}^J(t_4)}}{\braket{Z^J(t_3)\bar{Z}^J(t_4)}}=\underbrace{v_{\mathcal{O}}(t_1)v_{\mathcal{O}^\dagger}(t_2)\big\rvert_{\Psi_{\rm cl}}}_{\rm classical}+\underbrace{W_{\mathcal{O}\mathcal{O}^\dagger}(t_1,t_2)}_{\rm fluctuation}+O(g^0),
\end{align}
where we define the boundary-to-boundary propagator\footnote{The factor of $(2\Delta -d)^2$ that arises when relating the boundary-to-boundary propagator to the boundary limit of the bulk-to-bulk propagator was discussed in, for instance, \cite{freedman1999correlation}. In our case, $d=1$.}
\begin{align}\label{eq:schematic bndry-to-bndry}
    W_{\mathcal{O}\mathcal{O}^\dagger}(t_1,t_2)&\equiv (2\Delta -1)^2\lim_{\substack{z_1\to 0^+\\z_2\to 0^+}}\frac{2g}{z_1^{\Delta}}\frac{2g}{z_2^{\Delta}}G_{\mathcal{O}\mathcal{O}^\dagger}(t_1,z_1;t_2,z_2)
\end{align}
in terms of the bulk-to-bulk propagator
\begin{align}\label{eq:schematic bulk-to-bulk}
    G_{\mathcal{O}\mathcal{O}^\dagger}(t_1,z_1;t_2,z_2)\equiv \frac{\int \mathcal{D}\delta \Psi \text{exp}\left(-S_2[\delta \Psi]\right)\delta\Psi_{\mathcal{O}}(t_1,z_1)\delta\Psi_{\mathcal{O}^\dagger}(t_2,z_2)}{\int \mathcal{D}\delta \Psi \text{exp}\left(-S_2[\delta\Psi]\right)}.
\end{align}
Because $S_2$ is proportional to the string tension, the bulk-to-bulk propagator is proportional to its inverse, and the fluctuation piece in \eqref{eq:schematic 4 pt} is suppressed relative to the classical piece by $1/g$.
Thus, to determine the subleading correction to the large charge four-point functions using AdS/CFT, we need to determine the quadratic action of the fluctuations, compute the bulk-to-bulk propagators, and then send the two bulk points to the boundary.

We will compute the boundary-to-boundary propagators by first solving the Green's equations satisfied by the bulk-to-bulk propagators. This is the most technical step of the analysis and is the focus of Section~\ref{sec:series representations of 4pt functions}. The classical string is not homogeneous, unlike the AdS$_2$ string dual to the Wilson line without insertions. Nonetheless, since the classical string is symmetric under translations parallel to the boundary in global coordinates, we can take the Fourier transform with respect to the boundary global coordinate, in which case the Green's equations reduce to ODEs in the bulk global coordinate. The ODEs turn out to be of the Jacobi form of the Lam\'e equation, the solutions of which are known and given in terms of the theta functions. This lets us write explicit integral representations of the propagators. Moreover, we may write the boundary-to-boundary propagators as sums over the residues at the poles, which take particularly simple forms when written in terms of the fluctuation energies of the fluctuations. The series representations of the boundary-to-boundary propagators can be interpreted either as sums over stationary waves on the classical string, which lets us make contact with integrability in Section~\ref{sec:integrability}, or as sums over primaries in the the conformal block expansions of the four-point defect correlators, which lets us extract dCFT OPE data in Section~\ref{sec:extract OPE data}.

One could also study the subleading behavior of the two-point function that we normalize by in \eqref{eq:schematic 4 pt}, by evaluating a functional determinant that takes the schematic form
\begin{align}\label{eq:schematic 2-pt}
    \braket{Z^J(t_3)\bar{Z}^J(t_4)}&\propto \underbrace{e^{-S_{\rm tot}[\Psi_{\rm cl}]}}_{\rm classical}\underbrace{\left(\text{Det}\frac{\delta^2 S_{\rm tot}}{\delta \Psi\delta \Psi}\Big\rvert_{\Psi_{\rm cl}}\right)^{-\frac{1}{2}}}_{\rm fluctuation}\left(1+O(1/g)\right).
\end{align}
The classical contribution was computed in \cite{Giombi:2021zfb} and is given in eq.~(\ref{eq: Z^JZbar^J}). The calculation of the fluctuation determinant is a non-trivial problem
whose solution we will not pursue in this work.\footnote{In addition to needing to compute the functional determinants of the bosonic fluctuation operators, which would be complicated by two of the modes being coupled in the coordinates we use in Section~\ref{sec:quadratic action}, we would also need to include the contributions of the fermionic fluctuations.}
%Perhaps the fermionic contributions can be related to the bosonic contributions via the supersymmetry of the classical string.} 

The above sketch of the semiclassical analysis suppresses many details, including the standard steps of picking a suitable gauge and coordinates, redefining fields, simplifying the quadratic action, and keeping track of how these choices affect the dictionary given schematically in \eqref{eq:schematic 4 pt}-\eqref{eq:schematic bulk-to-bulk}. 
\begin{comment}
In addition, we will need to analyze the correlators with $Z$ and $\bar{Z}$ as light insertions differently from the correlators with $\Phi_i$ and $\mathbb{D}_a$ because we will see that, in our choice of coordinates, the fields dual to $Z$ and $\bar{Z}$ are coupled and solving for their propagators from their Green's equations is significantly more cumbersome. Our workaround will be to use the superconformal Ward identities to determine the $Z$ and $\bar{Z}$ correlators in terms of the $\Phi_i$ correlators, as discussed around \eqref{eq:RKeEazh2eB}-\eqref{eq:Kluz7kJ9dy}.
\end{comment}
In the remainder of this section, we will derive the quadratic action of the bosonic fluctuation modes in detail, identify the Green's equations satisfied by the bulk-to-bulk propagators, and formulate the precise dictionary between the defect correlators and the boundary-to-boundary propagators.

\subsection{Quadratic action for the fluctuations}\label{sec:quadratic action}

We begin by choosing coordinates for AdS$_5\times S^5$. We will use Euclidean signature throughout. First, we introduce the embedding coordinates $X^A$, $A=1,\ldots,6$ for AdS$_5$ and $Y^I$, $I=1,\ldots,6$ for $S^5$. These satisfy $\eta_{AB}X^AX^B=-1$ and $\delta_{IJ}Y^IY^J=1$, where $\eta_{AB}$ is the Minkowski metric tensor on $\mathbb{R}^{5,1}$ (with mostly plus convention),
%non-zero components $\eta_{11}=\eta_{22}=\ldots=\eta_{55}=1$ and $\eta_{66}=-1$, 
and $\delta_{IJ}$ is the standard Kronecker symbol on $\mathbb{R}^6$. 
%metric tensor on $\mathbb{R}^6$ with non-zero components $\delta_{11}=\delta_{22}=\ldots=\delta_{66}=1$. 
The metric on AdS$_5\times S^5$ may then be written:
\begin{equation}
\begin{aligned}
    &ds^2=ds_{\text{AdS}_5}^2+ds_{S^5}^2,\\
   &ds_{\text{AdS}_5}^2=\eta_{AB}dX^AdX^B\,,\qquad ds_{S^5}^2=\delta_{IJ}dY^IdY^J.
\end{aligned}
\end{equation}
Next, we parametrize the embedding coordinates in a way that is adapted to studying the fluctuations of the dual string. It will be convenient to foliate AdS$_5$ by AdS$_2\times S^2$ slices:
\begin{align}\label{eq:AdS5 embedding}
     X^a&=\frac{x^a}{1-\frac{1}{4}x^2},& (X^4,X^5,X^6)&=\frac{1+\frac{1}{4}x^2}{1-\frac{1}{4}x^2}\left(\sinh{\rho},\cosh{\rho}\sinh{\tau},\cosh{\rho}\cosh{\tau}\right).
\end{align}
Here $\rho,\tau\in \mathbb{R}$ are the bulk and Euclidean time coordinates on the AdS$_2$ slices and $x^a\in \mathbb{R}$, $a=1,2,3$, are three orthogonal coordinates with norm $x\equiv \sqrt{x^ax^a}$. If we decompose $x^a$ into radial and angular coordinates, then the two angular coordinates are coordinates on the $S^2$ slices and the radial coordinate $x$ parametrizes the different $S^2\times AdS_2$ slices. Similarly, it will be convenient to foliate $S^5$ by $S^3\times S^1$ slices:
\begin{align}\label{eq:S5 embedding}
    Y^i&=\frac{\cos{\bar{\theta}}\bar{y}^i}{1+\frac{1}{4}\bar{y}^2}, \hspace{1cm}Y^6=\frac{\cos{\bar{\theta}}(1-\frac{1}{4}\bar{y}^2)}{1+\frac{1}{4}\bar{y}^2}, &Y^4+iY^5&=\sin{\bar{\theta}}e^{i\bar{\phi}}.
\end{align}
Here $\bar{\phi}\in[0,2\pi)$ is the azimuthal angle on the $S^1$ slices corresponding to rotations in the $Y^4-Y^5$ plane, $\bar{y}^i\in \mathbb{R}$, $i=1,2,3$, are stereographic coordinates on the $S^3$ slices with norm $\bar{y}\equiv \sqrt{\bar{y}^i\bar{y}^i}$, and $\bar{\theta}\in[0,\pi)$ parametrizes the different $S^3\times S^1$ slices. We have written the coordinates on $S^5$ with bars to distinguish them from redefined coordinates that will appear later. In terms of the $x^a$, $\rho$, $\tau$ and $\bar{y}^i$, $\bar{\theta}$, $\bar{\phi}$ coordinates, the metrics on Euclidean AdS$_5$ and $S^5$ are
\begin{align}
ds_{\text{AdS}_5}^2&=\frac{\left(1+\frac{1}{4}x^2\right)^2}{\left(1-\frac{1}{4}x^2\right)^2}\left(d\rho^2+\cosh^2{\rho}d\tau^2\right)+\frac{dx^adx^a}{\left(1-\frac{1}{4}x^2\right)^2}\label{eq:EAdS5 metric},\\    ds_{S^5}^2&=d\bar{\theta}^2+\sin^2{\bar{\theta}}d\bar{\phi}^2+\cos^2{\bar{\theta}}\frac{d\bar{y}^jd\bar{y}^j}{\left(1+\frac{1}{4}\bar{y}^2\right)^2}\label{eq:S5 metric}.
\end{align}

Now we turn to the string in AdS$_5\times S^5$ that is dual to the Wilson line with $Z^J$ and $\bar{Z}^J$ inserted. We take the spacetime contour of the Wilson line to be a pair of antiparallel, antipodal lines located at the boundary of AdS$_5$ at $\rho\to \pm \infty$ (one may later perform a conformal transformation to the infinite straight line or circular loop). The contour at $x^a=0$ and $\rho\to +\infty$ runs in the positive $\tau$ direction, the contour at $x^a=0$ and $\rho\to -\infty$ runs in the negative $\tau$ direction, and $Z^J$ and $\bar{Z}^J$ are located at $\tau=-\infty$ and $\tau=\infty$, respectively. This configuration in global coordinates has manifest translational symmetry along $\tau$. The classical string dual to this operator was discussed in \cite{drukker2006small,Miwa2006HolographyOW,Gromov:2012eu} and reviewed in \cite{Giombi:2021zfb}. To summarize, the string forms a strip in AdS$_5$ that stretches between the two antipodal lines at $\rho\to \pm \infty$ and partly wraps an $S^2$ in $S^5$, carrying angular momentum $J$. Since the $S^5$ embedding coordinates $Y_I$ are dual to the scalars $\Phi_I$, and since the Wilson line couples to $\Phi_6$ and $Z$ and $\bar{Z}$ are chiral combinations of $\Phi_4$ and $\Phi_5$, the $S^2$ wrapped by the string is given by $(Y^4)^2+(Y^5)^2+(Y^6)^2=1$. With $\bar{y}^i=0$ in \eqref{eq:S5 embedding}, $\bar{\theta}$ and $\bar{\phi}$ are its polar and azimuthal angles. 

With $\rho$ and $\tau$ serving as the worldsheet coordinates, the classical dual string is given by:\footnote{Recall that we have already performed a Wick rotation to Euclidean AdS. The solution in Lorentzian signature is the same but with $\phi_{\rm cl}=\phi_0+\tau_L$, where $\tau_L$ is the Lorentzian time coordinate of global AdS.}
\begin{align}\label{eq:classical string solution}
    x^a&=0, & \bar{y}^i&=0, &\sin\bar{\theta}&=\sin\theta_{\rm cl}(\rho)\equiv \frac{c}{\cosh{\rho}}, & \bar{\phi}&=\phi_{\rm cl}(\tau)\equiv \phi_0-i\tau.
\end{align}
Here, the parameter $c\in[0,1)$, which we introduced for convenience in \eqref{eq: Z^JZbar^J}, determines the maximum polar angle of the string as well as the angular momentum of the string in the $Y^4-Y^5$ plane.  Fixing the angular momentum to be $J$ yields the condition \eqref{eq:J/g and c^2}. Finally, the parameter $\phi_0$ in \eqref{eq:classical string solution} is a modulus of the classical solution. This modulus played an important role in \cite{Giombi:2021zfb}, but a limited one in the present analysis. When mapping the string observables to the dCFT observables, one should integrate over the modulus to ensure that the string observables are dual to CFT observables in an R-charge eigenstate rather than a coherent state. The integration ensures that correlators not having equal numbers of $Z$ and $\bar{Z}$ are zero, and also gives rise to non-trivial combinatorial factors when the chiral primaries are non-zero linear combinations of both $Z$ and $\bar{Z}$. However, in the computation of the correlators in \eqref{eq:PhPhZ^JZb^J}-\eqref{eq:DDZ^JZb^J}, the integration over $\phi_0$ is trivial and we will ignore it. 

We are interested in the bosonic fluctuations about the classical solution in \eqref{eq:classical string solution}. We will work with the Nambu-Goto action and choose the static gauge such that $\rho$ and $\tau$ are not dynamical and serve as the worldsheet coordinates. For notational convenience, we package them into $\sigma^\mu\equiv (\rho,\tau)$, and let $\partial_\mu \equiv \partial/\partial \sigma^\mu$. We denote the AdS$_2$ metric by $h_{\mu\nu}$ and the metric induced on the classical string by $\gamma_{\mu\nu}$, which are explicitly
\begin{align}
    h_{\mu\nu}&\equiv \text{diag}\left(1,\cosh^2{\rho}\right),\\
    \gamma_{\mu\nu}&\equiv h_{\mu\nu}+\partial_\mu \theta_{\rm cl} \partial_\nu \theta_{\rm cl} + \sin^2{\theta_{\rm cl}}\partial_\mu \phi_{\rm cl}\partial_\nu \phi_{\rm cl}=\frac{\cosh^4{\rho}-c^2}{\cosh^2{\rho}}\text{diag}\left(\frac{1}{\cosh^2{\rho}-c^2},1\right).
\label{induced-metric}
\end{align}
We also note the inverse metric and the tensor density (here $\gamma\equiv \text{det}\gamma_{\mu\nu}$):
\begin{align}\label{eq:inverse metric gamma}
    \gamma^{\mu\nu}&= \frac{\cosh^2{\rho}}{\cosh^4{\rho}-c^2}\text{diag}\left(\cosh^2{\rho}-c^2,1\right),&
    \sqrt{\gamma}&= \frac{\cosh^4{\rho}-c^2}{\cosh^2{\rho}\sqrt{\cosh^2{\rho}-c^2}}.
\end{align}

The quadratic action of the bosonic fluctuation modes can be found by expanding the Nambu-Goto action
\begin{align}
    S= 2g \int d^2\sigma \sqrt{\Gamma},
\label{NG-action}
\end{align}
around the classical solution. Here $2g=\frac{\sqrt{\lambda}}{2\pi}$ is the string tension, and the induced metric on the fluctuating string is:
\begin{align}\label{eq:fluctuating string induced metric}
    \Gamma_{\mu\nu}\equiv \frac{\left(1+\frac{1}{4}x^2\right)^2}{\left(1-\frac{1}{4}x^2\right)^2}h_{\mu\nu}+\frac{\partial_\mu x^a\partial_\nu x^a}{\left(1-\frac{1}{4}x^2\right)^2}+\partial_\mu \bar{\theta} \partial_\nu \bar{\theta} + \sin^2{\bar{\theta}}\partial_\mu \bar{\phi}\partial_\nu \bar{\phi}+\frac{\cos^2{\bar{\theta}}}{\left(1+\frac{1}{4}y^2\right)^2}\partial_\mu \bar{y}^i \partial_\nu \bar{y}^i.
\end{align}
The $x^a$ coordinates within AdS$_5$ are already suitable for the small fluctuation expansion. For the coordinates in $S^5$, it is convenient to define the fluctuation fields $y^i$, $\theta$ and $\phi$ as 
\begin{align}\label{eq:redefining fields}
    \bar{y}^i&= \frac{1}{f_y(\rho)}y^i, & \bar{\theta}&=\theta_{\rm cl}+\frac{1}{f_\theta(\rho)}\theta, & \bar{\phi}&=\phi_{\rm cl}+\frac{1}{f_\phi(\rho)}\phi\,,
\end{align}
where 
\begin{equation}
\label{f-factors}
f_y(\rho)\equiv \frac{\sqrt{\cosh^2{\rho}-c^2}}{\cosh{\rho}}\,,\quad 
f_\theta(\rho)\equiv \frac{\cosh{\rho}\sqrt{\cosh^2{\rho}-c^2}}{\sqrt{\cosh^4{\rho}-c^2}}\,,\quad 
f_\phi(\rho)\equiv \frac{c\cosh{\rho}}{\sqrt{\cosh^4{\rho}-c^2}}\,.
\end{equation}
The rescaling by the $\rho$ dependent factors is necessary in order to obtain canonical kinetic terms for the fluctuations. Plugging (\ref{eq:redefining fields}) into (\ref{NG-action}) and expanding to quadratic order in $x^a$, $y^j$, $\theta$, $\phi$, the final result for the quadratic action takes the form (see Appendix~\ref{eq:deriving quadratic action} for the detailed derivation):
\begin{align}\label{eq:quadratic action}
    S_2[x^a,y^i,\theta,\phi]\equiv 2g \int d^2\sigma \sqrt{\gamma}\left[\mathcal{L}^{xx}+\mathcal{L}^{yy}+\mathcal{L}^{\theta\phi}\right],
\end{align}
where 
\begin{align}
    \mathcal{L}^{xx}&=\frac{1}{2}\gamma^{\mu\nu}\partial_\mu x^a \partial_\nu x^a+\frac{1}{2}m_x^2(\rho)x^ax^a,\label{eq:1wSykOeZ3v}\\
\mathcal{L}^{yy}&=\frac{1}{2}\gamma^{\mu\nu}\partial_\mu y^i \partial_\nu y^i+\frac{1}{2}m_y^2(\rho) y^iy^i,\label{eq:kf1t2Nw4fy}
\end{align}
and the ``masses" are given by 
\begin{align}
    m_x^2(\rho)&\equiv \frac{\cosh^2{\rho}(2\cosh^2{\rho}-c^2)}{\cosh^4{\rho}-c^2}, &
m_y^2(\rho)&\equiv-\frac{c^2\left(\cosh^2{\rho}-2\right)}{\cosh^4{\rho}-c^2}.\label{eq:mass x and y}
\end{align}
Note that $m_x^2(\rho)\to 2$ and $m_y^2(\rho)\to 0$ as $c\to 0$ or $\rho\to \pm\infty$, which are the expected values of the masses for the fluctuations around the undeformed AdS$_2$ string. The Lagrangian $\mathcal{L}^{\theta\phi}$ for the coupled $\theta$, $\phi$ modes is slightly more complicated and takes the form 
\begin{align}\label{eq:LpWrSXGjF0}
    \mathcal{L}^{\theta\phi}&=\frac{1}{2}\gamma^{\mu\nu}\left(\partial_\mu \theta\partial_\nu \theta+\partial_\mu \phi\partial_\nu\phi\right)+\frac{m_{\theta\phi}^2}{2}\left(\theta^2+\phi^2\right)-is(\theta\partial_\tau \phi-\phi\partial_\tau \theta),
\end{align}
where the prefactor for the cross-term is
\begin{align}\label{eq:s theta phi}
    s&\equiv \frac{\cosh^2{\rho}(-\cosh^4{\rho}+c^2\cosh{2\rho})}{(\cosh^4{\rho}-c^2)^2},
\end{align}
and the mass is
\begin{align}\label{eq:mass theta phi}
m_{\theta\phi}^2&\equiv-\frac{1}{(\cosh^4{\rho}-c^2)^3}\biggr[\cosh^{10}{\rho}+c^2\cosh^6{\rho}(10-12\cosh^2{\rho}+\cosh^4{\rho})\\&\hspace{5.5cm}+c^4\cosh^2{\rho}(1-12\cosh^2{\rho}+10\cosh^4{\rho})+c^6\cosh^2{\rho}\biggr].\nonumber
\end{align}
%Let us make a few comments about the $\mathcal{L}^{\theta\phi}$ Lagrangian  (\ref{eq:LpWrSXGjF0}). Firstly, note that it is symmetric in the fluctuations $\theta$ and $\phi$, despite the original coordinates $\bar{\theta}$ and $\bar{\phi}$ being a priori rather different (they are the polar and azimuthal angles on the two-sphere partly wrapped by the classical string). Secondly, 
Note that $\theta$ and $\phi$ are coupled by the $\theta\overset{\leftrightarrow}{\partial_\tau}\phi\equiv\theta\partial_\tau \phi-\phi\partial_\tau \theta$ term and there does not appear to be a simple coordinate transformation to decouple them. Note also that, from the dCFT perspective, the $SO(5)$ R-symmetry of the Wilson line that is broken to $SO(3)$ by the insertion of $Z^J$ and $\bar{Z}^J$ should be restored when $J=c^2=0$. In terms of the dual string, this implies that there should be a choice of coordinates in which the five $S^5$ fluctuation modes appear the same when $c^2=0$. The restoration of the broken $SO(5)$ symmetry is not manifest in terms of the $y^i$, $\theta$ and $\phi$ fluctuation fields, because when $c^2\to 0$, then $m_y^2\to 0$ but $m_{\theta\phi}^2\to -1/\cosh^2{\rho}$ and $s\to -1/\cosh^2{\rho}$. The symmetry can be made manifest, at least at quadratic order, by rotating the combination $(\theta,\phi)$ by $i\tau$. This is done explicitly in Appendix~\ref{app:perturbative analysis}, where it is a useful first step in the perturbative analysis of the $\theta$ and $\phi$ modes.

\begin{figure}
    \centering
    \begin{minipage}{0.49\hsize}
\centering
\includegraphics[clip, height=7cm]{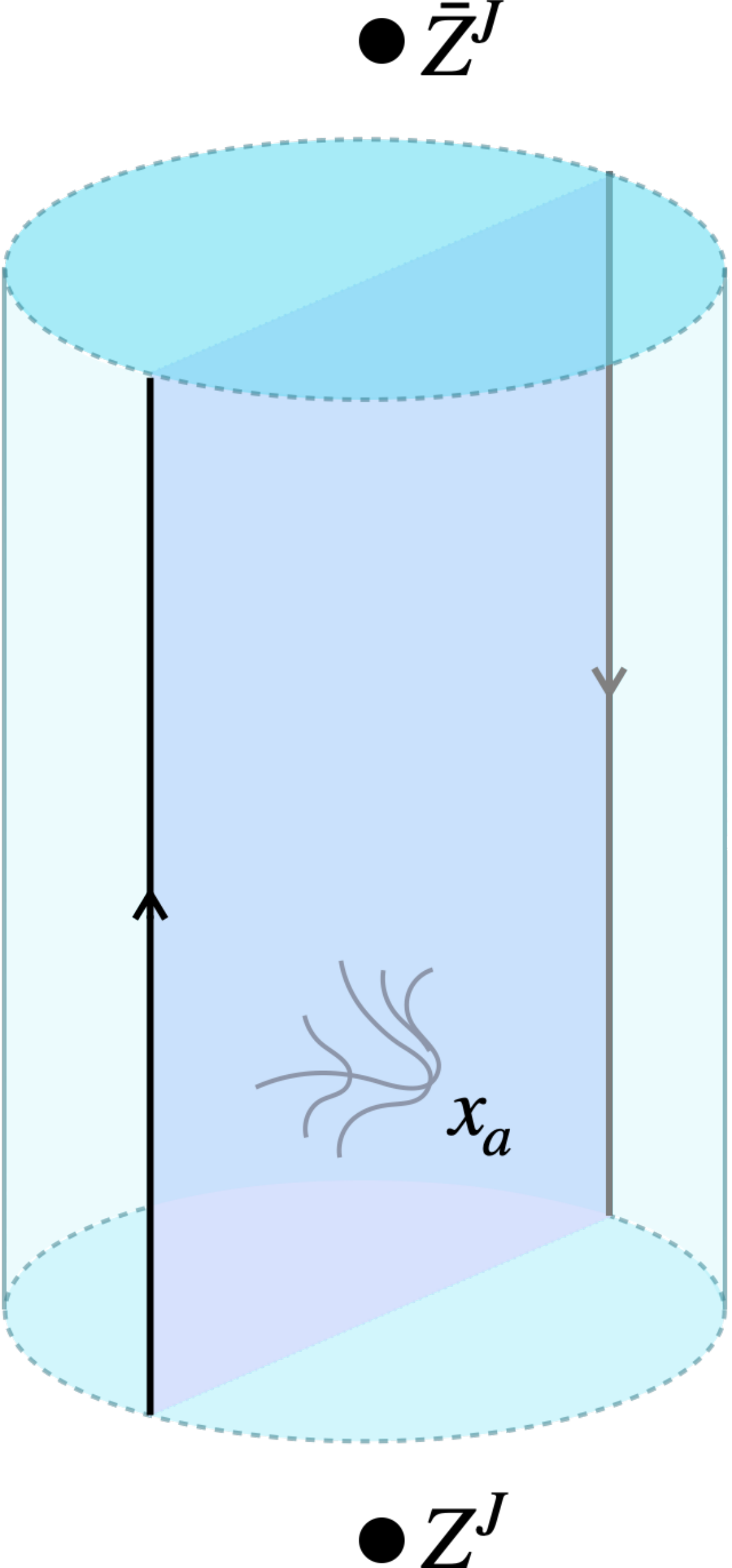}\\
\vspace{0.25cm}
{\bf a.} AdS$_5$ 
\end{minipage}
\begin{minipage}{0.49\hsize}
\centering
\vspace{1.5cm}
\includegraphics[clip, height=4cm]{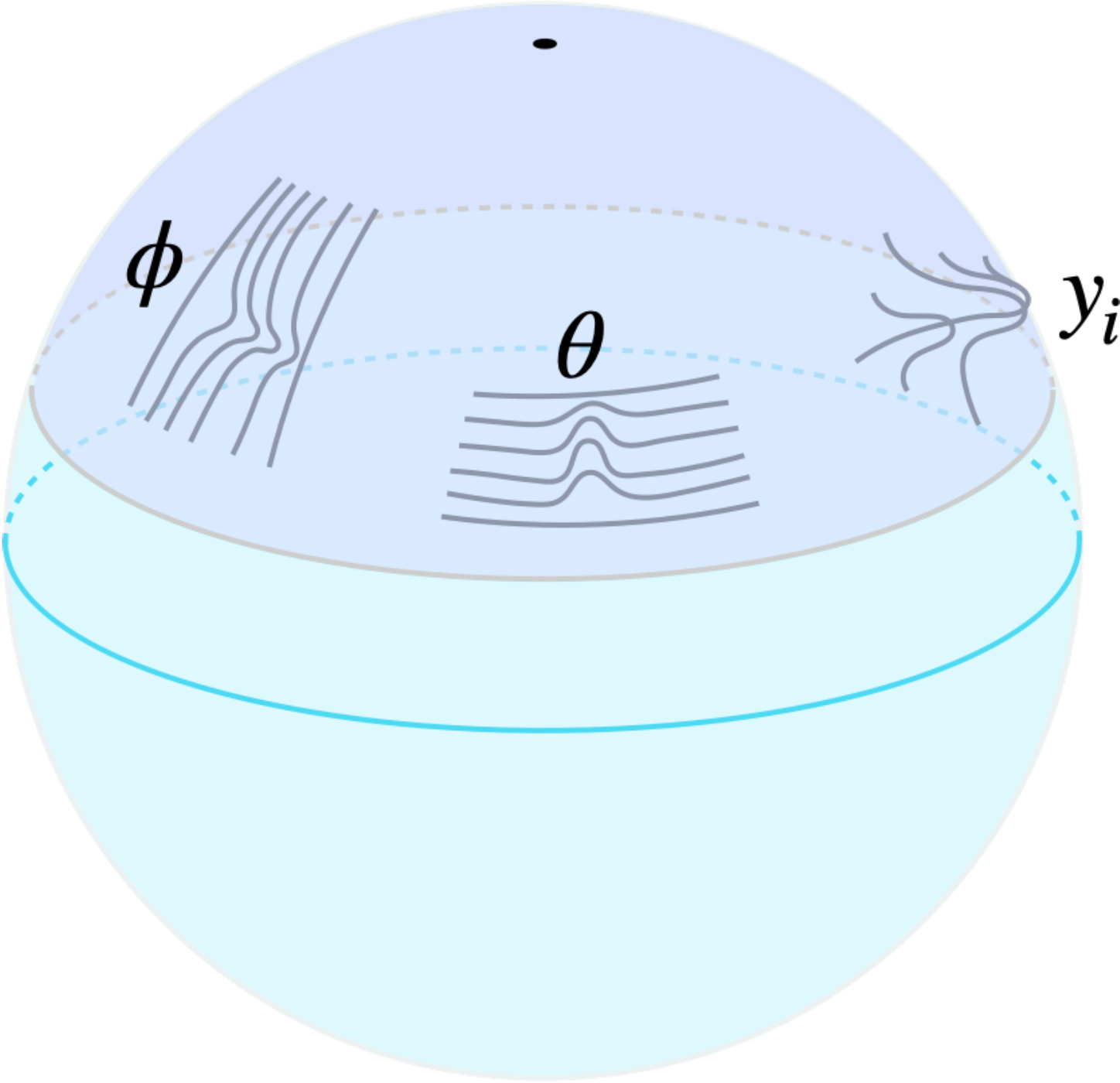}\\
\vspace{1.5cm}
{\bf b.} $S^5$
\end{minipage}    \caption{A qualitative sketch of the four types of bosonic fluctuations of the semiclassical string dual to the Wilson line with $Z^J$ and $\bar{Z}^J$. In the static gauge, there are \textbf{a.} $3$ fluctuation modes, labelled $x^a$, orthogonal to the AdS$_2$ strip formed by the classical string in AdS$_5$, and \textbf{b.} $1+1+3$ fluctuation modes, labelled $\phi$, $\theta$ and $y^i$, in the azimuthal, polar and orthogonal directions of the $S^2$ partially wrapped by the classical string in $S^5$.}
    \label{fig:fluctuations about classical string}
\end{figure}

To summarize, (\ref{eq:quadratic action}) is our final result for the quadratic action of fluctuations in the static gauge. The eight transverse bosonic modes $x^a$, $y^i$, $\theta$ and $\phi$, which are schematically depicted in Figure~\ref{fig:fluctuations about classical string}, can be viewed as fields propagating on an asymptotically AdS$_2$ background, where the deformation from AdS$_2$ corresponds to turning on the non-zero angular momentum. Note that there is a manifest $SO(3)\times SO(3)_R$ symmetry rotating the $x$ and $y$ coordinates, and the $x$, $y$ and $\theta/\phi$ coordinates are decoupled to this order. 

Let us finally remark that the expansion to quadratic order around a generic classical string solution in AdS$_5\times S^5$ was discussed using a rather general formalism in \cite{Forini:2015mca} (see also \cite{Drukker:2000ep}), working with the Polyakov action. 
%It should also be noted that an alternative to studying the quadratic fluctuations via the Nambu-Goto action in the static gauge is to implement the general formalism , which works with the Polyakov action in the conformal gauge. 
We have checked that the results given in \cite{Forini:2015mca} precisely agree with \eqref{eq:1wSykOeZ3v}, \eqref{eq:kf1t2Nw4fy} and \eqref{eq:LpWrSXGjF0}-\eqref{eq:mass theta phi}.\footnote{Our quadratic action in static gauge should be compared with the action for the transverse fluctuations given in eq. (3.35) of \cite{Forini:2015mca}.} 
%Thus, at least at the quadratic order, the Lagrangians for the bosonic fluctuations in the static and conformal gauges are the same.

\paragraph{Bulk-to-bulk propagators.} %The quadratic action in \eqref{eq:quadratic action} determines the bulk-to-bulk propagators for the four fluctuation modes, 

The bulk-to-bulk propagators for the fluctuation modes, $G_{yy}$, $G_{xx}$, $G_{\theta\theta}=G_{\phi\phi}$ and $G_{\phi\theta}=-G_{\theta\phi}$, satisfy Green's equations that follow from the quadratic action. For the $x$ and $y$ modes, they are\footnote{The Green's equation for the $y^i$ modes follows from evaluating the functional derivative in $0=\int \mathcal{D}y \frac{\delta}{\delta y^i(\rho,\tau)}\left(\text{exp}\left(-2g\int d^2\sigma \sqrt{\gamma}\mathcal{L}^{yy}\right)y^j(\rho',\tau')\right)$. Likewise for the $x^a$ modes.}
\begin{align}
    \left[\frac{1}{\sqrt{\gamma}}\partial_\mu\left(\sqrt{\gamma}\gamma^{\mu\nu}\partial_\nu\right)-m_x^2(\rho)\right]G_{xx}(\rho,\tau;\rho',\tau')&=-\frac{1}{2g\sqrt{\gamma}}\delta(\rho-\rho')\delta(\tau-\tau'),\label{eq:xx Green's eqn}\\
    \left[\frac{1}{\sqrt{\gamma}}\partial_\mu\left(\sqrt{\gamma}\gamma^{\mu\nu}\partial_\nu\right)-m_y^2(\rho)\right]G_{yy}(\rho,\tau;\rho',\tau')&=-\frac{1}{2g\sqrt{\gamma}}\delta(\rho-\rho')\delta(\tau-\tau').\label{eq:yy Green's eqn}
\end{align}
Meanwhile, since the $\theta$ and $\phi$ modes are coupled in the Lagrangian in \eqref{eq:LpWrSXGjF0}, the corresponding bulk-to-bulk propagators satisfy coupled Green's equations:\footnote{These follow, respectively, from evaluating $0=\int \mathcal{D}\theta \mathcal{D}\phi \frac{\delta}{\delta \theta(\rho,\tau)}\left(\text{exp}\left(-2g\int d^2\sigma \sqrt{\gamma}\mathcal{L}^{\theta\phi}\right)\theta(\rho',\tau')\right)$ and $0=\int \mathcal{D}\theta \mathcal{D}\phi \frac{\delta}{\delta \phi(\rho,\tau)}\left(\text{exp}\left(-2g\int d^2\sigma \sqrt{\gamma}\mathcal{L}^{\theta\phi}\right)\theta(\rho',\tau')\right)$.}
\begin{align}
    \left[\frac{1}{\sqrt{\gamma}}\partial_\mu\left(\sqrt{\gamma}\gamma^{\mu\nu}\partial_\nu\right)-m_{\theta\phi}^2(\rho)\right]G_{\theta\theta}+2is(\rho)\partial_\tau G_{\phi\theta}&=-\frac{1}{2g\sqrt{\gamma}}\delta(\rho-\rho')\delta(\tau-\tau'),\label{eq:thetatheta Green's eqn}\\
    \left[\frac{1}{\sqrt{\gamma}}\partial_\mu\left(\sqrt{\gamma}\gamma^{\mu\nu}\partial_\nu\right)-m_{\theta\phi}^2(\rho)\right]G_{\phi\theta}-2is(\rho)\partial_\tau G_{\theta\theta}&=0.\label{eq:phitheta Green's eqn}
\end{align}
These equations are accompanied by the boundary condition that each propagator vanishes at both boundaries of the strip: $G(\rho,\tau;\rho',\tau')\to 0$ as $\rho\to \pm\infty$. The normalization of the delta function on the RHS of \eqref{eq:xx Green's eqn}-\eqref{eq:thetatheta Green's eqn} makes it explicit that the bulk-to-bulk propagators scale with the inverse of the string tension.

The propagators have a number of useful symmetries that follow either from the quadratic action or the Green's equations. First, $G_{xx}$, $G_{yy}$ and $G_{\theta\theta}$ are all real, even under interchange of the bulk points (e.g., $G_{xx}(\rho,\tau;\rho',\tau')=G_{xx}(\rho',\tau';\rho,\tau)$), even under parity (e.g., $G_{xx}(\rho,\tau;\rho',\tau')=G_{xx}(-\rho,\tau;-\rho',\tau')=G_{xx}(\rho,-\tau;\rho',-\tau')$), and invariant under translations along $\tau$ (e.g., $G_{xx}(\rho,\tau;\rho',\tau')=G_{xx}(\rho,\tau+a;\rho',\tau'+a)$ for any $a\in \mathbb{R}$). By contrast, while $G_{\phi\theta}$ is also invariant under translations along $\tau$ and even under $\rho,\rho'\to -\rho,-\rho'$, it is imaginary, odd under interchange of the bulk points, and odd under $\tau,\tau'\to -\tau,-\tau'$. The difference in behavior of $G_{\phi\theta}$ can be traced to the $is\partial_\tau$ terms in \eqref{eq:thetatheta Green's eqn} and \eqref{eq:phitheta Green's eqn}.

\paragraph{From global to Poincar\'e coordinates.} 

So far we have studied the fluctuations of the dual string using coordinates on AdS$_5\times S^5$ in which the AdS$_2$ slices are parametrized by the global coordinates, $\rho,\tau$. In these coordinates, the translational symmetry of the classical string is manifest. However, Poincar\'e coordinates are perhaps a more familiar choice for stating the AdS/CFT dictionary between the defect four-point correlators in \eqref{eq:PhPhZ^JZb^J}-\eqref{eq:DDZ^JZb^J} and the bulk-to-bulk propagators and classical vertex operators. Thus, we will also parametrize the AdS$_2$ slices using Poincar\'e coordinates $t,z$ that are related to the global coordinates by\footnote{To study correlators with $Z^J$ at $t=0$ and $\bar{Z}^J$ at $t=\infty$, we should instead use Poincar\'e coordinates $t,z$ that are related to the global coordinates by $t+iz=e^\tau (\tanh{\rho}+i\sech{\rho})$. These are related to the coordinates in \eqref{eq:global to Poincare} by an $SL(2,\mathbb{R})$ transformation.}
\begin{align}\label{eq:global to Poincare}
    t+iz&\equiv \frac{e^\tau (\tanh{\rho}\pm i\sech{\rho})t_4 +t_3}{e^\tau(\tanh{\rho}\pm i\sech{\rho})+1}.
\end{align}
We take the upper sign if $t_3<t_4$ and the lower sign if $t_3>t_4$ so that $z\geq 0$. Going forward we assume $t_3<t_4$. The transformation indeed satisfies $d\rho^2+\cosh^2{\rho}d\tau^2=\frac{1}{z^2}(dt^2+dz^2)$.

Two properties of \eqref{eq:global to Poincare} are worth noting:  Firstly, $Z^J$ and $\bar{Z}^J$ are located at $t_3$ and $t_4$, respectively, on the AdS$_2$ boundary in accordance with \eqref{eq:PhPhZ^JZb^J}-\eqref{eq:DDZ^JZb^J}. This follows from sending $\tau\to -\infty$ and $\tau\to \infty$ in \eqref{eq:global to Poincare}. Secondly, to send the bulk point $\rho,\tau$ to the boundary point $t$, we
\begin{align}\label{eq:bulk to boundary}
    \text{fix }\tau&=\log\left|\frac{t-t_3}{t_4-t}\right|, &\text{and send }\rho&\to \eta\infty
\end{align}
where $\eta=1$ if $t_3<t<t_4$ and $\eta=-1$ otherwise. In this limit, $z$ is asymptotically
\begin{align}\label{eq:z as we approach boundary}
    z\sim \frac{2|t_4-t||t_3-t|}{t_4-t_3}e^{-\eta \rho}.
\end{align}

\subsection{Four-point functions as boundary-to-boundary propagators}\label{sec:diciontary}
We are now ready to state the precise dictionary between the defect correlators on the Wilson line and the propagators on the classical string.

First, we identify the fluctuation modes dual to the elementary insertions introduced in Section~\ref{sec: setup}. The six scalars, $\Phi_I$, in $\mathcal{N}=4$ SYM are dual to the $S^5$ embedding coordinates, $Y_I$. In particular, in the dCFT the $\Phi_i$, $i=1,2,3$, appearing in \eqref{eq:PhPhZ^JZb^J} are dual to $Y_i$, while $Z$ and $\bar{Z}$ are dual to $Y_4+iY_5=\sin{\bar{\theta}}e^{i\bar{\phi}}$ and $Y_4-iY_5=\sin{\bar{\theta}}e^{-i\bar{\phi}}$, respectively. Let us recall that the holographic correlators depend only on the asymptotic behavior of the fluctuation fields near the worldsheet boundary (see, e.g., \eqref{eq:string path integral}-\eqref{eq:schematic vertex operator}). Thus, given that $y_i\to 0$, $\bar{\theta}\to \bar{\theta}_{\rm cl}\to 1$ and $f_y\to 1$ as $\rho\to \pm \infty$, it follows from \eqref{eq:S5 embedding} and \eqref{eq:redefining fields} that $Y_i\sim y_i$ asymptotically. We may therefore equivalently take the fluctuation field dual to $\Phi_i$ to be $y_i$. Indeed, as can be seen from the Lagrangian in \eqref{eq:kf1t2Nw4fy}, the $y_i$ have an $SO(3)\subset SO(6)$ rotational symmetry and their masses asymptotically satisfy $m_y^2\to 0$ as $\rho\to \pm \infty$, whose dual statements in the dCFT are that the scalars $\Phi_i$ preserve an $SO(3)\subset SO(6)$ R-symmetry and have scaling dimension $\Delta=1$.\footnote{Recall that for a scalar, $m^2=\Delta(\Delta-d)$. In our case $d=1$.} We will also shortly discuss what the string observables look like when the fields dual to $Z$ and $\bar{Z}$ are expressed in terms of $\theta$ and $\phi$, which are related to $\bar{\theta}$ and $\bar{\phi}$ by the field redefinitions in \eqref{eq:redefining fields}. For now we note that the masses of the $\theta$ and $\phi$ fields satisfy $m_{\theta\phi}^2\to 0$ asymptotically near the boundary (see \eqref{eq:mass theta phi}), which matches the fact that $Z$ and $\bar{Z}$ have scaling dimension $\Delta=1$. Finally, the displacement operators $\mathbb{D}_a$ are dual to the AdS$_5$ embedding coordinates, $X_a$, $a=1,2,3$, that are transverse to the Wilson line on the boundary. Again because $x_a\to 0$ as $\rho\to \pm \infty$, it follows from \eqref{eq:AdS5 embedding} that $X_a\sim x_a$ asymptotically and therefore we may equivalently take the fluctuation field dual to $\mathbb{D}_a$ to be $x_a$. Indeed, as can be seen from the Lagrangian in \eqref{eq:1wSykOeZ3v}, the $x_a$ have an $SO(3)$ symmetry of rotations in AdS$_5$ about the classical string and their masses satisfy $m_x^2\to 2$ as $\rho\to \pm \infty$, whose dual statements in the dCFT are that the $\mathbb{D}_a$ have an $SO(3)$ symmetry of rotations in $\mathbb{R}^4$ about the Wilson line and have scaling dimension $\Delta=2$.

The vertex operators dual to $\Phi_i$, $\mathbb{D}_a$, $Z$ and $\bar{Z}$ are therefore:
\begin{align}
    v_{\Phi_i}(t)&\equiv \lim_{z\to 0}\frac{2g}{z}y_i(z,t),&v_{\mathbb{D}_a}(t)&\equiv \lim_{z\to 0}\frac{2g}{z^2}x_a(z,t),\label{eq:Phi and D vertex op}\\
    v_{Z}(t)&\equiv \lim_{z\to 0}\frac{2g}{z}\sin(\bar{\theta}(z,t))e^{i\bar{\phi}(z,t)}, &v_{\bar{Z}}(t)&\equiv \lim_{z\to 0}\frac{2g}{z}\sin(\bar{\theta}(z,t))e^{-i\bar{\phi}(z,t)}.\label{eq:vZ and vZb}
\end{align}
This makes \eqref{eq:schematic vertex operator} precise. Although the above vertex operators are defined using Poincar\'e coordinates, we will take advantage of the simplicity of the classical string in global coordinates and take the bulk-to-boundary limits using \eqref{eq:bulk to boundary} and \eqref{eq:z as we approach boundary}. 

\begin{figure}[t]
\centering
\begin{minipage}{0.24\hsize}
\centering
\includegraphics[clip, height=8cm]{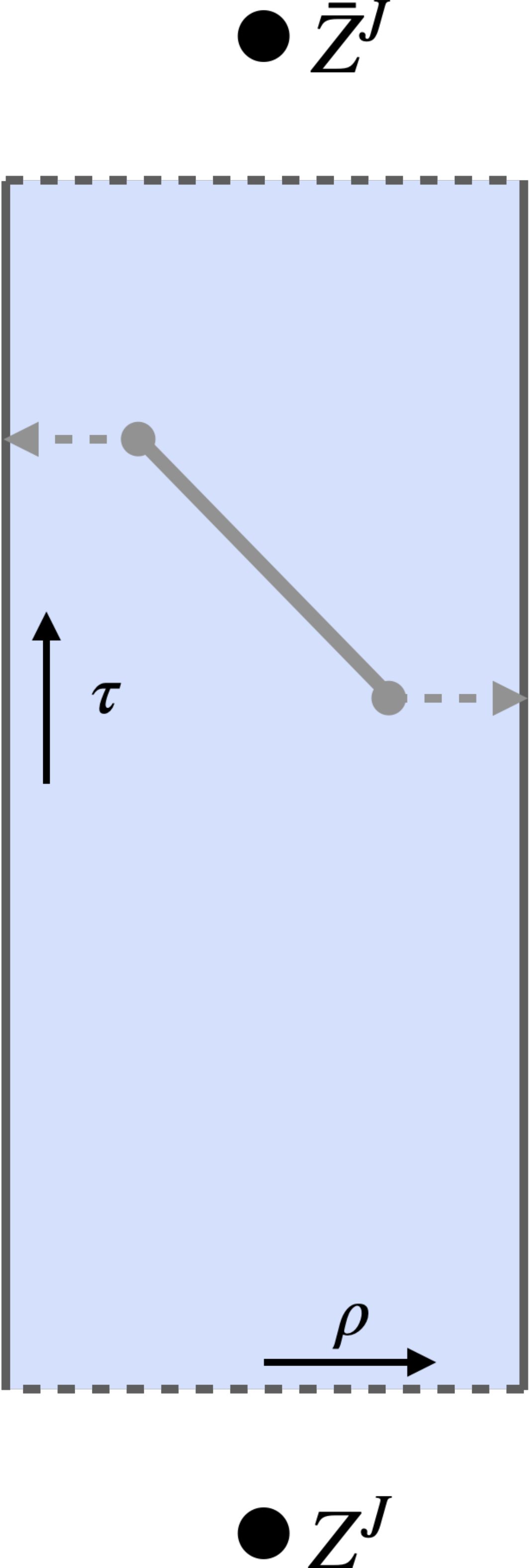}\\
\vspace{0.25cm}
{\bf a.}
\end{minipage}
\begin{minipage}{0.24\hsize}
\centering
\includegraphics[clip, height=8cm]{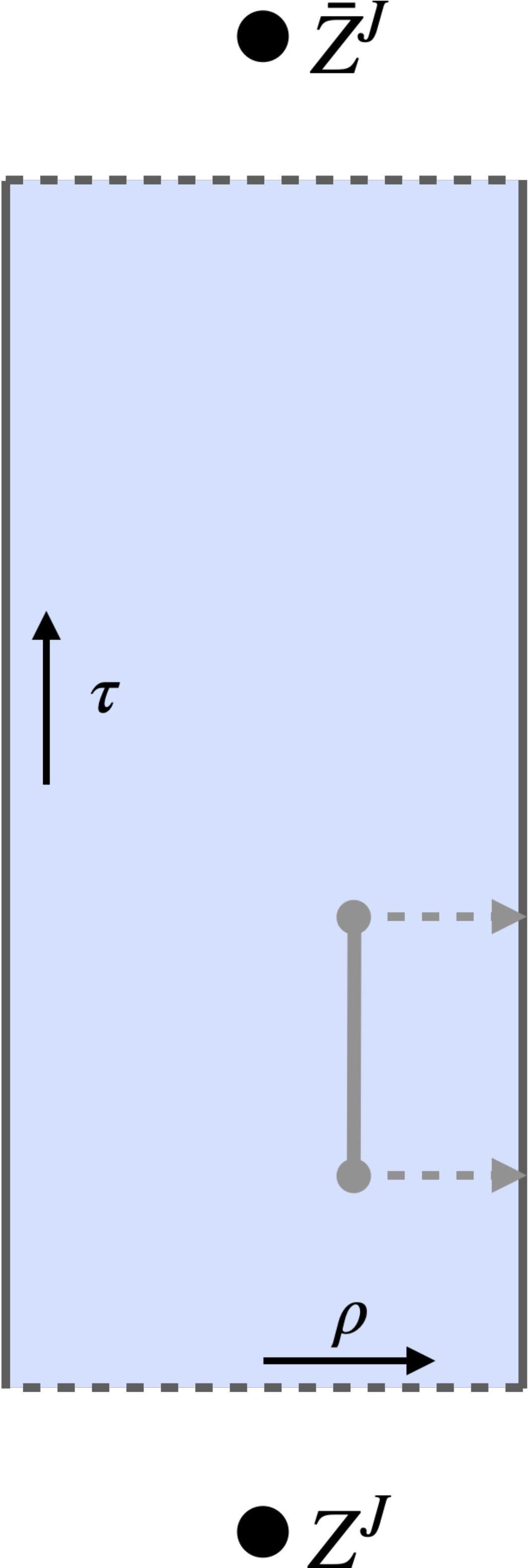}\\
\vspace{0.25cm}
{\bf b.} 
\end{minipage}
\begin{minipage}{0.49\hsize}
\centering
\includegraphics[clip, height=3cm]{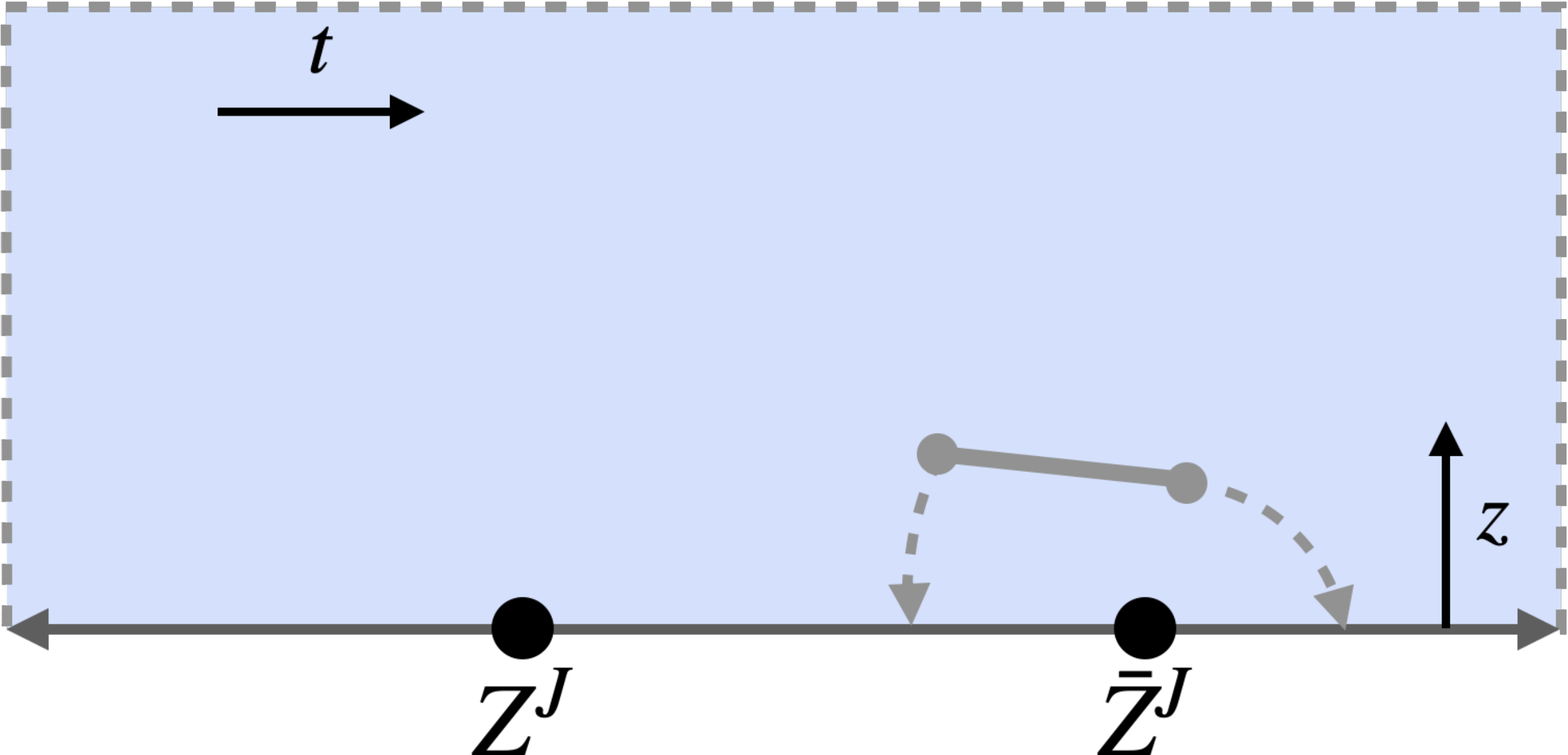}

\vspace{0.25cm}

{\bf c.} 

\vspace{0.25cm}

\includegraphics[clip, height=3cm]{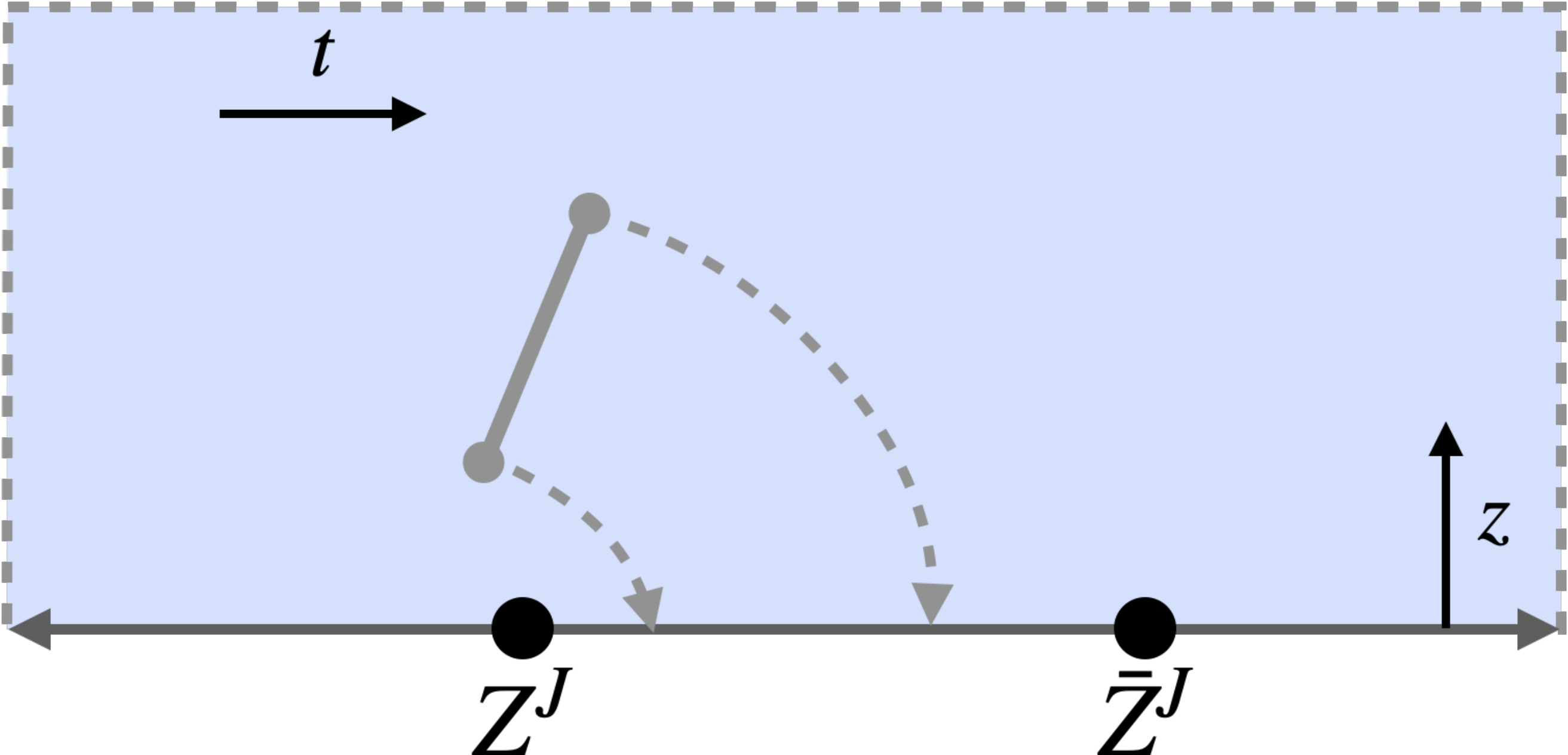}

\vspace{0.25cm}

{\bf d.} 
\vspace{0.25cm}
\end{minipage}
\caption{Sending the endpoints of the bulk-to-bulk propagators to the boundary in global (\textbf{a.} and \textbf{b.}) and in Poincar\'e (\textbf{c.} and \textbf{d.}) coordinates, in accordance with \eqref{eq:bulk to boundary}. When the bulk points are sent to opposite boundaries in global coordinates, as in \textbf{a.}, the light insertions on the Wilson line are separated by the heavy insertions, as in \textbf{c.}. In this case $\chi>1$. When the bulk points are sent to the same boundary in global coordinates, as in \textbf{b.}, the light insertions are between the two heavy insertions, as in \textbf{d.}. In this case $\chi<1$.} 
\label{fig:extrapolating bulk-to-boundary}
\end{figure}

Given \eqref{eq:classical string solution}, the vertex operators on the classical solution simplify to:
\begin{align}\label{eq:classical vertex operators}
    v_{\Phi_i}\rvert_{\Psi_{\rm cl}}&=v_{\mathbb{D}_a}\rvert_{\Psi_{\rm cl}}=0,&
    v_{Z}(t)\rvert_{\Psi_{\rm cl}}&=2gce^{i\phi_0}\frac{t_4-t_3}{(t_4-t)^2},&v_{\bar{Z}}(t)\rvert_{\Psi_{\rm cl}}&=2gce^{-i\phi_0}\frac{t_4-t_3}{(t_3-t)^2}.
\end{align}
These expressions for the classical vertex operators, which determine the leading large charge behavior of the defect correlators consisting of powers of $Z$ and $\bar{Z}$ in the background of $Z^J$ and $\bar{Z}^J$, are familiar from \cite{Giombi:2021zfb}. Meanwhile, the subleading large charge behavior due to the quadratic fluctuations about the classical solution are determined by sending the endpoints of the bulk-to-bulk propagators to the boundary as outlined in \eqref{eq:schematic 4 pt}-\eqref{eq:schematic bulk-to-bulk}. In particular, given that the vertex operators dual to $\Phi_i$ and $\mathbb{D}_a$ are zero on the classical solution, the leading contribution to the defect correlators in \eqref{eq:PhPhZ^JZb^J} and \eqref{eq:DDZ^JZb^J} are given by
\begin{align}
    \frac{\braket{\Phi_i(t_1)\Phi_j(t_2)Z^J(t_3)\bar{Z}^J(t_4)}}{\braket{Z^J(t_3)\bar{Z}^J(t_4)}}&=W_{yy}(t_1,t_2)\delta_{ij},\label{eq:yy bndy-to-bndy to Phi Phi defect fn}\\
    \frac{\braket{\mathbb{D}_a(t_1)\mathbb{D}_b(t_2)Z^J(t_3)\bar{Z}^J(t_4)}}{\braket{Z^J(t_3)\bar{Z}^J(t_4)}}&=W_{xx}(t_1,t_2)\delta_{ab},\label{eq:xx bndy-to-bndy to DD defect fn}
\end{align}
where the boundary-to-boundary propagator $W$ ($\equiv W_{yy},W_{xx}$) is related to the bulk-to-bulk propagator $G$ ($\equiv G_{yy},G_{xx}$) by
\begin{align}\label{eq:yy/xx bndy-to-bndy}
    W(t_1,t_2)&\equiv (2\Delta-1)^2\lim_{\substack{z_1\to 0^+\\z_2\to 0^+}}\frac{2g}{z_1^\Delta}\frac{2g}{z_2^\Delta }G(t_1,z_1;t_2,z_2)\nonumber\\&=\frac{(2\Delta -1)^2}{2^{2\Delta-2}}\frac{g^2}{t_{12}^{2\Delta}}\frac{\chi^{2\Delta}}{|1-\chi|^\Delta}\lim_{\substack{\rho\to \eta_1 \infty\\\rho'\to \eta_2\infty}}e^{\eta_1\Delta \rho}e^{\eta_2\Delta \rho'}G(\rho,\tau(t_1);\rho',\tau(t_2)).
\end{align}
In the second line, we have sent the bulk point to the boundary in accordance with \eqref{eq:bulk to boundary} (see Figure~\ref{fig:extrapolating bulk-to-boundary}) and written the ratios of distances on the Wilson line coming from \eqref{eq:z as we approach boundary} in terms of the conformal ratio $\chi$ defined in \eqref{eq:chi}. Because of the translational symmetry along $\tau$, $G$ depends only on the difference between $\tau(t_1)$ and $\tau(t_2)$, which reduces to
\begin{align}\label{eq:tau1-tau2}
    \tau(t_1)-\tau(t_2)=\log\left|\frac{t_{13}}{t_{14}}\frac{t_{24}}{t_{23}}\right|=-\log|1-\chi|.
\end{align}
Thus, the boundary-to-boundary propagator has the same conformal form as the defect correlators in \eqref{eq:PhPhZ^JZb^J}-\eqref{eq:DDZ^JZb^J}.

Finally, we express the vertex operators dual to the light insertions $Z$ and $\bar{Z}$ in terms of the rescaled fluctuation fields $\theta$ and $\phi$. We begin by expanding $v_Z$ in \eqref{eq:vZ and vZb} to linear order in $\theta$ and $\phi$:\footnote{Higher orders in $\theta$ do not contribute because $\theta/f_\theta\sim e^{\mp \rho}$ as $\rho\to \pm \infty$. On the other hand, because $\phi/f_\phi\to 1$ as $\rho\to \pm\infty$, the vertex operator has a series of corrections involving higher orders of $\phi$, $e^{\tau+i\phi_0}\lim_{z\to 0}\frac{2g}{z}\left(-\frac{\phi^2}{2f_\phi}-\frac{i\phi^3}{3!f_\phi^2}+\ldots\right)$, which are not suppressed. There may be a more convenient choice of coordinates than the one in \eqref{eq:EAdS5 metric}-\eqref{eq:S5 metric} that avoids this undesirable behavior. However, practically speaking, at the order in the large charge expansion that we are considering, we may simply ignore the higher order terms in the vertex operators. Contractions between more than one pair of copies of $\phi$ in different vertex operators involve at least two bulk-to-bulk propagators and are suppressed by $1/g$. Furthermore, while there is a contribution at the order of interest in the large charge expansion involving the self-contraction between the two copies of $\phi$ in the term $\sim \lim_{z\to 0}\frac{\phi^2}{zf_\phi}$, it yields a constant that can be absorbed into the definition of the vertex operator.}
\begin{align}
    v_{Z}(t)&=\lim_{z\to 0}\frac{2g}{z}\left[\sin{\theta_{\rm cl}}e^{i\phi_{\rm cl}}+\frac{\cos{\theta_{\rm cl}}}{f_\theta}e^{i\phi_{\rm cl}}\theta+i\frac{\sin{\theta_{\rm cl}}}{f_\phi}e^{i\phi_{\rm cl}}\phi+\ldots\right]\nonumber\\&=v_Z(t)\rvert_{\Phi_{\rm cl}}+e^{\tau(t)+i\phi_0}\lim_{z\to 0}\frac{2g }{z}\left(\theta+i\phi\right)+\ldots.
\end{align}
To get to the second line, we used that $\cos{\theta_{\rm cl}}/f_\theta\to 1$ and $\sin{\theta_{\rm cl}}/f_\phi\to 1$ as $\rho\to \pm \infty$, which follow from \eqref{eq:classical string solution} and \eqref{f-factors}. Likewise, $v_{\bar{Z}}$ in \eqref{eq:vZ and vZb} expanded to linear order is
\begin{align}
    v_{\bar{Z}}(t)&=v_{\bar{Z}}(t)\rvert_{\Psi_{\rm cl}}+e^{-\tau(t)-i\phi_0}\lim_{z\to 0}\frac{2g }{z}\left(\theta-i\phi\right)+\ldots.
\end{align}
Thus, the holographic expression for the leading and first subleading terms in the large charge expansion of the defect correlators in \eqref{eq:ZZbZ^JZb^J}-\eqref{eq:ZbZZ^JZb^J} is:
\begin{align}
    \frac{\braket{Z(t_1)\bar{Z}(t_2)Z^J(t_3)\bar{Z}^J(t_4)}}{\braket{Z^J(t_3)\bar{Z}^J(t_4)}}&=v_{Z}(t_1)\rvert_{\Psi_{\rm cl}}v_{\bar{Z}}(t_2)\rvert_{\Psi_{\rm cl}}+2e^{\tau(t_1)-\tau(t_2)}(W_{\theta\theta}(t_1,t_2)+iW_{\phi\theta}(t_1,t_2)),\label{eq:th-th and th-ph bndy-to-bndy to ZZb defect fn}\\
    \frac{\braket{\bar{Z}(t_1)Z(t_2)Z^J(t_3)\bar{Z}^J(t_4)}}{\braket{Z^J(t_3)\bar{Z}^J(t_4)}}&=v_{\bar{Z}}(t_1)\rvert_{\Psi_{\rm cl}}v_{Z}(t_2)\rvert_{\Psi_{\rm cl}}+2e^{\tau(t_2)-\tau(t_1)}(W_{\theta\theta}(t_1,t_2)-iW_{\phi\theta}(t_1,t_2)),\label{eq:th-th and th-ph bndy-to-bndy to ZbZ defect fn}
\end{align}
where the boundary-to-boundary propagator $W$ ($\equiv W_{\theta\theta},W_{\phi\theta}$) is again given in terms of the bulk-to-bulk propagator $G$ ($\equiv G_{\theta\theta},G_{\phi\theta}$) by \eqref{eq:yy/xx bndy-to-bndy} with $\Delta=1$. 

Finally, we note from \eqref{eq:classical vertex operators} that the classical pieces in \eqref{eq:th-th and th-ph bndy-to-bndy to ZZb defect fn}-\eqref{eq:th-th and th-ph bndy-to-bndy to ZbZ defect fn} are explicitly:
\begin{align}\label{eq:classical ZZb}
    v_{Z}(t_1)\rvert_{\Psi_{\rm cl}}v_{\bar{Z}}(t_2)\rvert_{\Psi_{\rm cl}}&=\frac{4g^2c^2}{t_{12}^2}\frac{\chi^2}{(1-\chi)^2},&
    v_{\bar{Z}}(t_1)\rvert_{\Psi_{\rm cl}}v_{Z}(t_2)\rvert_{\Psi_{\rm cl}}&=\frac{4g^2c^2}{t_{12}^2}\chi^2.
\end{align}

\section{Computing the Green's functions} 
\label{sec:series representations of 4pt functions}
%boundary-to-boundary propagators}

The last step in determining the defect four-point correlators is to compute the boundary-to-boundary propagators. We begin by presenting a general way to express the bulk propagators as integrals over Fourier modes conjugate to the global boundary coordinate, send the bulk points to the boundary, and rewrite the resulting integral representation of the boundary propagator as a series. We then apply the procedure to first compute $W_{yy}$, which lets us determine $W_{\theta\theta}$ and $W_{\theta\phi}$ via the superconformal Ward identities, and then compute $W_{xx}$.

\subsection{Integral and series representations of the propagators}\label{sec:integral and series reps of propagators}
Taking advantage of the symmetry of the classical string in \eqref{eq:classical string solution} under translations in the global time coordinate $\tau$, we write the $yy$ and $xx$ bulk-to-bulk propagators in their Fourier representations: 
\begin{align}\label{eq:KAKPSBQcaL}
    G(\rho,\tau;\rho',\tau')\equiv \frac{1}{4\pi g}\int_{-\infty}^\infty dk e^{ik(\tau-\tau')} g(\rho,\rho';k).
\end{align}
Here, $G\equiv G_{yy}$ and $g(\rho,\rho';k)\equiv g_{yy}(\rho,\rho';k)$ or $G\equiv G_{xx}$ and $g(\rho,\rho';k)\equiv g_{xx}(\rho,\rho';k)$. 

Noting the explicit form of the d'Alembertian on the classical string worldsheet in global coordinates,
\begin{align}
    \partial_\mu\left(\sqrt{\gamma}\gamma^{\mu\nu}\partial_\nu\right)=\frac{\partial}{\partial\rho}\left(\sqrt{\cosh^2{\rho}-c^2}\frac{\partial}{\partial \rho}\right)+\frac{1}{\sqrt{\cosh^2{\rho}-c^2}}\frac{\partial^2}{\partial \tau^2},
\end{align}
and substituting the Fourier representations of $G_{yy}$ and $G_{xx}$ into \eqref{eq:yy Green's eqn} and \eqref{eq:xx Green's eqn}, we find that $g_{yy}$ and $g_{xx}$ satisfy
\begin{align}\label{eq:zC572KjSOm}
    \left(\frac{d}{d\rho}\left(\sqrt{\cosh^2{\rho}-c^2}\frac{d}{d\rho}\right)-\frac{k^2}{\sqrt{\cosh^2{\rho}-c^2}}-\sqrt{\gamma}m^2(\rho)\right)g(\rho,\rho';k)=-\delta(\rho-\rho'),
\end{align}
where $m^2\equiv m_y^2$ or $m^2\equiv m_x^2$, as appropriate. We additionally impose the boundary condition $g(\rho,\rho';k)\to 0$ as $\rho\to \pm\infty$, which gives the standard asymptotic behavior $g\sim e^{\mp \Delta \rho}$.\footnote{Eq.~\eqref{eq:zC572KjSOm} reduces to $[e^{\mp\rho}\frac{d}{d\rho}\left(e^{\pm\rho} \frac{d}{d\rho}\right)-m^2\rvert_{\rho\to \pm \infty}]g=0$ in the regime $\rho\to \pm \infty$.} We also note the symmetries of $g$: $g(\rho,\rho';k)=g(\rho',\rho;k)=g(-\rho,-\rho';k)=g(\rho,\rho';-k)$. These properties follow from \eqref{eq:zC572KjSOm} and/or the discussion after \eqref{eq:phitheta Green's eqn}. It follows that $G$ only depends on $\tau$ and $\tau'$ through the combination $|\tau-\tau'|$.

The equation for $g(\rho, \rho';k)$ simplifies if we change variables from $\rho$ to the compactified coordinate
\begin{align}\label{eq:B5pnopgBFw}
    r\equiv \int_0^\rho \frac{\sqrt{1-c^2}d\rho'}{\sqrt{\cosh^2{\rho'}-c^2}}= -i\mathbb{F}\left(i\rho\Big\rvert\frac{1}{1-c^2}\right),
\end{align}
where $\mathbb{F}(x|m)$ is the incomplete elliptic integral of the first kind. Equivalently, $\cosh{\rho}= \text{cn}\left(ir|\frac{1}{1-c^2}\right)$, where $\text{cn}(x|m)$ is the Jacobi elliptic function.\footnote{Using \eqref{eq: Jacobi imag/real transformation n}, we can put the elliptic function in the form $\text{cn}\left(ir|\frac{1}{1-c^2}\right)=\text{dc}(r/\sqrt{1-c^2}|c^2)$, which makes it clearer that $\rho\mapsto r$ is a real map.} The asymptotic behavior of $r$ as $\rho\to \pm \infty$ will play an important role in what follows. It is given by
\begin{align}\label{eq:5kUeXpf7ju}
    r&\sim \pm\left(r_{m}+\bar{r}_me^{\mp \rho}+O(e^{\mp 2\rho})\right),&\text{where }r_{m}\equiv \sqrt{1-c^2}\mathbb{K}(c^2)\text{ and } \bar{r}_m\equiv -2\sqrt{1-c^2}.
\end{align}
Note that the change of variables in (\ref{eq:B5pnopgBFw}) is essentially the one that puts the induced metric (\ref{induced-metric}) in the conformal gauge form (up to rescaling by a constant factor).\footnote{See 
Appendix E of \cite{Gromov:2012eu} for the form of the solution in conformal gauge.}

After the change of variables, \eqref{eq:zC572KjSOm} becomes
\begin{align}
\left(\frac{d^2}{dr^2}-\frac{k^2}{1-c^2}-\frac{\sqrt{\cosh^2{\rho}-c^2}\sqrt{\gamma}m^2}{1-c^2}\biggr\rvert_{\cosh{\rho}\to \text{cn}\left(ir|\frac{1}{1-c^2}\right)}\right)g(r,r';k)&=-\frac{\delta(r-r')}{\sqrt{1-c^2}}.\label{eq:EfEiIZcZdN}
\end{align}
The solution to \eqref{eq:EfEiIZcZdN} may be written in the following piece-wise form:
\begin{align}\label{eq:XhqlLN6ozu}
    g(r,r';k)\equiv a(k) \left[g^R(r;k)g^L(r';k)\theta(r-r')+g^L(r;k)g^R(r';k)\theta(r'-r)\right].
\end{align}
Here, $g^{L}$ and $g^R$ solve \eqref{eq:EfEiIZcZdN} without the delta source term and satisfy the boundary conditions $g^R\to 0$ as $r\to r_{m}$ and $g^L\to 0$ as $r\to-r_m$. The normalization $a(k)$ is fixed by needing to reproduce the prefactor of the delta function in \eqref{eq:EfEiIZcZdN}:\footnote{Note that $a(k)$ is independent of $r$, since $\frac{d}{dr}a(k)^{-1}=0$ because of \eqref{eq:kZIdmPDcsW}-\eqref{eq:8txKFSC2oh}. Furthermore, there is a ``gauge'' freedom to rescale $g^R$ or $g^L$ by some $\lambda\neq 0$ because we impose a single homogeneous boundary condition on these solutions. Because $a(k)$ is rescaled by a compensating factor of $1/\lambda$, $g(r,r';k)$ is unaffected.}
\begin{align}\label{eq:DwWKnP1Xbu}
    a(k)&\equiv -\frac{1}{\sqrt{1-c^2}}\left(\frac{dg^R(r;k)}{dr}g^L(r;k)-\frac{dg^L(r;k)}{dr}g^R(r;k)\right)^{-1}.
\end{align}

The explicit forms of the homogeneous differential equations that $g^L$ and $g^R$ solve for the two different fluctuation modes are:
\begin{align}
    \left(-\frac{d^2}{dr^2}+\frac{2c^2}{1-c^2}\text{cn}\left(ir|\frac{1}{1-c^2}\right)^{-2}\right)g^{L/R}_{yy}(r;k)&=\frac{c^2-k^2}{1-c^2}g_{yy}^{L/R}(r;k),\label{eq:kZIdmPDcsW}\\
    \left(-\frac{d^2}{dr^2}+\frac{2}{1-c^2}\text{cn}\left(ir|\frac{1}{1-c^2}\right)^2\right)g^{L/R}_{xx}(r;k)&=\frac{c^2-k^2}{1-c^2}g_{xx}^{L/R}(r;k).\label{eq:8txKFSC2oh}
\end{align}
We can always choose $g^L$ and $g^R$ to satisfy
\begin{align}\label{eq:g^(R/L) properties}
    g^R(r;k)=g^R(r;-k)=g^L(-r;k).
\end{align}
The second equality can be imposed because the masses are even under $\rho\to -\rho$ or, equivalently, $\text{cn}(ir|\frac{1}{1-c^2})$ is even under $r\to -r$. Finally, we note that the boundary conditions impose the following asymptotic behavior:
\begin{align}\label{eq:gR/gL asymptotic behavior}
    g^R\sim (r-r_m)^\Delta\hspace{0.5cm}\text{as } r\to r_m,&& g^L\sim (r+r_m)^\Delta \hspace{0.5cm}\text{as } r\to -r_m.
\end{align}

\begin{figure}
    \centering
    \includegraphics[scale=0.35]{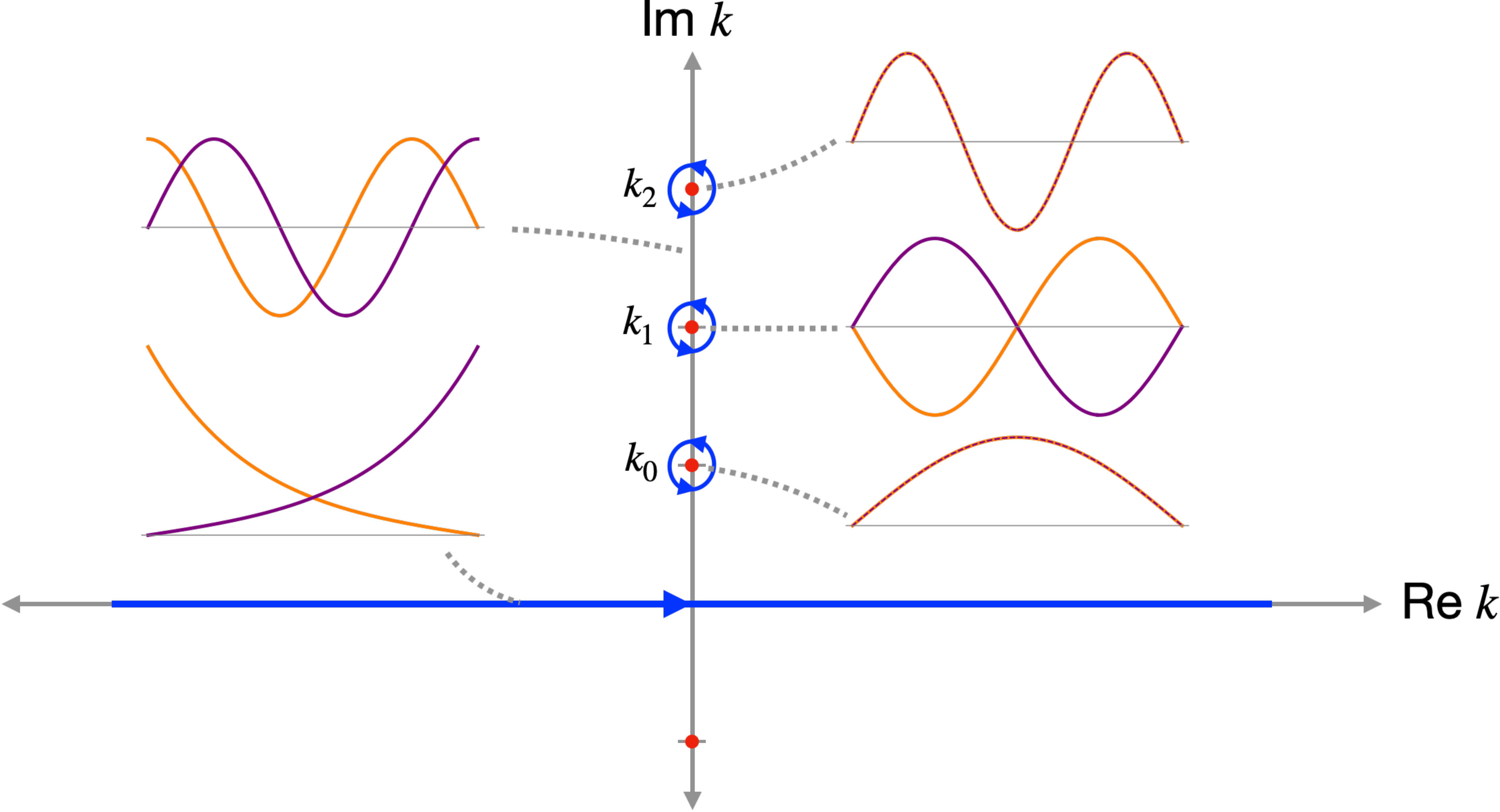}
    \caption{The contour in the Fourier integral representation of the bulk and boundary propagators runs along the real $k$ axis (shown in blue). We can close it at infinity in the upper half plane and pick up the residues at the poles, $k_n$, $n=0,1,2,\ldots$, along the positive imaginary $k$ axis. The insets sketch the solutions $g^R(r;k)$ (orange) and $g^L(r;k)$ (purple) at a few representative values of $k$, including at the poles ($k_0$, $k_1$ and $k_2$), on the imaginary axis away from the poles, and on the real axis.}
    \label{fig:integral to sum of residues}
\end{figure}

Eq.~\eqref{eq:KAKPSBQcaL} and \eqref{eq:XhqlLN6ozu} give an explicit integral representation for the $G_{xx}$ and $G_{yy}$ in terms of solutions to the ODEs in \eqref{eq:kZIdmPDcsW}-\eqref{eq:8txKFSC2oh}. The next step is to convert the integral in \eqref{eq:KAKPSBQcaL} into a sum of residues. First, we replace $\tau-\tau'$ by $|\tau-\tau'|$ in \eqref{eq:KAKPSBQcaL} (see the comment below \eqref{eq:zC572KjSOm}) and close the contour in the upper half-plane. There is no contribution from the arc at infinity because of the $e^{ik|\tau-\tau'|}$ term as long as $\tau\neq \tau'$. Next, we note that $g(r,r';k)$ has poles where $1/a(k)$, the Wronskian of $g^R(r;k)$ and $g^L(r;k)$, is zero. For these values of $k$, $g^R$ and $g^L$ are linearly dependent and define a single solution that vanishes at both $r=r_m$ and $r=-r_m$. Eqs.~\eqref{eq:kZIdmPDcsW} and \eqref{eq:8txKFSC2oh} take the form of the time-independent Schr\"odinger equation with real potentials $V_y(r)\equiv \frac{2c^2}{1-c^2}\text{cn}\left(ir\rvert \frac{1}{1-c^2}\right)^{-2}$ and $V_x(r)\equiv \frac{2}{1-c^2} \text{cn}\left(ir\rvert \frac{1}{1-c^2}\right)^2$ (and $\hbar^2/2/m=1$). Therefore, the solutions $g^L\propto g^R$ at the poles are naturally interpreted as bound states of a one-dimensional particle with energy $\mathcal{E}=\frac{c^2-k^2}{1-c^2}$ moving in a potential $V(r)$ ($\equiv V_y(r),V_x(r)$) with hard walls at $r=\pm r_m$ (because of the boundary conditions on $g^{R}$ and $g^L$). As we will verify explicitly in Sections~\ref{sec:integral and series scalar 4-pt} and \ref{sec:integral and series displacement 4-pt}, the poles of both $g_{yy}(\rho,\rho';k)$ and $g_{xx}(\rho,\rho';k)$ lie on the imaginary $k$ axis. Alternatively, this follows from the fact that the energy eigenvalues are necessarily real\footnote{The operator $-\frac{d^2}{dr^2}+V(r)$ is Hermitian with respect to the $L^2$ norm on the interval $[-r_m,r_m]$.} and obey the bound $\mathcal{E}>\frac{c^2}{1-c^2}$, which means the bound states only exist for $k^2<0$. The lower bound on $\mathcal{E}$ is obvious for $V_x(r)$, which attains the minimum value $\frac{2}{1-c^2}$ at $r=0$, and can be verified numerically for $V_{y}(r)$. The behavior of $g^R$ and $g^L$ at and away from the poles of $g(r,r;k)$ is illustrated in Figure~\ref{fig:integral to sum of residues}.

%The energy eigenvalues in these potentials have the lower bound $\mathcal{E}>\frac{c^2}{1-c^2}$,\footnote{The energies of the bound states are necessarily real because $-\frac{d^2}{dr^2}+V(r)$ is Hermitian with respect to the $L^2$ norm on the interval $[-r_m,r_m]$.}\textsuperscript{,}\footnote{This bound is easy to see for $V_x(r)$, which attains its minimum value of $\frac{2}{1-c^2}$ at $r=0$ (and $2>c^2$) and diverges to $+\infty$ at $r=\pm r_m$. By contrast, the bound is not analytically obvious for $V_y(r)$, which attains its maximum value $\frac{2c^2}{1-c^2}$ at $r=0$ and minimum value $0$ at $r=\pm r_m$, but it can be verified numerically.} which implies that the bound states only exist for $k^2<0$. We conclude that the poles of $g(\rho,\rho';k)$ lie on the imaginary axis. 

Therefore, labelling the poles $k_n\in i\mathbb{R}^+$, $n=0,1,2,\ldots$, we can write the bulk-to-bulk propagator as
\begin{align}\label{eq:2IkB8mXq6j}
    G(r,\tau;r',\tau')= \sum_{n=0}^\infty e^{ik_n|\tau-\tau'|} \left(\frac{1}{4\pi g}\oint_{k_n} dk\text{ }a(k)\right) g^{R}(r;k_n)g^{L}(r';k_n).
\end{align}
We used \eqref{eq:XhqlLN6ozu} and the fact that $g^R(r;k_n)g^L(r';k_n)=g^L(r;k_n)g^R(r';k_n)$, which follows because $g^L\propto g^R$ at $k=k_n$. We have also assumed that the $k_n$ are simple poles of $g(r,r';k)$, so we can evaluate $g^{R}$ and $g^L$ at $k_n$ and pull them out of the contour integral.

To convert the series for the bulk-to-bulk propagator into a series for the boundary-to-boundary propagator, we send $\rho\to \eta_1\infty$ and $\rho'\to \eta_2\infty$ (which sends $r\to\eta_1 r_m$ and $r'\to \eta_2r_m$) in accordance with \eqref{eq:yy/xx bndy-to-bndy}. We can do so term-by-term in the series because each $g^R(r;k_n)$ or $g^L(r;k_n)$ vanishes at both $r=\pm r_m$ and combines with the divergent factor $e^{\eta\Delta \rho}$ to yield a finite result in terms of $k_n$. There are two distinct cases to consider: $\eta_1=\eta_2$, when the two bulk points approach the same boundary, and $\eta_1=-\eta_2$, when the two bulk points approach opposite boundaries. These two cases are related in a simple way, as we can see by again exploiting the analogy with the Schr\"odinger equation in one dimension. Because the parities of the energy eigenstates in an even potential alternate (with the ground state being even), it follows that $g^R(r;k_n)=g^L(-r;k_n)=(-1)^{n}g^R(-r;k_n)$. Therefore,
\begin{align}\label{eq:g^R g^L different boundaries}
    \lim_{\rho\to \eta\infty}e^{\eta \Delta \rho}g^R(r(\rho);k_n)=\lim_{\rho\to -\eta\infty}e^{-\eta \Delta \rho}g^L(r(\rho);k_n)=\eta^{n}\lim_{\rho\to \infty}e^{\Delta \rho}g^R(r(\rho);k_n).
\end{align}
This is useful because it is easier in practice to evaluate the first limit on the LHS of \eqref{eq:g^R g^L different boundaries} when $\eta=1$. This is because the limit of $e^{\Delta \rho}g^R(r(\rho);k)$ as $\rho\to \infty$ is finite for any $k$ while the limit as $\rho\to -\infty$ is finite only if $k=k_n$.

Combining \eqref{eq:yy/xx bndy-to-bndy} and \eqref{eq:2IkB8mXq6j}, and applying \eqref{eq:tau1-tau2} and \eqref{eq:g^R g^L different boundaries}, we finally arrive at the following series representation for the boundary-to-boundary propagator:
\begin{align}\label{eq:dRNhIUWDdm}
    W(t_1,t_2)&=\frac{(2\Delta -1)^2}{2^{2\Delta -2}}\frac{g}{t_{12}^{2\Delta}}\frac{\chi^{2\Delta}}{|1-\chi|^{\Delta}}\sum_{n=0}^\infty \biggr[(-\eta_1\eta_2)^{n} e^{ik_n |\log|1-\chi||}\\&\hspace{6cm}\times\left(\frac{1}{4\pi} \oint_{k_n} dk a(k)\right)\left(\lim_{\rho\to \infty}e^{\Delta \rho}g^R(r(\rho);k_n)\right)^2\biggr].\nonumber
\end{align}
Note that $\eta_1\eta_2=\text{sgn}(1-\chi)$. 

It is also possible to send the bulk points to the boundary without writing the Fourier integral as a sum of residues. This is mainly useful if we can interchange the limit and the integral, which requires that the two bulk points be sent to opposite boundaries. For instance, if we first send $\rho\to \infty$ inside the integrand of \eqref{eq:KAKPSBQcaL}, then the step function in \eqref{eq:XhqlLN6ozu} sets $g(r,r';k)=a(k)g^R(r;k)g^L(r';k)$ and we must subsequently send $\rho'\to -\infty$ because $g^L(r';k)$ does not vanish at $\rho'\to \infty$ for real $k$. We find the following integral representation for the boundary-to-boundary propagator, which is valid when $\chi>1$:
\begin{align}\label{eq:bndy-to-bndy integral rep}
    W(t_1,t_2)&=\frac{(2\Delta-1)^2}{2^{2\Delta} \pi}\frac{g}{t_{12}^{2\Delta}}\frac{\chi^{2\Delta}}{(\chi-1)^\Delta}\int_{-\infty}^\infty dk e^{ik\log(\chi-1)}a(k) \left(\lim_{\rho\to \infty} e^{\Delta \rho} g^R(r(\rho);k)\right)^2.
\end{align}
In certain cases, this representation is more useful than \eqref{eq:dRNhIUWDdm}. 

\subsection{Computing \texorpdfstring{$W_{yy}$, $W_{\theta\theta}$ and $W_{\phi\theta}$}{Wyy, Wthth, and Wphith}}\label{sec:integral and series scalar 4-pt}

We now implement the analysis developed in the previous section to find the boundary-to-boundary propagator $W_{yy}$. Via \eqref{eq:yy bndy-to-bndy to Phi Phi defect fn} and \eqref{eq:PhPhZ^JZb^J}, this determines the leading large charge behavior of the defect correlator $\braket{\Phi\Phi Z^J\bar{Z}^J}/\braket{Z^J\bar{Z}^J}$ and of $G_1(\chi)$. Using the superconformal Ward identities, we will then also determine $G_{Z\bar{Z}}(\chi)$ and $G_{\bar{Z}Z}(\chi)$, which are equivalent to $W_{\theta\theta}$ and $W_{\phi\theta}$.

\subsubsection{Computing \texorpdfstring{$W_{yy}$}{Wyy}}\label{sec:computing G1}

The key to computing $W_{yy}$ analytically is to recognize that \eqref{eq:kZIdmPDcsW} can be put in the Jacobi form of the Lam\'e differential equation. This ODE appeared previously in studies of one-loop corrections to the energies of ``elliptic'' classical strings in AdS$_5\times S^5$ \cite{Beccaria:2010ry,Beccaria:2010zn,Forini:2010ek,Drukker:2011za}.\footnote{These are strings whose solutions can be written simply in terms of the Jacobi elliptic functions, and include the rotating folded string \cite{Beccaria:2010ry}, pulsating strings \cite{Beccaria:2010zn}, the string incident on anti-parallel lines on the boundary \cite{Forini:2010ek}, and a two-parameter family of strings incident on contours that interpolate between a circle and antiparallel lines \cite{Drukker:2011za}.} We summarize the equation in Appendix~\ref{app:special functions} along with conventions and identities for the elliptic integrals, Jacobi elliptic functions, and the theta functions. These special functions appear prominently throughout this section.

In order to put \eqref{eq:kZIdmPDcsW} into the Jacobi form of the Lam\'e equation, we rewrite it in terms of the new coordinate
\begin{align}\label{eq:0rquXvXaWd}
    \sigma&\equiv \frac{cr}{\sqrt{1-c^2}}+\ellK\left(\frac{1}{c^2}\right)+i\ellK\left(1-\frac{1}{c^2}\right).
\end{align}
Using the identities in \eqref{eq:IW5SfbL8zO}-\eqref{eq:8QBvEot7x6} and \eqref{eq: Jacobi imag/real transformation sn}-\eqref{eq: Jacobi imag/real transformation n}, we can simplify \eqref{eq:kZIdmPDcsW} to
\begin{align}\label{eq:auOOUFIO4A}
    \left[-\partial_\sigma^2+\frac{2}{c^2}\text{sn}^2\left(\sigma|\frac{1}{c^2}\right)\right]g_{yy}^{R/L}(\sigma;k)&=\left(1-\frac{k^2}{c^2}\right)g^{R/L}_{yy}(\sigma;k).
\end{align}
This matches the form of the Lam\'e equation given in \eqref{eq:ES9YxM59dJ}, if we identify the parameter $m$ and the eigenvalue $\Lambda$ to be:
\begin{align}\label{eq:42PG4Ezr6I}
    m&\equiv \frac{1}{c^2},& \Lambda &\equiv 1-\frac{k^2}{c^2}.
\end{align}

Because they are ubiquitous in the following discussion, it is convenient to introduce the following standard shorthand:
\begin{align}
    \ellK&\equiv \ellK\left(\frac{1}{c^2}\right)=c\ellK(c^2)-ic\ellK(1-c^2),\label{eq:def K}\\
    \ellK'&\equiv \ellK\left(1-\frac{1}{c^2}\right)=c\ellK(1-c^2).\label{eq:def K prime}
\end{align}
The second way of writing $\ellK$ and $\ellK'$ follows from \eqref{eq:E5NuDWFxub}-\eqref{eq:M8haq6dJl2}. Since $\ellK(c^2)$ and $\ellK(1-c^2)$ are positive real numbers for $c^2\in(0,1)$, it follows that $\ellK'$ and $\ellK+i\ellK'$ are also positive real numbers, but $\ellK$ is complex. It will be convenient to work with $\ellK$ and $\ellK'$ and convert to explicitly real expressions only at the end. 

We also note that as $r$ runs from $-r_m$ to $r_m$, $\sigma$ in \eqref{eq:0rquXvXaWd} runs from $\sigma_-$ to $\sigma_+$ along the real axis, where
\begin{align}
    \sigma_-&\equiv 0, &\sigma_+&\equiv 2\ellK+2i\ellK'.
\end{align}

As reviewed in Appendix~\ref{app:special functions}, the solutions to the Lam\'e equation are known in terms of theta functions. In particular, two linearly independent solutions to \eqref{eq:auOOUFIO4A} are
\begin{align}\label{eq:Gd7e84k9xY}
    f^\pm(\sigma;\alpha)\equiv \frac{H(\pm \sigma+\alpha|m)}{\Theta(\sigma|m)}e^{\mp\sigma Z(\alpha|m)}.
\end{align}
Here $H$, $\Theta$ and $Z$ are defined in \eqref{eq: H, Th, Z}, and $\alpha$ is related to $k$ by the transformation $\text{sn}\left(\alpha\bigr\rvert m\right)=\sqrt{1+k^2}$ or, equivalently,
\begin{align}\label{eq:k and alpha}
    k\equiv i \text{cn}\left(\alpha\rvert m\right). 
\end{align}

\begin{figure}[t]
\centering
\begin{minipage}{0.49\hsize}
\centering
\includegraphics[clip, height=8cm]{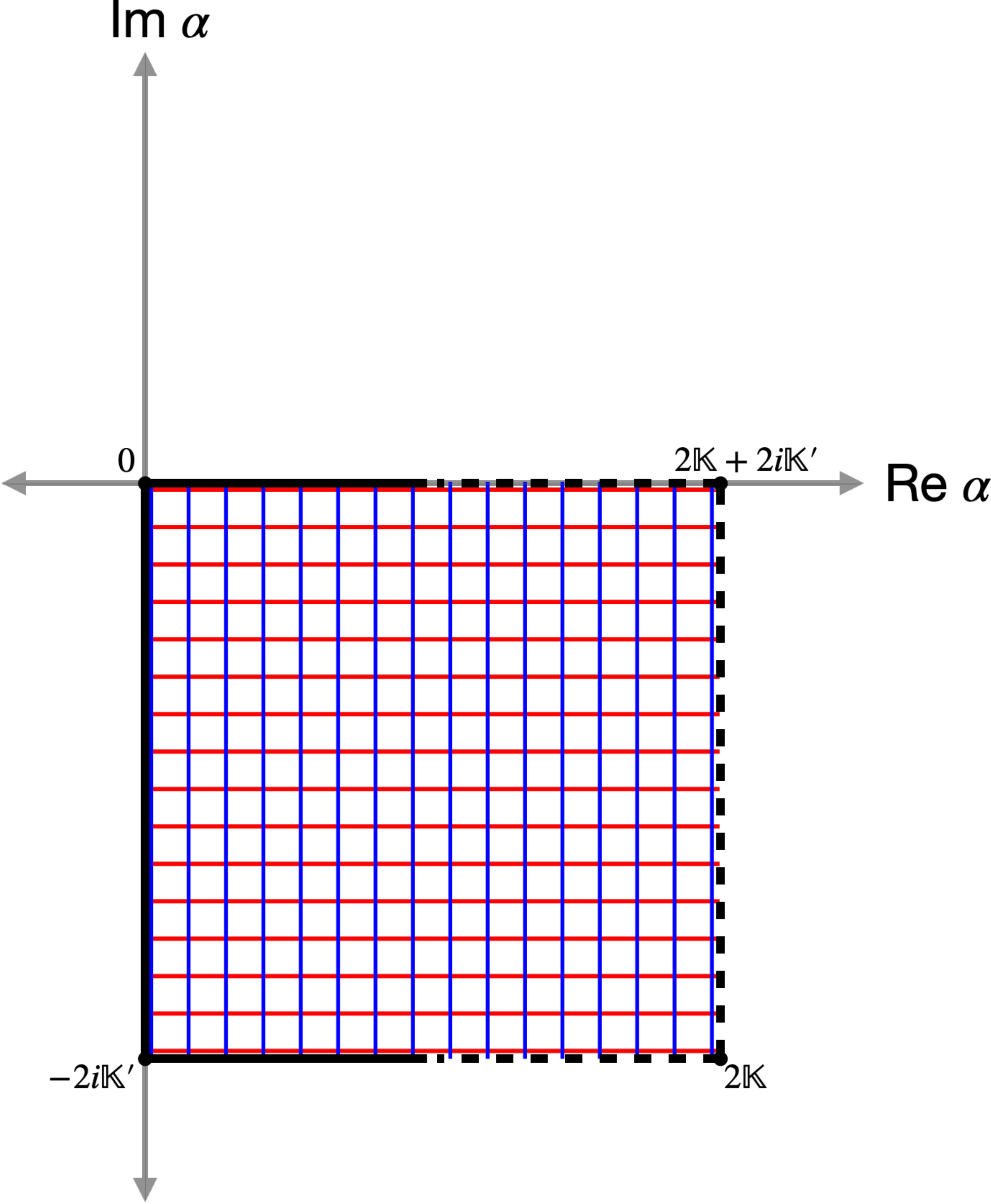}\\
\vspace{0.25cm}
{\bf a.} Fundamental unit cell in $\alpha$ plane
\end{minipage}
\begin{minipage}{0.49\hsize}
\centering
\vspace{1cm} 
\includegraphics[clip, height=6cm]{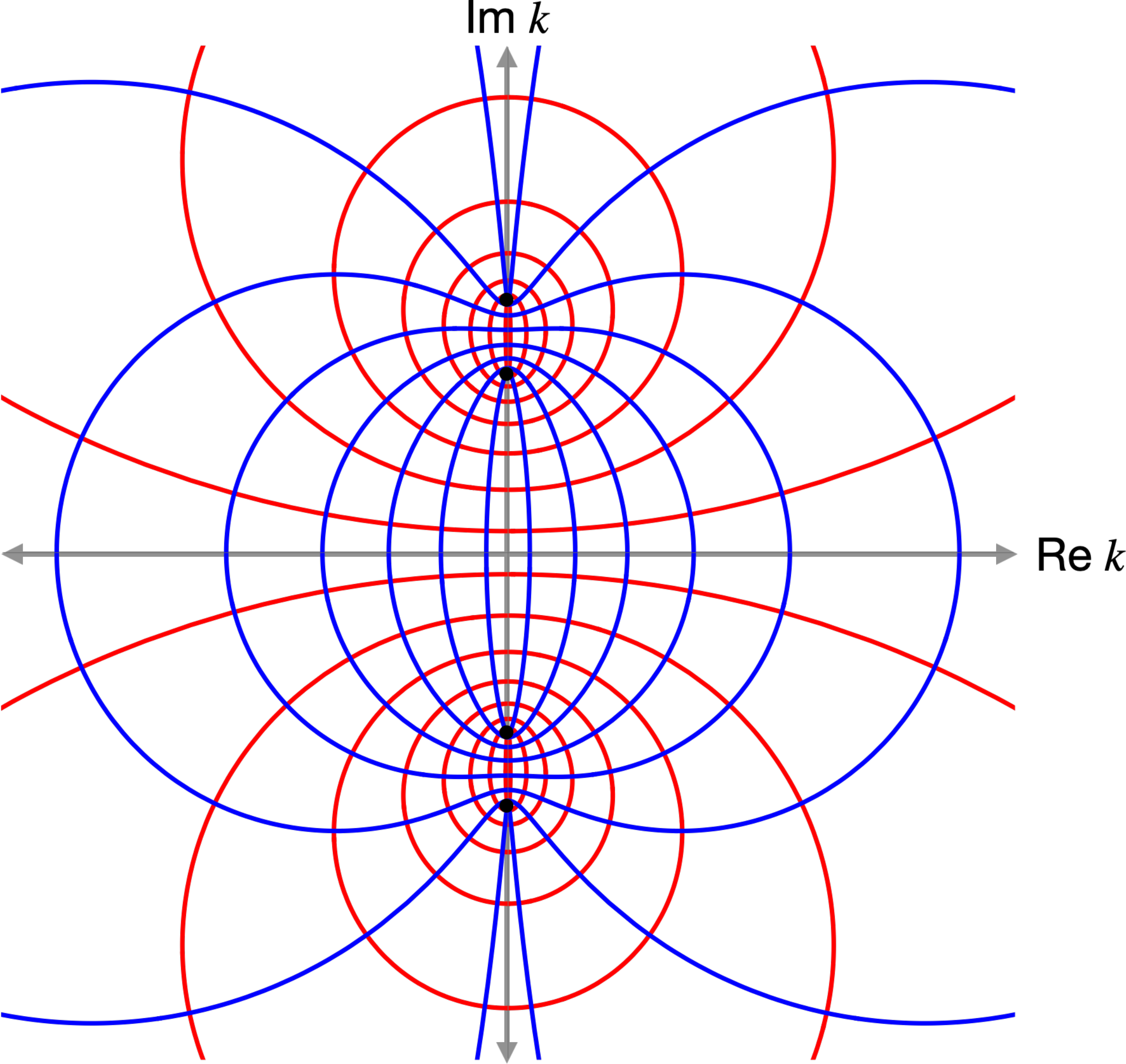}\\
\vspace{1.25cm} 

{\bf b.} Image of fundamental unit cell in $k$ plane
\end{minipage}
\caption{To express the propagator $G_{yy}$ explicitly as a Fourier integral involving theta functions, we need to change the integration variable from $k$, the wavenumber conjugate to global time $\tau$, to $\alpha$ via \eqref{eq:k and alpha}. The $\alpha$ plane is an infinite cover of the $k$ plane. The fundamental unit cell in the $\alpha$ plane is depicted in \textbf{a.} and its image in the $k$ plane is depicted in \textbf{b.} Recall from \eqref{eq:def K}-\eqref{eq:def K prime} that $\ellK'$ and $\ellK+i\ellK'$ are real.}
\label{fig:mapping alpha unit cell to k}
\end{figure}

\begin{figure}[t]
\centering
\begin{minipage}{0.49\hsize}
\centering
\includegraphics[clip, height=8cm]{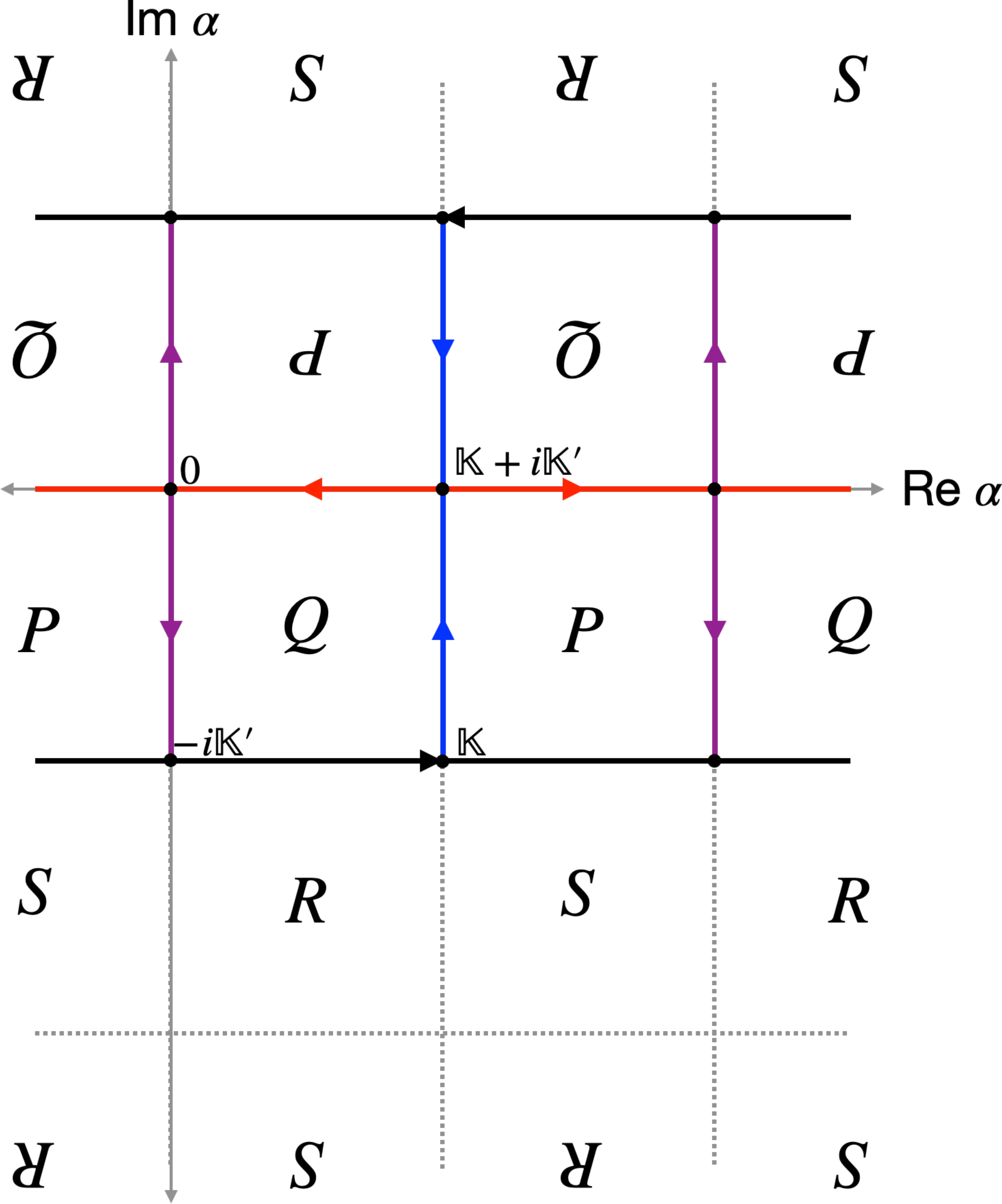}\\
\vspace{0.25cm}
{\bf a.} Periodicity of unit cells in $\alpha$ plane
\end{minipage}
\begin{minipage}{0.49\hsize}
\centering
\includegraphics[clip, height=8cm]{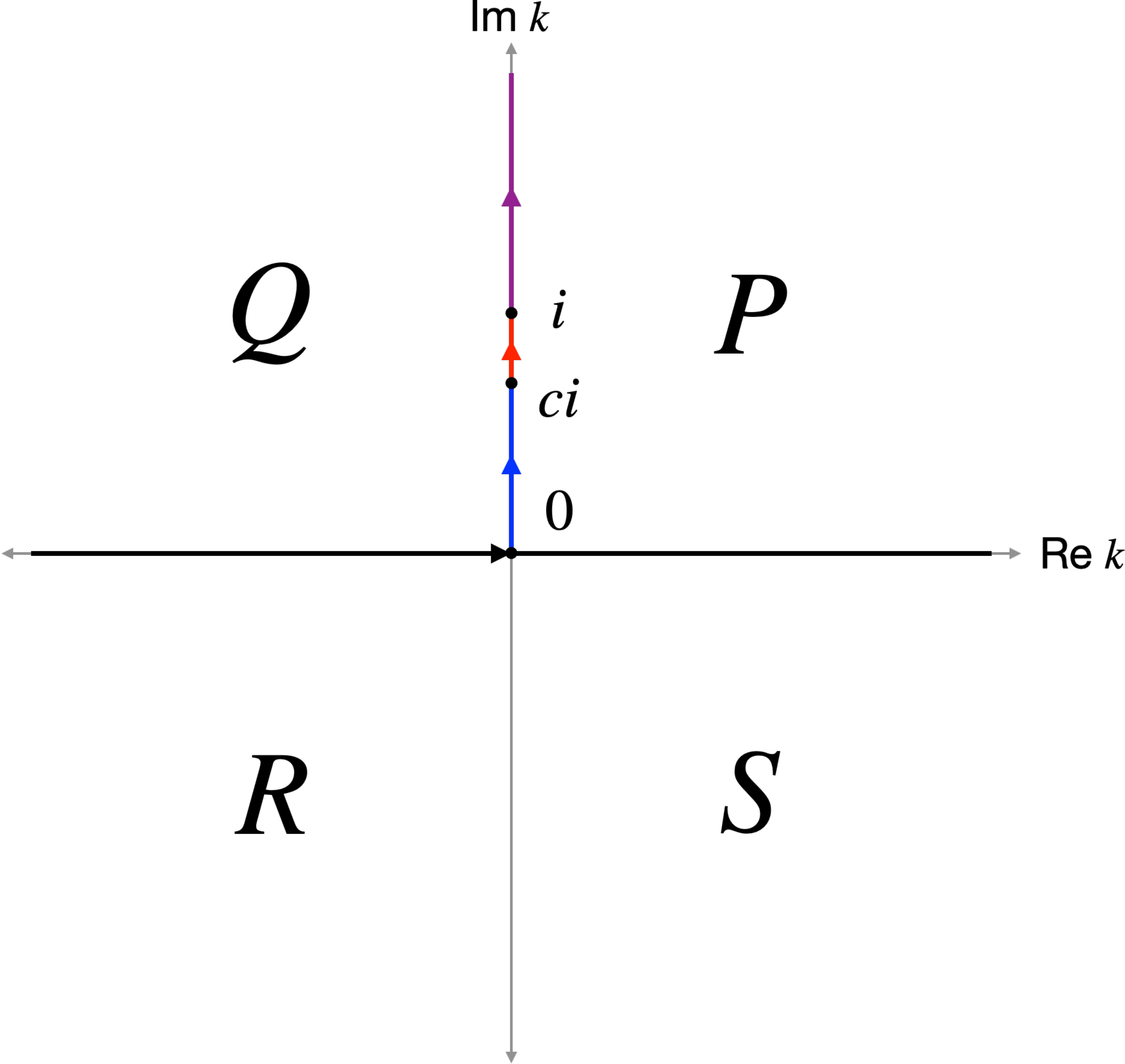}\\
\vspace{.25cm} 

{\bf b.} Image of $\alpha$ unit cells in $k$ plane
\end{minipage}
\caption{The Jacobi elliptic function relating $k$ and $\alpha$ in \eqref{eq:k and alpha} is doubly periodic, satisfying $\text{cn}\left(\alpha+2\ellK+2i\ellK'\rvert m\right)=\text{cn}\left(\alpha\rvert m\right)$, $\text{cn}\left(\alpha+2i\ellK'\rvert m\right)=-\text{cn}\left(\alpha\rvert m\right)$, as well as $\text{cn}\left(-\alpha\rvert m\right)=\text{cn}\left(\alpha\rvert m\right)$. Thus, the unit cells in the $\alpha$ plane are arranged as shown in \textbf{a.} Each grouping of four neighboring tiles labelled `Q', `P', `R' and `S', which are the pre-images of the four quadrants in the $k$ plane as shown in \textbf{b.}, is one unit cell. The intervals $(-\infty,0)$ and $(0,\infty)$ on the real axis and $(0,ci)$, $(ci,i)$ and $(i,+\infty i)$ in the $k$ plane are indicated in \textbf{b.} using black, blue, red and purple directed line segments, respectively. Their pre-images in the $\alpha$ plane are likewise indicated in \textbf{a.}} 
\label{fig:mapping alpha to k}
\end{figure}

Because the Jacobi elliptic functions are doubly periodic (see \eqref{eq:s5C1p1Pc5z}), the $\alpha$ complex plane is an infinite cover of the $k$ complex plane. We will see that this means the argument in Section~\ref{sec:integral and series reps of propagators} allowing us to convert the integral representation of the boundary-to-boundary propagator into a series representation can essentially be replicated in the $\alpha$ plane, but with some modification.\footnote{One could in principle work entirely in the $k$ plane instead, but it is more convenient to write the intermediate expressions for the bulk-to-bulk propagators using $\alpha$ and the Jacobi theta functions and then convert to expressions for the boundary-to-boundary propagators involving $k$ only at the end.} We will take the ``fundamental unit cell'' in the $\alpha$ plane to be the rectangle with vertices at $\alpha=0$, $\alpha=2\ellK+2i\ellK'$, $\alpha=-2i\ellK'$ and $\alpha=2\ellK$. A representative set of vertical and horizontal lines in the unit cell is depicted in Figure~\ref{fig:mapping alpha unit cell to k}~\hyperref[fig:mapping alpha unit cell to k]{a}, and its image in the $k$ plane is depicted in Figure~\ref{fig:mapping alpha unit cell to k}~\hyperref[fig:mapping alpha unit cell to k]{b}. The periodic placement of the other copies of the unit cell in the $\alpha$ plane is shown in Figure~\ref{fig:mapping alpha to k}. In particular, it will be useful to note the pre-images of the real and positive imaginary axes of the $k$ plane in the $\alpha$ unit cell: As $k$ runs from $-\infty$ to $\infty$ along the real axis (see Figure~\ref{fig:mapping alpha to k}~\hyperref[fig:mapping alpha to k]{b}), $\alpha$ runs along the line segment from $-i\ellK'$ to $i\ellK'+2\ellK$ (see Figure~\ref{fig:mapping alpha to k}~\hyperref[fig:mapping alpha to k]{a}), and as $k$ runs from $0$ to $ci$ to $i$ to $i\infty$ along the imaginary axis (see Figure~\ref{fig:mapping alpha to k}~\hyperref[fig:mapping alpha to k]{b}), $\alpha$ runs along the line segments from $\ellK$ to $\ellK+i\ellK'$ to $0$ to $-i\ellK'$ (see Figure~\ref{fig:mapping alpha to k}~\hyperref[fig:mapping alpha to k]{a}).

Using \eqref{eq:Gd7e84k9xY}, $g_{yy}^{R/L}$ can be written
\begin{align}
    g_{yy}^R(\sigma;\alpha)&=\frac{f^+(\sigma;\alpha)}{f^+(\sigma_+;\alpha)}-\frac{f^-(\sigma;\alpha)}{f^-(\sigma_+;\alpha)},\label{eq:hIhWEB0Kcn}\\
    g_{yy}^L(\sigma;\alpha)&=\frac{f^-(\sigma;\alpha)}{f^-(\sigma_-;\alpha)}-\frac{f^+(\sigma;\alpha)}{f^+(\sigma_-;\alpha)}.\label{eq:27Vo18ONzZ}
\end{align}
These manifestly satisfy the boundary conditions $g_{yy}^R(\sigma_+;\alpha)=g_{yy}^L(\sigma_-;\alpha)=0$. The parity and quasi-periodicity of the theta functions also imply that $g^{R/L}_{yy}(\sigma;\alpha)=g^{L/R}_{yy}(\sigma_+-\sigma;\alpha)$ and $g^{L/R}_{yy}(\sigma; \alpha+2\ellK a+2i\ellK'b)=g^{L/R}_{yy}(\sigma;\alpha)$, for $a,b\in \mathbb{Z}$, which are analogous to \eqref{eq:g^(R/L) properties}. 

Next, we compute the boundary limits of $g_{yy}^R(\sigma;\alpha)$ and $g_{yy}^L(\sigma;\alpha)$. First, we note the behavior of $\sigma$ near the end points, which follows from \eqref{eq:5kUeXpf7ju} and \eqref{eq:0rquXvXaWd}:
\begin{align}
    \sigma&\sim \sigma_{\pm}\mp 2ce^{\mp\rho},&\text{ as }\rho&\to \pm\infty.
\end{align} Second, \eqref{eq:gR/gL asymptotic behavior} implies that $g^R_{yy}\sim \sigma_+-\sigma$ as $\sigma\to \sigma_+$ and $g^L_{yy}\sim \sigma-\sigma_-$ as $\sigma\to \sigma_-$ (up to multiplicative factors independent of $\sigma$). Thus, we find:
\begin{align}\label{eq:xlJwDGHpIG}
    \lim_{\rho\to\infty}e^\rho g_{yy}^R(\sigma;\alpha)&=\lim_{\rho\to-\infty}e^{-\rho}g_{yy}^L(\sigma;\alpha)=-2c\frac{dg^R_{yy}}{d\sigma}\biggr\rvert_{\sigma=\sigma_+}=-4c (V(\alpha|m)-Z(\alpha|m)).
\end{align}
Here, $V(u|m)$ is defined in \eqref{eq:V}. Furthermore, the normalization becomes
\begin{align}
    \frac{1}{a_{yy}(\alpha)}&=-c\frac{dg_{yy}^R(\sigma_+;\alpha)}{d\sigma}g_{yy}^L(\sigma_+;\alpha)=4c(V(\alpha|m)-Z(\alpha|m))\sinh\left(\frac{\pi i\alpha}{\ellK}+\sigma_+Z(\alpha|m)\right).
\end{align}
To simplify this result, we used the fact that $a_{yy}(\alpha)$ is independent of $\sigma$ to evaluate the Wronskian at $\sigma=\sigma_+$, a point at which $g_{yy}^R=0$ and $\frac{dg_{yy}^R}{d\sigma}$ is given by \eqref{eq:xlJwDGHpIG}.

Using $\frac{dk}{d\alpha}= -i\text{sn}\left(\alpha|m\right)\text{dn}\left(\alpha|m\right)$ (note \eqref{eq:k and alpha} and \eqref{eq:dcn/dz}) and \eqref{eq:AO7XFN9xh0} to express $V(\alpha|m)-Z(\alpha|m)$ in terms of the Jacobi elliptic functions, we can write the boundary limit of the bulk-to-bulk propagator as the following integral:
\begin{align}\label{eq:mkhjfKkrkv}
    \lim_{\substack{\rho\to\infty\\\rho'\to-\infty}}e^{\rho-\rho'}G_{yy}(\rho,\tau;\rho',\tau')= -\frac{ci}{\pi g}\int_{-i\ellK'}^{i\ellK'+2\ellK}d\alpha \frac{\text{cn}\left(\alpha|m\right)\text{dn}^2\left(\alpha|m\right)\text{exp}\left(-\text{cn}\left(\alpha|m\right)|\tau-\tau'|\right)}{\sinh\left(F(\alpha)\right)},
\end{align}
where
\begin{align}\label{eq:F(alpha)}
    F(\alpha)\equiv\frac{\pi i \alpha}{\ellK}+2(\ellK+i\ellK')Z(\alpha|m).
\end{align}
This is equivalent to \eqref{eq:bndy-to-bndy integral rep} except we changed the integration variable from $k$ to $\alpha$.

\begin{figure}[!t]
\centering
\begin{minipage}{0.33\textwidth}
\centering
\includegraphics[clip, width=5cm]{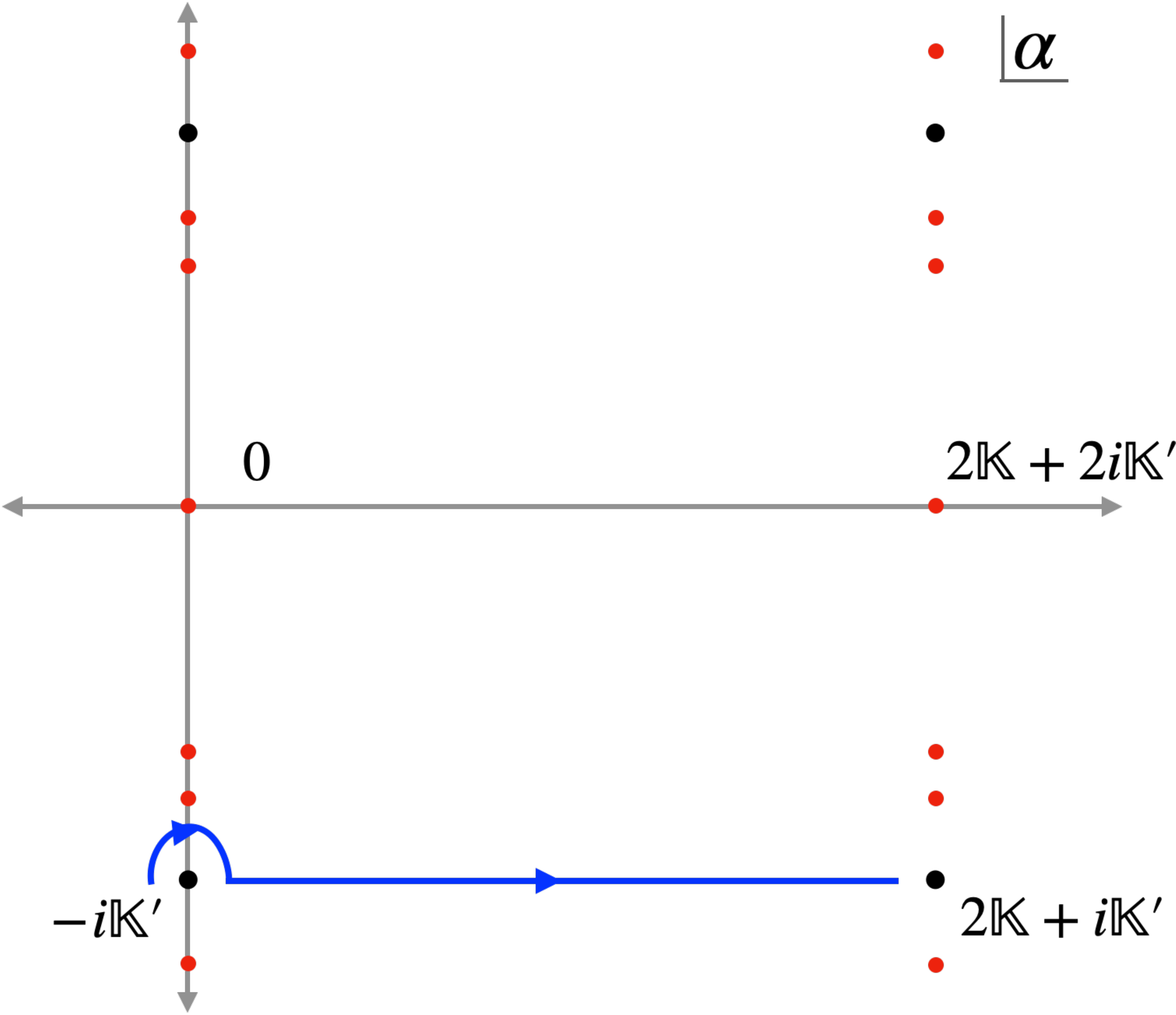}\\
{\bf a.}
\vspace{0.5cm}
\end{minipage}\hfill
\begin{minipage}{0.33\textwidth}
\centering
\includegraphics[clip, width=5cm]{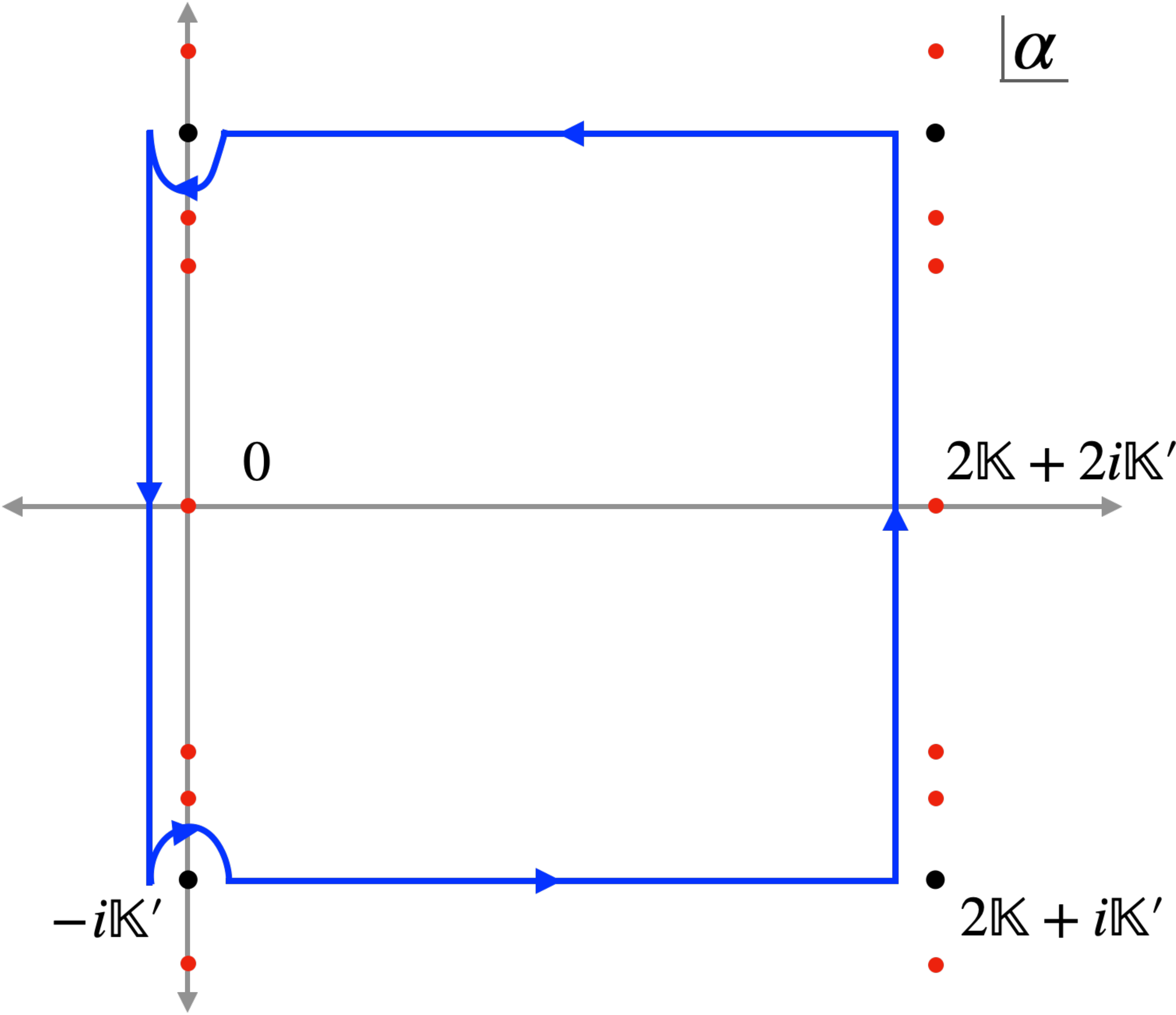}\\
{\bf b.}
\vspace{0.5cm}
\end{minipage}\hfill
\begin{minipage}{0.33\textwidth}
\centering
\includegraphics[clip, width=5cm]{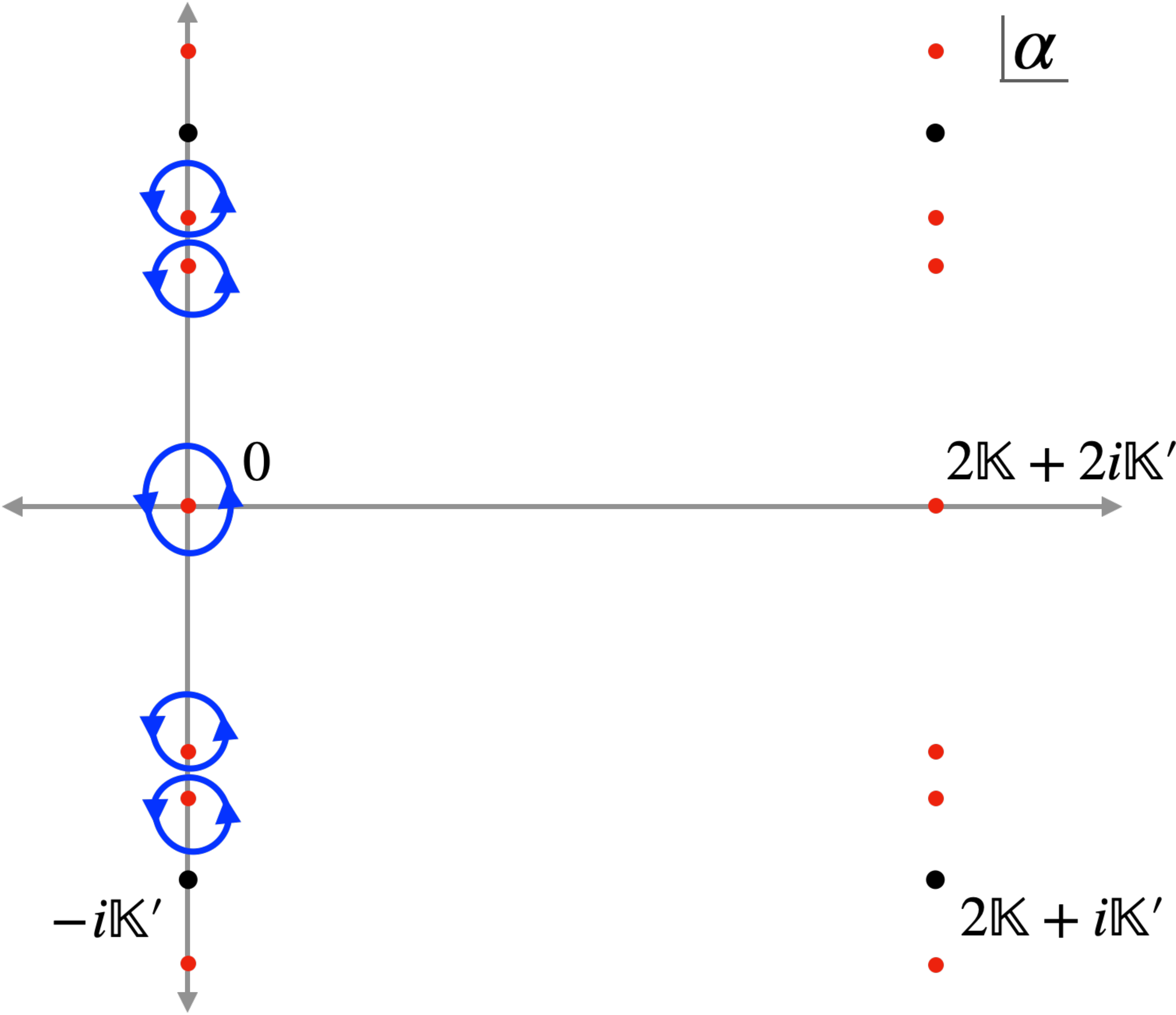}\\
{\bf c.}
\vspace{0.5cm}
\end{minipage} \\
\begin{minipage}{0.33\textwidth}
\centering
\includegraphics[clip, width=5cm]{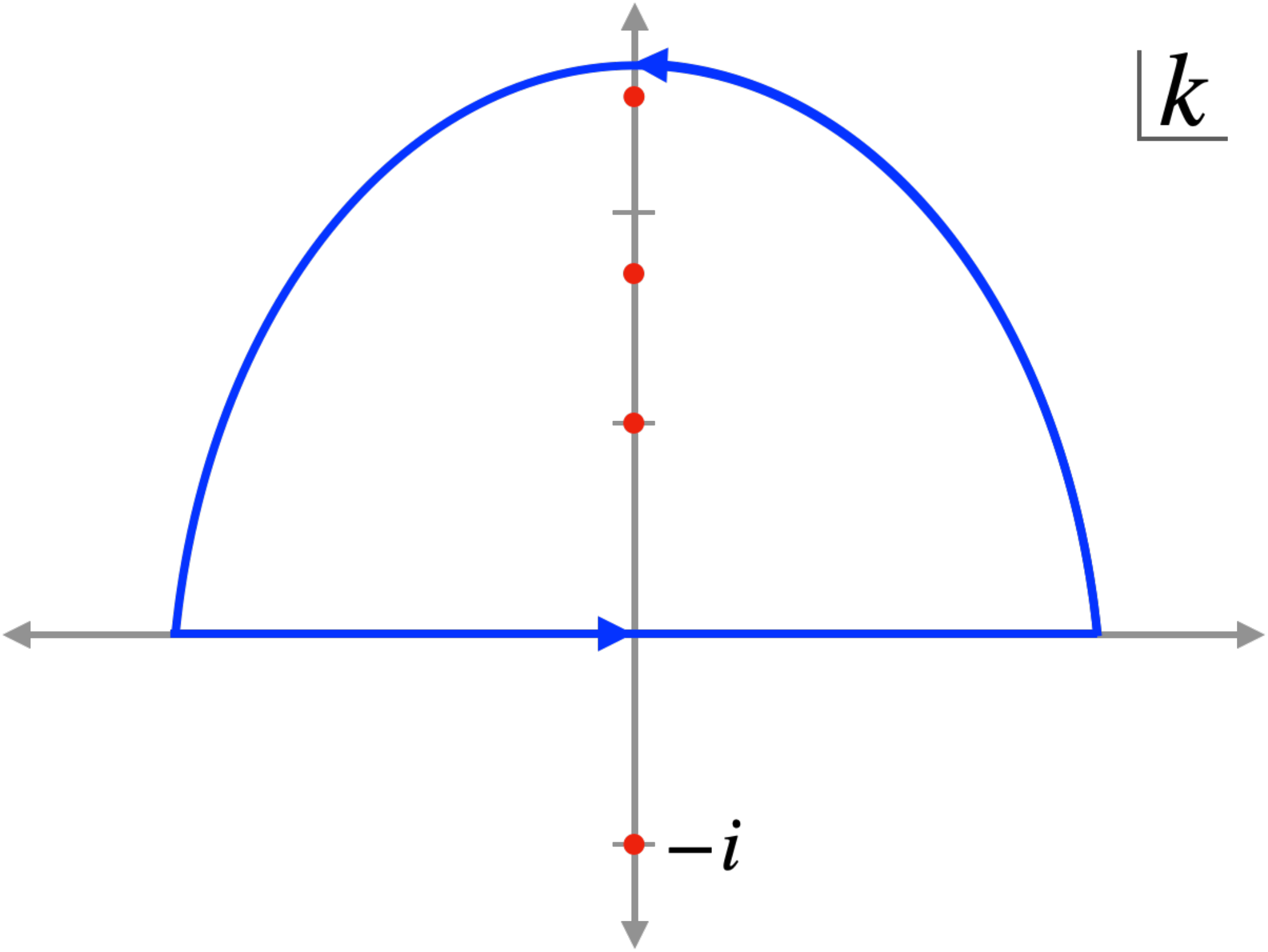}\\
{\bf d.}
\vspace{0.5cm}
\end{minipage}\hfill
\begin{minipage}{0.33\textwidth}
\centering
\includegraphics[clip, width=5cm]{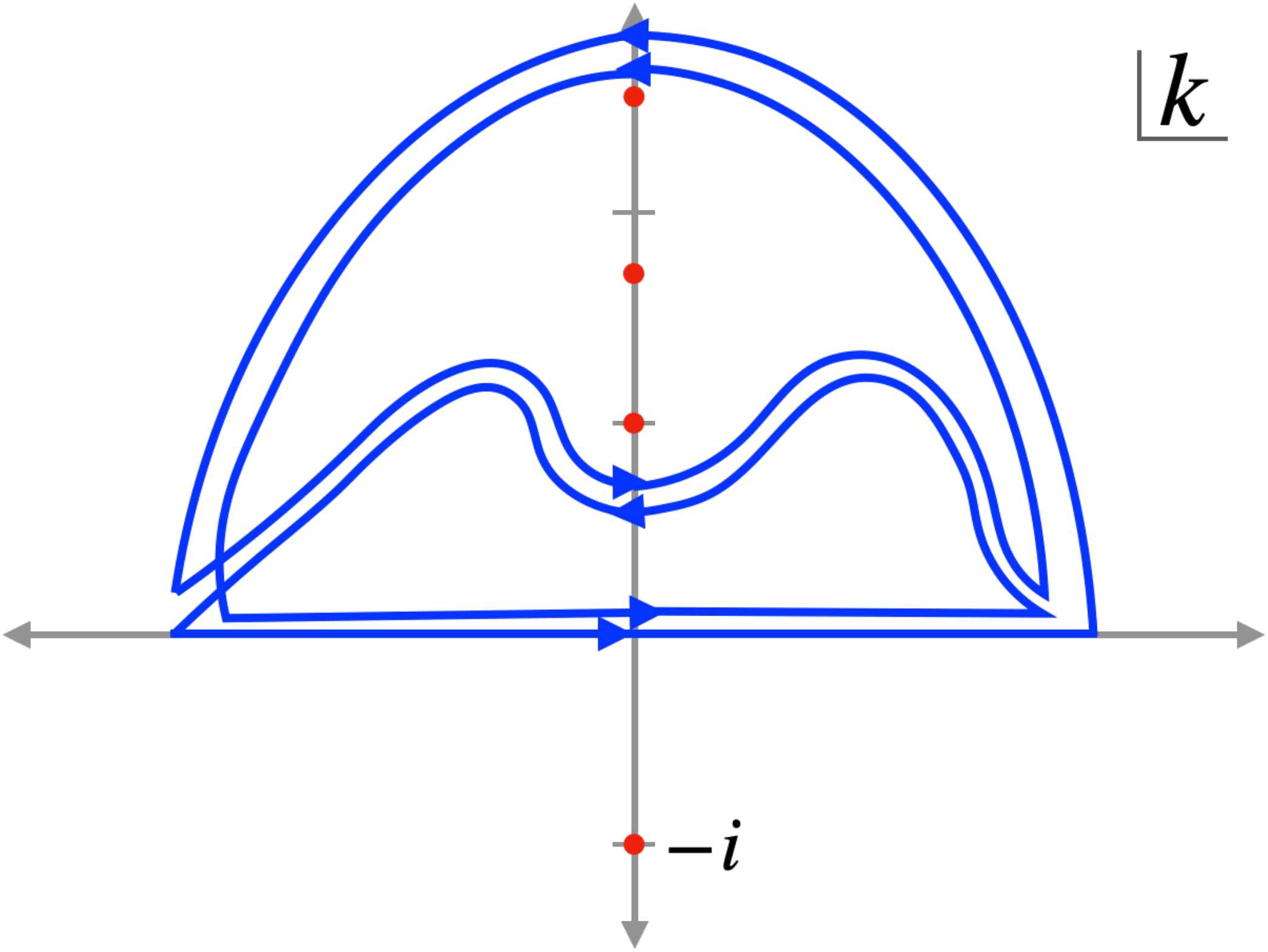}\\
{\bf e.}
\vspace{0.5cm}
\end{minipage}\hfill
\begin{minipage}{0.33\textwidth}
\centering
\includegraphics[clip, width=5cm]{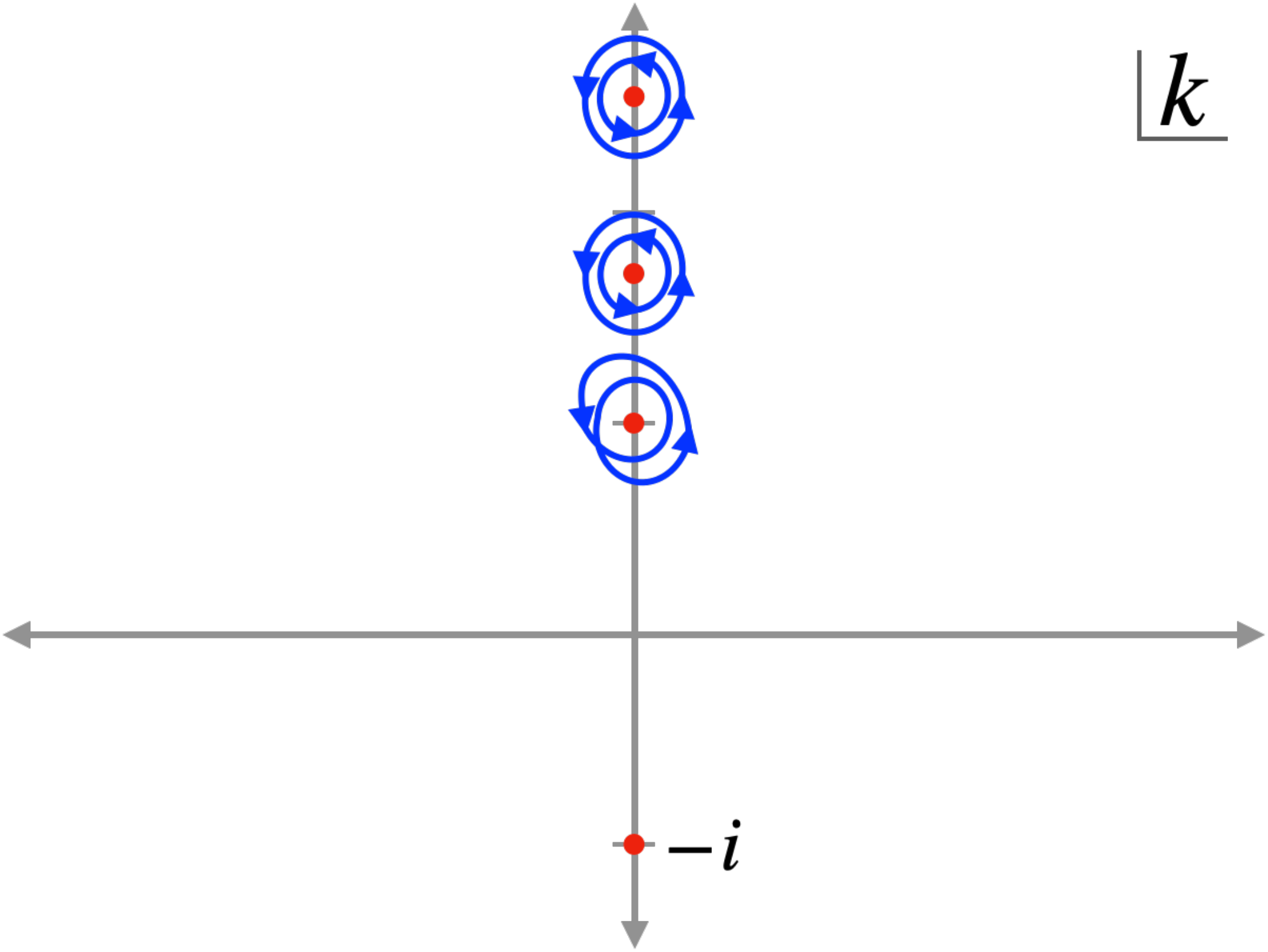}\\
{\bf f.}
\vspace{0.5cm}
\end{minipage}
\caption{Closing the integration contour in \eqref{eq:mkhjfKkrkv} and picking up the residues in both the $\alpha$ plane (\textbf{a.} - \textbf{c.}) and the $k$ plane (\textbf{d.} - \textbf{f.}). In \textbf{d.}, the integration contour along the real axis in the $k$ plane is closed with an arc at infinity. The lift of this closed contour to the $\alpha$ plane via the map in \eqref{eq:k and alpha} is not closed (\textbf{a.}). Using the periodic identification of the $\alpha$ plane (see Figure~\ref{fig:mapping alpha to k}), the contour can be closed as in \textbf{b.} at the cost of doubling the integral. The doubling is clear from the image of the new contour in the $k$ plane (\textbf{e.}). Finally, in the $\alpha$ plane we can pick up the residues of the poles on the interval $[-i\ellK',i\ellK']$ along the imaginary axis (\textbf{c.}). This corresponds to picking up two copies of the residues along the positive imaginary axis in the $k$ plane (\textbf{f.}).} 
\label{fig:contour deformation in alpha and k planes}
\end{figure}

The next step is to close the integration contour and pick up the residues at the poles. The integration contour in the $k$ plane can be closed using an arc at infinity, as in Figure~\ref{fig:contour deformation in alpha and k planes}~\hyperref[fig:contour deformation in alpha and k planes]{d}, but the lifted contour in the $\alpha$ plane, shown in Figure~\ref{fig:contour deformation in alpha and k planes}~\hyperref[fig:contour deformation in alpha and k planes]{a}, is not closed. However, the periodicity of the map from $\alpha$ to $k$, illustrated in Figure~\ref{fig:mapping alpha to k}, allows us to close the lifted contour as in Figure~\ref{fig:contour deformation in alpha and k planes}~\hyperref[fig:contour deformation in alpha and k planes]{b} at the cost of doubling the value of the integral. In particular, the contribution from the top horizontal segment duplicates the contribution from the bottom horizontal segment, while the contributions from the right and left vertical segments cancel. One can also see that the integral over the closed contour in the $\alpha$ plane is twice the original integral by noting the image of the the closed contour in the $k$ plane, which is given in Figure~\ref{fig:contour deformation in alpha and k planes}~\hyperref[fig:contour deformation in alpha and k planes]{e}. 

\begin{comment}
\footnote{This can be seen concretely by noting the parity and periodicity of the functions in the integrand in \eqref{eq:mkhjfKkrkv}.  The parity and periodicity of $\text{cn}(\alpha\rvert\frac{1}{c^2})$ and $\text{dn}(\alpha|\frac{1}{c^2})$ are summarized in \eqref{eq:AyWOmIh3i6}-\eqref{eq:AzboAe67sX}. Similarly, $\sinh\left(\frac{\pi i \alpha}{\ellK}+2(\ellK+i\ellK')Z(\alpha)\right)$ is invariant under $\alpha\to \alpha+2\ellK$ and $\alpha+2i\ellK'$ and odd under $\alpha \to -\alpha$.}
\end{comment}

To complete the argument, we need to identify the poles of the integrand of \eqref{eq:mkhjfKkrkv} that lie inside the closed contour. 
The poles of $\text{cn}(\alpha|m)$ and $\text{dn}(\alpha|m)$, which are at $\alpha=i\ellK'+2a\ellK+2bi\ellK'$, $a,b\in\mathbb{Z}$, all lie outside the contour. Thus, the only poles that contribute are the zeros of $\sinh\left(F(\alpha)\right)$ located inside the contour, which lie on the imaginary axis between $-i\ellK'$ and $i\ellK'$. We denote them $\alpha_n$, $n\in \mathbb{Z}$, and they satisfy\footnote{More generally, $\sinh\left(F(\alpha)\right)$ has zeros at $\alpha=\alpha_n+2a\ellK+2bi\ellK'$, $a,b\in \mathbb{Z}$.}
\begin{align}\label{eq:7u5v07hm5W}
    F(\alpha_n)= n\pi i.
\end{align}
Because $Z(z|m)$ is odd, $\alpha_0=0$ and $\alpha_{-n}= -\alpha_n$.

The result of deforming the contour to individually encircle the poles at each $\alpha_n$ is depicted in Figure~\ref{fig:contour deformation in alpha and k planes}~\hyperref[fig:contour deformation in alpha and k planes]{c}, and its image in the $k$ plane is depicted in Figure~\ref{fig:contour deformation in alpha and k planes}~\hyperref[fig:contour deformation in alpha and k planes]{f}. Note that each pole in Figure~\ref{fig:contour deformation in alpha and k planes}~\hyperref[fig:contour deformation in alpha and k planes]{c} is encircled once while each pole in Figure~\ref{fig:contour deformation in alpha and k planes}~\hyperref[fig:contour deformation in alpha and k planes]{f} is encircled twice. This is because $\alpha_n$ and $\alpha_{-n}$ for $n>0$ are mapped to the same point in the $k$ plane, so the contours individually encircling the two poles in the $\alpha$ plane are mapped to two contours encircling the same pole in the $k$ plane. Meanwhile, the pole at $\alpha_0=0$ is special: because the map in \eqref{eq:k and alpha} is even and therefore identifies points positioned antipodally with respect to the origin in the $\alpha$ plane, the image of the closed loop encircling $\alpha=0$ with winding number $1$ is a closed loop encircling $k=i$ with winding number $2$. 

In order to evaluate the residues of \eqref{eq:mkhjfKkrkv} at $\alpha_n$, it is convenient to note the following elementary result: If $f(z)$ and $g(z)$ are analytic at $z_n$, $f(z_n)=n\pi i$, and $f'(z_n)\neq 0$, then
\begin{align}\label{eq:RKtXoUrJeH}
    \frac{1}{2\pi i}\oint_{z_n} dz\text{ }\frac{g(z)}{\sinh\left(f(z)\right)}&=(-1)^n\frac{g(z_n)}{f'(z_n)}.
\end{align}
Applying this result, we find that \eqref{eq:mkhjfKkrkv} can be written as
\begin{align}\label{eq:12DBd7NOYO}
    \lim_{\substack{\rho\to\infty\\\rho'\to-\infty}}e^{\rho-\rho'}G(\rho,\tau;\rho',\tau')&=\frac{c}{g} \sum_{n=-\infty}^\infty(-1)^n \frac{\text{cn}\left(\alpha_n|m\right)\text{dn}^2\left(\alpha_n|m\right)\text{exp}\left(-\text{cn}\left(\alpha_n|m\right)|\tau-\tau'|\right)}{F'(\alpha_n)}.
\end{align}
Conveniently, the identity \eqref{eq:LIA7TEVOWb} lets us write the derivative of $Z(\alpha|m)$, and therefore $F'(\alpha)$, in terms of the Jacobi elliptic functions. Additionally, we can use \eqref{eq:E5NuDWFxub}-\eqref{eq:aPTCYjwzVg} and \eqref{eq:Legendre identity} to express $\ellK$, $\ellK'$, $\ellE(\frac{1}{c^2})$ in terms of $\ellK(c^2)$ and $\ellE(c^2)$, and we ultimately find
\begin{align}\label{eq:wcmt1PTwK8}
    F'(\alpha)= \frac{2}{c}\left((1-c^2)\ellK(c^2)-\ellE(c^2)\right)+2c\ellK(c^2)\text{dn}^2\left(\alpha|m\right).
\end{align}

Eq.~\eqref{eq:mkhjfKkrkv} is the final integral representation and \eqref{eq:wcmt1PTwK8}, supplemented with \eqref{eq:7u5v07hm5W}, is the final series representation for the boundary limit of $G_{yy}$ expressed in terms of the $\alpha$ variables. Both expressions are perhaps more transparent when written in terms of the original $k$ variables, which gets rid of the elliptic functions and recasts the quantization condition in \eqref{eq:7u5v07hm5W} in a more friendly form. 

We first rewrite \eqref{eq:mkhjfKkrkv}. $F$ can be written as a function of $k$ using $F(k)=\pi i+\int_0^k dk\frac{d\alpha}{dk}F'(\alpha)$, and noting \eqref{eq:wcmt1PTwK8}. Furthermore, using \eqref{eq:k and alpha} and the identities in \eqref{eq:oX5Dm1hVuw}-\eqref{eq:wtEDZNracI}, all of the elliptic functions can be expressed in terms of $k$. Ultimately, we find
\begin{align}\label{eq:s4xpFVW5pB}
    W_{yy}(t_1,t_2)&=\frac{g}{\pi}\frac{1}{ t_{12}^2}\frac{\chi^2}{\chi-1}\int_{-\infty}^\infty dk \frac{k\sqrt{1-c^2+k^2}\cos(k\log(\chi-1))}{\sqrt{1+k^2}\sinh\left(2\int_0^k d\ell \frac{\ellK(c^2)\ell^2+\ellE(c^2)}{\sqrt{1+\ell^2}\sqrt{1-c^2+\ell^2}}\right)}.
\end{align}
We used \eqref{eq:yy/xx bndy-to-bndy} to relate $W_{yy}$ to the boundary limit of $G_{yy}$. Eq.~\eqref{eq:s4xpFVW5pB} is valid when $\chi>1$.

Next, we rewrite \eqref{eq:12DBd7NOYO}. We introduce the ``fluctuation energies'':
\begin{align}\label{eq:En alpha-n}
    E_n\equiv -ik_n\equiv \text{cn}\left(\alpha_n\rvert m\right),
\end{align}
which are defined implicitly via \eqref{eq:7u5v07hm5W}. Note that $E_0=1$ and $E_{-n}=E_n$. It is perhaps more illuminating to define the energies in terms of an integral quantization condition, which follows from writing \eqref{eq:7u5v07hm5W} as $n\pi i=\int_0^{\alpha_n} d\alpha F'(\alpha)=\int_1^{E_n} dE' \frac{d\alpha}{dE'}F'(\alpha)$ and using \eqref{eq:wcmt1PTwK8} and the Jacobi elliptic function identities in \eqref{eq:oX5Dm1hVuw}-\eqref{eq:wtEDZNracI} to simplify the result. This leads precisely to \eqref{eq:quantization condition}.

Again using \eqref{eq:oX5Dm1hVuw}-\eqref{eq:wtEDZNracI} to express the elliptic functions in \eqref{eq:12DBd7NOYO} in terms of $E_n$, we find that the boundary-to-boundary propagator finally reduces to:
\begin{align}\label{eq:nfZXit03Ij}
    W_{yy}(t_1,t_2)&= \frac{2g}{\pi}\frac{1}{t_{12}^2}\frac{\chi^2}{|1-\chi|}\sum_{n\in \mathbb{Z}}\text{sgn}(1-\chi)^n f(E_n)e^{-E_n|\log|1-\chi||},
\end{align}
where $f(E)$ is given in \eqref{eq:form factor}. In \eqref{eq:nfZXit03Ij}, we have replaced $(-1)^n$ in \eqref{eq:12DBd7NOYO} by $\text{sgn}(1-\chi)^n$ in accordance with the discussion around \eqref{eq:g^R g^L different boundaries}-\eqref{eq:dRNhIUWDdm}. Thus, the series representation in \eqref{eq:nfZXit03Ij} is valid for both $\chi>1$ and $\chi<1$. Finally, $G_1(\chi)=\frac{\pi}{2g}t_{12}^2W_{yy}(t_1,t_2)$ leads to \eqref{eq:G1 series}. 

\paragraph{Remarks about $G_1(\chi)$.} Let us make a few remarks about the final result for $G_1(\chi)$. Firstly, because $E_n=E_{-n}$, the positive and negative terms in \eqref{eq:nfZXit03Ij} can naturally be combined. In general, the $n=0$ term must be treated separately from the $n\geq 1$ terms. It is special both because $E_0=1$ is independent of $c^2$ (this corresponds to the fact that the lowest operator in the conformal block expansion of the scalar four-point functions is protected), and because $k_0$ is the only pole in the $k$ upper half plane that is mapped $1$-to-$1$ with its pre-image $\alpha_0$ in the $\alpha$ plane.

Secondly, as a test of our results, we can consider the behavior of the four-point functions when $c^2=J=0$, in which case there are no large charges on the Wilson line and the four-point functions reduce to two-point functions. (Likewise the classical string does not rotate in $S^5$ and the propagators for the $S^5$ and AdS$_5$ modes reduce to those of scalars with $m^2=0$ and $m^2=2$ on AdS$_2$). It follows in this case from \eqref{eq:fluctuation energy density} that $\rho(E)=1$ and $E_n=|n|+1$, and from \eqref{eq:form factor} that $f(1)=1$ and $f(E)=\frac{E}{2}$ for $E>1$.\footnote{There is an order of limits issue when evaluating $f(E)$ at $E=1$ and $c^2=0$. Namely, $\lim_{c\to 0}\lim_{E\to 1}f(E)=2\lim_{E\to 1}\lim_{c\to 0}f(E)=1$. Since $E_0=1$ for all $c$, the $n=0$ term in \eqref{eq:nfZXit03Ij} is to be evaluated at $E=1$ before we take $c\to 0$.} Therefore, the series representation of $W_{yy}(t_1,t_2)$ from \eqref{eq:nfZXit03Ij} becomes
\begin{align}\label{eq:Wyy c=0 from general c exp}
    W_{yy}(t_1,t_2)&=\frac{2g}{\pi} \frac{1}{t_{12}^2}\frac{\chi^2}{|1-\chi|}\sum_{n=0}^\infty \text{sgn}(1-\chi)^n(n+1)e^{-(n+1)|\log|1-\chi||}.
\end{align}
Given that $4\sum_{n=1}^\infty n e^{-nx}=\sinh^{-2}{\frac{x}{2}}$ and $4\sum_{n=1}^\infty (-1)^{n+1} ne^{-nx}=\cosh^{-2}{\frac{x}{2}}$, the series can be explicitly summed for any $\chi$. We arrive at the result
\begin{align}
    \braket{\Phi_i(t_1)\Phi_j(t_2)}=W_{yy}(t_1,t_2)\delta_{ij}=\frac{2g}{\pi}\frac{1}{t_{12}^2}\delta_{ij}.
\end{align}
This correctly reproduces the leading behavior of the scalar two-point function in \eqref{eq:phi-phi D-D two pt}. 

Thirdly, \eqref{eq:G1 series} is a compact representation of a function defined piecewise on $\chi\in(-\infty,0)$, $\chi\in(0,1)$, $\chi\in(1,2)$ and $\chi\in(2,\infty)$. Unpacking the absolute values yields
\begin{align}\label{eq:G1 piecewise}
    G_1(\chi)&=\theta(-\chi)G^{(-\infty,0)}_1(\chi) +\theta(\chi)\theta(1-\chi)G^{(0,1)}_1(\chi)\nonumber\\&+\theta(\chi-1)\theta(2-\chi) G^{(1,2)}_1(\chi) +\theta(\chi-2) G^{(2,\infty)}_1(\chi),
\end{align}
where $\theta(b-x)\theta(a-x)$ is $1$ if $a<x<b$ and $0$ otherwise and $G_1^I(\chi)$ denotes the series representation of $G_1(\chi)$ on the interval $I\subset \mathbb{R}$. Explicitly, we have
\begin{align}
    G^{(-\infty,0)}_1(\chi)&=\chi^2\sum_{n\in \mathbb{Z}} \frac{f(E_n)}{(1-\chi)^{E_n+1}}, & G_1^{(0,1)}(\chi)&=\chi^2\sum_{n\in \mathbb{Z}}f(E_n)(1-\chi)^{E_n-1},\label{eq:G1(chi) explicit 1}\\
    G_1^{(1,2)}(\chi)&=\chi^2\sum_{n\in \mathbb{Z}} (-1)^n f(E_n)(\chi-1)^{E_n-1}, & G_1^{(2,\infty)}(\chi)&=\chi^2\sum_{n\in \mathbb{Z}}(-1)^n\frac{f(E_n)}{(\chi-1)^{E_n+1}}.\label{eq:G1(chi) explicit 2}
\end{align}
From the discussion around Figure~\ref{fig:four-point configurations}, one would expect $G_1$ to behave piecewise on $\chi\in(-\infty,0)$, $\chi\in(0,1)$ and $\chi\in(1,\infty)$. The apparently special role of $\chi=2$ in \eqref{eq:G1 piecewise} is an artifact of the series representation. In particular, while the series representations $G_1^{(1,2)}(\chi)$ and $G_1^{(2,\infty)}(\chi)$ do not converge at $\chi=2$, the integral representation in \eqref{eq:s4xpFVW5pB} is perfectly smooth at $\chi=2$.\footnote{This is analogous to the fact that  $\sum_{n=0}^\infty x^n$ diverges, whereas $\frac{1}{1-x}$ is perfectly smooth, at $x=-1$.}  This is related to the convergence of the OPE as we will discuss shortly.

Finally, we comment on how \eqref{eq:G1 series} is consistent with $G_1(\chi)\to 1$ as $\chi\to 0$, which is imposed by the OPE limit in which the two light operators (or the two heavy operators) approach each other (in this limit, the leading exchanged operator is the identity). As $\chi\to 0$, the exponentially damping term in \eqref{eq:G1 series} is turned off and the series diverges due to the ``infinite tail'' consisting of terms with arbitrarily large values of $n$. The contribution of the tail is captured by an integral over $E$ weighted by the energy density $\rho(E)$. Since large values of $E$ dominate, we may replace the energy density and form factor by their asymptotic forms: $\rho(E)\sim \frac{2\ellK(c^2)}{\pi}$, $f(E)\sim \frac{\pi}{4}\frac{E}{\ellK(c^2)}$. Thus, as $\chi\to 0$,
\begin{align}\label{eq:asymptotic chi->0}
    \sum_{n\in \mathbb{Z}} f(E_n) e^{-E_n|\log(1-\chi)|}&\sim \int_{E_{\rm cutoff}}^\infty 2 \rho(E) f(E) e^{-E|\chi|}dE \sim \int_{E_{\rm cutoff}}^\infty E e^{-E|\chi|}dE\sim \frac{1}{\chi^2},
\end{align}
where the leading behavior does not depend on the precise value of $E_{\rm cutoff}\gg1$. Eq.~\eqref{eq:asymptotic chi->0} combines with the prefactor $\chi^2/|1-\chi|\sim \chi^2$ in \eqref{eq:G1 series} to yield $G_1(\chi)\sim 1$, as desired. The key input in this reasoning is the fact that $2\rho(E)f(E)\sim E$ asymptotically for large $E$. Likewise, $G_4(\chi)\to 1$ follows from \eqref{eq:G4 series} and $2\rho(E)f(E)(E^2-1)/6\sim E^3/6$ asymptotically. Our argument is similar in spirit to the general analysis in \cite{Qiao:2017xif} of how the consistency of the OPE in different channels constrains its large dimension asymptotics in a 1d CFT.
\paragraph{Convergence of the series representation.}
The divergence of the series in \eqref{eq:G1(chi) explicit 1}-\eqref{eq:G1(chi) explicit 2}  at $\chi=2$ can be understood in terms of the limited radius of convergence of the OPE in a CFT. We recall that the product of two operators $O_1(s_1)$ and $O_2(s_2)$ can be written as a convergent sum over primaries at a point $s$ only if there exists a sphere centered on $s$ that contains $O_1(s_1)$ and $O_2(s_2)$ and no other operators (see, e.g., \cite{Pappadopulo:2012jk,Hogervorst:2013sma}). As we discuss in greater detail in Section~\ref{sec:extract OPE data}, the series representations of  $G_1(\chi)$ are essentially conformal block expansions of the four-point function in the light-heavy channel, expanded around the insertion point of the heavy operator. To illustrate concretely how this is related to \eqref{eq:G1(chi) explicit 1}-\eqref{eq:G1(chi) explicit 2}, let us use the conformal symmetry to set $t_1=1$, $t_3=0$ and $t_4=\infty$, in which case $1-\chi=t_2$. Then, $G_1^{(-\infty,0)}(\chi)$ and $G_1^{(2,\infty)}(\chi)$ are sums over positive powers of $1/t_2$ and correspond to taking the OPE of $\Phi(t_2)$ with $\bar{Z}^J(\infty)$ centered at $\infty$, while $G_1^{(0,1)}(\chi)$ and $G_1^{(1,2)}(\chi)$ are sums over positive powers of $t_2$ and correspond to taking the OPE of $\Phi(t_2)$ with $Z^J(0)$ centered at $0$. Note that a zero-sphere (i.e., two points) centered at $0$ and enclosing only $\Phi(t_2)$ and $Z^J(0)$ exists only if $|t_2|<1$ because the sphere would otherwise also enclose $\Phi(1)$. Likewise, a zero-sphere centered at $\infty$ and enclosing only $\Phi(t_2)$ and $\bar{Z}^J(\infty)$ exists only if $|t_2|>1$. Therefore the OPE is indeed expected to diverge at $t_2=1,-1$ or $\chi=0,2$.
%}\footnote{\textcolor{red}{It may be a general feature of working with simplified conformal blocks in the large charge limit that the radius of convergence is smaller than what is expected for a general CFT.}}

We can comment a little more concretely about the analyticity of the four-point functions and the convergence of their series representations. The convergence of each of the series in \eqref{eq:G1(chi) explicit 1} and \eqref{eq:G1(chi) explicit 2} is determined by the growth of $E_n$ with $n$. For $c\in (0,1)$, the energy density $\rho(E)$ is sharply peaked at $E=1$ and flattens out as $E$ increases. More precisely, $\rho(E)\sim \text{constant}/\sqrt{E-1}$ near $E=1$\footnote{Explicitly, $\rho(E)\sim \frac{\sqrt{2}}{\pi}\frac{\ellK(c^2)-\ellE(c^2)}{c}\frac{1}{\sqrt{E-1}}$ near $E=1$.} and $\rho(E)\sim \frac{2\ellK(c^2)}{\pi}+O(1/E^2)$ for large $E$, from which it follows that $E_n\sim \frac{\pi}{2\ellK(c^2)}|n|+O(|n|^0)$ for large $|n|$. The asymptotically linear growth of $E_n$ with $|n|$ means that each $G_1^I(\chi)$ converges absolutely and is analytic on a subset of the complex plane that includes the real interval $I$.\footnote{Absolute convergence follows from the ratio test and analyticity follows from applying Morera's theorem.} Because the terms in the series consist of non-integer powers of $1-\chi$, the series are multi-valued and the principal sheet should be defined with a branch cut. For example, $G_1^{(-\infty,0)}(\chi)$ converges for all $\chi\in \mathbb{C}$ such that $|\chi-1|>1$ with a natural choice of branch-cut being the interval $\chi\in (2,\infty)$; $G_1^{(2,\infty)}(\chi)$ converges for all $|\chi-1|>1$ with branch-cut $\chi\in (-\infty,0)$; $G_1^{(0,1)}(\chi)$ converges for all $0<|\chi-1|<1$ with branch-cut $\chi\in(1,2)$; and $G_1^{(1,2)}(\chi)$ converges for all $0<|\chi-1|<1$ with branch-cut $\chi\in (0,1)$. 

Each series in \eqref{eq:G1(chi) explicit 1} and \eqref{eq:G1(chi) explicit 2} can be analytically continued beyond its domain of convergence. In particular, the integral in \eqref{eq:s4xpFVW5pB} provides the maximal extension of $G^{(1,2)}_1$ and $G^{(2,\infty)}$ and smoothly stitches together their disjoint domains of convergence, which lie inside and outside the unit disk centered at $\chi=1$. Meanwhile, the analytic continuation of $G_1^{(-\infty,0)}$ and $G_1^{(0,1)}$ yields two additional distinct multi-valued functions on the complex $\chi$ plane. These observations are in accordance with the general behavior discussed in Section~\ref{sec:preliminaries}. Similar comments apply to the series expressions for $G_{Z\bar{Z}}$, $G_{\bar{Z}Z}$ and $G_4$, which we turn to in Section~\ref{sec:computing G2 and G3} and \ref{sec:integral and series displacement 4-pt}.

\subsubsection{Computing \texorpdfstring{$G_{Z\bar{Z}}$ and $G_{\bar{Z}Z}$ from $G_1$}{GZZb and GZbZ from G1}}\label{sec:computing G2 and G3}
Next, we determine integral and series representations of the defect four-point functions in \eqref{eq:ZZbZ^JZb^J}-\eqref{eq:ZbZZ^JZb^J}, in which the light insertions are $Z$ and $\bar{Z}$. According to \eqref{eq:th-th and th-ph bndy-to-bndy to ZZb defect fn}-\eqref{eq:th-th and th-ph bndy-to-bndy to ZbZ defect fn}, the leading contribution in the large charge expansion is given by the classical vertex operators in \eqref{eq:classical ZZb} and the first subleading correction is determined by the boundary-to-boundary propagators $W_{\theta\theta}$ and $W_{\theta\phi}$. One could try to solve for these boundary-to-boundary propagators in the same way that we solved for $W_{yy}$. This approach is more cumbersome for the $\theta$ and $\phi$ propagators than for the $y$ propagators--- both because $m_{\theta\phi}^2$ is less simple than $m_{yy}^2$ and because the Green's equations solved by $G_{\theta\theta}$ and $G_{\phi\theta}$ are coupled--- but one can nonetheless make progress working perturbatively in small $c^2$, as we demonstrate in Appendix \ref{app:perturbative analysis}. 

To determine the subleading corrections to $G_{Z\bar{Z}}(\chi)$ and $G_{\bar{Z}Z}(\chi)$ (which are equivalent to $W_{\theta\theta}$ and $W_{\theta\phi}$) for general $c^2$, we will instead make use of the superconformal Ward identities. The general solutions to the first order differential equations (\ref{eq:J1cWCLsjEW}) can be written as
\begin{align}
    G_2(\chi)&=\frac{\chi^2}{(1-\chi)^2}\left(C_2(\chi_0)+\int_{\chi_0}^{\chi}  dx\left(\frac{1}{x}-1\right)\frac{dG_1}{dx}\right),\label{eq:RKeEazh2eB}\\
    G_3(\chi)&=\chi^2\left(C_3(\chi_0)-\int_{\chi_0}^{\chi} dx\frac{1}{x}\frac{dG_1}{dx}\right).\label{eq:Kluz7kJ9dy}
\end{align}
After integrating by parts (with $C_2(\chi_0)$ and $C_3(\chi_0)$ absorbing the boundary terms) and converting from $G_2, G_3$ to $G_{Z\bar{Z}},G_{\bar{Z}Z}$ using \eqref{eq:def GZZb GZbZ}, this becomes:
\begin{align}
G_{Z\bar{Z}}(\chi)&=\frac{\chi^2}{(1-\chi)^2}\left[C_2(\chi_0)+\frac{1-\chi}{\chi^2}G_1(\chi)+\int_{\chi_0}^\chi dx \frac{G_1(x)}{x^2}\right],\label{eq:RRMXBWM1a4}\\    G_{\bar{Z}Z}(\chi)&=\chi^2\left[C_3(\chi_0)+\frac{1-\chi}{\chi^2}G_1(\chi)-\int_{\chi_0}^\chi dx \frac{G_1(x)}{x^2}\right].\label{eq:627HvU4xyA}
\end{align}
The remaining integrals in \eqref{eq:RRMXBWM1a4} and \eqref{eq:627HvU4xyA} are easy to evaluate when $G_1$ is expressed using the integral representation in \eqref{eq:s4xpFVW5pB} or the series representations in \eqref{eq:G1(chi) explicit 1}-\eqref{eq:G1(chi) explicit 2}. 

Let's first determine the series representations of $G_{Z\bar{Z}}$ and $G_{\bar{Z}Z}$. Because the series representations of $G_1$ are defined piecewise, we will likewise consider the four cases, $\chi\in (-\infty,0)$, $(0,1)$, $(1,2)$ and $(2,\infty)$ separately. We pick $\chi_0=\infty$ when $\chi\in(-\infty,0)$ or $\chi\in(2,\infty)$ and $\chi_0=1$ when $\chi\in(0,1)$ or $\chi\in(1,2)$. Since $G_1(\chi)$ is finite as $\chi\to \infty$ or $\chi\to1$ (see \eqref{eq:G1(chi) explicit 1}-\eqref{eq:G1(chi) explicit 2}), it follows that for these values of $\chi_0$ the integration constants can be written
\begin{align}\label{eq:Ward integration constants}
    C_2(\chi_0)&=\lim_{\chi\to \chi_0}\frac{(1-\chi)^2}{\chi^2}G_{Z\bar{Z}}, &C_3(\chi_0)&=\lim_{\chi\to \chi_0}\frac{1}{\chi^2}G_{\bar{Z}Z}.
\end{align}
Furthermore, the limits $\chi\to 1$ and $\chi\to \infty$ correspond to $Z$ and $\bar{Z}$ becoming coincident with $Z^J$ and $\bar{Z}^J$ in \eqref{eq:ZZbZ^JZb^J} and \eqref{eq:ZbZZ^JZb^J}, which means that $C_2(\chi_0)$ and $C_3(\chi_0)$ can be expressed in terms of certain normalized OPE coefficients. 

In particular, the primary of lowest conformal dimension in the OPE of $Z$ and $Z^J$ (resp. $\bar{Z}$ and $\bar{Z}^J$) is $Z^{J+1}$ ($\bar{Z}^{J+1}$) and the primary of lowest conformal dimension in the OPE of $Z$ and $\bar{Z}^J$ ($\bar{Z}$ and $Z^J$) is $Z^{J-1}$ ($\bar{Z}^{J-1}$). Thus, the leading contributions to the OPE of $Z$ and $Z^J$ and of $\bar{Z}$ and $Z^J$ are
\begin{align}\label{eq:leading ZZ^J and ZbZ^J OPE}
    Z(t)Z^J(t')&\sim \frac{\mathcal{C}_{ZZ^J\bar{Z}^{J+1}}}{\mathcal{N}_{Z^{J+1}\bar{Z}^{J+1}}}Z^{J+1}(t'),  &\bar{Z}(t)Z^J(t')&\sim \frac{\mathcal{C}_{\bar{Z}Z^J\bar{Z}^{J-1}}}{\mathcal{N}_{Z^{J-1}\bar{Z}^{J-1}}}\frac{1}{(t-t')^2}Z^{J-1}(t')
\end{align}
as $t\to t'$, with subleading terms suppressed by positive powers of $t-t'$. The relations with $Z\leftrightarrow \bar{Z}$ similarly hold. Then, sending $t_1\to t_3$ and $t_2\to t_4$, in which case $\chi\sim \frac{t_{34}^2}{t_{13}t_{24}}$, or $t_1\to t_4$ and $t_2\to t_3$, in which case $1-\chi\sim \frac{t_{14}t_{23}}{t_{34}^2}$, in \eqref{eq:ZZbZ^JZb^J}-\eqref{eq:ZbZZ^JZb^J} and applying the OPEs in \eqref{eq:leading ZZ^J and ZbZ^J OPE}, we find the following four limiting values of $G_{Z\bar{Z}}$ and $G_{\bar{Z}Z}$:
\begin{align}
    &\lim_{\chi\to \infty}G_{Z\bar{Z}}(\chi)&&=\lim_{\chi\to 1}G_{\bar{Z}Z}(\chi)&&=\frac{\pi}{4g}\frac{\mathcal{C}_{ZZ^J\bar{Z}^{J+1}} \mathcal{C}_{\bar{Z}\bar{Z}^JZ^{J+1}}}{\mathcal{N}_{Z^{J+1}\bar{Z}^{J+1}}\mathcal{N}_{Z^J\bar{Z}^J}},\label{eq:caIPyliddz}\\
    &\lim_{\chi\to \infty}\frac{G_{\bar{Z}Z}(\chi)}{\chi^2}&&=\lim_{\chi\to 1}(1-\chi)^2G_{Z\bar{Z}}(\chi)&&=\frac{\pi}{4g}\frac{\mathcal{C}_{\bar{Z}Z^J\bar{Z}^{J-1}} \mathcal{C}_{Z\bar{Z}^JZ^{J-1}}}{\mathcal{N}_{Z^{J-1}\bar{Z}^{J-1}}\mathcal{N}_{Z^J\bar{Z}^J}}.\label{eq:tBTFRl0q5L}
\end{align}
Since the normalized OPE coefficient in \eqref{eq:tBTFRl0q5L} is equal to the one in \eqref{eq:caIPyliddz} up to a unit shift in the large charge, it is useful to note that $c^2(J-1)=c^2(J)-\frac{dc^2(J)}{dJ}+O(1/g^2)$ and $\frac{dc^2(J)}{dJ}=\frac{1-c^2}{2g\ellE(c^2)}$, which follows from \eqref{eq:J/g and c^2}. Meanwhile, the OPE coefficient in \eqref{eq:caIPyliddz} was determined previously in \cite{Giombi:2021zfb} and is given in \eqref{eq:OPE coeff from localization}. Consequently, the constants of integration in \eqref{eq:Ward integration constants} with $\chi_0=1$ and $\chi_0=\infty$ simplify to
\begin{align}\label{eq:DlrRjXE6OH}
    C_2(\infty)&=C_3(1)=g\pi c^2+\frac{\pi}{2}\frac{1-c^2}{\ellE(c^2)},&C_2(1)&=C_3(\infty)=g\pi c^2.
\end{align}

Substituting the appropriate series expression for $G_1$ from \eqref{eq:G1(chi) explicit 1}-\eqref{eq:G1(chi) explicit 2} and \eqref{eq:DlrRjXE6OH} for $C_2$ and $C_3$ in \eqref{eq:RRMXBWM1a4}-\eqref{eq:627HvU4xyA} determines $G_{Z\bar{Z}}$ and $G_{\bar{Z}Z}$ for any $\chi$. For $\chi\in (0,2)$, we find
\begin{align}
    G_{Z\bar{Z}}(\chi)&=\frac{g\pi c^2\chi^2}{(1-\chi)^2} +\frac{\chi^2}{|1-\chi|}\sum_{n\in \mathbb{Z}} \text{sgn}\left(1-\chi\right)^{n+1}f(E_n)\frac{E_n-1}{E_n}e^{-(E_n-1)|\log|1-\chi||},\label{eq:G2 series}\\
    G_{\bar{Z}Z}(\chi)&=\left(g\pi c^2+\frac{\pi}{2}\frac{1-c^2}{\mathbb{E}(c^2)}\right)\chi^2\label{eq:G3 series}\\&\hspace{2cm}+\frac{\chi^2}{|1-\chi|}\sum_{n\in \mathbb{Z}} \text{sgn}\left(1-\chi\right)^{n+1}f(E_n)\frac{E_n+1}{E_n}e^{-(E_n+1)|\log|1-\chi||}.\nonumber
\end{align}
For $\chi\in(-\infty,0)\cup (2,\infty)$, we find
\begin{align}
    G_{Z\bar{Z}}(\chi)&=\left(g\pi c^2+\frac{\pi}{2}\frac{1-c^2}{\mathbb{E}(c^2)}\right)\frac{\chi^2}{(1-\chi)^2}\label{eq:i2F9Wwzuch}\\&\hspace{2cm}+\frac{\chi^2}{|1-\chi|}\sum_n \text{sgn}\left(1-\chi\right)^{n+1}f(E_n)\frac{E_n+1}{E_n}e^{-(E_n+1)|\log|1-\chi||}.\nonumber\\
    G_{\bar{Z}Z}(\chi)&=g\pi c^2\chi^2 +\frac{\chi^2}{|1-\chi|}\sum_n \text{sgn}\left(1-\chi\right)^{n+1}f(E_n)\frac{E_n-1}{E_n}e^{-(E_n-1)|\log|1-\chi||},\label{eq:s7VYqCQAUD}
\end{align}
One readily checks that \eqref{eq:G2 series}-\eqref{eq:G3 series} and \eqref{eq:i2F9Wwzuch}-\eqref{eq:s7VYqCQAUD} satisfy the crossing condition in \eqref{eq:crossing Gs}.

We can also determine the integral representations of $G_{Z\bar{Z}}$ and $G_{\bar{Z}Z}$ by applying \eqref{eq:RRMXBWM1a4}-\eqref{eq:627HvU4xyA} and substituting \eqref{eq:s4xpFVW5pB} for $G_1(\chi)=\frac{\pi t_{12}^2}{2g}W_{yy}(t_1,t_2)$. We find
\begin{align}
    G_{Z\bar{Z}}(\chi)&=\frac{\chi^2}{(1-\chi)^2}\biggr[g\pi c^2+\frac{\pi}{4}\frac{1-c^2}{\ellE(c^2)}\nonumber\\&+\int_{-\infty}^\infty dk \frac{\sqrt{1-c^2+k^2}\left[-k\cos(k\log(\chi-1))+\sin(k\log(\chi-1))\right]}{2\sqrt{1+k^2}\sinh\left(2\int_0^k d\ell \frac{\ellK(c^2)\ell^2+\ellE(c^2)}{\sqrt{1+\ell^2}\sqrt{1-c^2+\ell^2}}\right)}\biggr],\label{eq:X8lEuKtW9m}\\
    G_{\bar{Z}Z}(\chi)&=\chi^2\biggr[g\pi c^2+\frac{\pi}{4}\frac{1-c^2}{\ellE(c^2)}\nonumber\\&+\int_{-\infty}^\infty dk \frac{\sqrt{1-c^2+k^2}\left[-k\cos(k\log(\chi-1))-\sin(k\log(\chi-1))\right]}{2\sqrt{1+k^2}\sinh\left(2\int_0^k d\ell \frac{\ellK(c^2)\ell^2+\ellE(c^2)}{\sqrt{1+\ell^2}\sqrt{1-c^2+\ell^2}}\right)}\biggr].\label{eq:AKRvj3tMJx}
\end{align}
Because $\int_{-\infty}^\infty \cos(ak) f(k) dk\to 0$ as $a\to \pm\infty$ and $\int_{-\infty}^\infty \frac{\sin(ak)}{k} f(k)\to \text{sgn}(a)\pi f(0)$ as $a\to \pm \infty$ for a smooth function $f(k)$, the integrals in \eqref{eq:X8lEuKtW9m} and \eqref{eq:AKRvj3tMJx} approach $- \frac{\pi}{4}\frac{1-c^2}{\ellE(c^2)}$ and $ \frac{\pi}{4}\frac{1-c^2}{\ellE(c^2)}$ as $\chi\to 1$ and approach $\frac{\pi}{4}\frac{1-c^2}{\ellE(c^2)}$ and $-\frac{\pi}{4}\frac{1-c^2}{\ellE(c^2)}$ as $\chi\to \infty$, respectively. Therefore, \eqref{eq:X8lEuKtW9m} and \eqref{eq:AKRvj3tMJx} reproduce \eqref{eq:caIPyliddz}-\eqref{eq:tBTFRl0q5L}, as required by the OPE limit.

\subsection{Computing \texorpdfstring{$W_{xx}$}{Wxx}}\label{sec:integral and series displacement 4-pt}
Finally, we implement the general analysis from Section~\ref{sec:integral and series reps of propagators} to find the boundary-to-boundary propagator $W_{xx}$. Via \eqref{eq:xx bndy-to-bndy to DD defect fn} and \eqref{eq:DDZ^JZb^J}, this determines the leading large charge behavior of defect correlators $\braket{\mathbb{D}\mathbb{D} Z^J\bar{Z}^J}/\braket{Z^J\bar{Z}^J}$ or, equivalently,  $G_4(\chi)$. The computation of $W_{xx}$ is very similar to that of $W_{yy}$. We will therefore suppress some of the details and will use tildes to distinguish between parallel quantities.

We begin by rewriting \eqref{eq:8txKFSC2oh} in terms of the new variable $\tilde{\sigma}\equiv ir$. Using the identity in \eqref{eq:oX5Dm1hVuw} to simplify the result, we find:
\begin{align}\label{eq:xNlIa8LCN5}
    \left[-\partial_{\tilde{\sigma}}^2+\frac{2}{1-c^2}\text{sn}^2\left(\tilde{\sigma}|\frac{1}{1-c^2}\right)\right]g_{xx}^{R/L}(\tilde{\sigma};k)&=\left(\frac{2+k^2-c^2}{1-c^2}\right)g^{R/L}_{xx}(\tilde{\sigma};k).
\end{align}
This is also in the Jacobi form of the Lam\'e equation given in \eqref{eq:ES9YxM59dJ} if we identify the parameter, $\tilde{m}$, and the eigenvalue, $\tilde{\Lambda}$, to be:
\begin{align}\label{eq:FFlen54N2I}
    \tilde{m}&\equiv \frac{1}{1-c^2}, & \tilde{\Lambda}&\equiv \frac{2+k^2-c^2}{1-c^2}.
\end{align}
It is again useful to introduce the following shorthand:
\begin{align}
\tilde{\ellK}&\equiv \ellK\left(\frac{1}{1-c^2}\right)=\sqrt{1-c^2}(\ellK(1-c^2)-i\ellK(c^2)),\\
\tilde{\ellK}'&\equiv \ellK\left(\frac{c^2}{c^2-1}\right)=\sqrt{1-c^2}\ellK(c^2).
\end{align}
The second way of writing $\tilde{\ellK}$ and $\tilde{\ellK}'$ follows from \eqref{eq:E5NuDWFxub}-\eqref{eq:M8haq6dJl2} and makes it clear that $\tilde{\ellK}'$ and $\tilde{\ellK}+i\tilde{\ellK}'$ are positive real numbers.

As the coordinate $r$ runs from $-r_m$ to $r_m$, the coordinate $\tilde{\sigma}$ runs from $\tilde{\sigma}_-\equiv -i\tilde{\ellK}'$ to $\tilde{\sigma}_+\equiv i\tilde{\ellK}'$ along the imaginary axis. Its behavior near the end points follows from \eqref{eq:5kUeXpf7ju}:
\begin{align}
    \tilde{\sigma}&\sim \tilde{\sigma}_{\pm} \mp 2i\sqrt{1-c^2}e^{\mp\rho},&&\text{as }\rho\to\pm \infty.
\end{align}

Having recognized \eqref{eq:xNlIa8LCN5} as the Jacobi form of the Lam\'e equation with the identification in \eqref{eq:FFlen54N2I}, we identify two linearly independent solutions to be
\begin{align}
    \tilde{f}_{\pm}(\tilde{\sigma};\tilde{\alpha})\equiv \frac{H(\pm \tilde{\sigma} + \tilde{\alpha}|\tilde{m})}{\Theta(\tilde{\sigma}|\tilde{m})}e^{\mp \tilde{\sigma} Z(\tilde{\alpha}|\tilde{m})}.
\end{align}
Here, $\tilde{\alpha}$ is related to $k$ by
\begin{align}\label{eq:k and tilde alpha}
    k\equiv -i\text{sn}\left(\tilde{\alpha}|\tilde{m}\right).
\end{align}

\begin{figure}[t]
\centering
\begin{minipage}{0.49\hsize}
\centering
\includegraphics[clip, height=6.3cm]{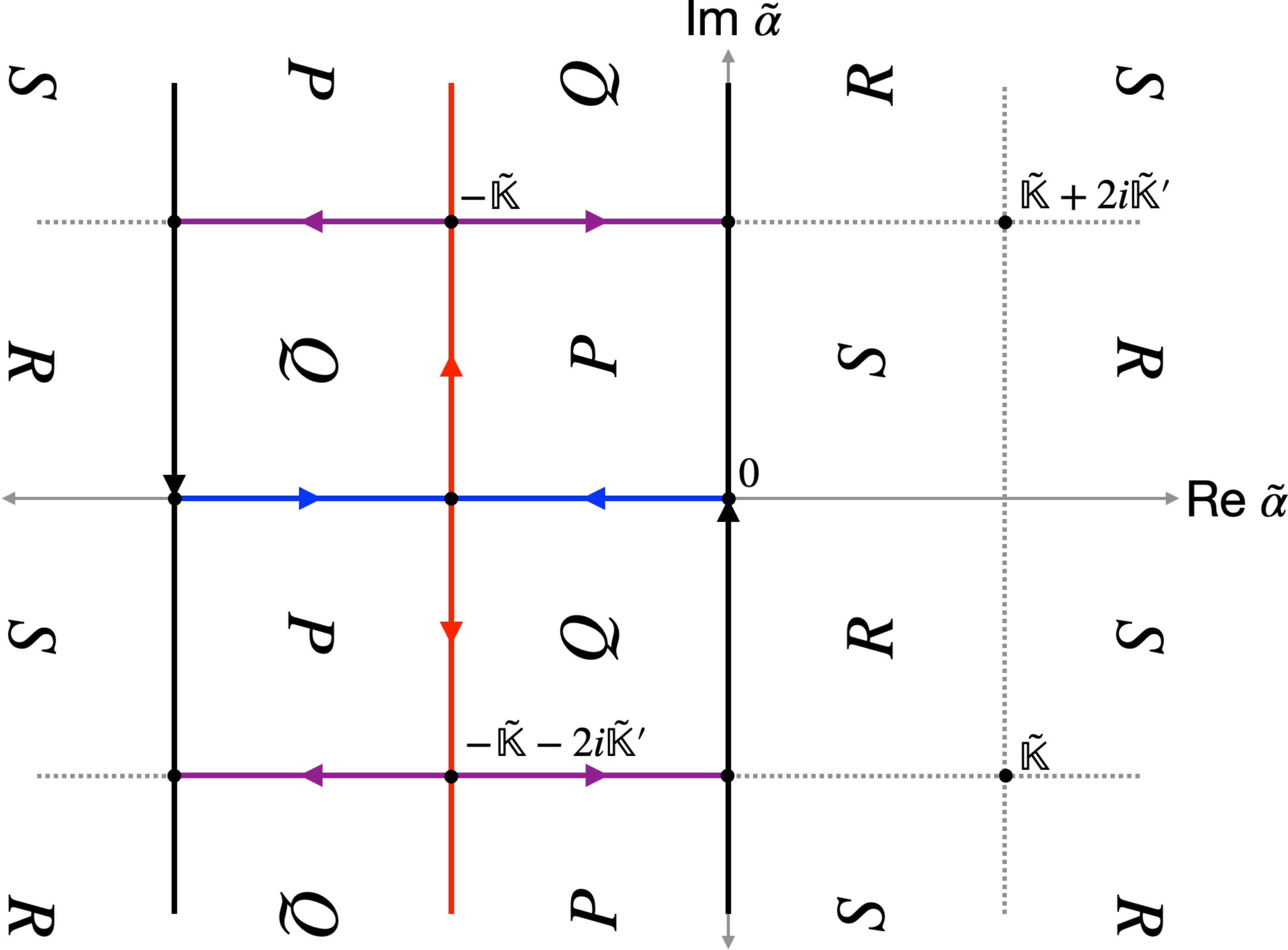}\\
\vspace{0.25cm}
{\bf a.} Periodicity of unit cells in $\tilde{\alpha}$ plane
\end{minipage}
\begin{minipage}{0.49\hsize}
\centering
\includegraphics[clip, height=6.5cm]{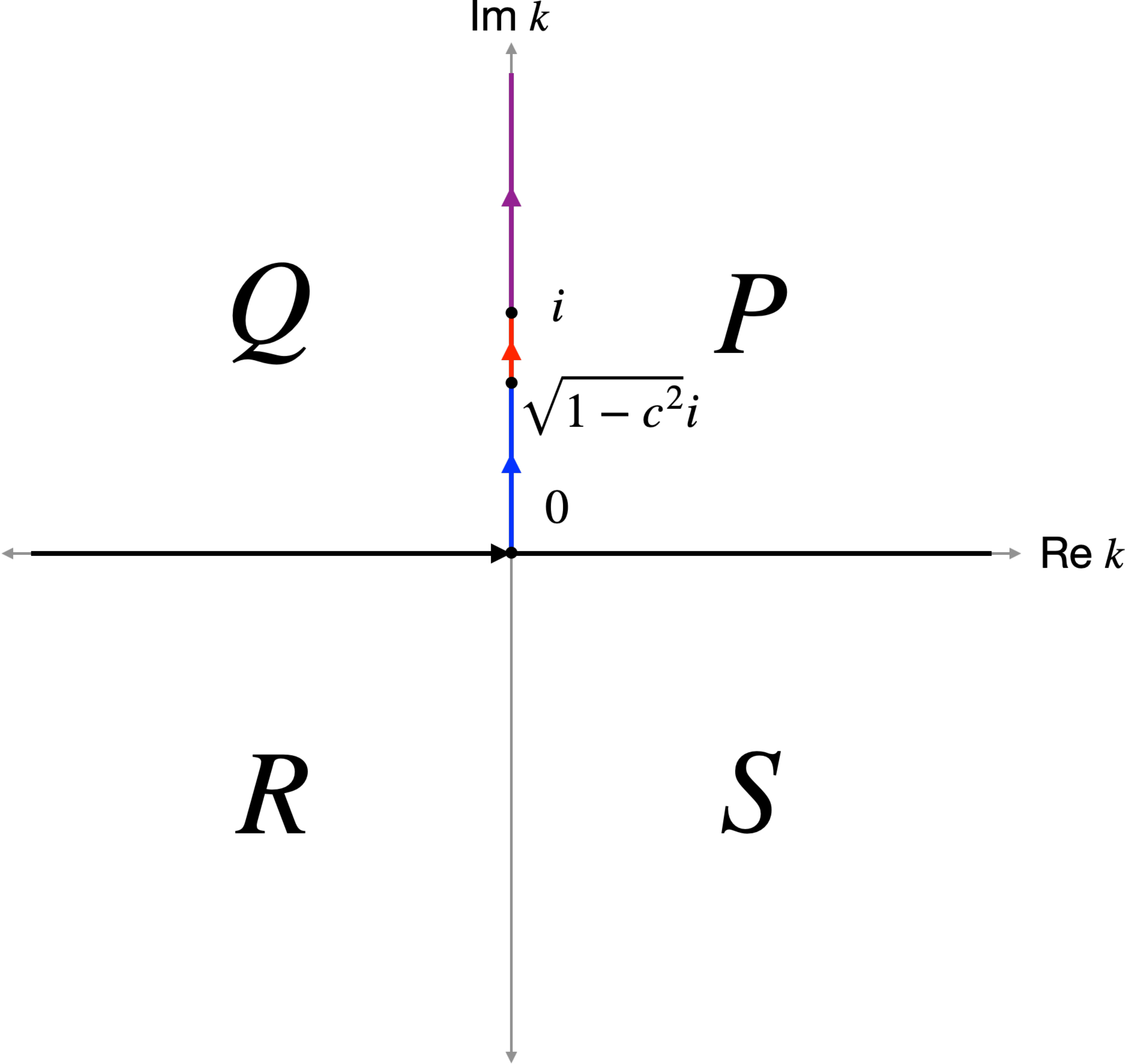}\\
\vspace{.05cm} 

{\bf b.} Image of $\tilde{\alpha}$ unit cells in $k$ plane
\end{minipage}
\caption{The Jacobi elliptic function relating $k$ and $\tilde{\alpha}$ in \eqref{eq:k and tilde alpha} is doubly periodic, satisfying $\text{sn}(\tilde{\alpha}+2\tilde{\ellK}+2i\tilde{\ellK}'\rvert\tilde{m})=-\text{sn}\left(\tilde{\alpha}\rvert \tilde{m}\right)$, $\text{sn}\left(\tilde{\alpha}+2i\ellK'\rvert \tilde{m}\right)=\text{sn}\left(\tilde{\alpha}\rvert \tilde{m}\right)$, as well as $\text{sn}\left(-\tilde{\alpha}\rvert \tilde{m}\right)=-\text{sn}\left(\tilde{\alpha}\rvert \tilde{m}\right)$. Thus, the unit cells in the $\tilde{\alpha}$ plane are arranged as shown in \textbf{a.} Each grouping of four neighboring tiles labelled `Q', `P', `R' and `S', which are the pre-images of the four quadrants in the $k$ plane as shown in \textbf{b.}, is one unit cell. The intervals $(-\infty,0)$ and $(0,\infty)$ on the real axis and $(0,\sqrt{1-c^2}i)$, $(\sqrt{1-c^2}i,i)$ and $(i,+\infty i)$ in the $k$ plane are indicated in \textbf{b.} using black, blue, red and purple directed line segments, respectively. Their pre-images in the $\tilde{\alpha}$ plane are likewise indicated in \textbf{a.}} 
\label{fig:mapping alpha tilde to k}
\end{figure}

Because of the double-periodicity of $\text{sn}\left(\tilde{\alpha}|\tilde{m}\right)$, the $\tilde{\alpha}$-plane is an infinite cover of the $k$ plane. See Figure~\ref{fig:mapping alpha tilde to k}. We take the fundamental unit cell to be the rectangle with vertices at $\tilde{\alpha}=-\tilde{\ellK}-2i\tilde{\ellK}'$, $\tilde{\alpha}=-\tilde{\ellK}$, $\tilde{\alpha}=\tilde{\ellK}+2i\tilde{\ellK}'$, and $\tilde{\alpha}=\tilde{\ellK}$. In particular, the pre-images of the real and positive imaginary axes are: As $k$ runs from $-\infty$ to $\infty$ along the real axis $\alpha$ runs from $-i\tilde{\ellK}'$ to $i\tilde{\ellK}'$ along the imaginary axis; as $k$ runs from $0$ to $\sqrt{1-c^2}i$ to $i$ to $+i\infty$ along the imaginary axis, $\tilde{\alpha}$ runs along the line segments from $0$ to $-\tilde{\ellK}-i\tilde{\ellK}'$ to $-\tilde{\ellK}$ to $i\tilde{\ellK}'$.

Next, we determine $g^{R/L}_{xx}$ as linear combinations of $\tilde{f}_{\pm}$. We cannot replicate \eqref{eq:hIhWEB0Kcn} and \eqref{eq:27Vo18ONzZ} because $\Theta(\tilde{\sigma}_{\pm}|\tilde{m})=0$. Instead, the appropriate linear combinations are
\begin{align}
    g^R_{xx}(\tilde{\sigma};\tilde{\alpha})&\equiv \frac{1}{\Theta(\tilde{\sigma}|\tilde{m})}\left[\frac{H(\tilde{\sigma}+\tilde{\alpha}|\tilde{m})}{H(\tilde{\sigma}_++\tilde{\alpha}|\tilde{m})}e^{(\tilde{\sigma}_+-\tilde{\sigma})Z(\tilde{\alpha}|\tilde{m})}-\frac{H(-\tilde{\sigma}+\tilde{\alpha}|\tilde{m})}{H(-\tilde{\sigma}_++\tilde{\alpha}|\tilde{m})}e^{(\tilde{\sigma}-\tilde{\sigma}_+)Z(\tilde{\alpha}|\tilde{m})}\right],\\
    g^L_{xx}(\tilde{\sigma};\tilde{\alpha})&\equiv \frac{1}{\Theta(\tilde{\sigma}|\tilde{m})}\left[\frac{H(-\tilde{\sigma}+\tilde{\alpha}|\tilde{m})}{H(-\tilde{\sigma}_-+\tilde{\alpha}|\tilde{m})}e^{(\tilde{\sigma}-\tilde{\sigma}_-)Z(\tilde{\alpha}|\tilde{m})}-\frac{H(\tilde{\sigma}+\tilde{\alpha}|\tilde{m})}{H(\tilde{\sigma}_-+\tilde{\alpha}|\tilde{m})}e^{(\tilde{\sigma}_--\tilde{\sigma})Z(\tilde{\alpha}|\tilde{m})}\right].
\end{align}
Let us check that these indeed satisfy the boundary condition $g^R_{xx}(\tilde{\sigma},\tilde{\alpha})\to 0$ as $\tilde{\sigma}\to \tilde{\sigma}_+$ (and therefore also $g^L_{xx}(\tilde{\sigma},\tilde{\alpha})\to 0$ as $\tilde{\sigma}\to \tilde{\sigma}_-$). We note the behavior of $\Theta(\tilde{\sigma}|\tilde{m})$ near $\tilde{\sigma}=\tilde{\sigma}_+$,
\begin{align}
    \Theta(\tilde{\sigma}|\tilde{m})=\frac{\pi}{2\tilde{\ellK}}\theta_4'\left(\frac{\pi \tilde{\tau}}{2},\tilde{q}\right)(\tilde{\sigma}-\tilde{\sigma}_+)+O(\tilde{\sigma}-\tilde{\sigma}_+)^2,
\end{align}
where $\tilde{\tau}\equiv \frac{i\tilde{\ellK}'}{\tilde{\ellK}}$ and $\tilde{q}\equiv e^{i\pi \tilde{\tau}}$. Furthermore, $\Theta$ is related to $H$ after a translation by $\pm i\tilde{\ellK}'$: 
\begin{align}
    H(\tilde{\sigma}\pm i \tilde{\ellK}'|\tilde{m})=\pm i \text{exp}\left(\mp \frac{i\pi}{2\tilde{\ellK}}\left(\tilde{\sigma}\pm \frac{i\tilde{\ellK}'}{2}\right)\right)\Theta(\tilde{\sigma}|\tilde{m}).
\end{align}
This allows us to express the $n$-fold derivatives of $H$ translated by half-periods (i.e., $H^{(n)}/H$ evaluated at $\tilde{\alpha}\pm i \tilde{\ellK}'$) as sums of the lower derivatives of $\Theta$ (i.e., $\Theta^{(m)}/\Theta$ evaluated at $\tilde{\alpha}$ for $0\leq m\leq n$). Recalling also that $Z=\Theta'/\Theta$, we ultimately find
\begin{align}
    g^{R}_{xx}(\tilde{\sigma};\tilde{\alpha})&=\frac{2\tilde{\ellK}}{3\pi}\frac{1}{\theta_4'\left(\frac{\pi \tilde{\tau}}{2},\tilde{q}\right)}\left[\frac{\Theta'''(\tilde{\alpha}|\tilde{m})}{\Theta(\tilde{\alpha}|\tilde{m})}-\frac{3\Theta''(\tilde{\alpha}|\tilde{m})}{\Theta(\tilde{\alpha}|\tilde{m})}Z(\tilde{\alpha}|\tilde{m})+2Z^3(\tilde{\alpha}|\tilde{m})\right](\tilde{\sigma}-\tilde{\sigma}_+)^2+O(\tilde{\sigma}-\tilde{\sigma}_+)^3\nonumber\\&=
    \frac{2\tilde{\ellK}}{3\pi}\frac{1}{\theta_4'\left(\frac{\pi \tilde{\tau}}{2},\tilde{q}\right)}Z''(\tilde{\alpha}|\tilde{m})(\tilde{\sigma}-\tilde{\sigma}_+)^2+O(\tilde{\sigma}-\tilde{\sigma}_+)^3.
\end{align}
This reproduces the expected behavior in \eqref{eq:gR/gL asymptotic behavior}.

Thus, sending $g_{xx}^R(\tilde{\sigma};\tilde{\alpha})$ and $g_{xx}^L(\tilde{\sigma};\tilde{\alpha})$ to opposite boundaries yields
\begin{align}
    \lim_{\rho\to\infty}e^{2\rho}g_{xx}^R(\tilde{\sigma};\tilde{\alpha})&=\lim_{\rho\to -\infty}e^{-2\rho}g_{xx}^L(\tilde{\sigma};\tilde{\alpha})=-\frac{8\tilde{\ellK}}{3\pi}\frac{1-c^2}{\theta_4'\left(\frac{\pi \tilde{\tau}}{2},\tilde{q}\right)}Z''(\tilde{\alpha}|\tilde{m}).
\end{align}
Furthermore, the normalization can be simplified to
\begin{align}
    a_{xx}(\tilde{\alpha})&=\frac{i}{\sqrt{1-c^2}}\left(2\frac{dg_{xx}^R(0;\tilde{\alpha})}{d\sigma}g_{xx}^R(0;\tilde{\alpha})\right)^{-1}\\&=-\frac{i\Theta(0|\tilde{m})^2e^{\frac{\pi \tilde{\ellK}'}{2\tilde{\ellK}}}}{4\sqrt{1-c^2}}\left(\frac{H(\tilde{\alpha}|\tilde{m})^2}{\Theta(\tilde{\alpha}|\tilde{m})^2}\left(V(\tilde{\alpha}|\tilde{m})-Z(\tilde{\alpha}|\tilde{m})\right)\sinh\left(\frac{i\pi\tilde{\alpha}}{\tilde{\ellK}}+2i\tilde{\ellK}'Z(\tilde{\alpha}|\tilde{m})\right)\right)^{-1},\nonumber
\end{align}
where we used the fact that $a_{xx}(\tilde{\alpha})$ is independent of $\tilde{\sigma}$ to evaluate the Wronskian at $\tilde{\sigma}=0$.

Using $\frac{dk}{d\tilde{\alpha}}=-i\text{cn}\left(\tilde{\alpha}\rvert\tilde{m}\right)\text{dn}\left(\tilde{\alpha}\rvert\tilde{m}\right)$ (note \eqref{eq:k and tilde alpha} and \eqref{eq:dsn/dz})
and \eqref{eq:vjHsgGxfAP}-\eqref{eq:AO7XFN9xh0} to express various combinations of the theta functions in terms of the Jacobi elliptic functions, we may write the boundary limit of the bulk-to-bulk propagators as the following integral:\footnote{To simplify the factors independent of $\tilde{\alpha}$ in front of the integral, we used $i\frac{2\tilde{\ellK}}{\pi}\frac{\theta_4(0,\tilde{q})}{\theta_4'(\frac{\pi \tilde{\tau}}{2},\tilde{q})}\tilde{q}^{-\frac{1}{4}}\tilde{m}^{\frac{1}{4}}=1$, which can be deduced from various identities in Ch. 1 of \cite{MR1007595}. 
%This follows from $\theta'(\pi \tau/2)=i/q^(1/4)\theta_2(0)\theta_3(0)\theta_4(0)$, $\ellK=\frac{\pi}{2}\theta_3(0)^2$ and $m=\theta_2(0)^4/\theta_3(0)^4$. These identities can be found in Chapter 1 of \cite{MR1007595}.
}

\begin{align}\label{eq:9pj5kESYtQ}
\lim_{\substack{\rho\to\infty\\\rho'\to-\infty}}e^{2\rho-2\rho'}G_{xx}(\rho,\tau;\rho',\tau')&=\frac{4\sqrt{1-c^2}}{9\pi g}\int_{-i\tilde{\ellK}'}^{i\tilde{\ellK}'} d\tilde{\alpha}\biggr[\text{exp}\left(\text{sn}\left(\tilde{\alpha}\rvert \tilde{m}\right)|\tau-\tau'|\right)\\&\hspace{3cm}\times\frac{\text{sn}\left(\tilde{\alpha}\rvert \tilde{m}\right)\text{cn}^2\left(\tilde{\alpha}\rvert \tilde{m}\right)\text{dn}^2\left(\tilde{\alpha}\rvert \tilde{m}\right)}{\sinh\left(\tilde{F}(\tilde{\alpha})\right)}\biggr],\nonumber
\end{align}
where
\begin{align}\label{eq:F-tilde(alpha-tilde)}
    \tilde{F}(\tilde{\alpha})&\equiv \frac{i\pi\tilde{\alpha}}{\tilde{\ellK}}+2i\tilde{\ellK}'Z(\tilde{\alpha}|\tilde{m}).
\end{align}
This is equivalent to \eqref{eq:bndy-to-bndy integral rep} except with a different parametrization of the integration variable.

\begin{figure}[!t]
\centering
\begin{minipage}{0.33\textwidth}
\centering
\includegraphics[clip, width=5cm]{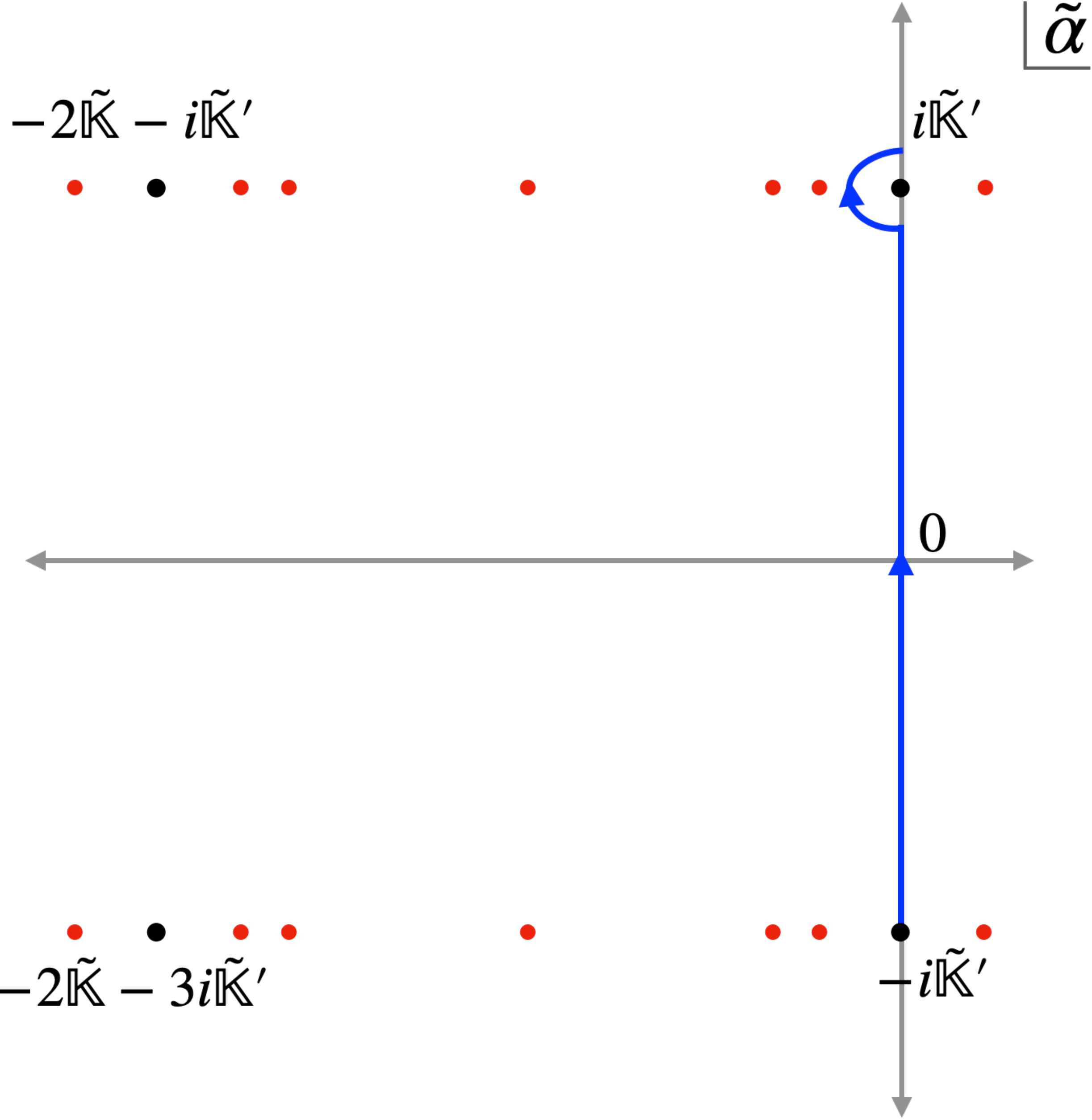}\\
{\bf a.}
\vspace{0.5cm}
\end{minipage}\hfill
\begin{minipage}{0.33\textwidth}
\centering
\includegraphics[clip, width=5cm]{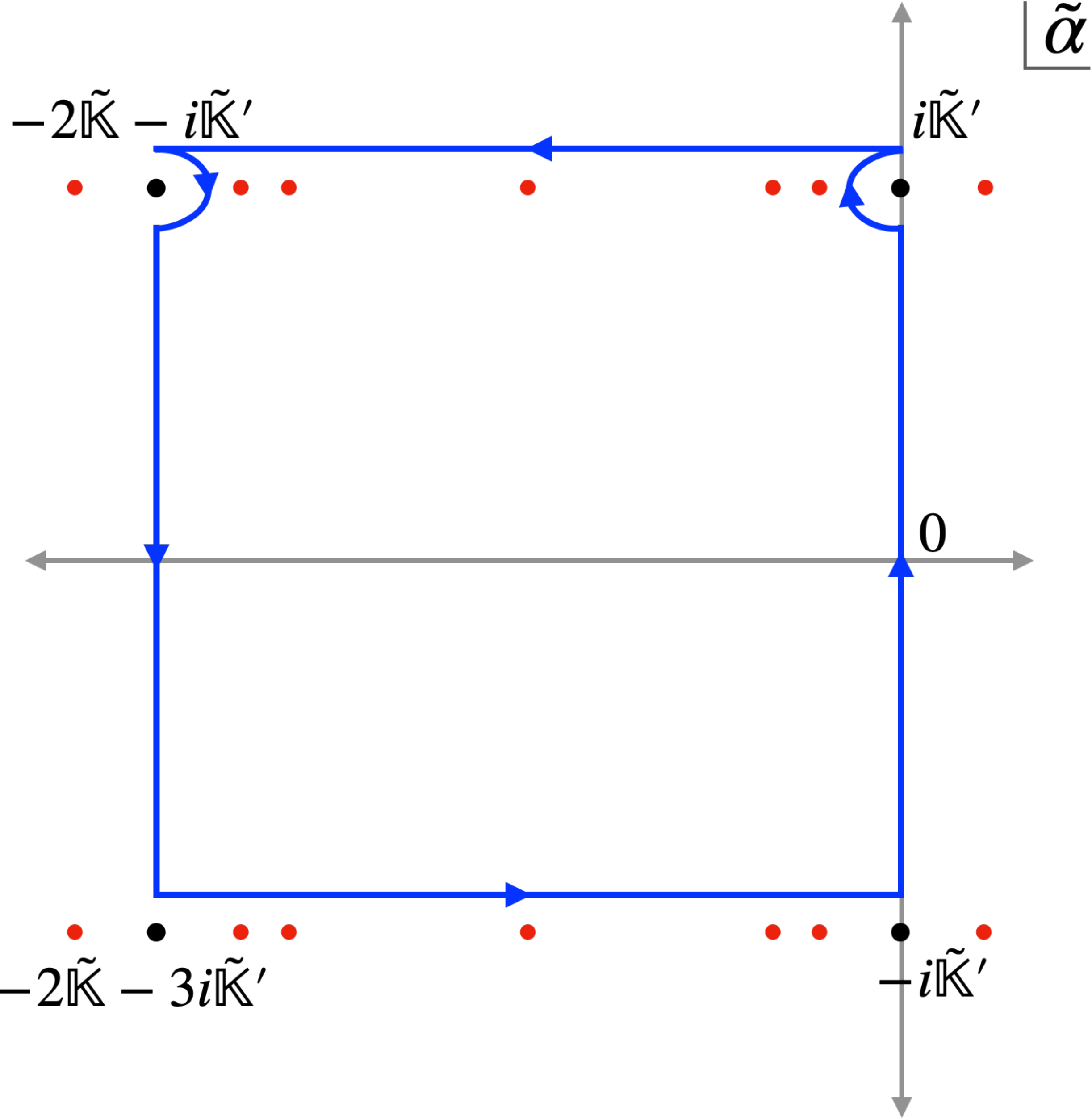}\\
{\bf b.}
\vspace{0.5cm}
\end{minipage}\hfill
\begin{minipage}{0.33\textwidth}
\centering
\includegraphics[clip, width=5cm]{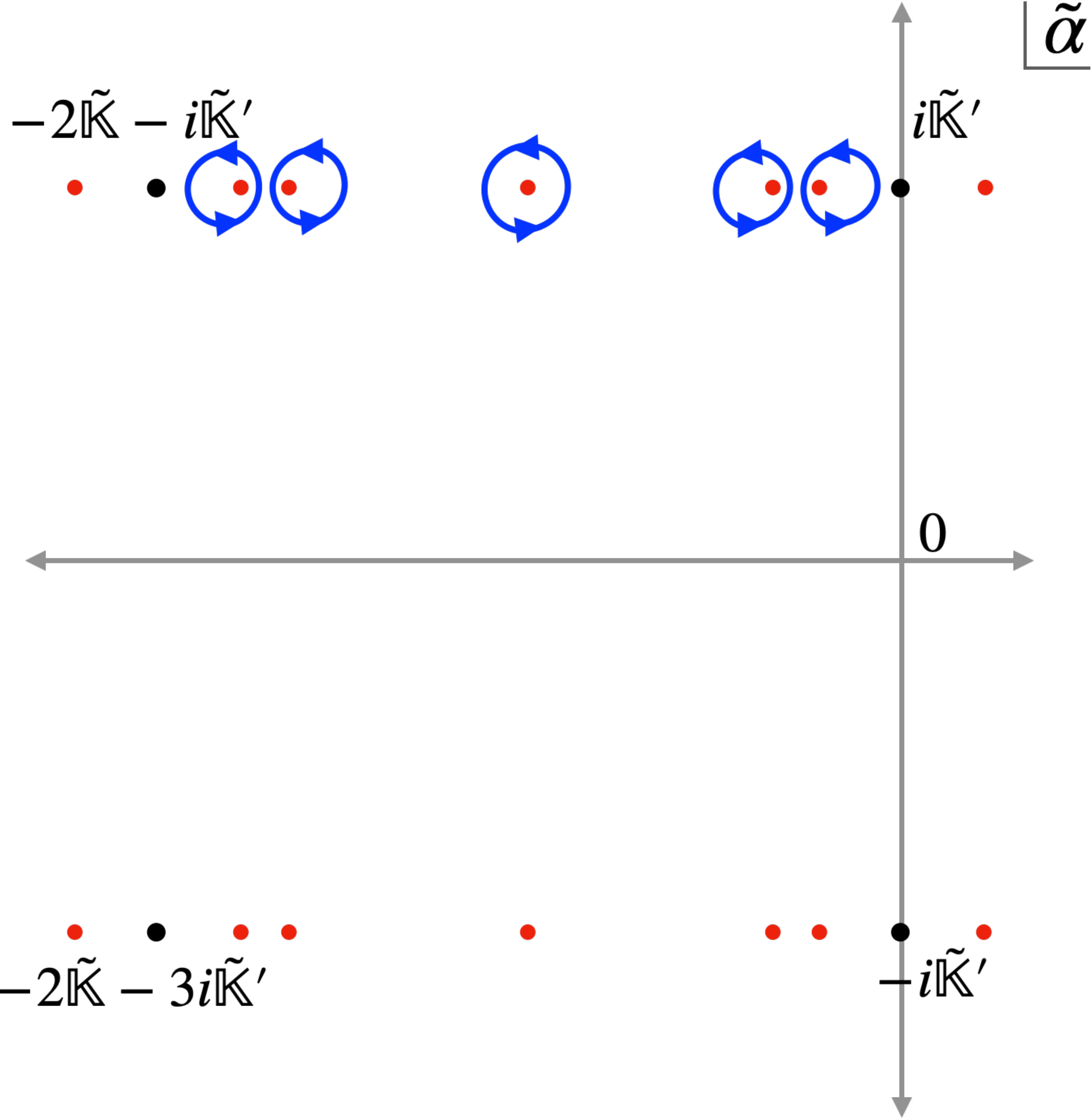}\\
{\bf c.}
\vspace{0.5cm}
\end{minipage} 
\caption{Closing the integration contour in \eqref{eq:9pj5kESYtQ} and picking up the residues in the $\tilde{\alpha}$ plane. The integration contour along the real axis in the $k$ plane can be closed in the usual way at infinity in the upper half-plane. The lift of the closed contour to the $\alpha$ plane via the map in \eqref{eq:k and tilde alpha} is depicted in \textbf{a.} and is not closed. Using the periodic identification of the $\tilde{\alpha}$ plane as depicted in Figure~\ref{fig:mapping alpha tilde to k}, the contour in \textbf{a.} can be closed as in \textbf{b.} at the cost of doubling the integral. We can then  pick up the residues at the poles on the interval $[-2\tilde{\ellK}-i\tilde{\ellK}',i\tilde{\ellK}']$ parallel to the real axis, as depicted in \textbf{c.}} 
\label{fig:contour deformation in alpha tilde and k planes}
\end{figure}

The next step is to close the contour in \eqref{eq:9pj5kESYtQ}. The image of the contour in the $k$ plane can be closed in the upper-half plane at infinity. The lift to the $\tilde{\alpha}$ plane, shown in Figure~\ref{fig:contour deformation in alpha tilde and k planes}~\hyperref[fig:contour deformation in alpha tilde and k planes]{a}., is not closed, but we can again use the periodicity of the map from $\tilde{\alpha}$ to $k$ to write \eqref{eq:mkhjfKkrkv} as one half of the integral over the closed contour shown in Figure~\ref{fig:contour deformation in alpha tilde and k planes}~\hyperref[fig:contour deformation in alpha tilde and k planes]{b}.

The poles of the integrand in \eqref{eq:9pj5kESYtQ} that lie inside the closed contour are the zeros of $\sinh(\tilde{F}(\tilde{\alpha}))$ that lie on the line segment between $i\tilde{\ellK}'$ and  $-2\tilde{\ellK}-i\tilde{\ellK}'$ (see Figure~\ref{fig:contour deformation in alpha tilde and k planes}~\hyperref[fig:contour deformation in alpha tilde and k planes]{c}), which we denote by $\tilde{\alpha}_n$, $n\in \mathbb{Z}$ and which satisfy
\begin{align}\label{eq:VQWwcAOEWM}
    \tilde{F}(\tilde{\alpha}_n)=(n-1)\pi i.
\end{align}
Here, we choose to define $\tilde{\alpha}_n$ by setting $\tilde{F}(\tilde{\alpha}_n)$ equal to $(n-1)\pi i$ instead of equal to $n\pi i$ because it is then possible to relate $\tilde{\alpha}_n$ to $\alpha_n$ by a linear transformation independent of $n$. In particular, we checked numerically that the following identity relating $F(z)$ defined in \eqref{eq:F(alpha)} and $\tilde{F}(z)$ defined in \eqref{eq:F-tilde(alpha-tilde)} appears to hold:
\begin{align}\label{eq:FF-tilde identity}
    F(i\ellK'z)-\tilde{F}\left(-\tilde{\ellK}-(\tilde{\ellK}+i\tilde{\ellK}')z\right)=\pi i,\hspace{1cm}\forall z\in \mathbb{C}.
\end{align}
Combined with \eqref{eq:7u5v07hm5W} and \eqref{eq:VQWwcAOEWM}, this identity implies that $\alpha_n$ and $\tilde{\alpha}_n$ are related by
\begin{align}\label{eq:CJ3lrOA5v5}
    \frac{\alpha_n}{i\ellK'}&=-\frac{\tilde{\alpha}_n+\tilde{\ellK}}{\tilde{\ellK}+i\tilde{\ellK}'}.
\end{align}
Namely, the positions of the poles in \eqref{eq:mkhjfKkrkv} along the line segment between $-i\ellK'$ and $i\ellK'$ in the $\alpha$ plane are equal to the positions of the poles of \eqref{eq:9pj5kESYtQ} along the line segment between $-2\tilde{\ellK}-i\tilde{\ellK}'$ and $i\tilde{\ellK}'$ in the $\tilde{\alpha}$ plane, as measured in units in which the two line segments have unit lengths. This can be also been seen qualitatively by comparing Figure~\ref{fig:contour deformation in alpha tilde and k planes} with Figure~\ref{fig:contour deformation in alpha and k planes}. 

As a consequence of \eqref{eq:CJ3lrOA5v5}, the quasi-energies defined in \eqref{eq:En alpha-n} are also equal to\footnote{This follows from substituting the expression for $\alpha_n$ in terms of $\tilde{\alpha}_n$ from \eqref{eq:CJ3lrOA5v5} into \eqref{eq:En alpha-n}, noting that $\frac{\tilde{\ellK}+i\tilde{\ellK}'}{\ellK'}=\frac{\sqrt{1-c^2}}{c}$, and using \eqref{eq: Jacobi imag/real transformation sn}-\eqref{eq: Jacobi imag/real transformation n} and \eqref{eq:IW5SfbL8zO}-\eqref{eq:8QBvEot7x6} to simplify the result.}
\begin{align}\label{eq:En tilde-alpha-n}
    E_n=-\text{sn}\left(\tilde{\alpha}_n|\tilde{m}\right).
\end{align}

Again using \eqref{eq:RKtXoUrJeH} to evaluate the residues of \eqref{eq:9pj5kESYtQ} at $\tilde{\alpha}_n$, we arrive at
\begin{align}\label{eq:4aMUyJTt40}
   &\lim_{\substack{\rho\to\infty\\\rho'\to-\infty}}e^{2\rho-2\rho'}G_{xx}(\rho,\tau;\rho',\tau')\\&=\frac{4\sqrt{1-c^2}}{9\pi g}\sum_{n=-\infty}^\infty(-1)^{n+1}\frac{\text{sn}\left(\tilde{\alpha}_n\rvert \tilde{m}\right)\text{cn}^2\left(\tilde{\alpha}_n\rvert \tilde{m}\right)\text{dn}^2\left(\tilde{\alpha}_n\rvert \tilde{m}\right)\text{exp}\left(\text{sn}\left(\tilde{\alpha}_n\rvert \tilde{m}\right)|\tau-\tau'|\right)}{\tilde{F}'(\tilde{\alpha}_n)}.\nonumber
\end{align}
Using the identity in \eqref{eq:LIA7TEVOWb}, we can write the derivative of $Z(\tilde{\alpha})|\tilde{m})$, and thus $\tilde{F}'(\tilde{\alpha})$, in terms of the Jacobi elliptic functions. Using \eqref{eq:E5NuDWFxub}-\eqref{eq:aPTCYjwzVg} to write $\tilde{\ellK}$, $\tilde{\ellK}'$ and $\ellE(\frac{1}{1-c^2})$ in terms of $\ellK(c^2)$ and $\ellE(c^2)$, we find
\begin{align}\label{eq:txtY7zZA0K}
    \tilde{F}'(z)&=2i\sqrt{1-c^2}\ellK(c^2)\left(\text{dn}\left(z|\tilde{m}\right)-1\right)+\frac{2i\ellE(c^2)}{\sqrt{1-c^2}}.
\end{align}
Eq.~\eqref{eq:9pj5kESYtQ} is the final integral representation and \eqref{eq:4aMUyJTt40}, supplemented with \eqref{eq:VQWwcAOEWM}, is the final series representation for the boundary limit of $G_{xx}$ expressed in terms of the $\tilde{\alpha}$ variables. It is again more transparent to express both in terms of the original $k$ variables. 

We first rewrite \eqref{eq:9pj5kESYtQ}. $\tilde{F}$ can be written as a function of $k$ using $\tilde{F}(k)=\int_0^k dk'\frac{d\tilde{\alpha}}{dk'}\frac{d\tilde{F}}{d\tilde{\alpha}}$ and noting \eqref{eq:txtY7zZA0K}. Using \eqref{eq:k and tilde alpha} and \eqref{eq:oX5Dm1hVuw}-\eqref{eq:wtEDZNracI}, all of the elliptic functions can be expressed in terms of $k$. Ultimately, we find
\begin{align}\label{eq:s00j8ye3cI}
    W_{xx}(t_1,t_2)&=\frac{g}{\pi t_{12}^4}\frac{\chi^4}{(1-\chi)^2}\int_{-\infty}^\infty dk \frac{k\sqrt{1+k^2}\sqrt{1-c^2+k^2}\cos(k\log(\chi-1))}{\sinh\left(2\int_0^k d\ell \frac{\ellK(c^2)\ell^2+\ellE(c^2)}{\sqrt{1+\ell^2}\sqrt{1-c^2+\ell^2}}\right)}.
\end{align}
This is valid for $\chi>1$.

Finally, we use \eqref{eq:oX5Dm1hVuw}-\eqref{eq:wtEDZNracI} and \eqref{eq:En tilde-alpha-n} to write all the elliptic functions in \eqref{eq:4aMUyJTt40} in terms of $E_n$. The boundary-to-boundary propagator simplifies to 
\begin{align}\label{eq:E2nGE9lYGN}
    W_{xx}(t_1,t_2)&= \frac{12g}{\pi}\frac{1}{t_{12}^4}\frac{\chi^4}{(1-\chi)^2}\sum_{n\in\mathbb{Z}}\text{sgn}(1-\chi)^{n+1}f(E_n)\frac{E_n^2-1}{6}e^{-E_n|\log|1-\chi||}.
\end{align}
Here $f(E)$ is the same function defined in \eqref{eq:form factor}. In the above expression, we have again replaced $(-1)^n$ in \eqref{eq:4aMUyJTt40} by $\text{sgn}(1-\chi)^n$ in accordance with the discussion around \eqref{eq:g^R g^L different boundaries}-\eqref{eq:dRNhIUWDdm}, and the resulting series representation for $W_{xx}$ holds for both $\chi>1$ and $\chi<1$. Note that because $E_0=1$, the $n=0$ term in the sum is zero. Using $G_4(\chi)=\frac{\pi}{12 g}t_{12}^4W_{xx}(t_1,t_2)$, this leads to \eqref{eq:G4 series}.

\paragraph{Remarks about $G_4(\chi)$.} Now we make two remarks about the final result for $G_4(\chi)$. We can check that the four-point function reduces to the two-point function of the displacement operators when $c=0$. In this case the series representation of $W_{xx}(t_1,t_2)$ from \eqref{eq:E2nGE9lYGN} becomes 
\begin{align}\label{eq:Wxx c=0 from general c exp}
    W_{xx}(t_1,t_2)&=\frac{2g}{\pi}\frac{1}{t_{12}^4}\frac{\chi^4}{(1-\chi)^2}\sum_{n=1}^\infty \text{sgn}(1-\chi)^{n+1}n(n+1)(n+2)e^{-(n+1)|\log|1-\chi||}.
\end{align}
Given that $\sum_{n=2}^\infty n(n^2-1) e^{-nx}=\frac{3}{8}\sinh^{-4}{\frac{x}{2}}$ and $\sum_{n=2}^\infty (-1)^nn(n^2-1) e^{-nx}=\frac{3}{8}\cosh^{-4}{\frac{x}{2}}$, we can explicitly sum the series for $\chi>1$ and $\chi<1$. In both cases, we arrive at the result
\begin{align}\label{eq:pMVATWnWqr}
    \braket{\mathbb{D}_a(t_1)\mathbb{D}_b(t_2)}&=W_{xx}(t_1,t_2)\delta_{ab}=\frac{12g}{\pi}\frac{1}{t_{12}^4}\delta_{ab}.
\end{align}
This reproduces the leading behavior of \eqref{eq:phi-phi D-D two pt}.

Next, comparing \eqref{eq:s4xpFVW5pB} and \eqref{eq:s00j8ye3cI} or \eqref{eq:nfZXit03Ij} and \eqref{eq:E2nGE9lYGN}, one sees that $G_1$ and $G_4$ satisfy:
\begin{align}
    G_4(\chi)&=-\frac{\chi^4}{6}\left(3\frac{d}{d\chi}+(\chi-1)\frac{d^2}{d \chi^2}\right)\frac{G_1(\chi)}{\chi^2}.
\end{align}
This identity, in analogy with \eqref{eq:J1cWCLsjEW}, should essentially be a Ward identity that relates the four-point functions in \eqref{eq:scalar large charge correlator} and \eqref{eq:displacement large charge correlator} that belong to the same superconformal multiplet.

\section{Extracting OPE data from the four-point functions}\label{sec:extract OPE data}

In this section, we extract defect OPE data of operators with large charge from the four-point defect correlators in \eqref{eq:scalar large charge correlator}-\eqref{eq:displacement large charge correlator}. The analysis in this case is particularly simple because the series representations of the defect correlators in \eqref{eq:G1 series}-\eqref{eq:G4 series} and \eqref{eq:G2 series}-\eqref{eq:s7VYqCQAUD} are already essentially in the form of conformal block expansions, and because the conformal blocks simplify when some of the operators have large dimensions. For a given four-point function, each term in the series corresponds to a different exchanged operator, the energies $E_n$ determine the anomalous dimensions, and the coefficients in the series determine the OPE coefficients. We also determine the small and large $\mathcal{J}$ behavior of the defect correlators and the OPE data.

\subsection{Extracting OPE data}\label{sec:extracting OPE data}

In a 1d CFT, the conformal block expansion of a four-point function in the $12\to 34$ channel is given by \cite{Dolan:2011dv}
\begin{align}\label{eq:4cFvvTr4hC}
    \braket{O_1(t_1)O_2(t_2)O_3(t_3)O_4(t_4)}&=\frac{1}{t_{21}^{\Delta_1+\Delta_2}t_{43}^{\Delta_3+\Delta_4}}\left(\frac{t_{42}}{t_{41}}\right)^{\Delta_1-\Delta_2}\left(\frac{t_{41}}{t_{31}}\right)^{\Delta_3-\Delta_4}\\&\times\sum_\Delta p_\Delta \chi^\Delta {_2F_1}(\Delta+\Delta_2-\Delta_1,\Delta+\Delta_3-\Delta_4,2\Delta,\chi).\nonumber
\end{align}
Here, we take the order of the operators to be $t_1<t_2<t_3<t_4$, in which case $0<\chi<1$. We will restrict our attention to four-point correlators in which $O_1^\dagger=O_3$ and $O_2^\dagger=O_4$ or $O_1^\dagger=O_4$ and $O_2^\dagger=O_3$, in which case we may choose a basis of primaries $O_k$ such that they satisfy the orthogonality condition $\braket{O_kO_{k'}^\dagger}\propto \delta_{kk'}$, and the OPEs in the $12\to 34$ channel take the form $O_1O_2\sim \sum_k \frac{\mathcal{C}_{O_1O_2O_k}}{\mathcal{N}_{O_kO_k^\dagger}}O_k$, $O_3O_4\sim \sum_k \frac{\mathcal{C}_{O_3O_4O_k}}{\mathcal{N}_{O_kO_k^\dagger}}O_k^\dagger$. The sum over $\Delta$ in \eqref{eq:4cFvvTr4hC} is then a sum over the primaries $O_k$ with $\Delta=\Delta_k$ and $p_\Delta = \mathcal{C}_{O_1O_2O_k}\mathcal{C}_{O_3O_4O_k^\dagger}/\mathcal{N}_{O_kO_k^\dagger}$.

We will study the conformal block expansions of the defect correlators in \eqref{eq:scalar large charge correlator}-\eqref{eq:displacement large charge correlator} in the ``light-heavy''$\to$ ``light-heavy'' channels, in which case the exchanged operators appearing in the OPEs of both ``light-heavy'' pairs of the external operators have large charge and large conformal dimension. When multiple of the external and exchanged operators have large conformal dimensions, the conformal blocks simplify. In particular, two asymptotic expansions of the hypergeometric function that will be useful are 
\begin{align}
    {_2F_1}(\alpha,\beta,L+\gamma,\chi)&=1+\frac{\alpha \beta}{L}\chi+O\left(\frac{1}{L^2}\right),\label{eq:large charge conformal block 1}\\
    {_2F_1}(L+\alpha,\beta,L+\gamma,\chi)&=\frac{1}{(1-\chi)^\beta}\left(1+\frac{\beta(\alpha-\gamma)}{L}\frac{\chi}{1-\chi}+O\left(\frac{1}{L^2}\right)\right).\label{eq:large charge conformal block 2}
\end{align}
The first expansion follows from the series definition ${_2F_1}(a,b,c,\chi)\equiv 1+\frac{ab}{c}z+\frac{a(a+1)b(b+1)}{c(c+1)}z^2+\ldots$, and is relevant when $\Delta=\Delta_1+O(1)=\Delta_4+O(1)$ are large in \eqref{eq:4cFvvTr4hC}. The second expansion can be found in, e.g., \cite{temme2003large} and is relevant when $\Delta= \Delta_2+O(1)=\Delta_4+O(1)$ are large.

\paragraph{OPE data from $\braket{Z^JZ\bar{Z}\bar{Z}^J}$ and $\braket{ZZ^J\bar{Z}\bar{Z}^J}$.} Consider the conformal block expansion of $\braket{(\epsilon_1\cdot \Phi(t_1))^J\epsilon_2\cdot \Phi(t_2)\epsilon_3\cdot \Phi(t_3)(\epsilon_4\cdot \Phi(t_4))^J}$ in the $12\to 34$ channel. It decomposes into three different channels distinguished by the $SO(5)$ irreducible representations (irreps) of the operators that appear in the OPE of $(\epsilon_1\cdot \Phi)^J$ and $\epsilon_2\cdot \Phi$ (and likewise in the OPE of $\epsilon_3\cdot \Phi$ and $(\epsilon_4\cdot \Phi)^J$).  In terms of Young tableaux, the decomposition of the tensor product of the rank $1$ and rank $J$ symmetric traceless representations into irreps is given by
\begin{align}
    \Yvcentermath2
    \yng(1)\otimes \overbrace{\yng(5)}^J=\overbrace{ \yng(6)}^{J+1}\oplus \overbrace{\yng(5,1)}^J\oplus \overbrace{\yng(4)}^{J-1},
\end{align}

The three $SO(5)$ channels can be disentangled and studied separately using, for instance, harmonic polynomials of $\xi$ and $\zeta$.\footnote{For instance, for the case $J=1$, the harmonic polynomials corresponding to the singlet, antisymmetric, and rank $2$ symmetric traceless representations are \cite{nirschl2005superconformal}: $Y=1$, $Y_{\tiny \yng(1,1)}=\xi-\zeta$, $Y_{\tiny\yng(2)}=\xi+\zeta-\frac{2}{5}$. We can therefore write the conformally invariant part of \eqref{eq:scalar large charge correlator} as 
\begin{equation}
    G_1+\xi G_3+\zeta G_2=(G_1+\frac{1}{5}G_3+\frac{1}{5}G_2)Y+\frac{1}{2}(G_3+G_2)Y_{\tiny \yng(2)}+\frac{1}{2}(G_3-G_2)Y_{\tiny \yng(1,1)}, 
\end{equation}
and then analyze each channel separately. The conformal block expansion of the three channels of the $J=1$ correlator was studied in \cite{giombi2017half}.} However, for simplicity we choose instead to focus on the four-point correlator $\braket{Z^J(t_1)Z(t_2)\bar{Z}(t_3)\bar{Z}^J(t_4)}$, which has only one channel because the operators appearing in the OPE of $Z^J$ and $Z$ and in the OPE of $\bar{Z}$ and $\bar{Z}^J$ all transform in the rank $J+1$ representation.\footnote{This follows from the fact that $Z^J(t_1)Z(t_2)\to e^{i(J+1)\theta}Z^J(t_1)Z(t_2)$ under the rotation $\Phi_4+i\Phi_5\to e^{i\theta}(\Phi_4+i\Phi_5)$. By contrast, the highest weight states in the rank $J-1$ symmetric traceless representation and the mixed representation will transform like $O_{\rm h.w.}\to e^{i(J-1)\theta}O_{\rm h.w.}$ and $O_{\rm h.w.}\to e^{iJ\theta}O_{\rm h.w.}$, respectively.} In principle, we do not lose anything with this restriction because the operators in the three channels are in the same superconformal multiplets, so their anomalous dimensions are the same and their OPE coefficients are related.

Relabelling $1\rightarrow 2\rightarrow 3\rightarrow 1$ in \eqref{eq:ZZbZ^JZb^J} and putting the resulting correlator in the form of \eqref{eq:4cFvvTr4hC} yields
\begin{align}
    \braket{Z^J(t_1)Z(t_2)\bar{Z}(t_3)\bar{Z}^J(t_4)}&=\frac{4g\mathcal{N}_{Z^J\bar{Z}^J}/\pi}{t_{21}^{J+1}t_{43}^{J+1}}\left(\frac{t_{42}t_{31}}{t_{41}^2}\right)^{J-1}\frac{\chi^{J+1}}{(1-\chi)^2}G_{Z\bar{Z}}\left(\frac{\chi-1}{\chi}\right).
\end{align}
Here $\chi\equiv \frac{t_{12}t_{34}}{t_{13}t_{24}}$, as in previous sections. Applying \eqref{eq:i2F9Wwzuch}, this may be explicitly written
\begin{align}\label{eq:Z^JZZbZb^J confblockexp}
    \braket{Z^J(t_1)Z(t_2)\bar{Z}(t_3)\bar{Z}^J(t_4)}&=\frac{4g\mathcal{N}_{Z^J\bar{Z}^J}/\pi}{t_{21}^{J+1}t_{43}^{J+1}}\left(\frac{t_{42}t_{31}}{t_{41}^2}\right)^{J-1}\biggr[\left(g\pi c^2 \left(1+\frac{2\chi}{J}\right)+\frac{\pi}{2}\frac{1-c^2}{\ellE(c^2)}\right)\chi^{J+1}\nonumber\\&+\sum_{n=1}^\infty 2f(E_n)\frac{E_n+1}{E_n}\chi^{E_n+J+1}+O(1/g)\biggr].
\end{align}
Here, we have combined the $n=0$ term in the sum in \eqref{eq:i2F9Wwzuch} with the term outside the sum (and recalled that $f(1)=\frac{g\pi c^2}{J}$) and also combined the $-n$ and $n$ terms.

Eq.~\eqref{eq:Z^JZZbZb^J confblockexp} is in the form of a conformal block expansion. Let us denote the primaries appearing in the OPE of $Z^J$ and $Z$ by $[Z^JZ]_n$, the operators appearing in the OPE of $\bar{Z}$ and $\bar{Z}^J$ by $[\bar{Z}\bar{Z}^J]_n$, and their dimension by $\Delta_n$. Here $n=0,1,2\ldots$ is (for the moment) simply a label to distinguish the different operators, ordered by increasing conformal dimension. For instance, the lowest primaries are $[Z^JZ]_0\equiv Z^{J+1}$ and $[\bar{Z}\bar{Z}^J]_0\equiv \bar{Z}^{J+1}$. We take the primaries to satisfy $[Z^JZ]_n^\dagger = [\bar{Z}\bar{Z}^J]_n$ and $\braket{[ZZ^J]_n[\bar{Z}\bar{Z}^J]_{n'}}\propto \delta_{nn'}$. 

We can now read off the OPE data of $[Z^JZ]_n$ and $[\bar{Z}\bar{Z}^J]_n$ by matching \eqref{eq:Z^JZZbZb^J confblockexp} to the general form of the conformal block expansion of $\braket{Z^JZ\bar{Z}\bar{Z}^J}$ given in \eqref{eq:4cFvvTr4hC}. Identifying $\Delta_1=\Delta_4=J$, $\Delta_2=\Delta_3=1$ and $\Delta=\Delta_n$, and simplifying the conformal blocks using \eqref{eq:large charge conformal block 1} with the identification $\alpha=\beta=\Delta_n+1-J$, $\gamma=2(\Delta_n-J)$ and $L=2J$, we arrive at
\begin{align}\label{eq:vxCVDhaIZJ}
    \braket{Z^J(t_1)Z(t_2)\bar{Z}(t_3)\bar{Z}^J(t_4)}&=\frac{1}{t_{21}^{J+1}t_{43}^{J+1}}\left(\frac{t_{42}t_{31}}{t_{41}^2}\right)^{J-1}\sum_{n=0}^\infty \frac{\mathcal{C}_{Z^JZ[Z^JZ]_n}\mathcal{C}_{\bar{Z}\bar{Z}^J[\bar{Z}\bar{Z}^J]_n}}{\mathcal{N}_{[Z^JZ]_n[\bar{Z}\bar{Z}^J]_n}}\chi^{\Delta_n}\\&\hspace{4cm}\times\left(1+\frac{(\Delta_n+1-J)^2}{2J}\chi+O(1/g^2)\right).\nonumber
\end{align}

We can now compare \eqref{eq:vxCVDhaIZJ} with \eqref{eq:Z^JZZbZb^J confblockexp} term-by-term, and read off the OPE data. To normalize the OPE coefficients by the norms of each of the operators, we note $\mathcal{N}_{Z\bar{Z}}=\frac{4g}{\pi}\left(1-\frac{3}{8\pi g}+\ldots\right)$, which follows from \eqref{eq:phi-phi D-D two pt}-\eqref{eq:n_1(g)}. Then, for the lowest operator, we find
\begin{align}\label{eq:Z^JZZ^(J+1) OPE data}
\Delta_0&=J+1, &\frac{|\mathcal{C}_{Z^JZZ^{J+1}}|^2}{\mathcal{N}_{Z\bar{Z}}\mathcal{N}_{Z^J\bar{Z}^J}\mathcal{N}_{Z^{J+1}\bar{Z}^{J+1}}}&=g\pi c^2+\frac{\pi}{2}\frac{1-c^2}{\ellE(c^2)}+\frac{3c^2}{8}+O(1/g).
\end{align}
This reproduces the OPE data that we used to fix the constants of integration of $G_{Z\bar{Z}}$ and $G_{\bar{Z}Z}$ in Section~\ref{sec:computing G2 and G3}. Meanwhile, the higher terms in the conformal block expansion produce the new OPE data:
\begin{align}\label{eq:Z^JZ[ZZ^J] OPE data}
    \Delta_n&=J+1+E_n, &\frac{|\mathcal{C}_{Z^JZ[Z^JZ]_n}|^2}{\mathcal{N}_{Z\bar{Z}}\mathcal{N}_{Z^J\bar{Z}^J}\mathcal{N}_{[ZZ^J]_n[\bar{Z}\bar{Z}^J]_n}}&=2f(E_n)\frac{E_n+1}{E_n},
\end{align}
where $n=1,2,3,\ldots$. In \eqref{eq:Z^JZZ^(J+1) OPE data} and \eqref{eq:Z^JZ[ZZ^J] OPE data}, we used $\mathcal{C}_{\bar{Z}\bar{Z}^J[\bar{Z}\bar{Z}^J]_n}=\mathcal{c}^*_{Z^JZ[Z^JZ]_n}$, which follows from reflection positivity of the dCFT. Because the simplified form of the conformal block in \eqref{eq:large charge conformal block 1} is valid when $\alpha,\beta,\gamma\ll L$, the OPE data in \eqref{eq:Z^JZ[ZZ^J] OPE data} is reliable as long as $\Delta_n-J\ll J\sim \sqrt{\lambda}$. In the strict large charge limit, where the charge $J$ and string tension $\frac{\sqrt{\lambda}}{2\pi}$ are taken to infinity, \eqref{eq:Z^JZ[ZZ^J] OPE data} gives the OPE data of an infinite tower of non-protected operators.

As a non-trivial check, we can also extract OPE data from $\braket{Z(t_1)Z^J(t_2)\bar{Z}(t_3)\bar{Z}^J(t_4)}$, in which we switch the relative order of $Z$ and $Z^J$. Relabelling $1\leftrightarrow 2$ in \eqref{eq:ZZbZ^JZb^J}, writing the correlator explicitly using \eqref{eq:i2F9Wwzuch}, and putting it in the form of \eqref{eq:4cFvvTr4hC}, we get
\begin{align}
    \braket{Z(t_1)Z^J(t_2)\bar{Z}(t_3)\bar{Z}^J(t_4)}&=\frac{4g \mathcal{N}_{Z^J\bar{Z}^J}/\pi}{t_{21}^{J+1}t_{43}^{J+1}}\left(\frac{t_{42}}{t_{31}}\right)^{1-J}\chi^{J+1}G_{Z\bar{Z}}(\chi^{-1}),\label{eq:3MbSdyqRZo}\\
    &=\frac{4g \mathcal{N}_{Z^J\bar{Z}^J}/\pi}{t_{21}^{J+1}t_{43}^{J+1}}\biggr[\left(g\pi c^2\left(1-\frac{2}{J}\frac{\chi}{1-\chi}\right)+\frac{\pi }{2}\frac{1-c^2}{\ellE(c^2)}\right)\frac{\chi^{J+1}}{(1-\chi)^2}\nonumber\\&+\sum_{n=1}^\infty (-1)^{n+1}\frac{2f(E_n)(E_n+1)}{E_n}\frac{\chi^{J+1+E_n}}{(1-\chi)^{E_n+2}}+O(1/g)\biggr].\label{eq:ZZ^JZbZb^J confblockexp}
\end{align}

We want to match \eqref{eq:ZZ^JZbZb^J confblockexp} to the general form of the conformal block expansion in \eqref{eq:4cFvvTr4hC}. We again denote the primaries appearing in the OPE of $Z$ and $Z^J$ by $[ZZ^J]_n\equiv [Z^JZ]_n$ and in the OPE of $\bar{Z}$ and $\bar{Z}^J$ by $[\bar{Z}\bar{Z}^J]_n$. Then, identifying $\Delta_1=\Delta_3=1$, $\Delta_2=\Delta_4=J$ and $\Delta=\Delta_n$ in \eqref{eq:4cFvvTr4hC}, which allows us to simplify the conformal blocks using \eqref{eq:large charge conformal block 2} if we set $L=2J$, $\alpha=\Delta_n-1-J$, $\beta=\Delta_n+1-J$, $\gamma=2(\Delta_n-J)$, we may write
\begin{align}\label{eq:6CQVIFAFGw}
    &\braket{Z(t_1)Z^J(t_2)\bar{Z}(t_3)\bar{Z}^J(t_4)}=\frac{1}{t_{21}^{J+1}t_{43}^{J+1}}\left(\frac{t_{42}}{t_{31}}\right)^{1-J}\sum_{n=0}^\infty \frac{\mathcal{C}_{ZZ^J[ZZ^J]_n}\mathcal{C}_{\bar{Z}\bar{Z}^J[\bar{Z}\bar{Z}^J]_n}}{\mathcal{N}_{[ZZ^J]_n[\bar{Z}\bar{Z}^J]_n}}\frac{\chi^{\Delta_n}}{(1-\chi)^{\Delta_n+1-J}}\nonumber\\&\ \hspace{6cm}\times \left(1-\frac{(\Delta_n+1-J)^2}{2J}\frac{\chi}{1-\chi}+O(1/g^2)\right).
\end{align}

We can now compare \eqref{eq:6CQVIFAFGw} with \eqref{eq:ZZ^JZbZb^J confblockexp} term-by-term. For the lowest operator, we find
\begin{align}\label{eq:ZZ^JZ^(J+1) OPE data}
    \Delta_0&=J+1, &\frac{\mathcal{C}_{ZZ^JZ^{J+1}}\mathcal{C}_{\bar{Z}\bar{Z}^J\bar{Z}^{J+1}}}{\mathcal{N}_{Z\bar{Z}}\mathcal{N}_{Z^J\bar{Z}^J}\mathcal{N}_{Z^{J+1}\bar{Z}^{J+1}}}&=g\pi c^2+\frac{\pi}{2}\frac{1-c^2}{\ellE(c^2)}+\frac{3c^2}{8}+O(1/g).
\end{align}
For the higher terms in the conformal block expansion, we find
\begin{align}\label{eq:ZZ^J[ZZ^J] OPE data}
    \Delta_n&=J+1+E_n, &\frac{\mathcal{C}_{ZZ^J[ZZ^J]_n}\mathcal{C}_{\bar{Z}\bar{Z}^J[\bar{Z}\bar{Z}^J]_n}}{\mathcal{N}_{Z\bar{Z}}\mathcal{N}_{Z^J\bar{Z}^J}\mathcal{N}_{[ZZ^J]_n[\bar{Z}\bar{Z}^J]_n}}&=(-1)^{n+1}\frac{2f(E_n)(E_n+1)}{E_n},
\end{align}
for $n=1,2,3,\ldots$. Eqs.~\eqref{eq:ZZ^JZ^(J+1) OPE data}-\eqref{eq:ZZ^J[ZZ^J] OPE data} match \eqref{eq:Z^JZZ^(J+1) OPE data}-\eqref{eq:Z^JZ[ZZ^J] OPE data} as long as $\mathcal{C}_{ZZ^J[Z^JZ]_0}=\mathcal{C}_{Z^JZ[ZZ^J]_0}$ and $\mathcal{C}_{ZZ^J[Z^JZ]_n}=(-1)^{n+1}\mathcal{C}_{Z^JZ[ZZ^J]_n}$, for $n=1,2,\ldots$. This is a consequence of parity, assuming $[Z^JZ]_n$ is odd for $n=2,4,\ldots$ and even otherwise. These parities of the exchanged operators are consistent with the observation that in the generalized free field limit (i.e., $g\to \infty$, which can be probed via the $J/g\to 0$ limit of the large charge results) composite primaries can be constructed out of $J+1$ copies of $Z$ and $0,2,3,4,\ldots$ derivatives (thus, the operators $[ZZ^J]_n$ for $n>0$ contain $n+1$ derivatives), but no primary can be constructed using a single derivative because $Z^J\partial Z\sim \partial Z^{J+1}$ is a descendent. See Appendix~\ref{sec:Finite charge OPE data} for more details.

The OPE data derived in this section come with a caveat. When extracting OPE data from the conformal block expansion of a four-point function, if one does not make sure to disentangle the contributions of operators with the same quantum numbers, the extracted OPE data will in general be averages over the degenerate operators.\footnote{Indeed, part of the motivation to focus on the $\braket{ZZ^J\bar{Z}\bar{Z}^J}$ correlator instead of the other scalar correlators is to reduce this issue of ``operator mixing.'' As discussed in \cite{giombi2017half} in the analysis of the $J=1$ four-point functions, at leading order in $1/g$ the operators on the Wilson line behave like generalized free fields, and there is a unique composite primary transforming in the rank $2$ symmetric traceless representation that can be constructed out of two copies of the elementary scalar field and a given even number of derivatives. Moreover, this composite operator does not mix with any other operators. However, as we discuss in Appendix~\ref{sec:Finite charge OPE data}, for $J>1$ there are multiple composite primaries transforming in the rank $J+1$ symmetric traceless representation that can be constructed out of $J+1$ copies of the elementary scalar field and a given number of derivatives, if the number of derivatives is sufficiently large.} Therefore, on general grounds we would expect the OPE data in \eqref{eq:Z^JZ[ZZ^J] OPE data} and \eqref{eq:ZZ^J[ZZ^J] OPE data} to be ``mixing-averaged.'' On the other hand, since we can precisely match the series representations of the four-point functions to the conformal block expansion, it is possible that the operator mixing is not relevant in the large charge limit. It would be good to understand this issue better.

\subsection{Defect correlators and OPE data at small and large \texorpdfstring{$\mathcal{J}$}{J/g}}\label{sec:small and large J/g}

In this section we examine the small $\mathcal{J}$ and large $\mathcal{J}$ behavior of the defect correlators derived in Section~\ref{sec:series representations of 4pt functions} and of the OPE data derived in Section~\ref{sec:extracting OPE data}. 

\paragraph{Small $\mathcal{J}$.} We first consider the defect correlators in the regime $1\ll J\ll g$, in which case $\mathcal{J}$ is small. The series in \eqref{eq:G1 series}-\eqref{eq:G4 series} and \eqref{eq:G2 series}-\eqref{eq:s7VYqCQAUD} allow us to determine with minimal effort the small $c^2$ expansions of the defect correlators, which can be converted to small $\mathcal{J}$ expansions using the Taylor series relating $\mathcal{J}$ and $c^2$ in this regime:
\begin{align}\label{eq:yxrxVwSqbk}
    \mathcal{J}&=\pi c^2 + \frac{3\pi c^4}{8}+\frac{15\pi c^6}{64}+O(c^8), &c^2&=\frac{1}{\pi}\mathcal{J}-\frac{3}{8\pi^2}\mathcal{J}^2+\frac{3}{64\pi^3}\mathcal{J}^3+O\left(\mathcal{J}^4\right).
\end{align}

We begin by determining the behavior of $E_n$ when $c^2$ is small. Qualitatively, the $E_n$ will be spaced approximately uniformly with unit separation, and deviations will be small in $c^2$ as long as $|n|$ is not too large. We can make this precise by determining the expansion of $E_n$ as a series in $c^2$. Recall that $E_0=1$ exactly for all $c$. Meanwhile, the divergence of $\rho(E)$ in \eqref{eq:fluctuation energy density} at $E=1$ makes determining the series expansion of $E_n$ for $n\neq 0$ directly from \eqref{eq:quantization condition} a bit subtle, because naively expanding $\rho(E)$ in small $c^2$ and then integrating from $1$ to $E_n$ order by order yields divergent integrals at each order. Therefore, we rewrite \eqref{eq:quantization condition} in terms of an integral that runs from $E_n$ to infinity, instead of from $1$ to $E_n$, in order to avoid the problematic point $E=1$:\footnote{To arrive at \eqref{eq:AsUXrW6fE5}, we used the following pair of results
\begin{align}
    \int_1^{\infty}\frac{dE}{\sqrt{E^2-1}\sqrt{E^2+c^2-1}}&=\ellK(1-c^2),&\int_1^\infty \left[\frac{E^2-1}{\sqrt{E^2-1}\sqrt{E^2+c^2-1}}-1\right]dE&=1-\ellE(1-c^2).
\end{align}
}
\begin{align}\label{eq:AsUXrW6fE5}
    E_n-\int_{E_n}^\infty \left[\frac{E^2-\ellE(c^2)/\ellK(c^2)}{\sqrt{E^2-1}\sqrt{E^2+c^2-1}}-1\right]dE&=\frac{\pi}{2\ellK(c^2)}(1+|n|).
\end{align}
For fixed $E_n>1$, it is now safe to expand the integrand in small $c^2$ and then integrate order by order, which yields rational functions of $E_n$ at each order. Writing $E_n=1+|n|+E_n^{(1)} c^2+E_n^{(2)} c^4+E_n^{(3)} c^6+\ldots$, we can then determine the $E_n^{(i)}$ by matching powers of $c^2$ on both sides of \eqref{eq:AsUXrW6fE5}. We ultimately find
\begin{align}
    E_0&=1,\label{eq:SKnSaQcLM1}\\
    E_n&=1+|n|-\frac{|n|+1}{4}c^2-\frac{(|n|+1)(5n^2+10|n|-4)}{64|n|(|n|+2)}c^4\nonumber\\&\hspace{1.76cm}-\frac{(|n|+1)(11n^2+22|n|-12)}{256|n|(|n|+2)}c^6+\ldots.\hspace{3cm}(n\neq 0)\label{eq:bb9j9wLlo6}
\end{align}
The above series expansion is reliable as long as $|n|\ll 1/c^2$.

Given the series expansion of $E_n$, it is straightforward to expand the summands in \eqref{eq:G1 series}-\eqref{eq:G4 series} and \eqref{eq:G2 series}-\eqref{eq:s7VYqCQAUD} in small $c^2$ and evaluate the sums order by order. For instance, the first two terms for the small $c^2$ expansion of the defect correlators are given explicitly by:
\begin{align}
    G_1(\chi)&=1+\frac{c^2}{4}\left[\frac{(-2+\chi+\chi^3)\log|1-\chi|}{\chi}-\frac{(2+\chi(\chi-2))(1-\chi+\chi^2\log|\chi|)}{(\chi-1)^2}\right]+O(c^4),\label{eq:vcByXaNCgz}\\
    G_2(\chi)&=\frac{g\pi c^2\chi^2}{(1-\chi)^2}+\frac{c^2}{4(\chi-1)^3}\biggr[(1-\chi)(1+\chi^2)(\chi+(1-\chi)^2)\log|1-\chi|\label{eq:X8H74HHQxk}\\&\hspace{1.5cm}+\chi^3(4-3\chi+\chi^2)\log|\chi|\biggr]+O(c^4),\nonumber\\
    G_3(\chi)&=g\pi c^2 \chi^2+\frac{c^2}{4(1-\chi)^2}\biggr[\chi(1-\chi)(1-2\chi+\chi^2)\label{eq:cF7APdqE0I}\\&\hspace{1.5cm}+(1-\chi)^3(1+\chi+2\chi^2)\log|1-\chi|+\chi^3(4-5\chi+2\chi^2)\log|\chi|\biggr]+O(c^4),\nonumber\\
    G_4(\chi)&=1+c^2\left[\left(\frac{1}{2}-\frac{1}{\chi}\right)\log|1-\chi|-1\right]+O(c^4).\label{eq:7k7MA0pcKL}
\end{align}

Using the same procedure, it is relatively easy to determine the first few higher order corrections to the defect correlators. Instead of providing explicit expressions, we summarize the type of functions that appear in the expansion up to order $c^{10}$, which may potentially be useful if one were to try to identify a suitable ansatz to ```bootstrap'' these or related correlators. At order $c^{2n}$, the $G_i(\chi)$ are sums of products of the polylogarithm function $\text{Li}_s$\footnote{Recall that rational functions and the logarithm are special cases with polylogarithmic order $0$ and $1$: $\text{Li}_0(z)=\frac{z}{1-z}$ and $\text{Li}_1(z)=-\log(1-z)$} such that the sum of the polylogarithmic orders of any of the products of polylogarithmic functions is less than or equal to $n$. For instance, for the first few terms in the small $c^2$ expansion, the combinations of polylogarithms of ``highest order'' are logarithms at order $c^2$, products of pairs of logarithms at order $c^4$, and terms schematically of the form $\text{Li}_3$, $\log\cdot \text{Li}_2$ and $\log^3$ appearing at order $c^6$. However, not every product of polylogarithms ``allowed'' by the above rule appears at a given order. For instance, there is no $\text{Li}_2$ term at $c^4$, and no $\text{Li}_4$ or $\text{Li}_2^2$ terms at $c^8$.

We can also determine the small $\mathcal{J}$ expansion of the OPE data of the operators exchanged in the light-heavy channel of $\braket{ZZ^J\bar{Z}\bar{Z}^J}$, which were determined for finite $\mathcal{J}$ in \eqref{eq:ZZ^JZ^(J+1) OPE data}-\eqref{eq:ZZ^J[ZZ^J] OPE data}. The first few terms in the expansions of the conformal dimensions are:
\begin{align}
    \Delta_0&=J+1,&\Delta_n&=J+\left(2+n-\frac{n+1}{4\pi}\mathcal{J}+O(\mathcal{J}^2)\right)+O(1/g),\label{eq:oSUHyieLIp}
\end{align}
The first few terms in the expansions of the OPE coefficients are:
\begin{align}
    \frac{\mathcal{C}_{ZZ^JZ^{J+1}}\mathcal{C}_{\bar{Z}\bar{Z}^J\bar{Z}^{J+1}}}{\mathcal{N}_{Z\bar{Z}}\mathcal{N}_{Z^J\bar{Z}^J}\mathcal{N}_{Z^{J+1}\bar{Z}^{J+1}}}&=g\left(\mathcal{J}-\frac{3}{8\pi}\mathcal{J}^2+O(\mathcal{J}^3)\right)+\left(1-\frac{3\mathcal{J}}{8\pi}+O(\mathcal{J}^2)\right)+O(1/g^2)\label{eq:RcZoCxIyG4}\\
    \frac{\mathcal{C}_{ZZ^J[ZZ^J]_n}\mathcal{C}_{\bar{Z}\bar{Z}^J[\bar{Z}\bar{Z}^J]_n}}{\mathcal{N}_{Z\bar{Z}}\mathcal{N}_{Z^J\bar{Z}^J}\mathcal{N}_{[ZZ^J]_n[\bar{Z}\bar{Z}^J]_n}}&=(-1)^{n+1}\left(n+2-\frac{(n+2)(2n-1)}{4\pi n}\mathcal{J}+O(\mathcal{J}^2)\right)+O(1/g).\label{eq:0IWgDZEzm3}
\end{align}
In \eqref{eq:oSUHyieLIp} and \eqref{eq:0IWgDZEzm3}, $n=1,2,3,\ldots$. Note that $\Delta_1=\Delta_0+2$ when $\mathcal{J}=0$, which again corresponds to the fact that there is no composite primary in the generalized free field limit constructed out of $J+1$ copies of $Z$ and a single derivative (i.e., the $n=1$ operator corresponds in the free limit to a primary built out of $Z$ and $Z^J$ and two derivatives).

Finally, we note that an alternative approach to probing the defect correlators in the $1\ll J\ll g$ regime is to first take $J$ finite and expand in $1/g$ (which is the usual strong coupling expansion of the Wilson line dCFT that can be studied using Witten diagrams on the AdS$_2$ dual string), and only then take $J$ to be large. We show in Appendix~\ref{sec:dyS0wZdrxG} that the results of the finite charge analysis match the large charge results in \eqref{eq:vcByXaNCgz}-\eqref{eq:0IWgDZEzm3}.

\paragraph{Large $\mathcal{J}$.}
The other limiting case of the defect correlators that is natural to consider is the regime $1\ll g\ll J$, in which case $\mathcal{J}$ is large. The series in \eqref{eq:G1 series}-\eqref{eq:G4 series} and \eqref{eq:G2 series}-\eqref{eq:s7VYqCQAUD} allow us to determine the leading behaviors of the defect correlators as $\mathcal{J}\to \infty$ or, equivalently, $c^2\to 1$. We note for convenience the asymptotic expansions relating $\mathcal{J}$ and $c^2$ in this regime:
\begin{align}
    \mathcal{J}&=-2\log(1-c^2)+4(2\log{2}-1)+O((1-c^2)\log(1-c^2)),\\
    c^2&=1-\frac{16}{e^2}e^{-\frac{\mathcal{J}}{2}}+O(\mathcal{J}e^{-\mathcal{J}}).
\end{align}

We first determine the behavior of the energies, $E_n$, when $\mathcal{J}\to \infty$. Because $\ellK(c^2)\to \infty$ as $c\to 1$, it follows from \eqref{eq:fluctuation energy density} that, in this limit, the $E_n$ condense into a continuum with a smooth distribution. The density of the $E_n$, normalized by the large charge, is asymptotically
\begin{align}\label{eq:ZN87g8mJ3y}
    \frac{\rho(E)}{\mathcal{J}}&\sim \frac{1}{2\pi}\frac{E}{\sqrt{E^2-1}}+\frac{2}{\pi}\frac{\sqrt{E^2-1}}{E}\frac{1}{\mathcal{J}}+O(e^{-\frac{\mathcal{J}}{2}}), &&\text{as }\mathcal{J}\to \infty.
\end{align}
The fine-grained behavior of the $E_n$ when $\mathcal{J}\to \infty$ can also be determined from the above density. Neglecting the exponentially suppressed term in $\rho(E)$, which is valid as long as $|n|\ll e^{\mathcal{J}}$, we can evaluate the integral in \eqref{eq:quantization condition} and the quantization condition  becomes 
\begin{align}
    \frac{\mathcal{J}+4}{2\pi}\sqrt{E_n^2-1}-\frac{2}{\pi}\arctan\left(\sqrt{E_n^2-1}\right)+O(e^{-\mathcal{J}/2})&=|n|.
\end{align}
This lets us determine $E_n$ in various regimes. When $|n|\ll \mathcal{J}$ (i.e., $E\approx 1$), the $E_n$ may be expanded in $1/\mathcal{J}$:
\begin{align}
    E_n&=1+\frac{2\pi^2 n^2}{\mathcal{J}^2}-\frac{2\pi^4 n^4}{\mathcal{J}^4}-\frac{64\pi^4 n^4}{3\mathcal{J}^5}+\frac{4\pi^6 n^6}{\mathcal{J}^6}+O(1/\mathcal{J}^7).
\end{align}
When $\mathcal{J}\ll |n|\ll e^{\mathcal{J}}$ (i.e., $E_n\gg 1$), the $E_n$ may be expanded in $1/|n|$:
\begin{align}
    E_n&=\frac{2\pi |n|}{4+\mathcal{J}}+\frac{2\pi}{4+\mathcal{J}}+\frac{\mathcal{J}-4}{4\pi |n|}+\frac{4-\mathcal{J}}{4\pi n^2}+O(1/|n|^3).
\end{align}
Finally, when $|n|\sim \mathcal{J}$ (i.e., $E_n\sim 1$), we may write $n=\eta \mathcal{J}$ and expand in $1/\mathcal{J}$:
\begin{align}
    E_n&=\sqrt{1+4\pi^2\eta^2}+\frac{8\pi \eta(-2\pi \eta+\arctan(2\pi \eta))}{\sqrt{1+4\pi^2\eta^2}}\frac{1}{\mathcal{J}}\label{eq:W2kzjjnFHW}\\&\hspace{2cm}+\frac{8(2\pi \eta-\arctan(2\pi\eta))(2\pi \eta-\arctan(2\pi \eta)+16\pi^3\eta^3)}{(1+4\pi^2\eta^2)^{\frac{3}{2}}}\frac{1}{\mathcal{J}^2}+O(1/\mathcal{J}^3).\nonumber
\end{align}
The leading term in \eqref{eq:W2kzjjnFHW} precisely matches the expression for the energy of fluctuations about the $c^2=1$ string in the BMN limit, which was derived in \cite{drukker2006small}. In particular, \cite{drukker2006small} found that the fluctuations have energies $\omega_p=\sqrt{1+p^2}$ and argued that the boundary conditions fix the momenta to be $p_n=\frac{n\sqrt{\lambda}}{2J}$ with $n\in \mathbb{Z}$. We note that in \eqref{eq:W2kzjjnFHW}, $2\pi\eta=\frac{2\pi n g}{J}=\frac{n\sqrt{\lambda}}{2J}.$ 

\begin{figure}[t]
\centering
\includegraphics[height=7cm]{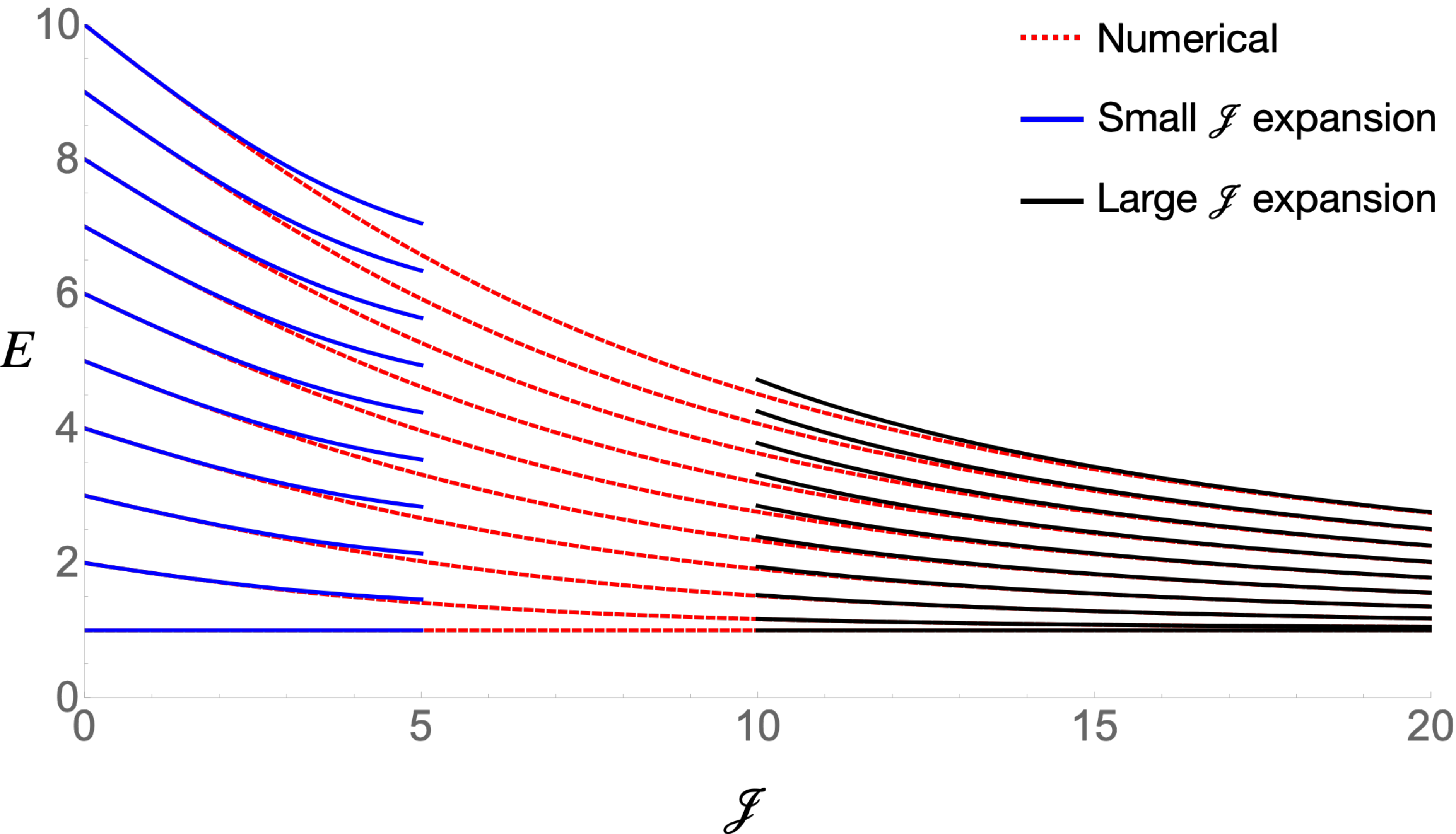}
\caption{$E_n$ is plotted for $n=0,1,2,\ldots,9$ and $\mathcal{J}\in [0,20]$. The red dashed curves depict the numerical values of $E_n$ computed using \eqref{eq:quantization condition}. The blue curves depict the small $c^2$ expansion given in \eqref{eq:bb9j9wLlo6}. The black curves depict the large $\mathcal{J}$ expansion given in \eqref{eq:W2kzjjnFHW}.}
\label{fig:En vs J}
\end{figure}

We can also study the four-point functions in the limit $\mathcal{J}\to \infty$. The condensation of the $E_n$ into a continuum makes it possible to convert the series representations of the $G_i(\chi)$ in \eqref{eq:G1 series}-\eqref{eq:G4 series} and \eqref{eq:G2 series}-\eqref{eq:s7VYqCQAUD} into simple integrals. In particular, we first note that $f(E)$ in \eqref{eq:form factor} asymptotically approaches 
\begin{align}
    f(E)&\to \frac{\pi E}{\mathcal{J}}-\frac{4\pi}{\mathcal{J}^2}\frac{E^2-1}{E}+O(e^{-\frac{\mathcal{J}}{2}}), &&\text{as }\mathcal{J}\to \infty,
\end{align}
and therefore $\rho(E)f(E)\to \frac{E^2}{2\sqrt{E^2-1}}$ is finite in the limit $\mathcal{J}\to \infty$. This is useful because, given a smooth function $g(E)$, the sum $\sum_{n\in \mathbb{Z}}f(E_n)g(E_n)$ approaches the integral $\int dE \frac{E^2}{\sqrt{E^2-1}}g(E)$ as $\mathcal{J}\to \infty$. When $\chi<1$ and the terms in the series are all positive, we can apply this to the series in \eqref{eq:G1 series} and \eqref{eq:G4 series} and find
\begin{align}
    G_1(\chi)&\to \frac{\chi^2}{1-\chi}\int_1^\infty dE \frac{E^2}{\sqrt{E^2-1}} e^{-E|\log(1-\chi)|}\nonumber\\&=\frac{\chi^2}{1-\chi}K_0''(|\log(1-\chi)|)\label{eq:XfKf2JJKZ8}\\
    G_4(\chi)&\to \frac{\chi^4}{(1-\chi)^2}\int_1^\infty dE\frac{E^2}{\sqrt{E^2-1}}\frac{E^2-1}{6}e^{-E|\log(1-\chi)|}\nonumber\\&=\frac{1}{6}\frac{\chi^4}{(1-\chi)^2}\biggr[K_0''''(|\log(1-\chi)|)-K_0''(|\log(1-\chi)|)\biggr].
\end{align}
Similarly, from the piecewise series representations of $G_{Z\bar{Z}}$ and $G_{\bar{Z}Z}$ in \eqref{eq:G2 series}-\eqref{eq:s7VYqCQAUD}, we also find
\begin{align}
    G_{Z\bar{Z}}(\chi)&\to g\pi \frac{\chi^2}{(1-\chi)^2}+\frac{\chi^2}{1-\chi}\int_1^\infty \frac{E^2}{\sqrt{E^2-1}}\frac{E\pm 1}{E}e^{-(E\pm 1)|\log(1-\chi)|}\nonumber\\&=g\pi \frac{\chi^2}{(1-\chi)^2}+\frac{\chi^2}{(1-\chi)^2}\biggr[K_0''(|\log(1-\chi)|)\mp K_0'(|\log(1-\chi)|)\biggr],\\
    G_{\bar{Z}Z}(\chi)&\to g\pi \chi^2+\frac{\chi^2}{1-\chi}\int_1^\infty \frac{E^2}{\sqrt{E^2-1}}\frac{E\mp 1}{E}e^{-(E\mp 1)|\log(1-\chi)|}\nonumber\\
    &=g\pi \chi^2+\chi^2\biggr[K_0''(|\log(1-\chi)|)\pm K_0'(|\log(1-\chi)|)\biggr],
\end{align}
where the upper (lower) sign is for $\chi<0$ ($0<\chi<1$). In these expressions, $K_0(z)$ is the modified Bessel function of the second kind.

By contrast, when $\chi>1$, the series become alternating and vanish in the continuum limit.\footnote{The different behavior of the $\mathcal{J}\to \infty$ limit of the $G_i(\chi)$ for $\chi<1$ and $\chi>1$ is analogous to the following simple example: Given a normalizable, smooth function $g:\mathbb{R}\to \mathbb{R}$ (e.g., $g(x)=e^{-x^2}$), the normalized sum of $g$ evaluated at uniformly spaced points separated by $1/N$ approaches either the integral of $g$ over $\mathbb{R}$ or zero as $N\to \infty$ depending on whether the sum is alternating: 
\begin{align}
\lim_{N\to \infty} \frac{1}{N}\sum_{n\in \mathbb{Z}}g\left(\frac{n}{N}\right)&= \int_{-\infty}^\infty g(x)dx, & \lim_{N\to \infty} \frac{1}{N}\sum_{n\in \mathbb{Z}}(-1)^n g\left(\frac{n}{N}\right)&=0.
\end{align}
} Thus, we find the following simplified asymptotic behavior as $\mathcal{J}\to \infty$:
\begin{align}\label{eq:h9fQgPQQXp}
    G_1(\chi),G_4(\chi)&\to 0, &
    G_{Z\bar{Z}}(\chi)&\to g\pi \frac{\chi^2}{(1-\chi)^2}, &
    G_{\bar{Z}Z}(\chi)&\to g\pi \chi^2. 
\end{align}
These can also be seen from the fact that the $\sinh$ term in the integrands in \eqref{eq:s4xpFVW5pB},\eqref{eq:X8lEuKtW9m}-\eqref{eq:AKRvj3tMJx} and \eqref{eq:Wxx c=0 from general c exp} diverges as $c^2\to 1$.

\begin{figure}[t]
\centering
\includegraphics[width=\hsize]{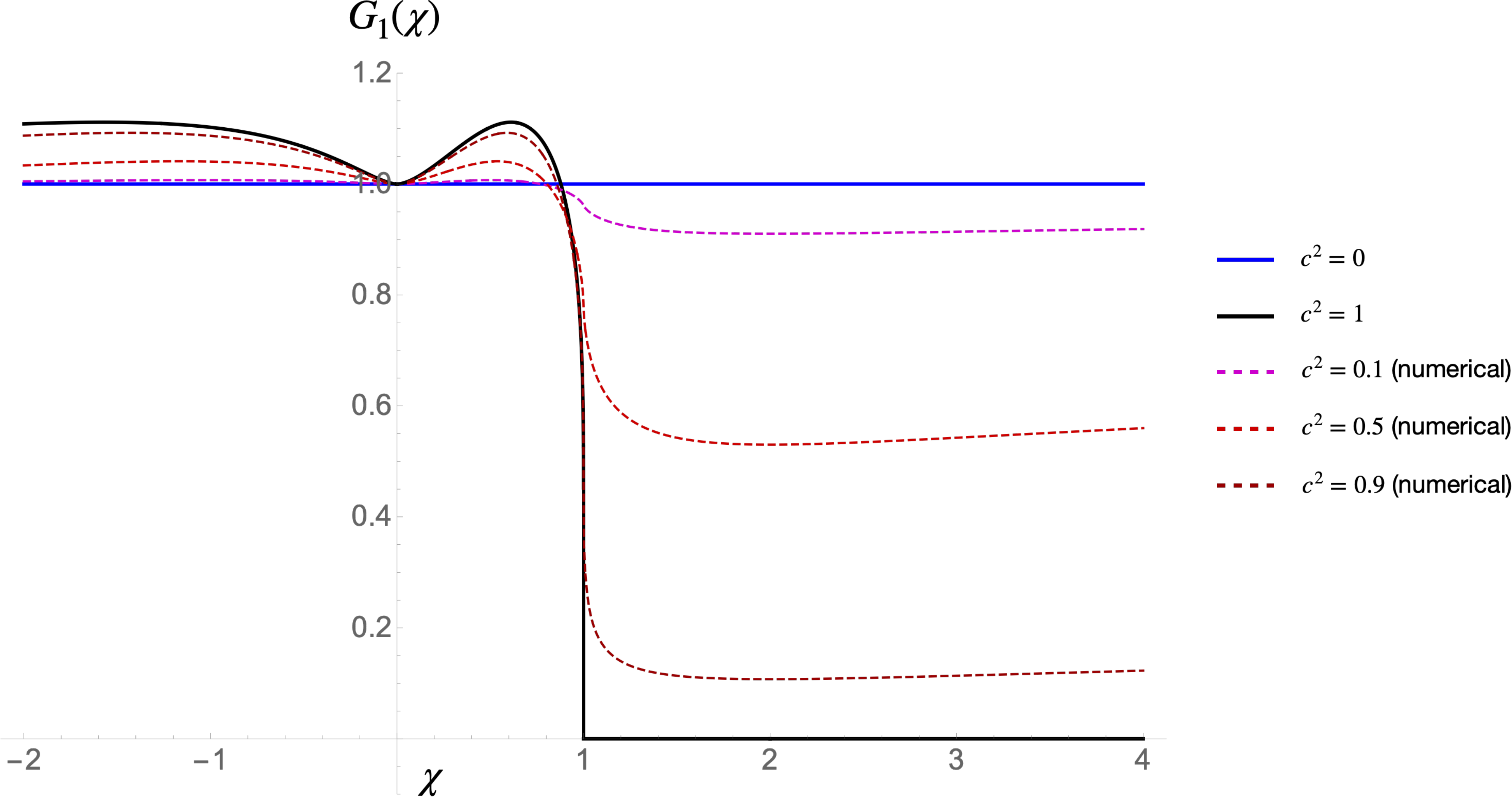}
\caption{$G_1(\chi)$ is plotted as a function of $\chi$ for representative values of $c^2$. The solid blue curve depicts $G_1(\chi)=1$, corresponding to the edge case $c^2=0$ (i.e., $\mathcal{J}=0$). The solid black curve depicts $G_1(\chi)$ for the edge case $c^2=1$ (i.e., $\mathcal{J}=\infty$), given by \eqref{eq:XfKf2JJKZ8} and \eqref{eq:h9fQgPQQXp}. The dashed curves depict $G_1(\chi)$ for $c^2=0.1,0.5,0.9$, evaluated numerically using the series representation in \eqref{eq:G1 series} for $\chi<1$ and using the integral representation in \eqref{eq:s4xpFVW5pB} for $\chi>1$. The numerical results on a small interval around $\chi=0$ are excluded due to slow convergence of the series. One can also plot $G_1(\chi)$ for $\chi>1$ using the series in \eqref{eq:s4xpFVW5pB}; the resulting curve matches the curve from the integral representation, but convergence is slow near $\chi=2$.}
\label{fig:G1 vs chi}
\end{figure}

\begin{figure}[t]
\centering
\includegraphics[width=\hsize]{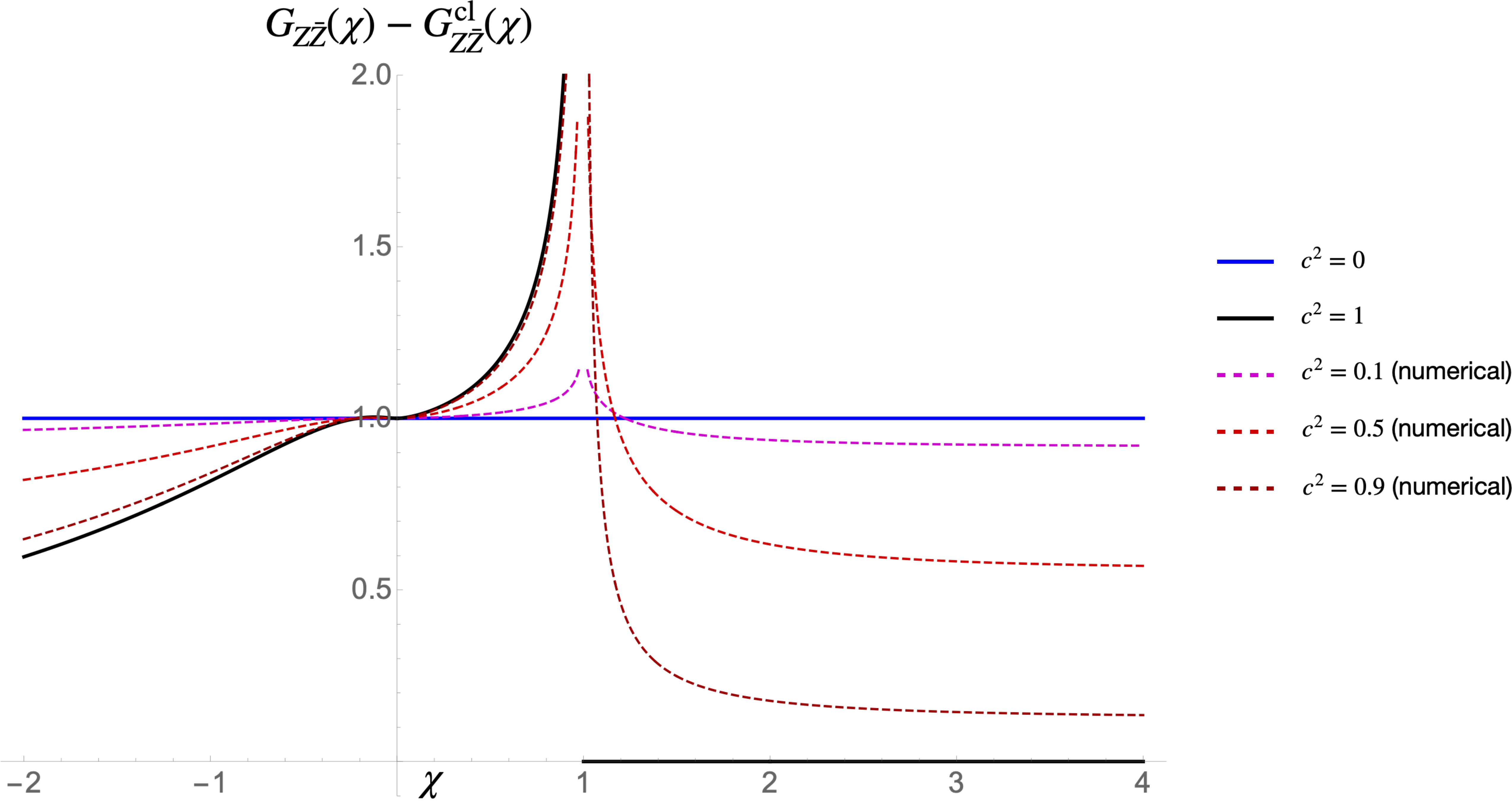}
\caption{The fluctuation component of $G_{Z\bar{Z}}(\chi)$ is plotted as a function of $\chi$ for representative values of $c^2$.}
\label{fig:G1+G2 vs chi}
\end{figure}

\begin{figure}[t]
\centering
\includegraphics[width=\hsize]{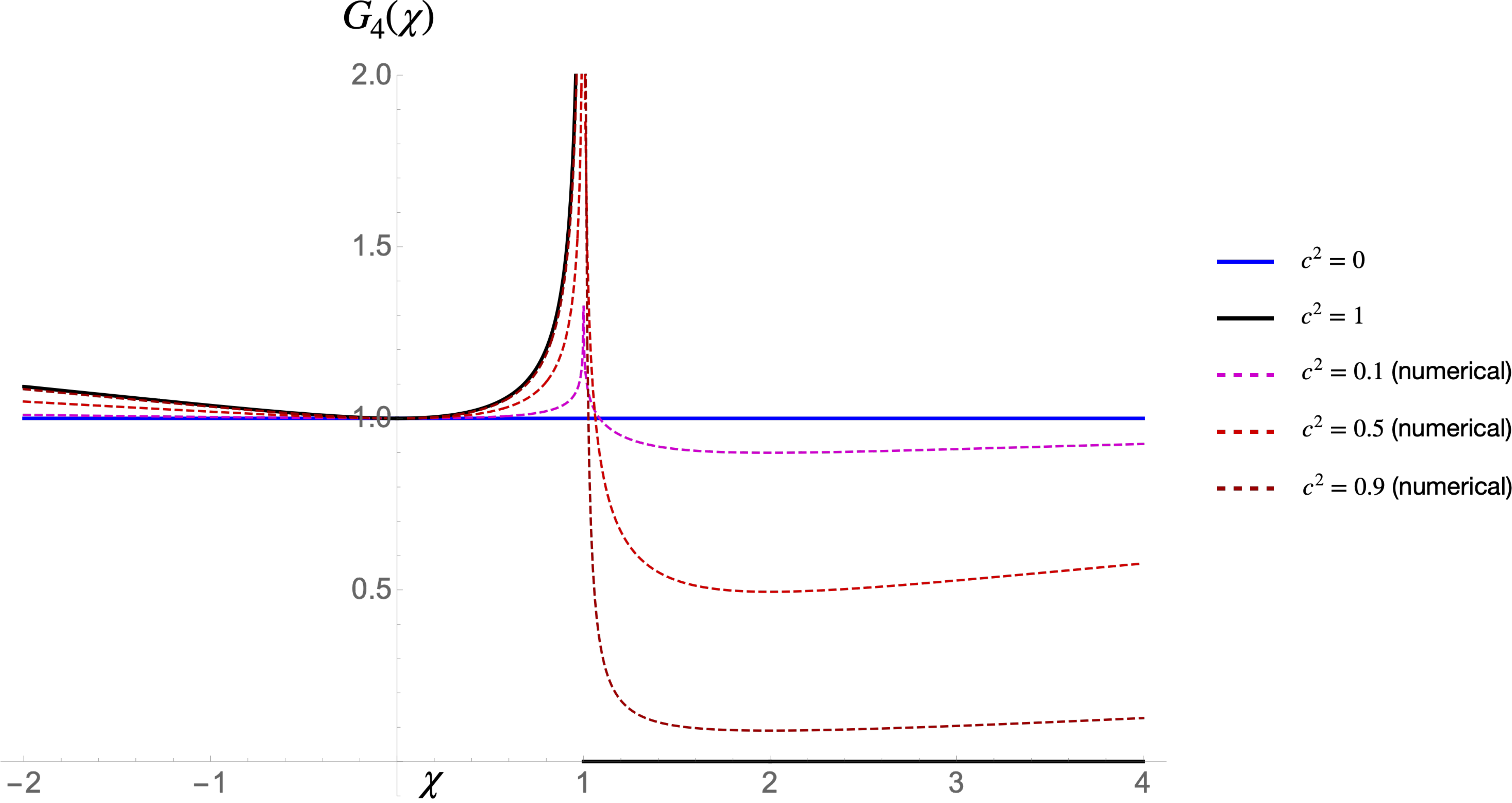}
\caption{$G_4(\chi)$ is plotted as a function of $\chi$ for representative values of $c^2$.}
\label{fig:G4 vs chi}
\end{figure}

Finally, we note the large $\mathcal{J}$ expansion of the OPE data of the operators exchanged in the light-heavy channel of $\braket{ZZ^J\bar{Z}\bar{Z}^J}$, which were determined for finite $\mathcal{J}$ in \eqref{eq:ZZ^JZ^(J+1) OPE data}-\eqref{eq:ZZ^J[ZZ^J] OPE data}. The leading terms in the expansions of the conformal dimensions are:
\begin{align}
    \Delta_0&=J+1,&\Delta_n&=J+\left(1+\sqrt{1+4\pi^2\eta^2}+O(1/\mathcal{J})\right)+O(1/g),\label{eq:bDdLaHv7rN}
\end{align}
The leading terms in the expansions of the OPE coefficients are:
\begin{align}
    \frac{\mathcal{C}_{ZZ^JZ^{J+1}}\mathcal{C}_{\bar{Z}\bar{Z}^J\bar{Z}^{J+1}}}{\mathcal{N}_{Z\bar{Z}}\mathcal{N}_{Z^J\bar{Z}^J}\mathcal{N}_{Z^{J+1}\bar{Z}^{J+1}}}&=g\pi \left(1+O(e^{-\frac{\mathcal{J}}{2}})\right)+\biggr(\frac{3}{8}+O(e^{-\frac{\mathcal{J}}{2}})\biggr)+O(1/g)\label{eq:QVFOEx8eRV}\\
    \frac{\mathcal{C}_{ZZ^J[ZZ^J]_n}\mathcal{C}_{\bar{Z}\bar{Z}^J[\bar{Z}\bar{Z}^J]_n}}{\mathcal{N}_{Z\bar{Z}}\mathcal{N}_{Z^J\bar{Z}^J}\mathcal{N}_{[ZZ^J]_n[\bar{Z}\bar{Z}^J]_n}}&=(-1)^{n+1}\left(\frac{2\pi}{\mathcal{J}}\left(1+\sqrt{1+4\pi^2\eta^2}\right)+O(1/\mathcal{J}^2)\right)+O(1/g).\label{eq:lmm9OYOjtR}
\end{align}
Eq.~\eqref{eq:bDdLaHv7rN} and \eqref{eq:lmm9OYOjtR} are valid for $\eta\equiv n/\mathcal{J}\sim 1$.

Having determined exact expressions for the energies and the four-point functions (which can be evaluated numerically with relative ease), as well as simplified analytic results when $\mathcal{J}$ is small or large, we close this section by summarizing our findings in a series of figures. Figure~\ref{fig:En vs J} illustrates the dependence of the first few energies $E_n$ on $\mathcal{J}$. We include the exact results evaluated numerically via the quantization condition \eqref{eq:quantization condition}, and compare them with the small $c^2$ expansion in \eqref{eq:bb9j9wLlo6} and the large $\mathcal{J}$ expansion in \eqref{eq:W2kzjjnFHW}. Furthermore, Figures~\ref{fig:G1 vs chi}-\ref{fig:G4 vs chi} depict $G_1(\chi)$, $G_{Z\bar{Z}}(\chi)$, and $G_4(\chi)$ for representative values of $c^2$, combining the simple analytic results for $c^2=0$ given in \eqref{eq:vcByXaNCgz}-\eqref{eq:7k7MA0pcKL} and for $c^2=1$ given in \eqref{eq:XfKf2JJKZ8}-\eqref{eq:h9fQgPQQXp} with the numerical implementation of the series representations given in \eqref{eq:G1 series}-\eqref{eq:G4 series} and \eqref{eq:G2 series}-\eqref{eq:s7VYqCQAUD}.

\section{Connections to integrability}\label{sec:integrability}
We now show that the results obtained in Section \ref{sec:series representations of 4pt functions} can be interpreted naturally in terms of the (semi-)classical integrability of the string sigma model. 
\subsection{Classical integrability, spectral curve, and quasi-momentum}
Let us first give a lightening  review of the classical integrability of the closed string. For more details, we refer the readers to the original papers \cite{Kazakov:2004qf,Beisert:2005bm,Gromov:2008ec} or a review \cite{Schafer-Nameki:2010qho}.

\paragraph{Spectral curve, quasi-momenta.}The Green-Schwarz string sigma model on AdS$_5\times S^5$ is known to be classically integrable. The simplest way to see it is to recast its equations of motion into the flatness of the {\it Lax connection}, which takes the following schematic form
\begin{align}\label{eq:Lax}
[\del_{\sigma}-J_{\sigma}\comma \del_{\tau}-J_{\tau}]=0\period
\end{align}
Here $J_{\sigma,\tau}(\sigma,\tau|x)$ is a $(4|4)\times (4|4)$ matrix-valued one-form built out of the worldsheet fields, and it depends on an extra complex parameter $x$, which we call the {\it spectral parameter}. Using the Lax connection, one can define a {\it monodromy matrix} along a nontrivial contour on the worldsheet, which is normally taken to be along the $\sigma$-direction:
\beq
\Omega (x)\equiv {\rm P}\exp\left(\int_{0}^{2\pi}J_{\sigma} d\sigma\right)\period
\eeq
Thanks to the flatness of the connection \eqref{eq:Lax}, the trace of the monodromy matrix does not depend on the local deformation of the contour and is therefore conserved for any value of $x$. In particular, when expanded in powers of $x$, it produces infinitely many conserved charges.

A more systematic way to compute the conserved charges and analyze the integrable structure is to consider the spectral curve defined by\footnote{Here ${\rm sdet}$ stands for the super-determinant.}
\beq
{\rm sdet}\left[y{\bf 1}-\Omega(x)\right]=0\period
\eeq
A standard way to parametrize the curve is to use the {\it quasi-momenta}, which are related to eigenvalues of $\Omega (x)$ by $y_i(x) =e^{ip_i (x)}$. More explicitly we have eight quasi-momenta
\beq
\Omega (x)\sim {\rm diag}\left(e^{i\hat{p}_1},e^{i\hat{p}_2},e^{i\hat{p}_3},e^{i\hat{p}_4}|e^{i\tilde{p}_1},e^{i\tilde{p}_2},e^{i\tilde{p}_3},e^{i\tilde{p}_4}\right)\comma
\eeq
where $\hat{p}_j$'s describe the dynamics in AdS$_5$ while $\tilde{p}_j$'s describe the dynamics in $S^5$. These eight quasimomenta together form an eight-sheeted covering of the complex $x$ plane, connected to each other by branch cuts (or poles\footnote{Precisely speaking, the cuts corresponding to fermionic excitations in \eqref{eq:polarizations} are infinitesimal: namely they are poles rather than branch cuts. For details, see \cite{Gromov:2008ec}. Here we are slightly abusing the notations and calling them ``branch cuts.''}). The branch cuts connect two of the eight sheets, and across each branch cut, the quasi-momenta have the following ``integer'' discontinuities:
\beq\label{eq:discontinuity}
p_I (x+i\epsilon)-p_J(x-i\epsilon)=2\pi n_{IJ}\comma\qquad n_{IJ}\in \mathbb{N}\period
\eeq 
Here $I$ and $J$ belong to the sets
\beq\label{eq:indices}
I\in \{\tilde{1},\tilde{2},\hat{1},\hat{2}\}\comma\qquad J\in \{\tilde{3},\tilde{4},\hat{3},\hat{4}\}\period
\eeq
As can be seen in \eqref{eq:discontinuity} and \eqref{eq:indices}, there are $16$ different kinds of branch cuts and they correspond to eight bosonic and eight fermionic excitations on the worldsheet:
\beq\label{eq:polarizations}
{\rm AdS}_5:\quad (\hat{i},\hat{j})\comma \qquad {\rm S}^5: \quad (\tilde{i},\tilde{j})\comma\qquad {\rm fermions}:  \quad (\tilde{i},\hat{j}) \,\,\text{or}\,\,(\hat{i},\tilde{j})\period
\eeq
In addition to branch cuts, the quasi-momenta have poles at $x=\pm1$ whose residues are correlated owing to the Virasoro constraints
\beq
\{\hat{p}_1,\hat{p}_2,\hat{p}_3,\hat{p}_4|\tilde{p}_1\,\tilde{p}_2,\tilde{p}_3,\tilde{p}_4\}\sim\frac{\{\alpha_{\pm},\alpha_{\pm},\beta_{\pm},\beta_{\pm}|\alpha_{\pm},\alpha_{\pm},\beta_{\pm},\beta_{\pm}\}}{x\pm 1}\period
\eeq
This property plays an important role when determining semi-classical corrections to the energy as we review below.

The main advantage of using the spectral curve and the quasi-momenta is that one can classify and construct them directly from their analytic properties without reference to explicit classical string solutions. Since the quasi-momenta encode all the higher conserved charges, it allows one to compute the quantum numbers of classical string solutions without constructing them explicitly. In particular, the global charges of the classical solution can be read off from the asymptotic behavior of the quasi-momenta at infinity:
\beq\label{eq:asymptps}
\pmatrix{c}{\hat{p}_1\\\hat{p}_2\\\hat{p}_3\\\hat{p}_4\\\tilde{p}_1\\\tilde{p}_2\\\tilde{p}_3\\\tilde{p}_4}=\frac{1}{2gx}\pmatrix{c}{+\Delta-S_1+S_2\\+\Delta+S_1-S_2\\-\Delta-S_1-S_2\\-\Delta+S_1+S_2\\+J_1+J_2-J_3\\+J_1-J_2+J_3\\-J_1+J_2+J_3\\-J_1-J_2-J_3}+O(1/x^2)\period
\eeq
Here $S_j$'s and $J_j$'s are angular momenta in AdS$_5$ and $S^5$ respectively. Using this, we can read off the conformal dimension of the classical solution in the following way:
\beq\label{eq:readingoffdelta}
\Delta=\lim_{x\to\infty}g x(\hat{p}_1+\hat{p}_2)\period
\eeq
\paragraph{Semi-classical fluctuations and quasi-energy.} Another advantage of the spectral curve is that it also encodes semi-classical corrections to the energy of classical string solutions. The details can be found in the original paper \cite{Gromov:2008ec}; here we sketch the outline of the derivation.

On the spectral curve, all the excitations on the worldsheet are described by branch cuts. Thus small perturbations, i.e.~semi-classical fluctuations, correspond to adding ``infinitesimal branch cuts''---namely poles---on the spectral curve. These poles also need to obey the analyticity requirements reviewed above; in particular, they can only be added at $x_n^{IJ}$ satisfying the following ``quantization'' condition:
\beq\label{eq:quantization}
p_I(x_n^{IJ})-p_J(x_n^{IJ})=2\pi n\qquad n\in\mathbb{N}\period
\eeq
After adding a pole, the quasi-momenta get shifted as $p_K(x)\to p_K(x)+\delta_{n}^{IJ}p_K(x)$. The residue of the pole is constrained to be
\beq
\delta_{n}^{IJ}p_I(x) \sim \pm \frac{\alpha (x_n^{IJ})}{x-x_n^{IJ}}\comma\qquad \alpha (x)=\frac{1}{g(1-1/x^2)}\period
\eeq
Physically, this constraint guarantees that the perturbation corresponds to adding one unit of quanta\fn{More precisely, it comes from requiring that the so-called filling fraction gets shifted by one. See \cite{Gromov:2008ec} for details.} on the worldsheet, rather than a composite of several quanta.

There are several other conditions satisfied by $\delta_n^{IJ}p_{K}$ which directly follow from the analyticity requirements of the quasi-momenta: For instance, $\delta_n^{IJ}p_I(x)$ should not modify the integer discontinuities \eqref{eq:discontinuity} of the existing cuts, and their residues at $x=\pm 1$ are correlated to be
\beq\label{eq:quasiflucpole}
\{\delta\hat{p}_1,\delta\hat{p}_2,\delta\hat{p}_3,\delta\hat{p}_4|\delta\tilde{p}_1\,\delta\tilde{p}_2,\delta\tilde{p}_3,\delta\tilde{p}_4\}\sim\frac{\{\delta\alpha_{\pm},\delta\alpha_{\pm},\delta\beta_{\pm},\delta\beta_{\pm}|\delta\alpha_{\pm},\delta\alpha_{\pm},\delta\beta_{\pm},\delta\beta_{\pm}\}}{x\pm 1}\period
\eeq
Because of this latter condition, the perturbation $\delta_n^{IJ}$ affects all the $p_K$'s, not just $p_I$ and $p_J$. In particular, any perturbation backreacts $\hat{p}_1$ and $\hat{p}_2$, and therefore changes the conformal dimension $\Delta $, which is determined by the asymptotics of $\hat{p}_{1,2}$.

In the actual computation, it is often convenient to compute the correction to $\Delta$ in two steps: first determine the correction to the quasi-momenta induced by the addition of a pole at an arbitrary position $x=y$ and later impose the quantization condition \eqref{eq:quantization} $y=x_n^{IJ}$. As a result of the first step, we obtain an expression for $\delta \Delta$ as a function of the position $y$:
\beq
q^{IJ}(y)\equiv \delta_{y}^{IJ}\Delta\period
\eeq
Here $\delta_y^{IJ}$ denotes a perturbation induced by an addition of a pole at $x=y$ that connects the $I$-th and $J$-th sheets. The function $q(y)$ is often called the {\it off-shell frequency} \cite{Gromov:2008ec} or the {\it quasi-energy} \cite{Dorey:2006zj} in the literature. In this paper, we adopt the latter convention.

\paragraph{$SU(2)$ two-cut solutions and the Wilson loop.} The explicit forms of the quasi-energies were written down in \cite{Gromov:2008ec} for the so-called $SU(2)$ symmetric two-cut solutions. The general $SU(2)$ solutions are dual to operators in $\mathcal{N}=4$ SYM which are made out of two complex scalar fields ($Z$ and $Y$), and describe string solutions that have nontrivial dynamics on $S^3$ inside $S^5$ but are point-like in AdS. In terms of the quasi-momenta, they are characterized by the fact that only $\tilde{p}_2$ and $\tilde{p}_3$ are connected by branch cuts outside the unit circle, while only $\tilde{p}_1=\tilde{p}_4$ are connected inside the unit circle. As a consequence, the quasi-momenta satisfy
\beq
\tilde{p}_2=-\tilde{p}_3\comma\qquad \tilde{p}_1=-\tilde{p}_4\comma\qquad \hat{p}_{1,2}=-\hat{p}_{3,4}=\frac{\Delta x}{2g(x^2-1)}\period
\eeq
The {\it symmetric two-cut} solutions are a subclass of such solutions which contain only two branch cuts that are symmetric around the origin (we take them to be on $[a, a^{\ast}]$ and $[-a^{\ast},-a]$ with $a^{\ast}$ being the complex conjugate of $a$). For such solutions, the quasi-energies for the $S^5$ excitations are given by
\beq\label{eq:2cut}
\begin{aligned}
q^{\tilde{2}\tilde{3}}(y)&=\frac{4}{b(1)+b(-1)}\left(\frac{b(y)}{y^2-1}-1\right)\comma\\
q^{\tilde{1}\tilde{4}}(y)&=-q^{\tilde{2}\tilde{3}}(1/y)+q^{\tilde{2}\tilde{3}}(0)\comma\\
q^{\tilde{2}\tilde{4}}(y)&=q^{\tilde{1}\tilde{3}}(y)=\frac{1}{2}\left(q^{\tilde{2}\tilde{3}}(y)+q^{\tilde{1}\tilde{4}}(y)\right)\comma
\end{aligned}
\eeq 
with $b(y)\equiv \sqrt{(y-a)(y-a^{\ast})(y+a)(y+a^{\ast})}$.

In \cite{Gromov:2012eu}, it was pointed out that the $S^3$ part of the quasi-momenta for the large charge insertions on the Wilson loop coincide with those of the $SU(2)$ two-cut solutions upon setting $a=e^{i\theta_0}$, where $\theta_0$ is related to the parameter $c$ used in sections \ref{sec:fluctuations about the classical string} and \ref{sec:series representations of 4pt functions} in the following way:
\beq
\sin \theta_0=c\period
\eeq
Performing the same substitution, we get the following quasi-energy $q^{\tilde{2}\tilde{3}}$
\beq\label{eq:substitution}
q^{\tilde{2}\tilde{3}}(y)\quad \to \quad \frac{1}{\sin\theta_0}\left(\frac{\sqrt{(y-e^{i\theta_0})(y-e^{-i\theta_0})(y+e^{i\theta_0})(y+e^{-i\theta_0})}}{y^2-1}-1\right)\period
\eeq
Note that the expressions for the quasi-energies in \eqref{eq:2cut} were derived for closed string solutions while the insertions on the Wilson loops correspond to open string solutions with nontrivial boundary conditions. Thus, the substitution \eqref{eq:substitution} is not guaranteed to give a correct answer for the Wilson loop (even though the quasi-momenta happen to coincide). Nevertheless, we will see below that the conformal dimensions of operators in the heavy-light channel on the Wilson loop defect CFT can be recast into a form almost identical to \eqref{eq:substitution}.
\subsection{Four-point functions and the spectral curve}
We now relate the results for the four-point function in Section \ref{sec:series representations of 4pt functions} to quantities on the spectral curve. For simplicity, we focus on $G_1$, the scalar four-point function corresponding to the $y$-excitation ($W_{yy}$). Generalizations to other four-point functions should be possible but we leave them for future investigations.
\paragraph{Quantization condition and the quasi-momenta.} In Section \ref{sec:series representations of 4pt functions}, we showed that $G_1$ can be expressed as a sum over contributions from poles $\alpha_n$ in the complex $\alpha$ plane satisfying the quantization condition,
\beq\label{eq:quantizationF}
F(\alpha_n)=n\pi i\comma\qquad \qquad F(\alpha)\equiv \frac{\pi i \alpha}{\mathbb{K}}+2(\mathbb{K}+i\mathbb{K}^{\prime})Z(\alpha|1/c^2)\comma
\eeq
with $\mathbb{K}= \mathbb{K}(\tfrac{1}{c^2})$ and $\mathbb{K}^{\prime}= \mathbb{K}(1-\tfrac{1}{c^2})$. We now show that this is equivalent to the quantization condition on the quasi-momentum.

The starting point of our analysis is the quasi-momentum $\tilde{p}_2=-\tilde{p}_3$ identified in \cite{Gromov:2012eu}, which coincides with the quasi-momentum for the $SU(2)$ symmetric two-cut solution with $a=e^{i\theta_0}$:
\beq\label{eq:GSquasi}
\begin{aligned}
\tilde{p}_2(x)&=\pi +2\mathbb{K}(c^2)\sqrt{\frac{1-x^{2}e^{2i\theta_0}}{-x^2+e^{2i\theta_0}}}\left(\frac{2i x\sin\theta_0}{x^2-1}-1\right)+\frac{4\mathbb{E}(c^2)}{\cos\theta_0}F_1-4\cos\theta_0 \mathbb{K}(c^2)F_2\comma\\
F_1&=i\mathbb{F}\left[\sin^{-1}\sqrt{\frac{(e^{2i\theta_0}+1)(e^{i\theta_0}-x)}{(e^{2i\theta_0}-1)(e^{i\theta_0}+x)}};-\tan^2\theta_0\right]\comma\\
F_2&=i\mathbb{E}\left[\sin^{-1}\sqrt{\frac{(e^{2i\theta_0}+1)(e^{i\theta_0}-x)}{(e^{2i\theta_0}-1)(e^{i\theta_0}+x)}};-\tan^2\theta_0\right]\period
\end{aligned}
\eeq
(Recall that $c$ and $\theta_0$ are related by $c=\sin\theta_0$.)
At first sight, the quasi-momentum \eqref{eq:GSquasi} seems unrelated to the function $F(\alpha)$ in \eqref{eq:quantizationF}, but using several identities of elliptic functions, one can rewrite it as follows\footnote{The simplest way to verify the equality between \eqref{eq:GSquasi} and \eqref{eq:GSquasi2} is to evaluate both numerically (e.g.~using Mathematica). Alternatively, one can prove the equivalence using identities that can be found in \cite{abramowitz1988handbook,NIST:DLMF,whittaker_watson_1996,stein2010complex,MR1007595}.}:
\beq\label{eq:GSquasi2}
\begin{aligned}
\tilde{p}_2(x)&=\frac{\pi}{\mathbb{K}}A(x)-2i (\mathbb{K}+i\mathbb{K}^{\prime})Z\left(A(x)|1/c^2\right)\comma\\
A(x)&=\mathbb{F}\left[\sin^{-1}\left(\frac{2i x\sin\theta_0}{1-x^2}\right);\frac{1}{c^2}\right]\period
\end{aligned}
\eeq 
In this form, it is clear that the quasi-momentum and the function $F$ are related by
\beq
F(A(x))=i\tilde{p}_2(x)\comma
\eeq
and the quantization condition \eqref{eq:quantizationF} coincides precisely with the quantization condition discussed in \cite{Gromov:2008ec}; $\tilde{p}_2(x)-\tilde{p}_3(x)=2n\pi$.

Thus the series representation of the four-point functions can be identified with a sum over contributions from points on the spectral curve satisfying the quantization condition. The basic physical reason of this coincidence is the following; the points on the spectral curve satisfying the quantization condition are known to correspond to normalizable fluctuations around the classical string solution \cite{Gromov:2008ec} while we saw in Section \ref{sec:series representations of 4pt functions} that the poles in the $\alpha$ plane correspond to bound states in the one-dimensional Schr\"odinger problem that describes linear fluctuations on the worldsheet. The two are basically the same concepts and therefore it is natural that they are identified. 
\paragraph{Conformal dimensions and quasi-energies.} Having identified the quantization conditions, we now consider the conformal dimensions of the operators exchanged in the heavy-light channel, which are given by
\beq
E_n\left(=\Delta_n-J\right)={\rm cn}(\alpha_n|1/c^2)\period
\eeq
Surprisingly, this takes a simple form when expressed in terms of the spectral curve using the relation $\alpha =A(x)$:
\beq\label{eq:openquasienergy}
E(x)\equiv {\rm cn}\left( A(x)\big| 1/c^2\right)=\frac{\sqrt{(x-e^{i\theta_0})(x-e^{-i\theta_0})(x+e^{i\theta_0})(x+e^{-i\theta_0})}}{x^2-1}\period
\eeq
The right hand side of this equation is almost identical to the expression for the quasi-energy $q^{\tilde{2}\tilde{3}}$ in \eqref{eq:substitution}. The differences are that $E(x)$ does not come with an overall factor $1/\sin\theta_0$ or the subtraction term $-1$.

Physically, the differences should come from the fact that the solution describing the Wilson loop has a nontrivial profile in AdS while that of the $SU(2)$ symmetric closed string is point-like. To derive \eqref{eq:openquasienergy} rigorously, one has to develop a spectral-curve description of open strings corresponding to the Wilson loop, which has not been done in the literature.\fn{The spectral curve for open strings satisfying a different boundary condition has been constructed in \cite{Bajnok:2013sza}.} Leaving this as an important future problem, here we present one possible scenario that would lead to \eqref{eq:openquasienergy}. Let us consider the following perturbation of the quasi-momentum $\tilde{p}_2$,
\beq
\delta \tilde{p}_2(x)=\frac{1}{b(x)}\left(-\frac{b(y)\alpha (y)}{x-y}+\frac{c^{\prime} b(1)}{x-1}+\frac{c^{\prime} b(1)}{x+1}+\frac{b(y)}{y}\right)+\frac{1}{2 g x}\comma
\eeq
with $b(x)=(x-e^{i\theta_0})(x+e^{i\theta_0})(x-e^{-i\theta_0})(x+e^{-i\theta_0})$ and
$c^{\prime}\equiv \frac{b(y)}{4g \sin\theta_0 (y^2-1)}$.
The perturbation adds a pole at $x=y$ with a required residue, satisfies all the properties reviewed above, and changes the charge $J_3$ by one unit (see \eqref{eq:asymptps}). This perturbation of $\tilde{p}_2$ backreacts on the AdS quasi-momenta $\hat{p}_2$ through the relation \eqref{eq:quasiflucpole}. To determine how precisely it backreacts, we need to know the unperturbed $\hat{p}_2$ describing the large charge insertions on the Wilson loop, which has not been worked out yet. Here we make an assumption\fn{This is motivated by the fact that the AdS part of the solution can be obtained by the analytic continuation of the sine-Gordon soliton, which describes the $S^3$-part of the solution (see the discussion in Section 2.2 of \cite{Correa:2012hh}).} that the unperturbed $\hat{p}_2$ also has two symmetric branch cuts $[e^{i\theta_0},e^{-i\theta_0}]$ and $[-e^{-i\theta_0},-e^{-i\theta_0}]$. Under this assumption, we write a general ansatz
\beq
\delta\hat{p}_2(x)=\frac{1}{b(x)}\left(\frac{c^{\prime} \, b(1)}{x-1}+\frac{c^{\prime}\, b(1)}{x+1}+\frac{b(y)}{y}+\tilde{c}x\right)\comma
\eeq
where $\tilde{c}$ is a constant which is to be determined. To determine $\tilde{c}$, we use the fact that $\hat{p}_2$ and $\hat{p}_1$ are related by the inversion transformation (see \cite{Gromov:2008ec})
\beq\label{eq:inversionp}
\hat{p}_1(x)=-\hat{p}_2(1/x) \comma
\eeq
and their asymptotics are both given by the conformal dimension $\Delta$, when the solution does not carry Lorentzian spins:
\beq
\hat{p}_1(x)(=-\hat{p}_2(1/x))\sim \hat{p}_2 (x)\sim \frac{\Delta}{2g x}\qquad \qquad (x\to \infty)\period
\eeq
Imposing this condition also on the perturbation $\delta\hat{p}_2(x)$, we can fix $\tilde{c}$ to be
\beq
\tilde{c}=\frac{b(y)}{2g(y^2-1)}\period
\eeq
We can then read off the correction to the conformal dimension from $\delta \tilde{p}_2$ by reading off the behavior at infinity:
\beq
\delta_y \Delta=\frac{b(y)}{y^2-1}=\frac{\sqrt{(y-e^{i\theta_0})(y-e^{-i\theta_0})(y+e^{i\theta_0})(y+e^{-i\theta_0})}}{y^2-1}\period
\eeq
This coincides with $E(x)$, suggesting that $E(x)$ can be interpreted as the quasi-energy. It is an important future problem to construct the spectral curve for open strings and rigorously show the equivalence of the two quantities.
\paragraph{Form factors as functions on the spectral curve.} We can also rewrite the coefficients of the series expansion, which we called the form factors in \eqref{eq:form factor}, as functions on the spectral curve. This can be achieved simply by substituting \eqref{eq:openquasienergy} into \eqref{eq:form factor}:
\begin{align}
\mathfrak{f}(x)\equiv& f(E(x))\\
=&\frac{\pi \sin^2 \theta_0}{4}\frac{(1+x^2)^2}{1-x^2}\frac{b(x)}{\mathbb{E}(\sin^2\theta_0)(x^{2}-1)^2-\mathbb{K}(\sin^2\theta_0) (x^{4}+1)+2\mathbb{K}(\sin^2\theta_0) \cos 2\theta_0 x^2}\period\nn
\end{align}
As shown above, the form factor $\mathfrak{f}$ is a product of a rational function in $x$ and $b(x)$, which gives rise to square-root branch cuts. In addition, it satisfies an ``inversion symmetry''
\beq
\mathfrak{f}(x)=-\mathfrak{f}(1/x)\comma
\eeq
which is reminiscent of the analogous relations for the quasi-momenta \eqref{eq:inversionp}. These features suggest that it  might be possible to directly compute $\mathfrak{f}(x)$ from the spectral curve. As we saw in the previous section, the form factors are essentially the structure constants of OPE. A method to compute the structure constants using the spectral curve and the classical integrability of the sigma model was developed for a different class of three-point functions, namely the ``heavy-heavy-heavy'' three-point functions of closed strings \cite{Janik:2011bd,Kazama:2011cp,Kazama:2013qsa,Kazama:2016cfl}. On the other hand, the three-point functions analyzed in this paper are ``heavy-heavy-light'' three-point functions \cite{Zarembo:2010rr,Costa:2010rz}. Conceptually, they are much simpler since one can compute them by perturbing the known solution that describes the two-point function\fn{By contrast, a direct computation of the heavy-heavy-heavy three-point functions would require finding a nontrivial saddle-point solution describing the three-point function, which is in general quite a difficult task. The works \cite{Janik:2011bd,Kazama:2011cp,Kazama:2013qsa,Kazama:2016cfl} circumvented this problem by judiciously using the integrability and developing a method that directly computes the three-point functions without constructing the saddle-point solution.}. Nevertheless, a systematic integrability approach\footnote{One possible way to develop such a framework is to relate the Green's functions to solutions to the so-called auxiliary linear problem: as briefly discussed in \cite{Vicedo:2008jy}, the solutions to the auxiliary linear problem are related to small perturbations of the classical solution. It would be worth trying to make the connection concrete in our setup.} based on the spectral curve has not been developed yet. Finding such a framework is an important future problem partly because it could shed a new light on the relation between the three-point functions and the (quantum) spectral curve, which has been a subject of active exploration in the past few years \cite{giombi2018exact,Giombi:2018hsx,Giombi:2020amn,Cavaglia:2018lxi,Cavaglia:2021mft}.

\section{Conclusions}\label{sec:vkOMSPdS3m}
In this paper and in \cite{Giombi:2021zfb}, we explored a connection between the Wilson loop defect CFT and the large charge expansion of CFT. 
In this second paper, we analyzed the $1/J$ corrections to the correlation functions of two large charge insertions with charge $J$ and two light insertions on the $1/2$-BPS Wilson loop. We computed them holographically by studying quantum fluctuations around a classical open string solution describing the large charge insertions. The final results are given by a series representation, which can be identified with the conformal block expansion in the heavy-light channel. We also found an interesting connection to classical integrability; each term in the series can be identified with a point on the spectral curve satisfying a quantization condition, and the conformal dimensions and the structure constants simplify when expressed in terms of the spectral parameter. 

There are several interesting future directions to explore, some of which we mention below. In this paper, we focused on correlation functions of operators inserted on the line defect. A natural generalization would be to study observables that also involve ``bulk" single-trace operators, and in particular to explore their large charge limits. The one-point functions of large charge operators in the presence of the Wilson line were studied in \cite{Zarembo:2002ph} (see also \cite{giombi2013correlators,Enari:2012pq}), and it would be interesting to extend those calculations by looking at correlation functions involving both bulk and defect insertions (these encode the bulk-defect OPE coefficients that are important physical data in dCFTs).\footnote{Some bulk-defect correlation functions with finite charges were studied in \cite{Giombi:2018hsx} from localization and string theory.} Both in this paper and in \cite{Giombi:2021zfb}, we focused on the Wilson loop in the fundamental representation, and another natural generalization would be study higher-rank representations. The correlation functions of light insertions in this setup have been analyzed in \cite{Giombi:2020amn} using holography and localization. Unlike the fundamental representation, the holographic duals are D-branes whose tension is proportional to $N$. Thus, an analog of what has been done in this paper would be to study the limit in which the charge of insertions goes to infinity with a ratio $J/N$ fixed. A D3 brane classical solution describing such a setup was found in \cite{Drukker:2006zk}. Following the same logic as in the analysis of the string, it would be interesting to study fluctuations around this D-brane solution. Also, the recent works \cite{Beccaria:2022bcr,Cuomo:2022xgw} demonstrated that, by taking a double-scaling limit in which one sends the rank of the representation to infinity while correlating it with the coupling constant of the theory, one can analytically study various properties of one-dimensional dCFT (in particular its renormalization group flow). This can be viewed as yet another dCFT analog of the large charge sector of CFT, different from what we studied in this paper, and it would be interesting to explore a possible connection.

On the integrability side, there are several important questions to be answered. An obvious one is to construct a classical spectral curve for open strings describing the Wilson loop (generalizing the result in \cite{Bajnok:2013sza}); this would allow us to rigorously establish the relation between the quasi-energy and the conformal dimension that we found in this paper. A more ambitious one is to develop a framework to compute the heavy-heavy-light structure constants directly from the spectral curve. As mentioned briefly at the end of Section \ref{sec:integrability}, one may be able to achieve this by using solutions to the auxiliary linear problem (or equivalently sections of Lax connections) and it would be interesting to explore this direction. 

Another possible direction is to reproduce the results in our paper from non-perturbative integrability approaches such as the hexagon formalism \cite{Kim:2017phs,Kiryu:2018phb,Basso:2015zoa,Fleury:2016ykk,Eden:2016xvg}. In particular, a sub-class of heavy-heavy-light three-point functions of single-trace operators at strong coupling were computed using the hexagon formalism in \cite{Basso:2019diw}, and it would be interesting to extend their analysis to insertions on the Wilson loop and make comparison with our semi-classical predictions.

As shown in \cite{giombi2018exact,Giombi:2021zfb}, the quasi-momentum and the spectral curve can also be computed by supersymmetric localization. A surprise that we found in this paper is that they also govern the spectrum and the structure constants of non-BPS excitations, which are not computable by localization. It is then interesting to see if the same holds also for rank-1 $\mathcal{N}=2$ SCFTs studied in \cite{Hellerman:2017sur,Bourget:2018obm,Grassi:2019txd}. Also there, the two-point functions of large charge BPS operators can be computed by supersymmetric localization and one obtains a matrix model and a spectral curve that describe a large charge double scaling limit in which the charge $J$ is sent to infinity while a product $g^2J$ is kept fixed (with $g$ being the Yang-Mills coupling constant). The focus of the works \cite{Hellerman:2017sur,Bourget:2018obm,Grassi:2019txd} was on the correlation functions of BPS operators, but it would be interesting to study non-BPS operators to see if the spectral curve obtained from the matrix model in \cite{Grassi:2019txd} encodes information about non-BPS operators as well. 

Finally, it is important to explore a ``closed-string analog'' of our analysis. In \cite{Buchbinder:2010ek}, the heavy-heavy-light-light four-point functions of single-trace operators  were studied at strong coupling (see also the analysis at weak coupling in \cite{Caetano:2011eb}). However they only computed the leading order results, which correspond to the classical limit of the string sigma model. It would be interesting to perform the computation at the next leader order by analyzing the quantum fluctuations, which may reveal an interesting connection to the spectral curve and semi-classical integrability as we saw in this paper.

\section*{Acknowledgements}
We thank Himanshu Khanchandani for helpful discussions. The work of SG and BO is supported in part by the US NSF under Grant No.~PHY-1914860. SK thanks the Kavli Institute for Theoretical Physics for hospitality during the completion of this project, where his work was supported in part by the National Science Foundation under Grant No. NSF PHY-1748958.
%\newpage 
\appendix 

\section{Special functions and the Lam\'e equation}\label{app:special functions}
Three related families of special functions feature prominently in the text, especially in Section~\ref{sec:series representations of 4pt functions}: the \textit{elliptic integrals} of the first and second kind, the \textit{Jacobi elliptic functions}, and the \textit{theta functions}. This appendix summarizes the conventions we adopt for these functions and collects some standard identities. We also summarize the Lam\'e equation and its solutions. Useful references include \cite{abramowitz1988handbook,NIST:DLMF,whittaker_watson_1996,stein2010complex,MR1007595}.

\paragraph{Elliptic integrals.} The \textit{incomplete elliptic integrals} of the first and second kind are given by
\begin{align}\label{eq:incompletel elliptic integrals}
    \mathbb{F}(\varphi|m)&\equiv \int_{0}^\varphi \frac{d\theta}{\sqrt{1-m\sin^2{\theta}}},&\mathbb{E}(\varphi|m)&\equiv \int_0^\varphi \sqrt{1-m\sin^2{\theta}}d\theta.
\end{align}
When $\varphi=\frac{\pi}{2}$, they become the \textit{complete elliptic integrals} of the first and second kind:
\begin{align}\label{eq:elliptic integrals}
    \ellK(m)&\equiv \mathbb{F}\left(\frac{\pi}{2}\Big|m\right),&\mathbb{E}(m)&\equiv \mathbb{E}\left(\frac{\pi}{2}\Big|m\right).
\end{align}
In \eqref{eq:incompletel elliptic integrals} and \eqref{eq:elliptic integrals}, $\varphi$ is the \textit{amplitude} and $m$ is the \textit{parameter}. The integrals in \eqref{eq:incompletel elliptic integrals} and \eqref{eq:elliptic integrals} are defined for $\varphi\in \mathbb{R}$ and $m\in [0,1)$, and can be analytically continued to complex $\varphi$ and $m$. In particular, it is useful to note the continuation of $\ellK(m)$ and $\ellE(m)$ to $m\in (-\infty,0)$ and $m\in (1,\infty)$, which is given by the relations
\begin{align}
    \ellK\left(1-\frac{1}{m}\right)&=\sqrt{m}\ellK(1-m),\label{eq:E5NuDWFxub}\\
    \ellK\left(\frac{1}{m}\right)&=\sqrt{m}\left[\ellK(m)-i\ellK(1-m)\right],\label{eq:M8haq6dJl2}\\
    \ellE\left(1-\frac{1}{m}\right)&=\frac{1}{\sqrt{m}}\ellE(1-m),\label{eq:0PodEedJMj}\\
    \ellE\left(\frac{1}{m}\right)&=\frac{1}{\sqrt{m}}\left[\ellE(m)+i\ellE(1-m)-(1-m)\ellK(m)-im\ellK(1-m)\right].\label{eq:aPTCYjwzVg}
\end{align}
In Mathematica, $\mathbb{F}(x|m)$ and $\ellE(x|m)$ are implemented by the commands $\texttt{EllipticF[x,m]}$ and $\texttt{EllipticE[x,m]}$. $\ellK(m)$ and $\ellE(m)$ are implemented by $\texttt{EllipticK[m]}$ and $\texttt{EllipticE[m]}$.

When $m$ is unambiguous, one often adopts the shorthand $\mathbb{K}\equiv \mathbb{K}(m)$, $\mathbb{K}'\equiv \mathbb{K}(1-m)$ and $\mathbb{E}\equiv \mathbb{E}(m)$, $\mathbb{E}'\equiv \mathbb{E}(1-m)$. It is also worth noting the Legendre identity:
\begin{align}\label{eq:Legendre identity}
    \ellE \ellK' +\ellE'\ellK-\ellK\ellK' = \frac{\pi}{2}.
\end{align}

\paragraph{Jacobi elliptic functions.}
The Jacobi elliptic functions are the inverses of the incomplete elliptic integrals. They also form a one-parameter family of doubly periodic meromorphic functions on $\C$ such that the defining parallelogram has corners at $s$: $z=0$, $c$: $z=\ellK$, $n$: $z=i\ellK$ and $d$: $z=\ellK+i\ellK$, where we adopt the shorthand from the previous section. The Jacobi elliptic function with a simple zero at corner $q$ and a simple pole at corner $p$ is denoted $\text{pq}(z|m)$, where $p$ and $q$ can be $s$, $c$, $n$ or $d$.\footnote{Uniqueness under such a construction requires a few additional conditions on $\text{pq}(z,m)$.} There are therefore twelve distinct elliptic functions. They are interrelated by the identities
\begin{align}
\text{pq}\left(z|m\right)&=\frac{1}{\text{qp}\left(z|m\right)}, && \text{pq}\left(z|m\right)=\frac{\text{pr}\left(z|m\right)}{\text{qr}\left(z|m\right)}, && \text{pq}\left(z|m\right)=\text{pr}\left(z|m\right)\text{rq}\left(z|m\right).
\end{align}
Consequently, all the elliptic functions can be expressed in terms of just three: $\text{sn}(z|m)$, $\text{cn}(z|m)$, and $\text{dn}(z|m)$. In Mathematica, these are implemented using $\texttt{JacobiSN[z,m]}$, $\texttt{JacobiCN[z,m]}$ and $\texttt{JacobiDN[z,m]}$.

The aforementioned inverse relationship between the elliptic functions and the elliptic integrals can be expressed
\begin{align}
    \sin{\varphi}&=\text{sn}\left(\mathbb{F}(\varphi|m)|m\right),&\cos{\varphi}&=\text{cn}\left(\mathbb{F}(\varphi|m)|m\right).
\end{align}
The aforementioned periodicity of the elliptic functions is given by (here, $a,b\in \mathbb{Z}$)
\begin{align}
    \text{sn}\left(z+2a \mathbb{K}+2b i \mathbb{K}' |m\right)&=(-1)^a \text{sn}\left(z |m\right),\label{eq:AyWOmIh3i6}\\
    \text{cn}\left(z+2a \mathbb{K}+2b i \mathbb{K}' |m\right)&=(-1)^{a+b} \text{cn}\left(z|m\right),\label{eq:s5C1p1Pc5z}\\
    \text{dn}\left(z+2a \mathbb{K}+2b i \mathbb{K}' |m\right)&=(-1)^{b} \text{dn}\left(z|m\right),\label{eq:AzboAe67sX}
\end{align}
Furthermore, $\text{sn}\left(-z\rvert m \right)=-\text{sn}\left(z\rvert m \right)$, $\text{cn}\left(-z\rvert m \right)=\text{cn}\left(z\rvert m \right)$ and $\text{dn}\left(-z\rvert m \right)=\text{dn}\left(z\rvert m \right)$.

Some additional useful results include: the derivatives of the elliptic functions,
\begin{align}
    \frac{d}{dz}\text{sn}\left(z|m\right)&=\text{cn}(z|m)\text{dn}(z|m),\label{eq:dsn/dz}\\\frac{d}{dz}\text{cn}\left(z|m\right)&=\text{sn}(z|m)\text{dn}(z|m),\label{eq:dcn/dz}\\ \frac{d}{dz}\text{dn}\left(z|m\right)&=-m\text{sn}(z|m)\text{cn}(z|m),\label{eq:ddn/dz}
\end{align}
two identities involving the squares of the elliptic functions,
\begin{align}
    \text{cn}^2\left(z|m\right)+\text{sn}^2\left(z|m\right)&=1,\label{eq:oX5Dm1hVuw}\\
    \text{dn}^2\left(z|m\right)+m\text{ sn}^2\left(z|m\right)&=1,\label{eq:wtEDZNracI}
\end{align}
and relations between different elliptic functions under translations by half of a period:
\begin{align}
    \text{sn}\left(z+\ellK|m\right)&=\text{cd}\left(z|m\right), &\text{sn}\left(z+i\ellK'|m\right)&=\frac{1}{\sqrt{m}}\text{ns}\left(z|m\right),\label{eq:IW5SfbL8zO}\\
    \text{cn}\left(z+\ellK|m\right)&=-\sqrt{1-m}\text{sd}\left(z|m\right),&\text{cn}\left(z+i\ellK'|m\right)&=-\frac{i}{\sqrt{m}}\text{ds}\left(z|m\right),\\
    \text{dn}\left(z+\ellK|m\right)&=\sqrt{1-m}\text{nd}\left(z|m\right), &\text{dn}\left(z+i\ellK'|m\right)&=-i\text{cs}(z|m).\label{eq:8QBvEot7x6}
\end{align}
Finally, we note that the elliptic functions with parameter $m$ are related to elliptic functions with parameter $1-m$ by the ``Jacobi imaginary transformations'' and to elliptic functions with parameter $1/m$ by the ``Jacobi real transformations":
\begin{alignat}{3}
\text{sn}(z|m)&=-i\text{sc}(iz|1-m),\quad &&=\frac{1}{\sqrt{m}}\text{sn}\left(\sqrt{m}z|\frac{1}{m}\right),\label{eq: Jacobi imag/real transformation sn}\\
\text{cn}(z|m)&=\text{nc}(iz|1-m),\quad &&=\text{dn}\left(\sqrt{m}z|\frac{1}{m}\right),\label{eq: Jacobi imag/real transformation cn}\\
\text{dn}(z|m)&=\text{dc}(iz|1-m),\quad &&=\text{cn}\left(\sqrt{m}z|\frac{1}{m}\right).\label{eq: Jacobi imag/real transformation n}
\end{alignat}

\paragraph{Theta functions.}
The theta functions are a family of four analytic functions of two complex variables: the argument $z\in \C$ and the half-period ratio $\tau$ satisfying $\operatorname{Im}(\tau)>0$. It is standard to also define the nome, $q\equiv e^{i\pi \tau}$, which satisfies $|q|<1$. In our conventions, the two theta functions appearing in the text are given explicitly by the series
\begin{align}
    \theta_1\left(z,q\right)&\equiv-i\sum_{n=-\infty}^\infty (-1)^{n}q^{(n+1/2)^2}e^{(2n+1)iz}, & \theta_4\left(z,q\right)&\equiv\sum_{n=-\infty}^\infty (-1)^{n}q^{n^2}e^{2niz}.
\end{align}
They are implemented in Mathematica using $\texttt{EllipticTheta[n,z,q]}$ with $n=1,4$. Their derivatives with respect to the argument are implemented using $\texttt{EllipticThetaPrime[n,z,q]}$.

Evidently, $\theta_1(-z,q)=-\theta_1(z,q)$ and $\theta_4(-z,q)=\theta_4(z,q)$. Furthermore, the theta functions are quasiperiodic in translations by $\pi$ and $\pi \tau$:
\begin{align}
    \theta_1(z+(a+b\tau)\pi,q)&=(-1)^{a+b}q^{-b^2}e^{-2ibz}\theta_1(z,q),\\
    \theta_4(z+(a+b\tau)\pi,q)&=(-1)^bq^{-b^2}e^{-2ibz}\theta_4(z,q).
\end{align}
Here $a,b \in \mathbb{Z}$. This means their logarithmic derivatives satisfy 
\begin{align}
    \frac{\theta_1'(z+(a+b\tau)\pi,q)}{\theta_1(z+(a+b\tau)\pi,q)}&=\frac{\theta_1'(z,q)}{\theta_1(z,q)}-2ib,\label{eq:1649587669}
    \\
    \frac{\theta_4'(z+(a+b\tau)\pi,q)}{\theta_4(z+(a+b\tau)\pi,q)}&=\frac{\theta_4'(z,q)}{\theta_4(z,q)}-2ib.\label{eq:4181514526}
\end{align}
Here the prime denotes differentiation with respect to the argument.

\paragraph{The Lam\'e equation.}
The Lam\'e differential equation in Jacobi form is \cite{whittaker_watson_1996}
\begin{align}\label{eq:ES9YxM59dJ}
    \left[-\frac{d^2}{dx^2}+2m\text{ sn}\left(x|m\right)^2\right]f(x)=\Lambda f(x).
\end{align}
Two linearly independent solutions are \cite{c73aa227856b43099500121c82d81bdb}:
\begin{align}\label{eq:mlMZiqzziA}
    f^{\pm}(x)\equiv \frac{H(x\pm \alpha|m)}{\Theta(x|m)}e^{\mp Z(\alpha|m)},
\end{align}
where $\alpha$ is related to $m$ via
\begin{align}
    \text{sn}(\alpha|m)&=\frac{1}{\sqrt{m}}\sqrt{1+m-\Lambda},
\end{align}
and, letting $\mathbb{K}\equiv \mathbb{K}(m)$, $\mathbb{K}'\equiv \mathbb{K}(1-m)$, $\mathbb{E}\equiv \mathbb{E}(m)$, and $q\equiv e^{-\frac{\pi \mathbb{K}'}{\mathbb{K}}}$, the Jacobi $H$, $\Theta$, and $Z$ functions can be defined in terms of the theta functions:
\begin{align}\label{eq: H, Th, Z}
    H(u|m)&\equiv \theta_1\left(\frac{\pi}{2}\frac{u}{\mathbb{K}},q\right), &
    \Theta(u|m)&\equiv \theta_4\left(\frac{\pi}{2}\frac{u}{\mathbb{K}},q\right), & Z(u|m)&\equiv \frac{\Theta'(u)}{\Theta(u)}= \frac{\pi}{2\mathbb{K}}\frac{\theta_4'\left(\frac{\pi u}{2\mathbb{K}},q\right)}{\theta_4\left(\frac{\pi u}{2\mathbb{K}},q\right)}.
\end{align}
Though not as standard, it is also convenient for us to also define
\begin{align}\label{eq:V}
    V(u|m)\equiv \frac{H'(u|m)}{H(u|m)}=\frac{\pi}{2\mathbb{K}}\frac{\theta_1'\left(\frac{\pi u}{2\mathbb{K}},q\right)}{\theta_1\left(\frac{\pi u}{2\mathbb{K}},q\right)}.
\end{align}
$Z(u|m)$ can be implemented in Mathematica using \texttt{JacobiZN[u,m]},\footnote{This is preferable to \texttt{JacobiZeta[JacobiAmplitude[u,m],m]}, which does not extend properly into the complex $u$ plane.} or simply by defining it in terms of $\theta_4$. Likewise, $\Theta$, $H$ and $V$ can be implemented in terms of $\theta_1$ and $\theta_4$. 

We note the parity of these functions: $H(u|m)=-H(-u|m)$, $\Theta(u|m)=\Theta(-u|m)$, $Z(u|m)=-Z(-u|m)$ and $V(u|m)=-V(-u|m)$. Due to \eqref{eq:1649587669}-\eqref{eq:4181514526}, $Z$ and $V$ have simple behavior under translations by $2\ellK$ and $2i\ellK'$:
\begin{align}
    V(u\pm 2\ellK|m)&=V(u|m), &V(u\pm 2i\ellK'|m)&=V(u|m)\mp\frac{\pi i}{\ellK}\\
    Z(u\pm 2\ellK|m)&=Z(u|m), &Z(u\pm 2i\ellK'|m)&=Z(u|m)\mp\frac{\pi i}{\ellK}.
\end{align}
Finally, there exist combinations and derivatives of $H$, $\Theta$, $Z$ and $V$ that are doubly periodic and can therefore be expressed in terms of the Jacobi elliptic functions, including
\begin{align}
    \frac{H(u|m)}{\Theta(u|m)}&=m^{\frac{1}{4}}\text{sn}\left(u|m\right),\label{eq:vjHsgGxfAP}\\
    Z'(u|m)&=\text{dn}^2\left(u|m\right)-\frac{\mathbb{E}}{\mathbb{K}},\label{eq:LIA7TEVOWb}\\
    V(u|m)-Z(u|m)&=\frac{\text{cn}\left(u|m\right)\text{dn}\left(u|m\right)}{\text{sn}\left(u|m\right)}.\label{eq:AO7XFN9xh0}
\end{align}
See \cite{MR1007595}\cite{whittaker_watson_1996}\cite{abramowitz1988handbook}, respectively.

\section{Details of the derivation of the quadratic action}\label{eq:deriving quadratic action}
This appendix fills in the steps to get from \eqref{NG-action}-\eqref{eq:fluctuating string induced metric} to \eqref{eq:quadratic action}. First, we define the fluctuations of $\bar{\theta}$ and $\bar{\phi}$ about their classical values: 
\begin{align}
    \tilde{\theta}&\equiv \bar{\theta}-\theta_{\rm cl}, & \tilde{\phi}&\equiv \bar{\phi}-\phi_{\rm cl}.
\end{align}
Then we expand $\Gamma_{\mu\nu}$ to quadratic order in $x^a$, $\bar{y}^j$, $\tilde{\theta}$ and $\tilde{\phi}$ and write it as a sum of five terms:
\begin{align}\label{eq:mF8y35FsCD}
    \Gamma_{\mu\nu}&=\gamma_{\mu\nu}+\Gamma_{\mu\nu}^\theta+\Gamma_{\mu\nu}^{xx}+\Gamma_{\mu\nu}^{yy}+\Gamma_{\mu\nu}^{\theta\phi}+\text{higher than quadratic}.
\end{align}
Here, $\Gamma_{\mu\nu}^{xx}$ is quadratic in $x^a$, $\Gamma_{\mu\nu}^{yy}$ is quadratic in $\bar{y}^i$, $\Gamma_{\mu\nu}^{\theta}$ is linear in $\tilde{\theta}$ and $\tilde{\phi}$, and $\Gamma_{\mu\nu}^{\theta\phi}$ is quadratic in $\tilde{\theta}$ and $\tilde{\phi}$.\footnote{Note, $\Gamma$ is not a Christoffel symbol and the symbols $x$, $y$, $\theta$ and $\phi$ are labels rather than indices in \eqref{eq:mF8y35FsCD}.} The higher order terms in \eqref{eq:mF8y35FsCD} that we neglect are interaction terms for the fluctuation fields, which are suppressed relative to the quadratic action by positive powers of $1/g$. 

$\Gamma_{\mu\nu}^{xx}$ and $\Gamma_{\mu\nu}^{yy}$ take simple explicit forms:
\begin{align}
    \Gamma_{\mu\nu}^{xx}&=\partial_\mu x^a\partial_\nu x^a+h_{\mu\nu}x^2,&
    \Gamma_{\mu\nu}^{yy}&=\frac{\cosh^2{\rho}-c^2}{\cosh^2{\rho}}\partial_\mu \bar{y}^i \partial_\nu \bar{y}^i,\label{eq:9sTzr6VK09}
\end{align}
while $\Gamma_{\mu\nu}^{\theta}$ and $\Gamma_{\mu\nu}^{\theta\phi}$ are more complicated:
\begin{align}
    \Gamma_{\mu\nu}^{\theta}&=\left(\begin{array}{cc}-\tilde{\theta} f_3+2if_4\partial_\tau \tilde{\phi} & if_4 \partial_\rho \tilde{\phi}-f_1 \partial_\tau \tilde{\theta}\\ if_4 \partial_\rho \tilde{\phi} -f_1 \partial_\tau \tilde{\theta}& -2f_1\partial_\rho \tilde{\theta} \end{array}\right),\label{eq:W2NrLRX7Wg}\\
    \Gamma_{\mu\nu}^{\theta\phi}&=\partial_\mu \tilde{\theta}\partial_\nu \tilde{\theta}+f_4\partial_\mu \tilde{\phi}\partial_\nu \tilde{\phi}+\left(\begin{array}{cc} -f_2 \tilde{\theta}^2 +2i\tilde{\theta}\partial_\tau \tilde{\phi} f_3& i\tilde{\theta} \partial_\rho \tilde{\phi} f_3\\i\tilde{\theta} \partial_\rho \tilde{\phi} f_3 & 0\end{array}\right).\label{eq:HlgnCwoL8M}
\end{align}
Here we have introduced the auxiliary functions,
\begin{align}
    f_1(\rho)&=\frac{c\sinh{\rho}}{\cosh{\rho}\sqrt{\cosh^2{\rho}-c^2}}, &
    f_2(\rho)&=\cos{2\theta_{\rm cl}}=\frac{\cosh^2{\rho}-2c^2}{\cosh^2{\rho}},\\
    f_3(\rho)&=\sin{2\theta_{\rm cl}}=\frac{2c\sqrt{\cosh^2{\rho}-c^2}}{\cosh^2{\rho}}, & f_4(\rho)&=\sin^2{\theta_{\rm cl}}=\frac{c^2}{\cosh^2{\rho}}.
\end{align}

Expanding the Nambu-Goto Lagrangian in \eqref{NG-action} to quadratic order in the fluctuations yields
\begin{align}\label{eq:wMZNYbKL4t}
    \mathcal{L}\equiv \frac{\sqrt{\Gamma}}{\sqrt{\gamma}}=1+\mathcal{L}^\theta+\mathcal{L}^{xx}+\mathcal{L}^{yy}+\mathcal{L}^{\theta\phi}+\text{higher than quadratic},
\end{align}
where the different terms are related to the different components of the metric by 
\begin{align}
    \mathcal{L}^\theta &\equiv \frac{1}{2}\gamma^{\mu\nu}\Gamma_{\mu\nu}^\theta,\hspace{2cm} \mathcal{L}^{xx}\equiv \frac{1}{2}\gamma^{\mu\nu}\Gamma_{\mu\nu}^{xx},\hspace{2cm}
    \mathcal{L}^{yy}\equiv\frac{1}{2}\gamma^{\mu\nu}\Gamma_{\mu\nu}^{yy},\label{eq:theta, x, y Lagrangian}\\
    \mathcal{L}^{\theta\phi}&\equiv\frac{1}{2}\gamma^{\mu\nu}\Gamma_{\mu\nu}^{\theta\phi}+\frac{1}{8}\left((\gamma^{\mu\nu}\Gamma_{\mu\nu}^\theta)^2-2\gamma^{\mu\alpha}\gamma^{\nu\beta}\Gamma_{\mu\nu}^\theta\Gamma_{\alpha\beta}^\theta\right).\label{eq:theta phi Lagrangian}
\end{align}

Let us consider each term in \eqref{eq:wMZNYbKL4t} in turn. Firstly, using integration-by-parts, we find $\mathcal{L}^\theta = 0$ up to boundary terms, as expected for the linear variation about an extremum. Secondly, the quadratic Lagrangian for the $x^a$ fluctuations is simply given by \eqref{NG-action}, and the quadratic Lagrangian for the $\bar{y}^i$ fluctuations is
\begin{align}
    \mathcal{L}^{yy}&=\frac{1}{2}\frac{\cosh^2{\rho}-c^2}{\cosh^2{\rho}}\gamma^{\mu\nu}\partial_\mu \bar{y}^i \partial_\nu \bar{y}^i.\label{eq:GvOyNn2SSU}
\end{align}
Since the kinetic term in \eqref{eq:GvOyNn2SSU} is not in the canonical form, we introduce the redefined fields $y^i\equiv f_y(\rho)\bar{y}^i$, where $f_y(\rho)$ is given in \eqref{f-factors}. Getting rid of the resulting $y^i\partial y^i=\frac{1}{2}\partial(y^iy^i)$ term using integration-by-parts yields the Lagrangian in \eqref{eq:kf1t2Nw4fy}. 

Finally, though $\mathcal{L}^{\theta\phi}$ is more cumbersome than the other terms, it can also be massaged into a relatively nice form. We first write \eqref{eq:theta phi Lagrangian} explicitly:
\begin{align}\label{eq:sBOKwjesKh}
\mathcal{L}^{\theta\phi}&=\frac{1}{2}\frac{\cosh^2{\rho}(\cosh^2{\rho}-c^2)}{\cosh^4{\rho}-c^2}\gamma^{\mu\nu}\partial_\mu \tilde{\theta} \partial_\nu \tilde{\theta}+\frac{1}{2}\frac{c^2\cosh^2{\rho}}{\cosh^4{\rho}-c^2}\gamma^{\mu\nu}\partial_\mu \tilde{\phi}\partial_\nu \tilde{\phi}\nonumber\\&
-\frac{ic^3\cosh{\rho}\sqrt{\cosh^2{\rho}-c^2}\sinh{\rho}}{(c^2-\cosh^4{\rho})^2}(\partial_\rho \tilde{\theta} \partial_\tau \tilde{\phi}-\partial_\tau \tilde{\theta} \partial_\rho \tilde{\phi})\nonumber\\&-\frac{ic(c^2-2\cosh^4{\rho})\sqrt{\cosh^2{\rho}-c^2}}{(\cosh^4{\rho}-c^2)^2}\tilde{\theta} \partial_\tau \tilde{\phi}\nonumber\\&
-\frac{(c^4-2c^2\cosh^4{\rho}+\cosh^6{\rho})}{2(c^2-\cosh^4{\rho})^2}\tilde{\theta}^2+\frac{c^2(\cosh^2{\rho}-c^2)\sinh{2\rho}}{4(c^2-\cosh^4{\rho})^2}\partial_\rho (\tilde{\theta}^2).
\end{align}
We can combine the second and third lines above using the product rule and integration-by-parts to get rid of the $\partial_\rho$ derivatives. Likewise, we use integration-by-parts to convert the $\partial_\rho \tilde{\theta}^2$ term in the fourth line into a $\tilde{\theta}$ mass term. Finally, to put the kinetic terms in the first line in canonical form, we introduce the redefined fields $\theta\equiv f_\theta(\rho)\tilde{\theta}$ and $\phi\equiv f_\phi(\rho)\tilde{\phi}$, where $f_\theta(\rho)$ and $f_\phi(\rho)$ are again given in \eqref{f-factors}. Using integration-by-parts one last time to get rid of the $\theta\partial \theta$ and $\phi\partial\phi$ terms, we finally arrive at (\ref{eq:LpWrSXGjF0}). 

\section{Four-point functions perturbatively at small \texorpdfstring{$\mathcal{J}$}{J/g}}\label{app:perturbative analysis}
In this Appendix, we study $W_{yy}$, $W_{xx}$, $W_{\theta\theta}$ and $W_{\phi\theta}$ by perturbatively solving the Green's equations in \eqref{eq:xx Green's eqn}-\eqref{eq:phitheta Green's eqn}, treating $c^2$ as a small parameter. This is in contrast to the discussion in Section~\ref{sec:small and large J/g}, where we expanded the finite $c^2$ series for the boundary propagators in \eqref{eq:G1 series}-\eqref{eq:G4 series} and \eqref{eq:G2 series}-\eqref{eq:s7VYqCQAUD} in small $c^2$. The analysis in this appendix serves as an independent check of the results in Section~\ref{sec:small and large J/g}. It also allows us to calculate $W_{\theta\theta}$ and $W_{\phi\theta}$ directly via the semiclassical analysis of the string discussed in Section~\ref{sec:fluctuations about the classical string}, instead of using the Ward identities and inputs from localization, as in Section~\ref{sec:computing G2 and G3}.

We expand the boundary propagator $W$ ($\equiv W_{yy},W_{xx},W_{\theta\theta},W_{\phi\theta}$) in $c^2$:
\begin{align}
    W(t_1,t_2)=W_{0}(t_1,t_2)+c^2W_{1}(t_1,t_2)+\ldots.
\end{align}
We will determine $W_{yy}$, $W_{\theta\theta}$ and $W_{\phi\theta}$ to linear order and $W_{xx}$ to zeroth order for simplicity.

\paragraph{Perturbative calculation of $W_{yy}$.} We begin by determining $W_{yy}$ to linear order in $c^2$. We apply the formalism in Section~\ref{sec:integral and series reps of propagators}, and in particular \eqref{eq:dRNhIUWDdm}, essentially without modification. The first step is to find $g_{yy}^R(r;k)$ from its differential equation, \eqref{eq:kZIdmPDcsW}, supplemented with the boundary condition $g_{yy}^R(r_m;k)=0$, where $r_m=\sqrt{1-c^2}\ellK(c^2)$. Then $a_{yy}(k)$ is found in terms of $g^R_{yy}(r;k)$ via \eqref{eq:DwWKnP1Xbu}.

We expand $g_{yy}^R$ in $c^2$: $g_{yy}^R(r;k)=g_{yy,0}^R(r;k)+c^2g_{yy,1}^R(r;k)+\ldots$. Expanding \eqref{eq:kZIdmPDcsW} to linear order in $c^2$ yields the following two differential equations solved by $g_{yy,0}^R$ and $g_{yy,1}^R$:
\begin{align}\label{eq:ODE c^0 c^2}
    \left(\frac{d^2}{dr^2}-k^2\right)g_{yy,0}^R(r;k)&=0,&
    \left(\frac{d^2}{dr^2}-k^2\right)g_{yy,1}^R(r;k)&=(k^2+\cos{2r})g_{yy,0}^R(r;k).
\end{align}
Furthermore, expanding $g_{yy}^R(r_m;k)=0$ to linear order in $c^2$ (note that $r_m=\frac{\pi}{2}-\frac{\pi}{8}c^2+\ldots$) yields the following boundary conditions for $g_{yy,0}^R$ and $g_{yy,1}^R$:
\begin{align}\label{eq:FUFcEZgW99}
    g_{yy,0}^R\left(\frac{\pi}{2};k\right)&=0,&g_{yy,1}^R\left(\frac{\pi}{2};k\right)-\frac{\pi}{8}\frac{dg_{yy,0}^R}{dr}\biggr\rvert_{r=\frac{\pi}{2}}=0.
\end{align}
The explicit solutions to \eqref{eq:ODE c^0 c^2} are:
\begin{align}
    g_{yy,0}^R(r;k)&=2\sinh\left(k\left(\frac{\pi}{2}-r\right)\right), \\g_{yy,1}^R(r;k)&=\frac{1}{8(1+k^2)}\biggr[2k\cosh\left(k\left(\frac{\pi}{2}-r\right)\right)\left((\pi-4r)(1+k^2)-2\sin{2r}\right)\\&\hspace{9cm}-8\cos^2{r}\sinh\left(k\left(\frac{\pi}{2}-r\right)\right)\biggr].\nonumber
\end{align}

Given these expressions, the boundary limit of $g^R_{yy}$ simplifies significantly:
\begin{align}
    \lim_{\rho\to \infty} e^{\rho}g^R_{yy}(r;k)&=\bar{r}_m \frac{dg^R_{yy}}{dr}\biggr\rvert_{r_m}=4k-\frac{2k}{1+k^2}c^2+\ldots.
\end{align}
We used \eqref{eq:5kUeXpf7ju} and noted that $\bar{r}_m=-2+c^2+\ldots$. Meanwhile, \eqref{eq:DwWKnP1Xbu} yields the normalization:
\begin{align}\label{eq:AQW0pCnZg0}
    a_{yy}(k)&=\frac{1}{4k\sinh{\pi k}}+\frac{c^2}{16}\left[\frac{2}{k(1+k^2)\sinh{\pi k}}-\frac{\pi \cosh{\pi k}}{\sinh^2{\pi k}}\right]+\ldots.
\end{align}
The poles of $a_{yy}(k)$ in the upper half plane are at $k_n=ni$ for $n=1,2,3,\ldots$. They are of degree $1$ at order $c^0$, and degree $2$ at order $c^2$, which reflects the fact that a simple pole can appear as a higher order pole when treated perturbatively (e.g., if $w=w_0+c^2w_1+\ldots$ then $\frac{1}{z-w}=\frac{1}{z-w_0}+\frac{w_1 c^2}{(z-w_0)^2}+\ldots$). Consequently, we need to pull the factor 
\begin{align}\label{eq:ZswDz4szas}
    &e^{ik|\tau|}a_{yy}(k)\left(\lim_{\rho\to \infty} e^{\rho}g^R_{yy}\right)^2=e^{ik|\tau|}\left[\frac{4k}{\sinh{\pi k}}-c^2\left(\frac{2k}{(1+k^2)\sinh{\pi k}}+\frac{\pi k^2\cosh{\pi k}}{\sinh^2{\pi k}}\right)+\ldots\right].
\end{align}
back inside the contour integral in order to apply \eqref{eq:dRNhIUWDdm}, which assumes the poles at $k_n$ are simple. Here, for ease of notation, we use 
\begin{align}
    \tau\equiv \tau(\chi)\equiv \log|1-\chi|.
\end{align}

Evaluating the residues of \eqref{eq:ZswDz4szas}, we find at order $c^0$:
\begin{align}
    2\pi i\underset{k=ni}{\text{Res}}\biggr[e^{ik |\tau|} a_{yy}(k)\left(\lim_{\rho\to \infty} e^{\rho}g^R_{yy}\right)^2\biggr\rvert_{c^0}\biggr]&=8(-1)^{n+1}ne^{-n|\tau|}
\end{align}
and at order $c^2$:
\begin{align}
    2\pi i\underset{k=i}{\text{Res}}\biggr[e^{ik |\tau|} a_{yy}(k)\left(\lim_{\rho\to \infty} e^{\rho}g^R_{yy}\right)^2\biggr\rvert_{c^2}\biggr]&=-3e^{-|\tau|}\label{eq:SXYVlyiBHu}\\
    2\pi i\underset{k=ni}{\text{Res}}\biggr[e^{ik |\tau|} a_{yy}(k)\left(\lim_{\rho\to \infty} e^{\rho}g^R_{yy}\right)^2\biggr\rvert_{c^2}\biggr]&=2(-1)^{n+1}\left[\frac{2n}{n^2-1}-2n+n^2|\tau|\right]e^{-n |\tau|}.\label{eq:mTtm3To8ag}
\end{align}
We need to treat the residue at $k=i$ separately from the other residues. 

Finally we expand $W_{yy}(t_1,t_2)=W_{yy,0}(t_1,t_2)+c^2W_{yy}(t_1,t_2)+\ldots$ and apply \eqref{eq:dRNhIUWDdm} at each order. Up to the index shift $n\mapsto n+1$, the $c^0$ result reproduces \eqref{eq:Wyy c=0 from general c exp}:
\begin{align}\label{eq:fmXxUbUWeZ}
    W_{yy,0}(t_1,t_2)&=\frac{2g}{\pi}\frac{1}{t_{12}^2}\frac{\chi^2}{|1-\chi|}\sum_{n=1}^\infty \text{sgn}(1-\chi)^{n+1}ne^{-n|\tau|}.
\end{align}
The $c^2$ correction is given by
\begin{align}\label{eq:ilBM8gsJnk}
    W_{yy,1}(t_1,t_2)&=\frac{g}{2\pi}\frac{1}{t_{12}^2}\frac{\chi^2}{|1-\chi|}\biggr[-\frac{3}{2}e^{-|\tau|}+\sum_{n=1}^\infty \text{sgn}(1-\chi)^{n+1}\biggr(\frac{2n}{n^2-1}-2n+n^2|\tau|\biggr)e^{-n|\tau|}\biggr].
\end{align}
The series can be summed explicitly for all $\chi$. Using $W_{yy}(t_1,t_2)=\frac{\pi}{2g}t_{12}^2G_1(\chi)$, we can compare the result with the $c^2$ term in \eqref{eq:vcByXaNCgz}, and we find perfect agreement. 

\paragraph{Perturbative calculation of $W_{xx}$.} Next, we determine $W_{xx}$ to zeroth order in $c^2$.\footnote{We study $W_{yy}$ to linear order and $W_{xx}$ only to zeroth order because the zeroth order ODE in \eqref{eq:ODE c^0 c^2} is nicer than \eqref{eq:byEHl3fVEY}, and makes the perturbative analysis simpler for $W_{yy}$ than $W_{xx}$.} We need to find $g_{xx}^R(r;k)$ from \eqref{eq:8txKFSC2oh}, supplemented with the boundary condition $g_{xx}^R(r_m;k)=0$. To zeroth order, the differential equation becomes
\begin{align}\label{eq:byEHl3fVEY}
    \left(\frac{d^2}{dr^2}-2\sec^2{r}-k^2\right) g_{xx,0}^{R}(r;k)&=0.
\end{align}
and the boundary condition is $g_{xx,0}^R(\frac{\pi}{2})=0$. Two linearly independent solutions to \eqref{eq:byEHl3fVEY} are $\frac{e^{\pm kr}}{2}\left(k\pm  \tan{r}\right)$, and the linear combination that satisfies the bounary condition is
\begin{align}\label{eq:g_xx^R c=0}
    g_{xx,0}^R(r;k)&=k\cosh\left(k\left(\frac{\pi}{2}-r\right)\right)-\sinh\left(k\left(\frac{\pi}{2}-r\right)\right)\tan(r).
\end{align}

The boundary limit of $g_{xx,0}^R$ simplifies to
\begin{align}
    \lim_{\rho\to \infty}e^{2\rho}g_{xx,0}^R(r;k)=\frac{\bar{r}_m^2}{2} \frac{d^2g_{xx,0}^R}{dr^2}\biggr\rvert_{r_m}=\frac{4}{3}k(1+k^2).
\end{align}
We used $r_m=\frac{\pi}{2}$ and $\bar{r}_m=-2$ to zeroth order. Moreover, \eqref{eq:DwWKnP1Xbu} yields the normalization
\begin{align}
    a_{xx,0}(k)&=\frac{1}{k(1+k^2)\sinh{k\pi}}.
\end{align}
Thus, the combination
\begin{align}\label{eq:7Q40YLp8aV}
    e^{ik|\tau|}a_{xx,0}(k)\left(\lim_{\rho\to \infty}e^{2\rho}g_{xx,0}^R \right)^2=\frac{16}{9}e^{ik|\tau|}\frac{k(1+k^2)}{\sinh{k\pi}}
\end{align}
has simple poles at $k_n=ni$, $n=2,3,4,\ldots$ in the upper half plane. It is analytic at $k=0,\pm i$ even though these are double poles of $a_{xx,0}(k)$. Thus, we can apply \eqref{eq:dRNhIUWDdm} with the slight modification that $n=2$ is the first term in the sum. Given the residue
\begin{align}
    2\pi i\underset{k=n i}{\text{Res}}\biggr[e^{ik |\tau|} a_{xx,0}(k)\left(\lim_{\rho\to \infty} e^{2\rho}g^R_{xx,0}\right)^2\biggr]&=\frac{32}{9}(-1)^{n}n(n^2-1)e^{-n|\tau|},
\end{align}
the sum for the zeroth order expression for $W_{xx}$ becomes
\begin{align}\label{eq:Wxx c=0}
    W_{xx}(t_1,t_2)&=\frac{2g}{\pi}\frac{1}{t_{12}^4}\frac{\chi^4}{(1-\chi)^2}\sum_{n=2}^\infty \text{sgn}(1-\chi)^{n} e^{-n|\log|1-\chi||}n(n^2-1).
\end{align}
This reproduces \eqref{eq:Wxx c=0 from general c exp}, up to the index shift $n\mapsto n+1$.

\paragraph{Peturbative calculation of $W_{\theta\theta}$ and $W_{\phi\theta}$.} Next, we turn to the computation of $W_{\theta\theta}$ and $W_{\phi\theta}$ to linear order in $c^2$. The analysis is conceptually similar to that of $W_{yy}$, but slightly more complicated because the $\theta$ and $\phi$ modes are coupled by the $-is \theta \overset{\leftrightarrow}{\partial }\phi$ term in the Lagrangian in \eqref{eq:LpWrSXGjF0}.

We begin by noting that the prefactor of the term coupling $\theta$ and $\phi$, which is given in \eqref{eq:s theta phi}, satisfies $s\to -\frac{1}{\cosh^2{\rho}}$ as $c\to 0$. Likewise the mass of the $\theta$ and $\phi$ fields, given in \eqref{eq:mass theta phi}, satisfies $m_{\theta\phi}^2\to -\frac{1}{\cosh^2{\rho}}$ as $c\to 0$, in contrast with the masses of the $x$ and $y$ fluctuations, which approach $m_{xx}^2\to 2$ and $m_{yy}^2\to 0$. This makes the direct analysis of the propagators of the $\theta$ and $\phi$ fields in \eqref{eq:LpWrSXGjF0} somewhat cumbersome even when $c^2=0$. Therefore, it is useful as a preliminary step to rotate the fields as follows:
\begin{align}\label{eq:rotated theta phi}
    \left(\begin{array}{c}\theta\\\phi\end{array}\right)&\to \left(\begin{array}{cc}\cos{i\tau} & \sin{i\tau} \\ -\sin{i\tau} & \cos{i\tau} \end{array}\right)\left(\begin{array}{c}\theta\\ \phi\end{array}\right),
\end{align}
This leaves invariant the form of the Lagrangian in \eqref{eq:LpWrSXGjF0}, as well as the Green's equations in \eqref{eq:phitheta Green's eqn}-\eqref{eq:thetatheta Green's eqn}, except that the coupling term $s$ and the mass $m_{\theta\phi}^2$ are transformed to
\begin{align}
    s&\to s+\gamma^{\tau\tau}&&=\frac{c^2\sinh^2{2\rho}}{2(\cosh^4{\rho}-c^2)^2},\label{eq:wP3XUQ0z4T}\\
    m_{\theta\phi}^2&\to  m_{\theta\phi}^2-\gamma^{\tau\tau}-2s&&=-\frac{c^2\cosh^2{\rho}}{(\cosh^4\rho-c^2)^3}\biggr[\cosh^4{\rho}(8-8\cosh^2{\rho}+\cosh^4{\rho})\label{eq:Ir6apX7pK}\\&&&\hspace{4cm}+2c^2(2-8\cosh^2{\rho}+5\cosh^4{\rho})+c^4\biggr]\nonumber.
\end{align}
These now have the desired property $s,m_{\theta\phi}^2\to 0$ as $c\to 0$. The rotation in \eqref{eq:rotated theta phi} also modifies the contribution of the $\theta$ and $\phi$ fluctuations in the dictionary in \eqref{eq:th-th and th-ph bndy-to-bndy to ZZb defect fn}-\eqref{eq:th-th and th-ph bndy-to-bndy to ZbZ defect fn} to 
\begin{align}
    \frac{\braket{Z(t_1)\bar{Z}(t_2)Z^J(t_3)\bar{Z}^J(t_4)}}{\braket{Z^J(t_3)\bar{Z}^J(t_4)}}&=v_{Z}(t_1)\rvert_{\Psi_{\rm cl}}v_{\bar{Z}}(t_2)\rvert_{\Psi_{\rm cl}}+2(W_{\theta\theta}(t_1,t_2)+iW_{\phi\theta}(t_1,t_2)),\label{eq:2pcuM8BXmq}\\
    \frac{\braket{\bar{Z}(t_1)Z(t_2)Z^J(t_3)\bar{Z}^J(t_4)}}{\braket{Z^J(t_3)\bar{Z}^J(t_4)}}&=v_{\bar{Z}}(t_1)\rvert_{\Psi_{\rm cl}}v_{Z}(t_2)\rvert_{\Psi_{\rm cl}}+2(W_{\theta\theta}(t_1,t_2)-iW_{\phi\theta}(t_1,t_2)).\label{eq:FerYQICfpw}
\end{align}

Now we can determine $W_{\theta\theta}$ and $W_{\phi\theta}$ to linear order in $c^2$ in essentially the same way that we determined $W_{yy}$. We first write $G_{\theta\theta}$ and $G_{\phi\theta}$ as Fourier integrals:
\begin{align}
    G_{\theta\theta}(\rho,\tau;\rho',\tau')&=\frac{1}{4\pi g}\int_{-\infty}^\infty dk e^{ik(\tau-\tau')}g_{\theta\theta}(\rho,\rho';k), \\G_{\phi\theta}(\rho,\tau;\rho',\tau')&=\frac{1}{4\pi g}\int_{-\infty}^\infty dk e^{ik(\tau-\tau')}g_{\phi\theta}(\rho,\rho';k).
\end{align}
Substituting these into the Green's equations in \eqref{eq:thetatheta Green's eqn}-\eqref{eq:phitheta Green's eqn} yields the following coupled equations for $g_{\theta\theta}(\rho,\rho';k)$ and $g_{\phi\theta}(\rho,\rho';k)$:
\begin{align}
    \left[\frac{d}{d\rho}\left(\sqrt{\cosh^2{\rho}-c^2}\frac{d}{d\rho}\right)-\frac{k^2}{\sqrt{\cosh^2{\rho}-c^2}}-\sqrt{\gamma}m_{\theta\phi}^2\right]g_{\theta\theta}-2k\sqrt{\gamma}sg_{\phi\theta}&=-\delta(\rho-\rho'),\label{eq:SUiyvwRUoD}\\
    \left[\frac{d}{d\rho}\left(\sqrt{\cosh^2{\rho}-c^2}\frac{d}{d\rho}\right)-\frac{k^2}{\sqrt{\cosh^2{\rho}-c^2}}-\sqrt{\gamma}m_{\theta\phi}^2\right]g_{\phi\theta}+2k\sqrt{\gamma}sg_{\theta\theta}&=0.\label{eq:xUF4HZJnud}
\end{align}
It is again sensible to change variables from $\rho$ to $r$ via \eqref{eq:B5pnopgBFw}. Then \eqref{eq:SUiyvwRUoD} and \eqref{eq:xUF4HZJnud} become
\begin{align}
    \left[\frac{d^2}{dr^2}-\frac{k^2}{1-c^2}-\frac{\sqrt{\cosh^2{\rho}-c^2}\sqrt{\gamma}m_{\theta\phi}^2}{1-c^2}\right]g_{\theta\theta}-\frac{2k\sqrt{\cosh^2{\rho}-c^2}\sqrt{\gamma}s}{1-c^2}g_{\phi\theta}&=-\frac{\delta(r-r')}{\sqrt{1-c^2}},\label{eq:zNpTPuiX2U}\\
    \left[\frac{d^2}{dr^2}-\frac{k^2}{1-c^2}-\frac{\sqrt{\cosh^2{\rho}-c^2}\sqrt{\gamma}m_{\theta\phi}^2}{1-c^2}\right]g_{\phi\theta}+\frac{2k\sqrt{\cosh^2{\rho}-c^2}\sqrt{\gamma}s}{1-c^2}g_{\theta\theta}&=0,\label{eq:naFY1mZecH}
\end{align}
where $\cosh{\rho}$ is to be replaced by $\text{cn}\left(ir|\frac{1}{1-c^2}\right)$.

Now consider expanding $g_{\theta\theta}$ and $g_{\phi\theta}$ in $c^2$. Because $s=O(c^2)$, it follows that $g_{\phi\theta}=O(c^2)$ and the term containing $g_{\phi\theta}$ in \eqref{eq:zNpTPuiX2U} is $O(c^4)$ and can therefore be ignored when we are interested in $W_{\theta\theta}$ and $W_{\phi\theta}$ to linear order in $c^2$. Thus, to linear order, $g_{\theta\theta}$ is effectively decoupled from $g_{\phi\theta}$, and \eqref{eq:zNpTPuiX2U} reduces to the same form as \eqref{eq:EfEiIZcZdN} except with $m_{yy}^2$ replaced by $m_{\theta\phi}^2$. We will therefore first determine $W_{\theta\theta}$ to linear order in $c^2$ using the same approach as with $W_{yy}$. We will then return to $W_{\phi\theta}$, the details of whose analysis are slightly different.

To linear order in $c^2$, it is valid to write
\begin{align}
    g_{\theta\theta}(r,r';k)=a_{\theta\theta}(k)(g_{\theta\theta}^R(r;k)g_{\theta\theta}^L(r';k)\theta(r-r')+g_{\theta\theta}^L(r;k)g_{\theta\theta}^R(r';k)\theta(r'-r))
\end{align}
where $a_{\theta\theta}(k)\equiv -\frac{1}{\sqrt{1-c^2}}\left(2\frac{dg^R_{\theta\theta}(r;k)}{dr}g^R_{\theta\theta}(-r;k)\right)^{-1}$. Here $g_{\theta\theta}^R$ satisfies 
\begin{align}\label{eq:Hebcu4Z9hN}
    \left[\frac{d^2}{dr^2}-\frac{k^2}{1-c^2}-\frac{\sqrt{\cosh^2{\rho}-c^2}\sqrt{\gamma}m_{\theta\phi}^2}{1-c^2}\biggr\rvert_{\cosh{\rho}\to\text{cn}\left(ir|\frac{1}{1-c^2}\right)}\right]g^R_{\theta\theta}(r;k)&=0
\end{align}
and obeys the boundary condition $g_{\theta\theta}^R(r_m)=0$, and by symmetry $g_{\theta\theta}^L(r;k)=g_{\theta\theta}^R(-r;k)$. Next, writing $g_{\theta\theta}^R(r;k)=g_{\theta\theta,0}^R(r;k)+c^2g_{\theta\theta,1}^R(r;k)+\ldots$ and expanding \eqref{eq:Hebcu4Z9hN} to linear order in $c^2$ yields the two differential equations solved by $g_{\theta\theta,0}^R(r;k)$ and $g_{\theta\theta,1}^R(r;k)$:
\begin{align}
    \left(\frac{d^2}{dr^2}-k^2\right)g_{\theta\theta,0}^R&=0,&
    \left(\frac{d^2}{dr^2}-k^2\right)g_{\theta\theta,1}^R&=(k^2-\cos{4r})g_{\theta\theta,0}^R.
\end{align}
Here, we used \eqref{eq:Ir6apX7pK} and \eqref{eq:inverse metric gamma}. Furthermore, expanding $g_{\theta\theta}^R(r_m)=0$ to linear order in $c^2$ yields the following boundary conditions for $g_{\theta\theta,0}^R$ and $g_{\theta\theta,1}^R$, which are the same as in \eqref{eq:FUFcEZgW99}:
\begin{align}
    g_{\theta\theta,0}^1\left(\frac{\pi}{2};k\right)&=0,&g_{\theta\theta,1}^R\left(\frac{\pi}{2};k\right)-\frac{\pi}{8}\frac{dg_{\theta\theta,0}^R}{dr}\biggr\rvert_{r=\frac{\pi}{2};k}=0.
\end{align}

The solutions to these differential equations and boundary conditions are:
\begin{align}
    g_{\theta\theta,0}^R(r;k)&=2\sinh\left(k\left(\frac{\pi}{2}-r\right)\right)\label{eq:xlz8cMEe4m}\\
    g_{\theta\theta,1}^R(r;k)&=\frac{1}{8(4+k^2)}\biggr[2k\cosh\left(k\left(\frac{\pi}{2}-r\right)\right)\left((\pi-4r)(4+k^2)+\sin{4r}\right)\\&\hspace{9cm}-8\sin^2{2r}\sinh\left(k\left(\frac{\pi}{2}-r\right)\right)\biggr]\nonumber
\end{align}
Given these expressions, the boundary limit of $g^R_{\theta\theta}$ simplifies to:
\begin{align}
    \lim_{\rho\to \infty} e^{\rho}g^R_{\theta\theta}(r;k)&=\bar{r}_m \frac{dg^R_{\theta\theta}}{dr}\biggr\rvert_{r_m}=4k-\frac{2k}{4+k^2}c^2+\ldots.
\end{align}
Furthermore, \eqref{eq:DwWKnP1Xbu} yields the normalization
\begin{align}
    a_{\theta\theta}(k)&=\frac{1}{4k\sinh{\pi k}}+\frac{c^2}{16}\left[\frac{2}{k(4+k^2)\sinh{\pi k}}-\frac{\pi \cosh{\pi k}}{\sinh^2{\pi k}}\right]+\ldots.
\end{align}
As with $a_{yy}(k)$ in \eqref{eq:AQW0pCnZg0}, the poles of $a_{\theta\theta}(k)$ in the upper half-plane are at $k_n=ni$ for $n=1,2,3,\ldots$ and are of degree $1$ at order $c^0$ but of order $1$ or $2$ at order $c^2$. Consequently, in order to apply \eqref{eq:dRNhIUWDdm} to determine $W_{\theta\theta}$, we need to pull the factor 
\begin{align}\label{eq:y54edhy54edg}
    &e^{ik|\tau|}a_{\theta\theta}(k)\left(\lim_{\rho\to \infty} e^{\rho}g^R_{\theta\theta}\right)^2=e^{ik|\tau|}\left[\frac{4k}{\sinh{\pi k}}-c^2\left(\frac{2k}{(4+k^2)\sinh{\pi k}}+\frac{\pi k^2\cosh{\pi k}}{\sinh^2{\pi k}}\right)+\ldots\right].
\end{align}
back inside the contour integral. Evaluating the contour integral around the pole at $k_n$ amounts to computing the residues of \eqref{eq:y54edhy54edg}. At order $c^0$, we find
\begin{align}
    2\pi i\underset{k=n i}{\text{Res}}\biggr[e^{ik |\tau|} a_{\theta\theta}(k)\left(\lim_{\rho\to \infty} e^{\rho}g^R_{\theta\theta}\right)^2\biggr\rvert_{c^0}\biggr]&=8(-1)^{n+1}ne^{-n|\tau|}
\end{align}
and at order $c^2$:
\begin{align}
    2\pi i\underset{k=2i}{\text{Res}}\biggr[e^{ik |\tau|} a_{\theta\theta}(k)\left(\lim_{\rho\to \infty} e^{\rho}g^R_{\theta\theta}\right)^2\biggr\rvert_{c^2}\biggr]&=\frac{3}{2}\left(5-4|\tau|\right)e^{-2|\tau|}\label{eq:44a3RftUk8}\\
    2\pi i\underset{k=n i}{\text{Res}}\biggr[e^{ik |\tau|} a_{\theta\theta}(k)\left(\lim_{\rho\to \infty} e^{\rho}g^R_{\theta\theta}\right)^2\biggr\rvert_{c^2}\biggr]&=2(-1)^{n+1}\left[\frac{2n}{n^2-4}-2n+n^2|\tau|\right]e^{-n |\tau|}.\label{eq:fDZZfISjLQ}
\end{align}
We need to treat the residue at $k=2i$ separately from the other residues. 

Finally we write $W_{\theta\theta}(t_1,t_2)=W_{\theta\theta,0}(t_1,t_2)+c^2W_{\theta\theta,1}(t_1,t_2)+\ldots$ and apply \eqref{eq:dRNhIUWDdm} at each order. The $c^0$ result is
\begin{align}\label{eq:0b6hM12vzf}
    W_{\theta\theta,0}(t_1,t_2)=W_{yy,0}(t_1,t_2),
\end{align}
(see \eqref{eq:fmXxUbUWeZ}) and the $c^2$ result is
\begin{align}\label{eq:OYcCNSoZMG}
W_{\theta\theta,1}(t_1,t_2)&=\frac{g}{2\pi}\frac{1}{t_{12}^2}\frac{\chi^2}{|1-\chi|}\biggr[\text{sgn}(1-\chi)\left(-\frac{15}{4}+3|\tau|\right)e^{-2|\tau|}\nonumber\\&+\sum_{n=1,n\neq 2}^\infty \text{sgn}(1-\chi)^{n+1}\left(\frac{2n}{n^2-4}-2n+n^2|\tau|\right)e^{-n |\tau|}\biggr].
\end{align}
The series can be summed explicitly for all $\chi\in \mathbb{R}$. 

The last step is to compute $g_{\phi\theta}$ to order $c^2$. Since there is no $c^0$ term, we can write $g_{\phi\theta}=g_{\phi\theta,1}c^2+\ldots$. The linear term satisfies the boundary condition $g_{\phi\theta,1}(\pm\frac{\pi}{2},r';k)=0$ and the equation 
\begin{align}\label{eq:PNp3UKAx2i}
    \left(\frac{\partial^2}{\partial r^2}-k^2\right)g_{\phi\theta,1}(r,r';k)&=-k\sin^2(2r)g_{\theta\theta,0}(r,r';k).
\end{align}
Given that $g_{\theta\theta,0}(r,r';k)=a_{\theta\theta,0}(k)(g_{yy,0}^R(r;k)g_{yy,0}^R(-r';k)\theta(r-r')+g_{yy,0}^R(r';k)g_{yy,0}^R(-r;k)\theta(r'-r))$ with $g_{yy,0}^R(r;k)=2\sinh\left(k\left(\frac{\pi}{2}-r\right)\right)$ and $a_{\theta\theta,0}(k)=\frac{1}{4k\sinh{\pi k}}$, we consider the ansatz
\begin{align}\label{eq:g_(phitheta,1) ansatz}
    g_{\phi\theta,1}(r,r';k)=R(r,r';k)\theta(r-r')+L(r,r';k)\theta(r'-r).
\end{align}
It follows that
$R$ and $L$ satisfy
\begin{align}
    \left(\frac{\partial^2}{\partial r^2}-k^2\right)R(r,r';k)&=-\csch(\pi k)\sinh\left(k\left(\frac{\pi}{2}+r'\right)\right)\sin^2(2r)\sinh\left(k\left(\frac{\pi}{2}-r\right)\right),\\
    \left(\frac{\partial^2}{\partial r^2}-k^2\right)L(r,r';k)&=-\csch(\pi k)\sinh\left(k\left(\frac{\pi}{2}-r'\right)\right)\sin^2(2r)\sinh\left(k\left(\frac{\pi}{2}+r\right)\right),
\end{align}
and the boundary conditions $R(\frac{\pi}{2},r';k)=L(-\frac{\pi}{2},r';k)=0$. To ensure that the derivatives of the step functions do not produce delta functions in \eqref{eq:PNp3UKAx2i}, $R$ and $L$ also satisfy $R(r',r';k)=L(r',r';k)$ and $\partial_r R(r',r';k)=\partial_rL(r',r';k)$. These conditions uniquely determine $R$ and $L$. They are given explicitly by:
\begin{align}
    R(r,r';k)&=\biggr[\frac{1}{2k}\left(1-\frac{k^2\cos(4r)}{4+k^2}\right)\sinh\left(k\left(\frac{\pi}{2}-r\right)\right)+\left(r-\frac{k^2\sin(4r)}{4(4+k^2)}\right)\cosh\left(k\left(\frac{\pi}{2}-r\right)\right)\nonumber\\&-\frac{\pi}{2}\csch(\pi k)\sinh\left(k\left(\frac{\pi}{2}+r\right)\right)\biggr]\frac{\sinh\left(k\left(\frac{\pi}{2}+r'\right)\right)}{4k\sinh{\pi k}}+(r\leftrightarrow -r'),\label{eq:prRliJQwHv}\\
    L(r,r';k)&=R(r',r;k).\label{eq:VAdyqSAJ2j}
\end{align}
Note that $R(r,r';-k)=-R(r,r';k)$, which means $g_{\phi\theta,1}(r,r';-k)=-g_{\phi\theta,1}(r,r';k)$. This reflects the fact that $G_{\phi\theta}(\rho,\tau;\rho',\tau')$ changes sign when $\tau,\tau'\to -\tau,-\tau'$, as discussed below \eqref{eq:phitheta Green's eqn}. 

Next, we apply \eqref{eq:yy/xx bndy-to-bndy}, which tells us
\begin{align}\label{eq:npycTN4Fss}
    W_{\phi\theta,1}(t_1,t_2)&=\frac{g}{4\pi}\frac{1}{t_{12}^2}\frac{\chi^2}{|1-\chi|}\text{sgn}(\tau)\int_{-\infty}^\infty dk e^{ik|\tau|}\lim_{\substack{\rho\to \eta_1\infty \\ \rho'\to\eta_2\infty}}e^{\eta_1\rho+\eta_2\rho'}g_{\phi\theta,1}(r,r';k).
\end{align}
where $\eta_1=1$ if $t_3<t_1<t_4$ and $\eta_1=-1$ otherwise, and analogously for $\eta_2$. We have used that $g_{\phi\theta,1}$ is odd under $k\to -k$ to pull out the sign of $\tau$ from the exponential. 

Thus we need to evaluate the boundary limits of $g_{\phi\theta,1}(r,r';k)$. We note that when $c=0$, \eqref{eq:B5pnopgBFw} reduces to $r=\rm{arcsec}(\cosh{\rho})$, whose asymptotic behavior is $r\sim \frac{\pi}{2}-2e^{-\rho}$ as $\rho\to \infty$ and $r\sim -\frac{\pi}{2}+2e^\rho$ as $\rho\to -\infty$. Likewise for $r'$. Thus, given \eqref{eq:g_(phitheta,1) ansatz} and \eqref{eq:prRliJQwHv}-\eqref{eq:VAdyqSAJ2j}, the boundary limits are found to be
\begin{align}
    \lim_{\substack{\rho\to \infty \\ \rho'\to-\infty}}e^{\rho-\rho'}g_{\phi\theta,1}(r,r';k)&= \lim_{\substack{\rho\to -\infty \\ \rho'\to\infty}}e^{\rho'-\rho}g_{\phi\theta,1}(r,r';k)\label{eq:58d0zfA36R}\\&\hspace{4cm}=k \pi \csch(k \pi)^2\cosh(k\pi)-\frac{4}{4+k^2}\csch{\pi k}\nonumber\\
    \lim_{\substack{\rho\to \infty \\ \rho'\to\infty}}e^{\rho+\rho'}g_{\phi\theta,1}(r,r';k)&=\lim_{\substack{\rho\to -\infty \\ \rho'\to-\infty}}e^{-\rho-\rho'}g_{\phi\theta,1}(r,r';k)\label{eq:JcyYhkaVhJ}\\&\hspace{4cm}=-k \pi \csch(k \pi)^2+\frac{4}{4+k^2}\csch{\pi k}\cosh(k\pi)\nonumber
\end{align}
Note that the order of limits in \eqref{eq:JcyYhkaVhJ} does not matter.

Finally, we close the contour in \eqref{eq:npycTN4Fss} at infinity in the upper half-plane and deform it to pick up the residues at the poles at $k=ni$, $n=1,2,3,\ldots$. 

\begin{align}
    2\pi i\underset{k=2i}{\text{Res}}\biggr[e^{ik|\tau|}\lim_{\substack{\rho\to \eta_1\infty \\ \rho'\to\eta_2\infty}}e^{\eta_1\rho+\eta_2\rho'}g_{\phi\theta,1}\biggr]&=i\eta_1\eta_2 e^{-2|\tau|}\left(-\frac{3}{2}+6|\tau|\right),\label{eq:j93M4dPUUw}\\
    2\pi i\underset{k=ni}{\text{Res}}\biggr[e^{ik|\tau|}\lim_{\substack{\rho\to \eta_1\infty \\ \rho'\to\eta_2\infty}}e^{\eta_1\rho+\eta_2\rho'}g_{\phi\theta,1}\biggr]&=i(\eta_1\eta_2)^{n+1}2e^{-n|\tau|}\left(\frac{n^2}{4-n^2}+n|\tau|\right).\label{eq:Yv0hghwOtj}
\end{align}
We need to treat the residue at $k=2i$ separately from the residues at $k=ni$, $n=1,3,4,5,\ldots$. The simple relation between the residues when $\rho$ and $\rho'$ are sent to opposite boundaries (i.e., $\eta_1=-\eta_2$) and the residues when $\rho$ and $\rho'$ are sent to the same boundary (i.e., $\eta_1=\eta_2$) results from the fact that \eqref{eq:58d0zfA36R} and \eqref{eq:JcyYhkaVhJ} differ by multiplicative factors of $\cosh{\pi k}$ and $\cosh(n\pi i)=(-1)^n$.

Substituting \eqref{eq:j93M4dPUUw}-\eqref{eq:Yv0hghwOtj} into \eqref{eq:npycTN4Fss}, we finally find
\begin{align}\label{eq:FZXxEOjkOQ}
W_{\phi\theta,1}(t_1,t_2)&=\frac{ig}{2\pi}\frac{1}{t_{12}^2}\frac{\chi^2}{|1-\chi|}\text{sgn}(\tau)\biggr[\text{sgn}(1-\chi)\left(-\frac{3}{4}+3|\tau|\right)e^{-2|\tau|}\\&\hspace{5cm}+\sum_{n=1,n\neq 2}^\infty \text{sgn}(1-\chi)^{n+1}\left(\frac{n^2}{4-n^2}+n|\tau|\right)e^{-n |\tau|}\biggr].\nonumber
\end{align}
The series can be summed explicitly for all $\chi\in \mathbb{R}$.

To summarize, starting from the quadratic Lagrangian for the $\theta$ and $\phi$ fluctuation modes, given in \eqref{eq:LpWrSXGjF0}, we were able to determine the boundary-to-boundary propagators for the $\theta$ and $\phi$ modes to linear order in $c^2$. When these propagators are combined with the classical vertex operators given in \eqref{eq:classical ZZb}, the dictionary in \eqref{eq:2pcuM8BXmq} and \eqref{eq:FerYQICfpw} determines the defect correlators in \eqref{eq:ZZbZ^JZb^J}-\eqref{eq:ZbZZ^JZb^J} to be:
\begin{align}
    \frac{\braket{Z(t_1)\bar{Z}(t_2)Z^J(t_3)\bar{Z}^J(t_4)}}{\braket{Z^J(t_3)\bar{Z}^J(t_4)}}&=\frac{4g^2 c^2}{t_{12}^2}\frac{\chi^2}{(1-\chi)^2}+2\biggr[W_{\theta\theta,0}(t_1,t_2)\nonumber\\&+(W_{\theta\theta,1}(t_1,t_2)+iW_{\phi\theta,1}(t_1,t_2))c^2+O(c^4)\biggr]+O(g^0),\\
    \frac{\braket{\bar{Z}(t_1)Z(t_2)Z^J(t_3)\bar{Z}^J(t_4)}}{\braket{Z^J(t_3)\bar{Z}^J(t_4)}}&=\frac{4g^2 c^2}{t_{12}^2}\chi^2+2\biggr[W_{\theta\theta,0}(t_1,t_2)\nonumber\\&+(W_{\theta\theta,1}(t_1,t_2)-iW_{\phi\theta,1}(t_1,t_2))c^2+O(c^4)\biggr]+O(g^0).
\end{align}
where $W_{\theta\theta,0}$,  $W_{\theta\theta,1}$ and $W_{\phi\theta,1}$ are given in \eqref{eq:0b6hM12vzf}-\eqref{eq:OYcCNSoZMG} and \eqref{eq:FZXxEOjkOQ}. When the series are summed explicitly and the defect correlators are translated to $G_{Z\bar{Z}}$ and $G_{\bar{Z}Z}$ using \eqref{eq:ZZbZ^JZb^J}-\eqref{eq:ZbZZ^JZb^J}, we find perfect agreement with \eqref{eq:vcByXaNCgz}-\eqref{eq:cF7APdqE0I}.

We were also able to extend the perturbative analysis in this section to order $c^4$, but with considerably more effort. We spare the reader the details, but the results for $W_{yy,2}$, $W_{\theta\theta,2}$ and $W_{\phi\theta,2}$, were again in perfect agreement with the perturbative results using the much simpler approach discussed in Section~\ref{sec:small and large J/g}.

\section{Four-point functions and OPE data at finite charge}\label{sec:dyS0wZdrxG}

In this appendix, we study the four-point correlators in \eqref{eq:scalar large charge correlator}-\eqref{eq:displacement large charge correlator} and the OPE data that can be extracted from them in the planar, strongly coupled regime ($N\to \infty$, $g\gg 1$) with $J$ finite. The ``finite charge'' results are valid in the parameter range $1\sim J\ll g$ whereas the large charge results discussed in the body of the paper are valid in the parameter range $1\ll J\sim g$. When $1\ll J\ll g$, the two regimes overlap and can be compared.

\subsection{Finite charge four-point functions}

The correlation functions of light scalar and displacement operators in the strongly coupled regime of the Wilson line dCFT can be computed holographically by studying the fluctuations about the AdS$_2$ classical string dual to the Wilson line. The fluctuations are weakly coupled and governed by a tower of interactions in the AdS$_2$ bulk suppressed by $1/g$ \cite{giombi2017half}. At leading order the defect operators behave like generalized free fields, and the Witten diagrams involve only Wick contractions of the defect operators using the AdS$_2$ boundary-to-boundary propagators. At the first sub-leading order, the correlators receive contributions from the one-loop self-energy corrections to the boundary-to-boundary propagators and from four-point contact diagrams. The self energy corrections can be deduced indirectly using results from localization \cite{giombi2018exact} while the contact diagrams can be evaluated by integrating over the position in the bulk of the appropriate interaction vertices connecting four boundary-to-bulk propagators \cite{giombi2017half}. The four-point functions of four elementary (i.e., $J=1$, non-composite) scalars, four elementary displacement operators, and two elementary scalars and two elementary displacement operators were discussed in \cite{giombi2017half}. The two- and three-point functions of higher finite charge composite scalars was discussed in \cite{giombi2018exact}. The present discussion is a straightforward adaptation of those analyses.  

\begin{figure}[t]
\centering
\begin{minipage}{0.48\textwidth}
\centering
\includegraphics[clip, height=3cm]{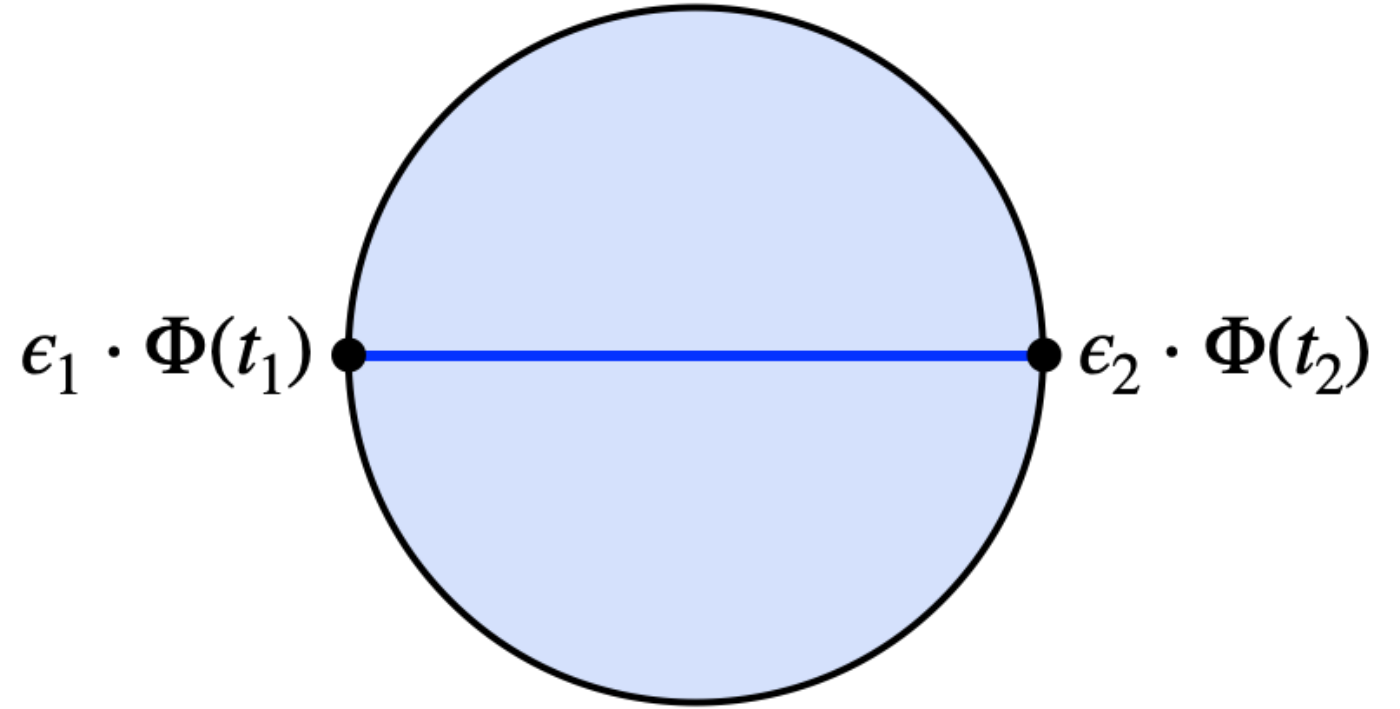}\\
{\bf a. $(12)$}
\end{minipage}
\begin{minipage}{0.48\textwidth}
\centering
\includegraphics[clip, height=3cm]{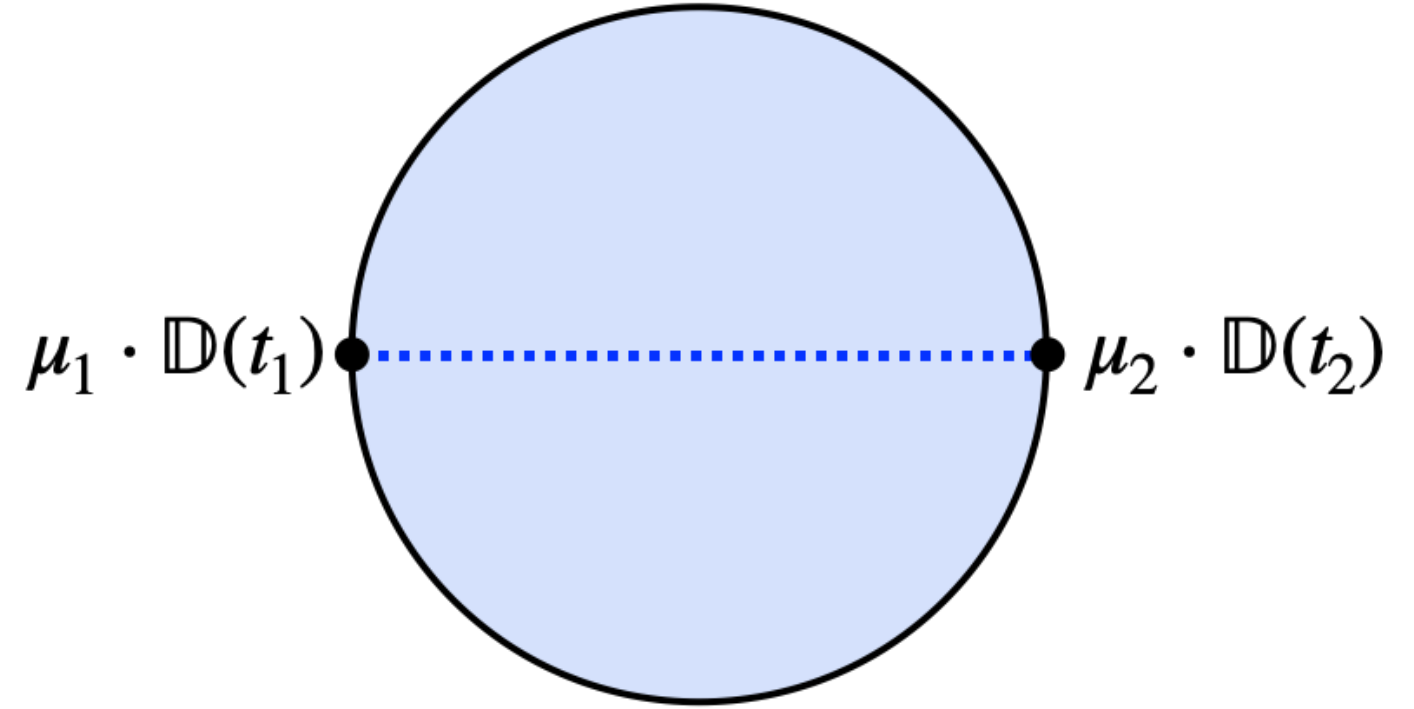}\\
{\bf b. $[12]$}
\end{minipage} \\ 

\vspace{0.5 cm}

\begin{minipage}{0.48\textwidth}
\centering
\includegraphics[clip, height=3cm]{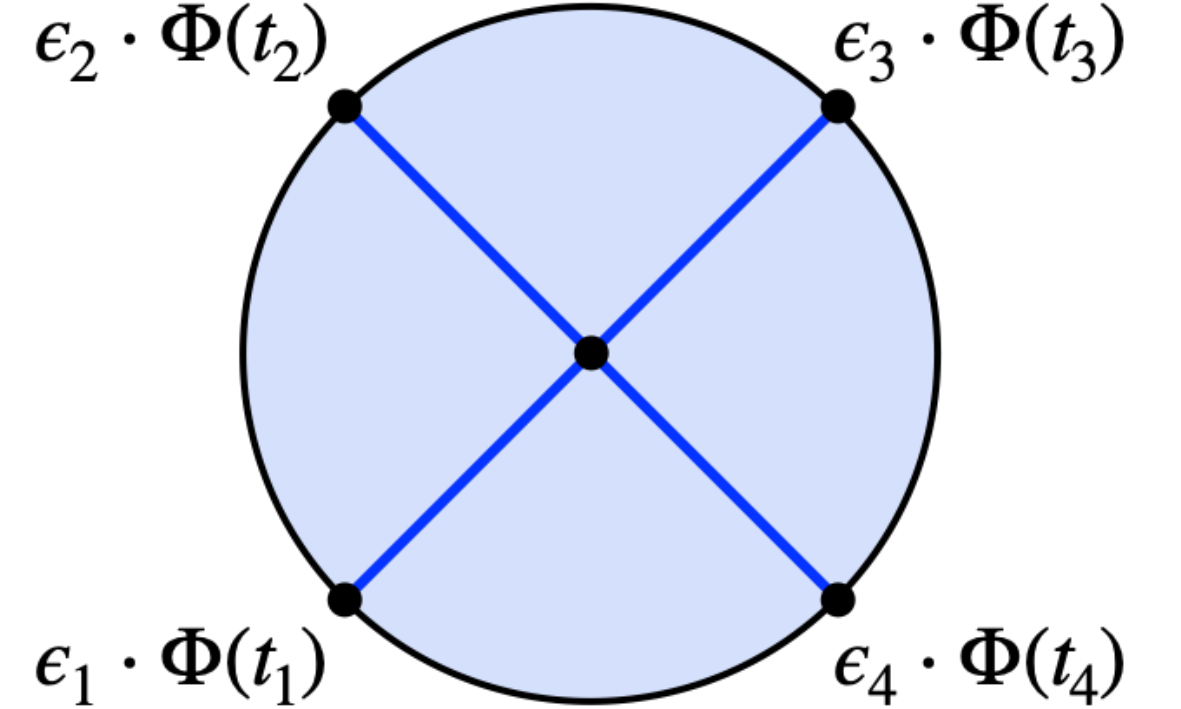}\\
{\bf c. $(1234)$}
\end{minipage}
\begin{minipage}{0.48\textwidth}
\centering
\includegraphics[clip, height=3cm]{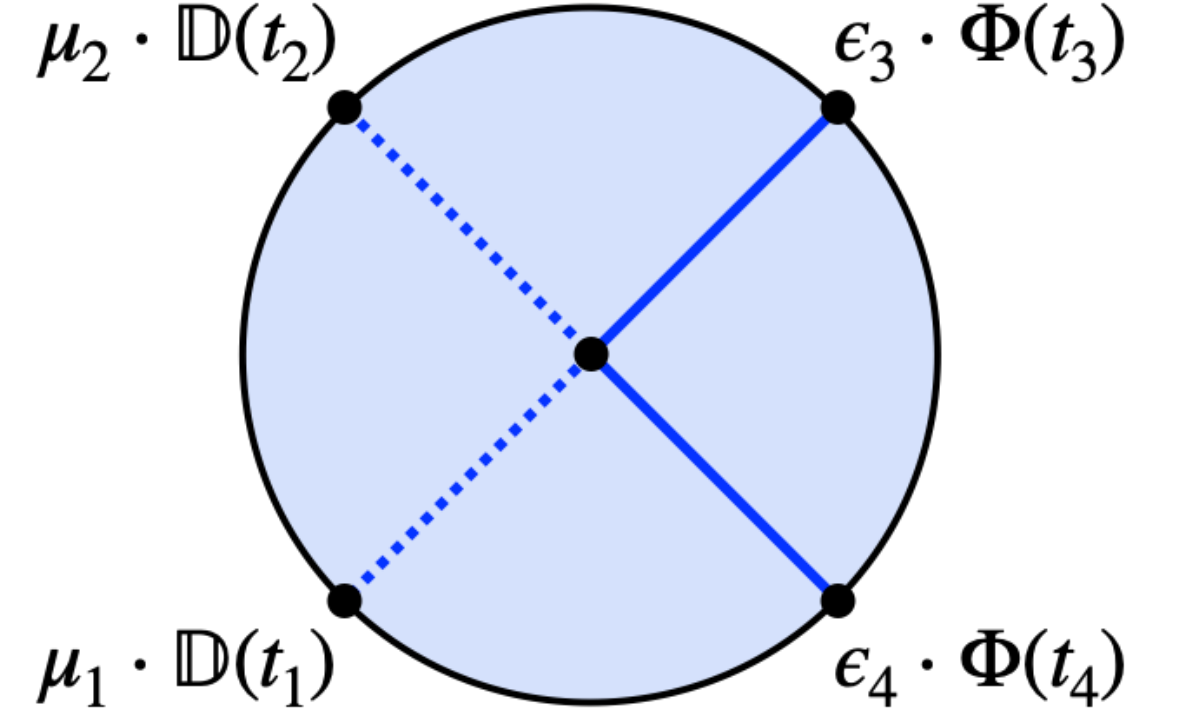}\\
{\bf d. $[12|34)$}
\end{minipage}
\caption{Four elementary Witten diagrams that we use to compute the two-point function with two charge $J$ scalars as well as the four-point functions with two charge $J$ scalars and two unit scalars or two unit displacement operators. \textbf{a.} scalar boundary-to-boundary propagator with one-loop correction (not depicted). \textbf{b.} displacement boundary-to-boundary propagator with one-loop correction (not depicted). \textbf{c.} four-point contact diagram with four legs incident on scalars. \textbf{d.} four-point contact diagram with two legs incident on scalars and two legs incident on displacment operators.} 
\label{fig:elementary Witten diagrams}
\end{figure}

The Witten diagrams contributing to the correlators in \eqref{eq:scalar large charge correlator}-\eqref{eq:displacement large charge correlator} to subleading order in $1/g$ consist of four ``building blocks.'' The first two are the ``dressed'' (i.e., one-loop corrected) boundary-to-boundary propagators that connect two elementary scalars or two elementary displacement operators. These are given in \eqref{eq:phi-phi D-D two pt}. We will write them using the abbreviated notation
\begin{align}
    (12)&\equiv \braket{\epsilon_1\cdot \Phi(t_1)\epsilon_2\cdot \Phi(t_2)}=\frac{2g}{\pi}\frac{\epsilon_1\cdot \epsilon_2}{t_{12}^2}\left(1-\frac{3}{8\pi g}+O(1/g^2)\right),\label{eq:scalar bdy-to-bdy propagator}
    \\\relax [12]&\equiv \braket{\mu_1\cdot \mathbb{D}(t_1)\mu_2\cdot \mathbb{D}(t_2)}=\frac{12g}{\pi}\frac{\mu_1\cdot \mu_2}{t_{12}^4}\left(1-\frac{3}{8\pi g}+O(1/g^2)\right).\label{eq:displacement bdy-to-bdy propagator}
\end{align}
and represent them graphically using solid and dashed curves, respectively, as illustrated in Figure~\ref{fig:elementary Witten diagrams}~\hyperref[fig:elementary Witten diagrams]{a} and Figure~\ref{fig:elementary Witten diagrams}~\hyperref[fig:elementary Witten diagrams]{b}. 

The third and fourth building blocks are the contact diagram with four external legs connected to elementary scalars and the contact diagram with two external legs connected to elementary scalars and two legs connected to elementary displacement operators. Due to the interaction vertex in the bulk, these are suppressed by $1/g$. We write the contact diagrams using the abbreviated notation,
\begin{align}
    (1234)&\equiv \frac{(12)(34)}{4\pi g}\Lambda(\chi,\xi,\zeta),&[12|34)&\equiv \frac{[12](34)}{4\pi g}G_{\mathbb{D}\Phi}(\chi),\label{eq:4pt contact diagrams}
\end{align}
and depict them graphically as in Figure~\ref{fig:elementary Witten diagrams}~\hyperref[fig:elementary Witten diagrams]{c} and Figure~\ref{fig:elementary Witten diagrams}~\hyperref[fig:elementary Witten diagrams]{d}.
Here, 
\begin{align}
    \Lambda(\chi,\xi,\zeta)&\equiv\biggr[G_S^{(1)}(\chi)-\frac{2}{5}G_T^{(1)}(\chi)+\xi\left(G_T^{(1)}(\chi)+G_A^{(1)}(\chi)\right)+\zeta \left(G_T^{(1)}(\chi)-G_A^{(1)}(\chi)\right)\biggr],
\end{align}
and the functions $G_S^{(1)}(\chi)$, $G_T^{(1)}(\chi)$ and $G_A^{(1)}(\chi)$ are given explicitly in \cite{giombi2017half}. Likewise, $G_{\mathbb{D}\Phi}(\chi)$ also follows from the results therein and is given explicitly by
\begin{align}\label{eq:x969Tukd9K}
    G_{\mathbb{D}\Phi}(\chi)\equiv -4\left(1+\left(\frac{1}{\chi}-\frac{1}{2}\right)\log|1-\chi|\right).
\end{align}

Because $(\epsilon_3\cdot \Phi(t_3))^J$ and $(\epsilon_4\cdot \Phi(t_4))^J$ are composite operators, there will be also be diagrams in which one pair of legs of the scalar four-point contact diagram are incident on the same point, or two pairs of legs are incident on two points.\footnote{By contrast, there are no contributions from self-contracting boundary-to-boundary propagators because $\epsilon_i^2=\mu_i^2=0$.} When $\epsilon_4=\epsilon_1$ and $t_4\to t_1$, then $\xi=1$, $\zeta=0$, $\chi\to 1$, and therefore $\Lambda\to-3$. Thus, the scalar contact diagrams with one or two pairs of legs coincident simplify to
\begin{align}
    (1123)&=-\frac{3}{4\pi g}(12)(13), & (1122)=-\frac{3}{4\pi g}(12)^2.
\end{align}

\begin{figure}
\centering
\includegraphics[clip, height=3cm]{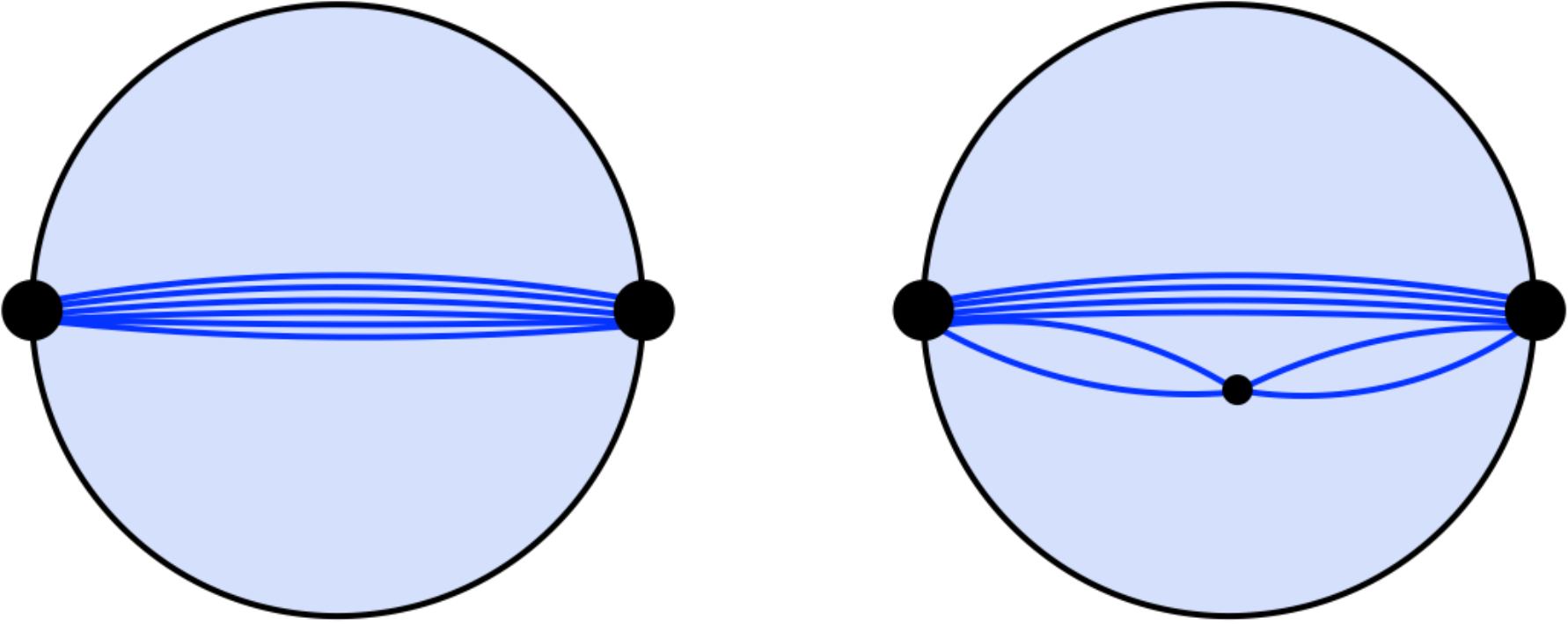}
\caption{The Witten diagrams contributing to the two-point function $\braket{(\epsilon_3\cdot \Phi(t_3))^J(\epsilon_4\cdot \Phi(t_4))^J}$ to first subleading order in $1/g$ have two distinct topologies, corresponding to the two terms in \eqref{eq:HH 1}.} 
\label{fig:HH Witten diagrams}
\end{figure}

Now we have everything necessary to determine the four-point correlators in \eqref{eq:scalar large charge correlator}-\eqref{eq:displacement large charge correlator}, as well as the ``heavy-heavy'' two-point function they are normalized by, to subleading order. Let's start with the heavy-heavy two-point function. The contributing Witten diagrams belong to one of two classes distinguished by their topologies, which are depicted in Figure~\ref{fig:HH Witten diagrams}. Keeping track of the combinatorial factors, we find
\begin{align}
    \braket{(\epsilon_3\cdot \Phi(t_3))^J(\epsilon_4\cdot \Phi(t_4))^J}&= J!(34)^J +\binom{J}{2}^2(J-2)!(3344)(34)^{J-2}\label{eq:HH 1}\\&=\left(\frac{2g}{\pi}\frac{\epsilon_3\cdot \epsilon_4}{t_{34}^2}\right)^JJ! \left(1-\frac{3J(J+1)}{16\pi g}+O(1/g^2)\right).\label{eq:HH 2}
\end{align}

\begin{figure}
\centering
\includegraphics[clip, height=6cm]{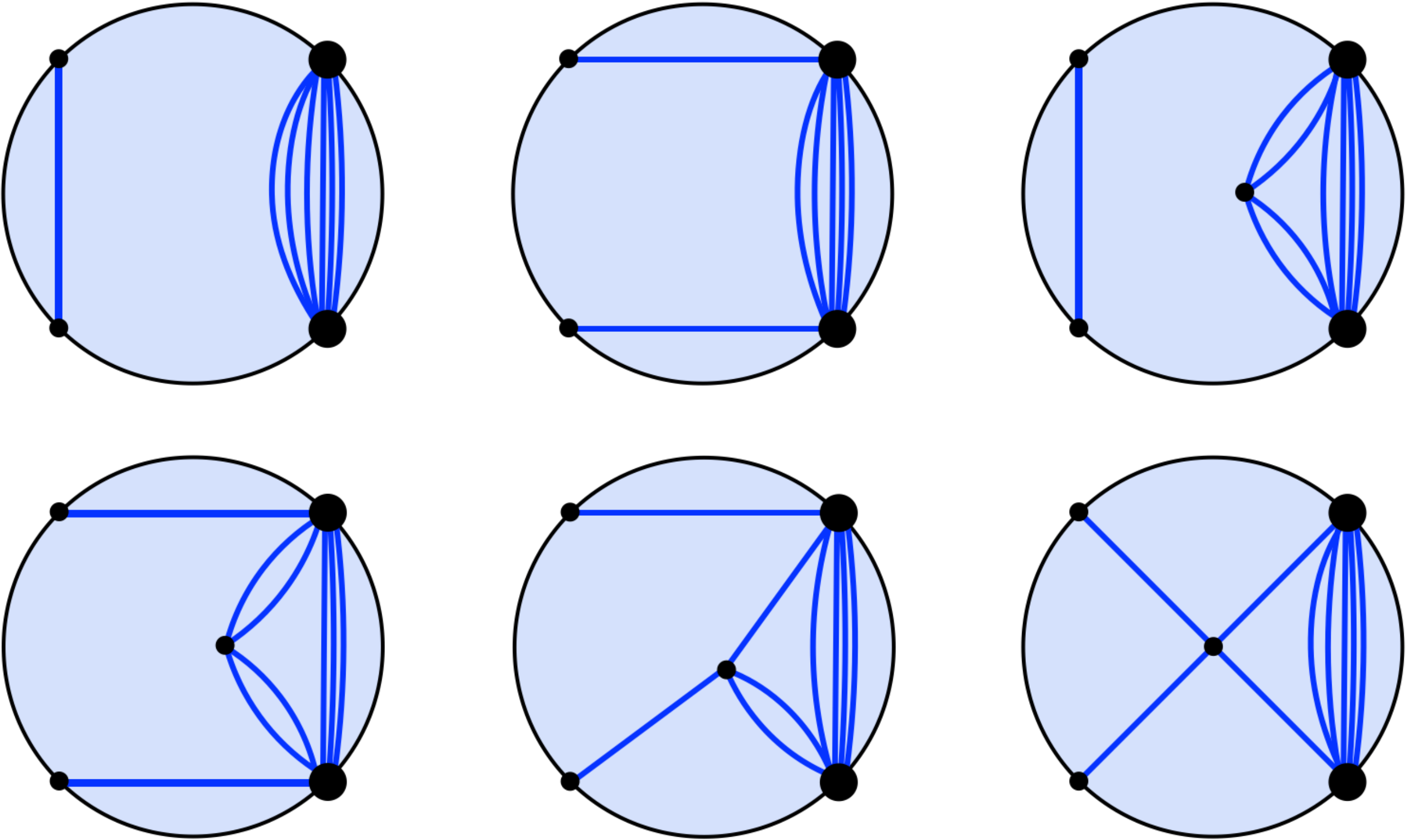}
\caption{The Witten diagrams contributing to the four-point function $\braket{\epsilon_1\cdot \Phi(t_1)\epsilon_2\cdot \Phi(t_2)(\epsilon_3\cdot \Phi(t_3))^J(\epsilon_4\cdot \Phi(t_4))^J}$ to first subleading order in $1/g$ have six distinct topologies. There are eleven topologies ($=1+2+1+2+4+1$) if we distinguish between configurations with the light and heavy operators permuted. These correspond to the eleven terms in \eqref{eq:PhPhHH}.} 
\label{fig:PhPhHH Witten diagrams}
\end{figure}

Second, the Witten diagrams contributing to the four-point function in \eqref{eq:scalar large charge correlator} belong to one of six classes distinguished by their topologies, which are depicted in Figure~\ref{fig:PhPhHH Witten diagrams}. Permuting the two light operators and the two heavy operators, and keeping track of the appropriate combinatorial factors, we find
\begin{align}
    &\braket{\epsilon_1\cdot \Phi(t_1)\epsilon_2\cdot \Phi(t_2)(\epsilon_3\cdot \Phi(t_3))^J(\epsilon_4\cdot \Phi(t_4))^J}=J!\left[(12)(34)+J(13)(24)+J(14)(23)\right](34)^{J-1}\nonumber\\&\hspace{2cm}+\binom{J}{2}^2(J-2)!\left[(12)(34)+(J-2)(13)(24)+(J-2)(14)(23)\right](3344)(34)^{J-3}\nonumber\\&\hspace{2cm}+\binom{J}{2}J!\left[(1344)(23)+(2344)(13)+(1334)(24)+(2334)(14)\right](34)^{J-2}\nonumber\\&\hspace{2cm}+J!J(1234)(34)^{J-1}.\label{eq:PhPhHH}
\end{align}
The first line above captures the contributions from the first two topologies in Figure~\ref{fig:PhPhHH Witten diagrams}; the second line captures the contributions from the third and fourth topologies; the third line captures the contributions from the fifth topology; and the last line captures the contribution from the sixth topology.

Given the explicit expressions for the propagators and the contact diagram in \eqref{eq:scalar bdy-to-bdy propagator} and \eqref{eq:4pt contact diagrams}, the expression in \eqref{eq:PhPhHH} simplifies to
\begin{align}
    &\braket{\epsilon_1\cdot \Phi(t_1)\epsilon_2\cdot \Phi(t_2)(\epsilon_3\cdot \Phi(t_3))^J(\epsilon_4\cdot \Phi(t_4))^J}=J!\left(\frac{2g}{\pi}\right)^{J+1}\frac{\epsilon_1\cdot \epsilon_2}{x_{12}^2}\frac{(\epsilon_3\cdot \epsilon_4)^J}{x_{34}^{2J}}\biggr[1-\frac{3(J^2+J+2)}{16\pi g}\nonumber\\&\hspace{3cm}+J\left(\xi \chi^2+\frac{\zeta\chi^2}{(1-\chi)^2}\right)\left(1-\frac{3J(J+3)}{16\pi g}\right)+\frac{J}{4\pi g}\Lambda+O(1/g^2)\biggr].
\end{align}
Thus, given \eqref{eq:HH 2}, the normalized four-point function is
\begin{align}
    &\frac{\braket{\epsilon_1\cdot\Phi(t_1)\text{ }\epsilon_2\cdot \Phi(t_2)(\epsilon_3\cdot\Phi(t_3))^J(\epsilon_4\cdot \Phi(t_4))^J}}{\frac{2g}{\pi}\frac{\epsilon_1\cdot \epsilon_2}{t_{12}^2}\braket{(\epsilon_3\cdot\Phi(t_3))^J(\epsilon_4\cdot \Phi(t_4))^J}}\\&\hspace{3cm}=1-\frac{3}{8\pi g}+J\left(\xi \chi^2+\frac{\zeta \chi^2}{(1-\chi)^2}\right)\left(1-\frac{3J}{8\pi g}\right)+\frac{J}{4\pi g}\Lambda(\xi,\zeta,\chi)+O(1/g^2),\nonumber
\end{align}
and the three conformally invariant functions reduce to
\begin{align}
    G_1(\chi)&=1-\frac{3}{8\pi g}+\frac{J}{4\pi g}\left(G_S^{(1)}-\frac{2}{5}G_T^{(1)}\right)+O(1/g^2),\label{eq:0ngGpoeYzk}\\
    G_2(\chi)&=\frac{J\chi^2}{(1-\chi)^2}\left(1-\frac{3J}{8\pi g}\right)+\frac{J}{4\pi g}(G_T^{(1)}-G_A^{(1)})+O(1/g^2),\label{eq:Ty4BinBhU5}\\
    G_3(\chi)&=J\chi^2\left(1-\frac{3J}{8\pi g}\right)+\frac{J}{4\pi g}(G_T^{(1)}+G_A^{(1)})+O(1/g^2).\label{eq:Ieh8qqkwO0}
\end{align}
We can thus identify the first few terms in the double expansion of the conformal functions in $1/g$ and $\mathcal{J}=J/g$. Using \eqref{eq:yxrxVwSqbk}, we can also write the result as an expansion in $1/g$ and $c^2$. We find:
\begin{align}
    G_1(\chi)&=\left(1+\frac{\mathcal{J}}{4\pi }\left(G_S^{(1)}-\frac{2}{5}G_T^{(1)}\right)+O(\mathcal{J}^2)\right)+\frac{1}{g}\left(-\frac{3}{8\pi}+O(\mathcal{J})\right)+O(1/g^2)\nonumber\\&=\left(1+\frac{c^2}{4}\left(G_S^{(1)}-\frac{2}{5}G_T^{(1)}\right)+O(c^4)\right)+\frac{1}{g}\left(-\frac{3}{8\pi }+O(c^2)\right)+O(1/g^2)\label{eq:l10mb4DJQx}\\
    G_2(\chi)&=\frac{g\chi^2}{(1-\chi)^2}\left(\mathcal{J}-\frac{3\mathcal{J}^2}{8\pi}+O(\mathcal{J}^3)\right)+\left(\frac{\mathcal{J}}{4\pi}\left(G_T^{(1)}-G_A^{(1)}\right)+\mathcal{J}^2\right)+O(1/g)\nonumber\\&=\frac{g\pi\chi^2}{(1-\chi)^2}\left(c^2+0c^4+O(c^6)\right)+\left(\frac{c^2}{4}\left(G_T^{(1)}-G_A^{(1)}\right)+O(c^4)\right)+O(1/g).\label{eq:D1q58K74wn}\\
    G_3(\chi)&=g\chi^2\left(\mathcal{J}-\frac{3\mathcal{J}^2}{8\pi}+O(\mathcal{J}^3)\right)+\left(\frac{\mathcal{J}}{4\pi}\left(G_T^{(1)}+G_A^{(1)}\right)+O(\mathcal{J}^2)\ldots\right)+O(1/g)\nonumber\\&=g\pi\chi^2\left(c^2+0c^4+O(c^6)\right)+\left(\frac{c^2}{4}\left(G_T^{(1)}+G_A^{(1)}\right)+O(c^4)\right)+O(1/g)\label{eq:P8FIswyodt}
\end{align}
Given the explicit forms of $G_S^{(1)}(\chi)$, $G_T^{(1)}(\chi)$ and $G_A^{(1)}(\chi)$ from \cite{giombi2017half}, we easily check that \eqref{eq:l10mb4DJQx}-\eqref{eq:P8FIswyodt} precisely match the \eqref{eq:vcByXaNCgz}-\eqref{eq:cF7APdqE0I} in the overlapping terms.

\begin{figure}
\centering
\includegraphics[clip, height=3cm]{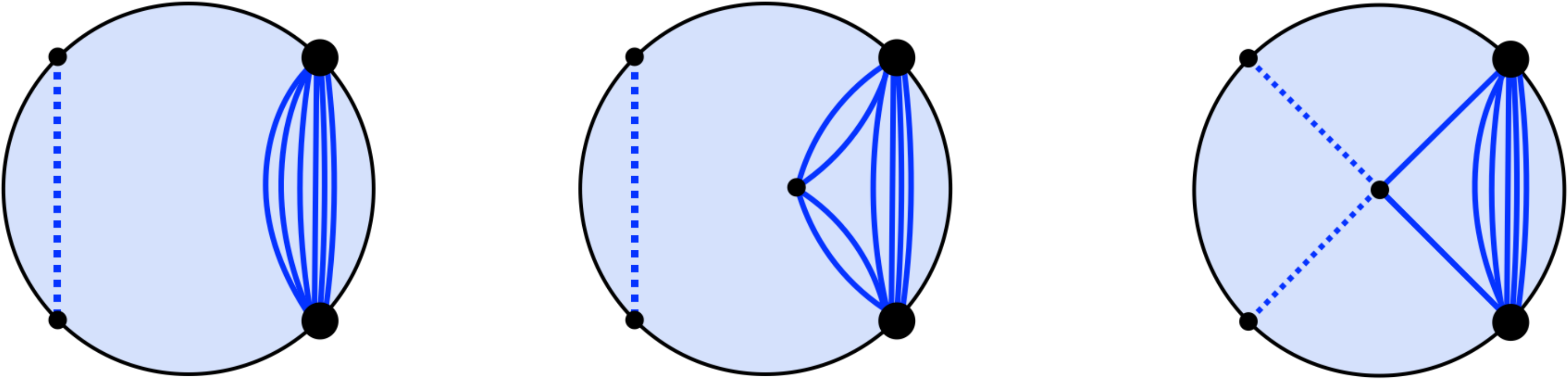}
\caption{The Witten diagrams contributing to the four-point function $\braket{\mu_1\cdot \mathbb{D}(t_1)\mu_2\cdot \mathbb{D}(t_2)(\epsilon_3\cdot \Phi(t_3))^J(\epsilon_4\cdot \Phi(t_4))^J}$ to first subleading order in $1/g$ have three distinct topologies corresponding to the three terms in \eqref{eq:TLMgM4Hxbp}.} 
\label{fig:DDHH Witten diagrams}
\end{figure}

Finally, the Witten diagrams contributing to the four-point function in \eqref{eq:displacement large charge correlator} belong to one of three classes distinguished by their topologies, which are depicted in Figure~\ref{fig:DDHH Witten diagrams}. Explicitly, we have
\begin{align}
    &\braket{\mu_1\cdot \mathbb{D}(t_1)\mu_2\cdot\mathbb{D}(t_2)(\epsilon_3\cdot\Phi(t_3))^J(\epsilon_4\cdot \Phi(t_4))^J}\nonumber\\&=J![12](34)^J+\binom{J}{2}^2(J-2)![12](3344)(34)^{J-2}+J!J[12|34)(34)^{J-1}\label{eq:TLMgM4Hxbp}\\&=6J!\left(\frac{2g}{\pi}\right)^{J+1}\frac{\mu_1\cdot \mu_2}{t_{12}^4}\frac{(\epsilon_3\cdot \epsilon_4)^J}{t_{34}^{2J}}\biggr[1-\frac{3(J^2+J+2)}{16\pi g}+\frac{J}{4\pi g}G_{\mathbb{D}\Phi}(\chi)+\ldots\biggr].\label{eq:L66U2nnaPq}
\end{align}
The conformal function $G_4(\chi)$ is therefore:
\begin{align}
    G_4(\chi)&=\frac{\braket{\mu_1\cdot \mathbb{D}(t_1)\mu_2\cdot\mathbb{D}(t_2)(\epsilon_3\cdot\Phi(t_3))^J(\epsilon_4\cdot \Phi(t_4))^J}}{\frac{12g}{\pi}\frac{\mu_1\cdot \mu_2}{t_{12}^4}(\epsilon_3\cdot\Phi(t_3))^J(\epsilon_4\cdot \Phi(t_4))^J}=1-\frac{3}{4\pi g}+\frac{J}{4\pi g}G_{\mathbb{D}\Phi}(\chi)+O(1/g^2),
\end{align}
which can also be written as an expansion in $1/g$ and $\mathcal{J}$ or $1/g$ and $c^2$:
\begin{align}
    G_4(\chi)&=\left(1+\frac{\mathcal{J}}{4\pi }G_{\mathbb{D}\Phi}(\chi)+O(\mathcal{J}^2)\right)+\frac{1}{g}\left(-\frac{3}{4\pi}+O(\mathcal{J})\right)+O(1/g^2)\nonumber\\&=\left(1+\frac{c^2}{4}G_{\mathbb{D}\Phi}(\chi)+O(c^4)\right)+\frac{1}{g}\left(-\frac{3}{4\pi}+O(c^2)\right)+O(1/g^2).
\end{align}
Given \eqref{eq:x969Tukd9K}, this precisely matches \eqref{eq:7k7MA0pcKL} in the overlapping terms. 

\subsection{Finite charge OPE data}\label{sec:Finite charge OPE data} Next we extract OPE data from the finite charge four-point functions. Because the large charge and finite charge four-point functions agree in the regime of overlapping validity, the corresponding OPE data must necessarily agree as well. Nonetheless, the procedure we use in the present section to extract the finite charge OPE data is conceptually different from what we used in Section~\ref{sec:extract OPE data}, and therefore provides additional insight and serves as a nice check of the large charge results.

Like in Section~\ref{sec:extract OPE data}, we focus on the four-point function $\braket{ZZ^J\bar{Z}\bar{Z}^J}$ in the $12\to 34$ channel. (For brevity, we will not also consider $\braket{Z^JZ\bar{Z}\bar{Z}^J}$). From \eqref{eq:3MbSdyqRZo} combined with \eqref{eq:0ngGpoeYzk} and \eqref{eq:Ty4BinBhU5}, we have
\begin{align}
    \braket{Z(t_1)Z^J(t_2)\bar{Z}(t_3)\bar{Z}^J(t_4)}&=\frac{\mathcal{N}_{Z\bar{Z}}\mathcal{N}_{Z^J\bar{Z}^J}}{t_{21}^{J+1}t_{43}^{J+1}}\left(\frac{t_{42}}{t_{31}}\right)^{1-J}\chi^{J+1}\biggr[1+\frac{J}{(1-\chi)^2}+\frac{J}{4\pi g}\biggr(-\frac{3(J-1)}{2(1-\chi)^2}\nonumber\\&+\left(G_S^{(1)}+\frac{3}{5}G_T^{(1)}-G_A^{(1)}\right)\biggr\rvert_{\chi^{-1}}\biggr)+O(1/g^2)\biggr].\label{eq:wjNw4KhCRA}
\end{align}
We can again analyze this correlator using the conformal block expansion given in \eqref{eq:4cFvvTr4hC}.  With $J$ finite, we work perturbatively in $1/g$ instead of in $1/J$. 

\paragraph{Constructing primaries out of generalized free fields.} In the strong coupling limit, $g\to \infty$, the Wilson line dCFT reduces to a generalized free field theory. Since $Z$ and $\bar{Z}$ behave like generalized free fields with unit conformal dimension, we can explicitly construct the conformal primaries appearing in the OPEs of $Z$ and $Z^J$ (and likewise in the OPE of $\bar{Z}$ and $\bar{Z}^J$) using the conformal algebra. Even though we can deduce OPE data using the conformal block expansion without needing to know the precise form of the exchanged operators, constructing at least the first few operators explicitly makes the analysis more transparent and illustrates some of the assumptions we made about the operators appearing in the conformal block expansions in Section~\ref{sec:extract OPE data}.  

In $1d$, the conformal algebra is given by $[K,P]=2D$, $[D,P]=P$, $[D,K]=-K$. Since $Z$ is a primary of dimension $1$, it satisfies $DZ=Z$ and $KZ=0$, and its descendants are $P^mZ$ for $m=1,2,3,\ldots$. We can construct additional primaries by taking linear combinations of the different ways that $m$ copies of $P$ can act on $J+1$ copies of $Z$.\footnote{The construction (and counting) of primaries out of more than two copies of generalized free fields was discussed in, for instance, Appendix A of \cite{Roumpedakis:2016qcg}} If $J$ is arbitrarily large, the number of different ways is equal to the number of ways, denoted $p_m$, to partition the positive integer $m$ into a sum of positive integers. The counting of partitions has a nice generating function: $\sum_{m=0}^\infty p_m q^m=\prod_{m=1}^\infty \frac{1}{1-q^m}=1+q+2q^2+3q^3+5q^4+7q^5+\ldots$. 
\begin{comment}Explicitly, the different operators with $m$ copies of $P$ acting on $J+1$ copies of $Z$ for the first few values of $m$ are given by:
\begin{align}
    &m=0:&&Z^{J+1},\nonumber\\
    &m=1:&&Z^JPZ,\nonumber\\
    &m=2:&&Z^{J}P^2Z,&&Z^{J-1}(PZ)^2\nonumber\\
    &m=3:&&Z^J P^3Z,&&Z^{J-1}(P^2Z)(PZ),&&Z^{J-2}(PZ)^3\nonumber\\
    &m=4:&&Z^J P^4Z,&&Z^{J-1}(P^3Z)(PZ),&&Z^{J-1}(P^2Z)^2,&&Z^{J-2}(P^2Z)(PZ)^2,&&Z^{J-3}(PZ)^4.\nonumber
\end{align}
\end{comment}
Since each operator can be written as a linear combination of primaries and descendants, it follows that the $p_m$-dimensional vector space spanned by the operators constructed out of $m$ copies of $P$ acting on $J+1$ copies of $Z$ has a $(p_m-p_{m-1})-$dimensional subspace spanned by primaries and a $p_{m-1}$-dimensional subspace spanned by descendants of the lower-$m$ primaries. The subspace of primaries is equivalently the space of solutions to $K\mathcal{O}=0$ at a given $m$.

For the sake of illustration, let us construct the first few primaries. We will denote them by $[Z^{J+1}]^0_{m,i}$ where $m$ denotes the number of copies of $P$, $i$ is a label to distinguish between degenerate primaries, if the space of solutions to $K\mathcal{O}=0$ has dimension greater than one, and the superscript $0$ indicates that these are the primaries in the generalized free field theory. Using $DP^mZ=(\Delta+m)P^mZ$ and $KP^mZ=m(m+2\Delta-1)P^{m-1}Z$, which follow from the conformal algebra, one can show that the following are conformal primaries
\begin{align}
    [Z^{J+1}]_0&=Z^{J+1},\label{eq:xlXoEqFNPz}\\
    [Z^{J+1}]_2&=Z^JP^2Z-\frac{3}{2}Z^{J-1}(PZ)^2,\\
    [Z^{J+1}]_3&=Z^{J}P^3Z-6Z^{J-1}(P^2Z)(PZ)+6Z^{J-2}(PZ)^3\\
    [Z^{J+1}]_{4,1}&=Z^JP^4Z-10 Z^{J-1}(P^3Z)(PZ)+10Z^{J-1}(P^2Z)^2\\
    [Z^{J+1}]_{4,2}&=Z^{J-1}(P^2Z)^2-3Z^{J-2}(P^2Z)(PZ)^2+\frac{9}{4}Z^{J-3}(PZ)^4.\label{eq:geql6usUHA}
\end{align}
In particular, they satisfy $K[Z^{J+1}]_{m,i}=0$ and $D[Z^{J+1}]_{m,i}=J+1+m$. These have not been normalized in any particular way and $[Z^{J+1}]_{4,1}$ and $[Z^{J+1}]_{4,2}$ form an arbitrary basis for the $2$-dimensional space of $m=4$ primaries.

If we were to continue the construction for larger $m$, we would find that the primaries with $m>4$ obey the following basic patterns: 
\begin{itemize}
    \item There is one conformal primary constructed out of $J+1$ copies of $Z$ (likewise $\bar{Z}$) at $m=0$, $m=2$, and $m=3$. There is more than one conformal primary constructed out of $J+1$ copies of $Z$ (likewise $\bar{Z}$) at each $m\geq 4$. There is no conformal primary at $m=1$.
    \item The dimensions of the primaries are $\Delta_{[Z^{J+1}]_{m,i}}=\Delta_{[\bar{Z}^{J+1}]_{m,i}}=J+1+m$. Moreover, under parity $[Z^{J+1}]_{m,i}$ and $[\bar{Z}^{J+1}]_{m,i}$ are odd if $m$ is odd and even if $m$ is even. This follows because $P\mathcal{O}=-i\partial \mathcal{O}$.
    \item We may choose the basis of primaries constructed out of $\bar{Z}$ so that they satisfy $[Z^{J+1}]_{m,i}^\dagger=[\bar{Z}^{J+1}]_{m,i}$. This follows from $[P,Z]^\dagger=-[P,\bar{Z}]$. Furthermore, applying a Gram-Schmidt-like procedure, using $\braket{Z(t_1)\bar{Z}(t_2)}=\frac{4g}{\pi}\frac{1}{t_{21}^2}$ and Wick contractions to evaluate the inner product $\braket{[Z^{J+1}]_{m,i}[\bar{Z}^{J+1}]_{m',i'}}$, we can construct an orthonormal basis that satisfies $\braket{[Z^{J+1}]_{m,i}[Z^{J+1}]_{m',i'}}=\braket{[\bar{Z}^{J+1}]_{m,i}[\bar{Z}^{J+1}]_{m',i'}}=0$ and $\braket{[Z^{J+1}]_{m,i}[\bar{Z}^{J+1}]_{m',i'}}=\delta_{mm'}\delta_{ii'}$.
\end{itemize}

\paragraph{Extracting OPE data.} Let us now return to the four-point function in \eqref{eq:wjNw4KhCRA}. We will denote the conformal primaries contributing to the conformal block expansion by $[Z^{J+1}]_{m,i}$, which we have labelled based on the property that they reduce to the generalized free-field primaries at strong coupling: $[Z^{J+1}]_{m,i}\to [Z^{J+1}]^0_{m,i}$ as $g\to \infty$. In analogy with degenerate perturbation theory in quantum mechanics, this means choosing a basis $[Z^{J+1}]^0_{m,i}$ for each degenerate subspace with $m\geq 4$ that diagonalizes the interaction.

We expand the conformal dimensions and OPE coefficients of the primaries in $1/g$:
\begin{align}
    \Delta_{m,i}&\equiv \Delta_{[Z^{J+1}]_{m,i}}=J+1+m+\frac{\gamma_{m,i}}{4\pi g}+O(1/g^2),\\
    a_{m,i}&\equiv \frac{\mathcal{C}_{ZZ^J[Z^{J+1}]_{m,i}}\mathcal{C}_{\bar{Z}\bar{Z}^J[\bar{Z}^{J+1}]_{m,i}}}{\mathcal{N}_{Z\bar{Z}}\mathcal{N}_{Z^J\bar{Z}^J}\mathcal{N}_{[Z^{J+1}]_{m,i}[\bar{Z}^{J+1}]_{m,i}}}=a_{m,i}^0+\frac{a_{m,i}^1}{4\pi g}+O(1/g^2).
\end{align}
Invoking \eqref{eq:4cFvvTr4hC}, we can write the conformal block expansion of \eqref{eq:wjNw4KhCRA} and expand in $1/g$:
\begin{align}
    &\braket{Z(t_1)Z^J(t_2)\bar{Z}(t_3)\bar{Z}^J(t_4)}=\frac{\mathcal{N}_{Z\bar{Z}}\mathcal{N}_{Z^J\bar{Z}^J}}{t_{21}^{J+1}t_{43}^{J+1}}\left(\frac{t_{42}}{t_{31}}\right)^{1-J}\sum_{m=0}^\infty \sum_{i\geq 1}a_{m,i}\chi^{\Delta_{m,i}}F_{\Delta_{m,i},J-1}(\chi)\nonumber\\&=\frac{\mathcal{N}_{Z\bar{Z}}\mathcal{N}_{Z^J\bar{Z}^J}}{t_{21}^{J+1}t_{43}^{J+1}}\left(\frac{t_{42}}{t_{31}}\right)^{1-J}\sum_{m=0}^\infty \sum_{i\geq 1}\biggr[a_{m,i}^0\chi^{J+1+m}F_{J+1+m,J-1}(\chi)\label{eq:lH2ZFUnUL8}\\&\hspace{2cm}+\frac{\chi^{J+1+m}}{4\pi g}\left(a_{m,i}^0\gamma_{m,i}\log{\chi}+a_{m,i}^1+a_{m,i}^0\gamma_{m,i}\partial_m\right)F_{J+1+m,J-1}(\chi)+O(1/g^2)\biggr],\nonumber
\end{align}
where we introduced $F_{h,a}(\chi)\equiv {_2F_1}(h+a,h-a,2h,\chi)$ for shorthand. 

By comparing the expansions in \eqref{eq:lH2ZFUnUL8} and \eqref{eq:wjNw4KhCRA} and disentangling the different conformal blocks, we can extract information about $a_{m,i}^0$, $\gamma_{m,i}$ and $a_{m,i}^1$. Because $[Z^{J+1}]_{m,i}$ for each $i\geq 1$ have the same conformal dimension at leading order, the conformal block expansion does not distinguish between the different operators, and the OPE data that we can extract are the ``operator-averaged'' OPE coefficients and anomalous dimensions:
\begin{align}
    \bar{a}_m^0&\equiv \sum_{i\geq 1}a_{m,i}^0,&&\bar{a}_m^1\equiv \sum_{i\geq 1} a_{m,i}^1, &&\bar{\gamma}_{m}\equiv \frac{1}{\bar{a}_m^0}\sum_{i\geq 1}a_{m,i}^0\gamma_{m,i}.
\end{align}
For simplicity, we will focus on $\bar{a}_m^0$ and $\bar{\gamma}_m$. To disentangle the conformal blocks, it is useful to note that they satisfy the following orthogonality relation:
\begin{align}\label{eq:FbWbQjm5HP}
    \oint \frac{dz}{2\pi i}\frac{1}{z^2}z^{\Delta+m}F_{\Delta+m,a}(z)z^{1-\Delta -m'}F_{1-\Delta-m',a}(z)=\delta_{mm'}.
\end{align}
This holds for any $a\in \mathbb{R}$, $\Delta=1,2,3,\ldots$ and $n,n'=0,1,2,\ldots$. We also note
\begin{align}
    \oint \frac{dz}{2\pi i}\frac{1}{z^2}z^{-J-m}F_{-J-m,J-1}(z) z^{J+1}&=(-1)^m\frac{\Gamma(2J+m)\Gamma(m+2)\Gamma(m+2J+1)}{\Gamma(2J)\Gamma(m+1)\Gamma(2m+2J+1)},\label{eq:i3jJYovZz9}\\
    \oint \frac{dz}{2\pi i}\frac{1}{z^2}z^{-J-m}F_{-J-m,J-1}(z)\frac{z^{J+1}}{(1-z)^2}&=\frac{\Gamma(m+2)^2\Gamma(m+2J+1)}{\Gamma(m+1)\Gamma(2m+2J+1)}.\label{eq:NEtszcxVfE}
\end{align}

Finally, comparing \eqref{eq:lH2ZFUnUL8} with \eqref{eq:wjNw4KhCRA} at leading order, we have
\begin{align}
    \sum_{m=0}^\infty \bar{a}_{m}^0 \chi^{J+1+m}F_{J+1+m,J-1}(\chi)&=\chi^{J+1}\left(1+\frac{J}{(1-\chi)^2}\right).
\end{align}
Applying the orthogonality condition \eqref{eq:FbWbQjm5HP} and \eqref{eq:i3jJYovZz9}-\eqref{eq:NEtszcxVfE} yields
\begin{align}
    \bar{a}_{m}^0&=\left(J\Gamma(2+m)\Gamma(2J)+(-1)^m\Gamma\left(2J+m\right)\right)\frac{(m+1)\Gamma\left(m+2J+1\right)}{\Gamma(2J)\Gamma(2m+2J+1)}.\label{eq:z7Rr1ljXaB}
\end{align}
Note that the RHS evaluates to zero $m=1$, reflecting the fact that there is no primary with $m=1$, as per the discussion around \eqref{eq:xlXoEqFNPz}-\eqref{eq:geql6usUHA}.

Next, if compare the term at order $1/g$ that is proportional to $\log{\chi}$, we have
\begin{align}
    \sum_{m=0}^\infty \bar{a}_m^0\bar{\gamma}_m \chi^{J+1+m}F_{J+1+m,J-1}(\chi)&=\frac{J\chi^{J+1}}{4\pi g}\left[G_S^{(1)}(\chi^{-1})+\frac{3}{5}G_T^{(1)}(\chi^{-1})-G_A^{(1)}(\chi^{-1})\right]_{\log{\chi}}
\end{align}
The prefactors of the $\log{\chi}$ terms in $G_S^{(1)}$, $G_T^{(1)}$ and $G_A^{(1)}$ are rational functions that can be read off the explicit expressions in \cite{giombi2017half}. Using the orthogonality of the conformal blocks, we find
\begin{align}\label{eq:}
    \bar{\gamma}_{m}&=\frac{J}{\bar{a}_m^0}\oint \frac{d\chi}{2\pi i}\frac{1}{\chi^2}\chi^{-J-m}F_{-J-m,J-1}(\chi)\chi^{J+1}\frac{2\chi^2(3+\chi(\chi-3))}{(\chi-1)^3},
\end{align}
By evaluating the above contour integral analytically for a range of $J$ and $m$, we deduce the following simple expression for $\bar{\gamma}_m$:
\begin{align}
    \bar{\gamma}_0&=0,&\bar{\gamma}_m&=-\frac{m(m+1)}{2}-Jm, \text{ for } m=2,3,\ldots.\label{eq:rA2TIqPXMU}
\end{align}

Finally, we can compare the finite charge OPE data with the large charge OPE data. First, we identify the operators we denoted $[ZZ^J]_n$ in Section~\ref{sec:extracting OPE data} with the operators $[Z^{J+1}]_m$ we defined in this section such that $n=0$ corresponds to $m=0$, and $n\geq 1$ corresponds to $m=n+1$. Then, expanding the finite charge OPE coefficients in $1/J$, we find for the $m=0$ and $m\geq 2$ operators:
\begin{align}
    \bar{a}_0 &=J+1+O(1/g)=g\left(\mathcal{J}+O(\mathcal{J}^2)\right)+(1+O(\mathcal{J}))+O(1/g).\\\
    \bar{a}_m&=\left[(-1)^m (m+1) +O(1/J)\right]+O(1/g).
\end{align}
Identifying $m=n+1$ for $n\geq 1$, we see that these expressions match \eqref{eq:RcZoCxIyG4}-\eqref{eq:0IWgDZEzm3} in the overlapping terms.

We can similarly expand the conformal dimensions in $1/g$ and $1/J$, we find for the $m=0$ and $m\geq 2$ operators:
\begin{align}
    \Delta_0&=J+1+O(1/g^2)\label{eq:pNnKUE8NKS}\\&=g\left(\mathcal{J}+0\mathcal{J}^2+O(\mathcal{J}^3)\right)+(1+0\mathcal{J}+O(\mathcal{J}^2))+1/g(0+O(\mathcal{J}))+O(1/g^2)\nonumber\\
    \Delta_{n}&=J+1+m-\frac{Jm}{4\pi g}-\frac{m(m+1)}{8\pi g}+O(1/g^2)\label{eq:aOWwcRHg2M}\\&=g\left(\mathcal{J}+0\mathcal{J}^2+O(\mathcal{J}^3)\right)+(m+1-\frac{m\mathcal{J}}{4\pi}+O(\mathcal{J}^2))+\frac{1}{g}\left(-\frac{m(m+1)}{8\pi}+O(\mathcal{J})\right)+O(1/g^2),\nonumber
\end{align}
We find agreement with \eqref{eq:bDdLaHv7rN}, including the absence of an operator with leading dimension $\Delta=J+2$. The fact that the agreement in the overlapping terms is exact reinforces the suggestion in Section~\ref{sec:extract OPE data} that operator mixing may not be relevant for the OPE data in the large charge limit.

\bibliographystyle{ssg}
\bibliography{mybib}

\end{document}